\documentclass[12pt,a4paper]{book}

\makeatother

\usepackage{geometry}
\usepackage[toc,page]{appendix}
\RequirePackage{ifpdf} 
\usepackage{amsmath} 
\usepackage{mathtools}
\usepackage{textalpha}

\usepackage{lipsum}%% a garbage package you don't need except to create examples.
\usepackage{fancyhdr}

\usepackage{pstricks}
\usepackage[final]{pdfpages} 
\usepackage{ifpdf} 
\usepackage{slashed}

\usepackage{hyperref}
\usepackage{xcolor}
\hypersetup{
  colorlinks   = true, %Colours links instead of ugly boxes
  urlcolor     = magenta, %Colour for external hyperlinks
  linkcolor    = blue, %Colour of internal links
  citecolor    = purple %Colour of citations
}

\usepackage{array}
\newcolumntype{L}[1]{>{\raggedright\let\newline\\\arraybackslash\hspace{0pt}}m{#1}}
\newcolumntype{C}[1]{>{\centering\let\newline\\\arraybackslash\hspace{0pt}}m{#1}}
\newcolumntype{R}[1]{>{\raggedleft\let\newline\\\arraybackslash\hspace{0pt}}m{#1}}

\usepackage{color} 
\usepackage{graphics}

\usepackage{etoolbox} 
\usepackage{fixmath}

\usepackage{caption} 
\usepackage{subcaption} 
\usepackage{amsfonts}

\usepackage{multirow}
\usepackage{epstopdf}
\usepackage{upgreek}

\usepackage{relsize}
\usepackage{float}

\usepackage{tabularx}
\usepackage{tabu}

\usepackage{blindtext}
\usepackage{quotchap}
\usepackage{times}

\DeclareUnicodeCharacter{2212}{-}

\usepackage{graphicx}
\usepackage{setspace}
\usepackage{aurical}
\usepackage{calligra}
\usepackage[T1]{fontenc}
\usepackage{txfonts}
\usepackage{notoccite} % If you have \cite commands in \section-like commands, or in \caption, the
\usepackage[backend=biber,sorting=none,style=chem-acs,doi=false,abbreviate=false,isbn=false,url=false,eprint=true]{biblatex}
\usepackage{tikz}
\usetikzlibrary{positioning,arrows}
\usetikzlibrary{decorations.pathmorphing}
\usetikzlibrary{decorations.markings}
\usetikzlibrary{shapes.geometric}
\usetikzlibrary{trees,shadows}
\usepackage{forest}
\usepackage{endnotes}
\tikzset{
    vector/.style={decorate, decoration={snake}, draw},
    provector/.style={decorate, decoration={snake,amplitude=2.5pt}, draw},
    antivector/.style={decorate, decoration={snake,amplitude=-2.5pt}, draw},
    fermion/.style={draw=black,
      postaction={decorate},decoration={markings,mark=at position .55
        with {\arrow[draw=black]{>}}}}, 
    fermionbar/.style={draw=black, postaction={decorate},
                       decoration={markings,mark=at position .55 with {\arrow[draw=black]{<}}}},
    fermionnoarrow/.style={draw=black},
    gluon/.style={decorate, draw=black,decoration={coil,amplitude=4pt, segment length=6pt}},
    scalar/.style={dashed,draw=black,
      postaction={decorate},decoration={markings,mark=at position .55
        with {\arrow[draw=black]{>}}}}, 
    scalarbar/.style={dashed,draw=black,
      postaction={decorate},decoration={markings,mark=at position .55
        with {\arrow[draw=black]{<}}}}, 
    scalarnoarrow/.style={dashed,draw=black},
    electron/.style={draw=black,
      postaction={decorate},decoration={markings,mark=at position .55
        with {\arrow[draw=black]{>}}}}, 
    bigvector/.style={decorate, decoration={snake,amplitude=4pt}, draw}
} 
\definecolor{headercolor}{gray}{0.65} % chapter numbers will be semi transparent .5 .55 .6 .0
\usepackage{fancyhdr}
\fancypagestyle{plain}{%
\fancyhf{} % clear all header and footer fields 
\fancyfoot[C]{\fontfamily{ppl}\selectfont \small\bfseries \thepage} % except the center 
 
}
\renewcommand{\chaptermark}[1]{\markboth{#1}{}}

\usepackage{emptypage}

\definecolor{halfgray}{gray}{0.55} % chapter numbers will be semi transparent .5 .55 .6 .0
\newfont{\chapNumFont}{eurb10 scaled 7000}
\newfont{\chapTitFont}{pplr9d}
\usepackage[explicit]{titlesec}
\titleformat{\section}[hang]{\bfseries\Large}{\fontfamily{ppl}\selectfont \thesection}{15pt}{\fontfamily{ppl}\selectfont #1}
\titleformat{\subsection}[hang]{\bfseries\large}{\fontfamily{ppl}\selectfont \thesubsection}{15pt}{\fontfamily{ppl}\selectfont #1}
\titleformat{\subsubsection}[hang]{\bfseries}{\fontfamily{ppl}\selectfont \thesubsubsection}{15pt}{\fontfamily{ppl}\selectfont #1}
\titleformat{\chapter}[block]%
{\Huge}{\raggedleft{\color{halfgray}\chapNumFont\thechapter}}{20pt}%
{\raggedright{\fontfamily{ppl}\selectfont #1}}

\usepackage{titletoc}
\titlecontents{chapter}[1.4em]{\vspace{8pt}}{\fontfamily{ppl}\selectfont\bfseries \contentslabel{1.4em}}%
{\fontfamily{ppl}\selectfont\bfseries \hspace*{-1.4em}}{\titlerule*[1pc]{.}\fontfamily{ppl}\selectfont\bfseries\contentspage}
\titlecontents{section}[4em]{}{\fontfamily{ppl}\selectfont\small \contentslabel{2.2em}}{}{\titlerule*[1pc]{.}\fontfamily{ppl}\selectfont\small\contentspage}
\titlecontents{subsection}[7em]{}{\fontfamily{ppl}\selectfont \contentslabel{2.6em}}{}{\hfill\contentspage}

\newcommand{\ab}[1]{{\color{black}  #1}}

\newcommand{\mk}[1]{{\color{black} #1 }}%comments by Kumar
\newcommand{\thr}[1]{{\color{black}  #1}}
\newcommand{\asr}[1]{{\color{black}  #1}}
\newcommand{\thrasr}[1]{{\color{black}  #1}}

\begin{document}
\begin{doublespace}

\allowdisplaybreaks[4]
\unitlength1cm

\newcommand{\dis}[1]{\mathbold{#1}}
\newcommand{\overbar}[1]{mkern-1.5mu\overline{\mkern-1.5mu#1\mkern-1.5mu}\mkern
1.5mu}

\def\D{{\cal D}}
\def\g{\overline {\cal G}}
\def\gm{\gamma}
\def\ep{\epsilon}
\def\zo{\overline{z}_1}
\def\zt{\overline{z}_2}
\def\zob{\overline{z}_1}
\def\ztb{\overline{z}_2}
\def\C{{C}}
\def\C{{C}}
\def\Aob{\overline A_1^I}
\def\Atb{\overline A_2^I}
\def\Athb{\overline A_3^I}
\def\Afb{\overline A_4^I}
\def\Ao{A_1^I}
\def\At{A_2^I}
\def\Ath{A_3^I}
\def\fo{f_1^I}
\def\ft{f_2^I}
\def\fth{f_3^I}
\def\Af{A_4^I}
\def\Bo{B_1^I}
\def\Bt{B_2^I}
\def\Bth{B_3^I}
\def\Dob{\overline D_1^I}
\def\Dobd{\overline D_{d,1}^I}
\def\Dtb{\overline D_2^I}
\def\Dtbd{\overline D_{d,2}^I}
\def\Dthb{\overline D_3^I}
\def\Dthbd{\overline D_{d,3}^I}
\def\btob{\overline \beta_1}
\def\bttb{\overline \beta_2}
\def\btthb{\overline \beta_3}
\def\lfr{\log\left({\mu_F^2 \over \mu_R^2}\right)}
\def\lfrt{\log^2\left({\mu_F^2 \over \mu_R^2}\right)}
\def\lfrtt{\log^3\left({\mu_F^2 \over \mu_R^2}\right)}
\def\lqr{\log\left({q^2 \over \mu_R^2}\right)}
\def\lqrt{\log^2\left({q^2 \over \mu_R^2}\right)}
\def\lqrtt{\log^3\left({q^2 \over \mu_R^2}\right)}
\def\lmt{\log\left({\mu_R^2 \over m_t^2}\right)}
\def\lmts{\log^2\left({\mu_R^2 \over m_t^2}\right)}
\def\lw{\ln(1-\omega)}
\def\w{\omega}
\def\one{\ln(w)}
\def\two{\ln^2(w)}
\def\three{\ln^3(w)}
\def\four{\ln^4(w)}
\def\five{\ln^5(w)}
\def\six{\ln^6(w)}
\def\aLqf{\log\left({q^2 \over \mu_F^2}\right)} 
\def\aLqftwo{\log^2\left({q^2 \over \mu_F^2}\right)} 
\def\aLqfthree{\log^3\left({q^2 \over \mu_F^2}\right)} 
\def\M{{\cal M}}
\def\ep{\epsilon}
\def\unM{\hat{\cal M}}
\def\unas{ \left( \frac{\hat{a}_s}{\mu^{\epsilon}} S_{\epsilon} \right) }
\def\rnM{{\cal M}}
\def\rnas{ \left( a_s  \right) }
\def\b0{\beta_0}
\def\cD{{\cal D}}
\def\cC{{\cal C}}
\def\ca{\text{\tiny C}_\text{\tiny A}}
\def\cf{\text{\tiny C}_\text{\tiny F}}

\def\spt{(s+t)}
\def\spu{(s+u)}
\def\tpu{(t+u)}

\pagenumbering{roman}
%\scalefont{1.1}
%\hspace{0pt}
%\vfill
\begin{center}
%
%{\Huge {\bf Precision QCD at the LHC}}
{\Large {\bf Radiative corrections and threshold resummed predictions to pseudoscalar Higgs boson production in QCD}}
%via gluon fusion
%
%\vskip 0.50cm
%
\\{\bf {\em By}}
\vskip 0.0cm
{\bf {\normalsize ARUNIMA BHATTACHARYA}}
\vskip 0.0cm
{\bf {\normalsize PHYS05201704009}}
\vskip 0.3cm
{\bf {\normalsize Saha Institute of Nuclear Physics, Kolkata}}
\vskip 0.5cm
{\textit{\normalsize A thesis submitted to the
\vskip 0.05cm
Board of Studies in Physical Sciences
\vskip 0.05cm
In partial fulfillment of requirements
\vskip 0.05cm
for the Degree of 
}}
\vskip 0.05cm
{\bf {\normalsize DOCTOR OF PHILOSOPHY}}
\vskip 0.1cm
{\textit{of}}
\vskip 0.1cm
{\bf {\normalsize HOMI BHABHA NATIONAL INSTITUTE}}
\vskip 0.7cm
\includegraphics[scale=0.56,keepaspectratio=true]{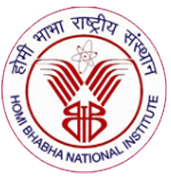}
% university.png: 234x215 px, 72dpi, 8.26x7.58 cm, bb=0 0 234 215
%
\vskip 0.5cm
{\bf {\normalsize February 2023}}
\vfill
\end{center}
\hspace{0pt}
%\footnotetext{Version 4}
%
\vfill
\clearpage
\newpage
\
\newpage
%
%~ \vskip 0.7cm
%
\includepdf[pages=-]{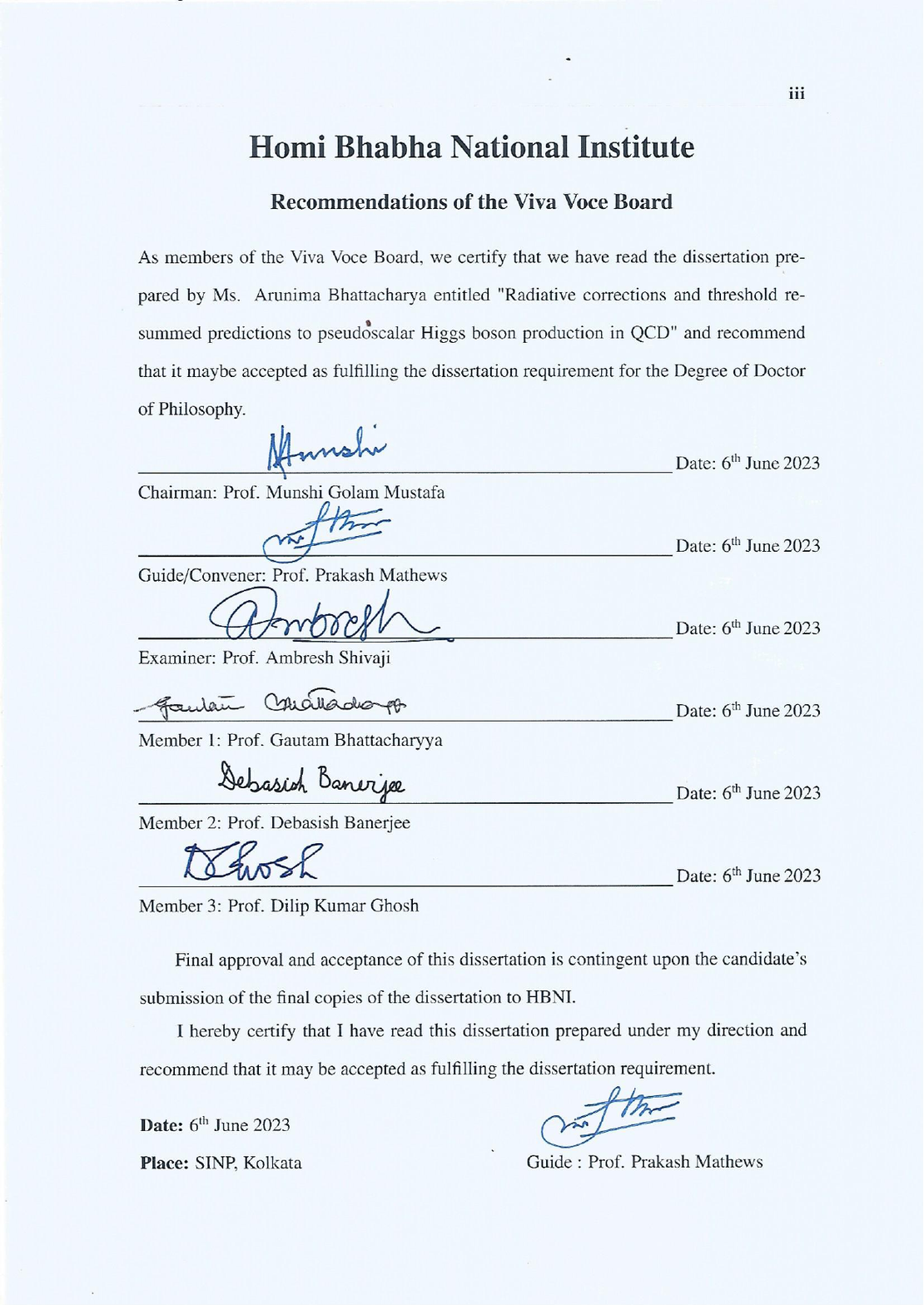}
\newpage
\
\newpage
\centerline{{\bf {\large STATEMENT BY AUTHOR}}}
\vskip 1.00cm
\doublespacing
This dissertation has been submitted in partial fulfillment of requirements for an advanced degree at Homi Bhabha National Institute (HBNI) and is deposited in the Library to be made available to borrowers under rules of the HBNI.
\vskip 0.6cm
Brief quotations from this dissertation are allowable without special
permission, provided that accurate acknowledgement of source is made.
Requests for permission for extended quotation from or reproduction of
this manuscript in whole or in part may be granted by the Competent
Authority of HBNI when in his or her judgement the proposed use of the
material is in the interests of scholarship. In all other instances,
however, permission must be obtained from the author.

\vskip 1.5cm

$~$\hspace{11.0cm}Signature

\vspace{5 mm}

$~$\hspace{10.2cm}Arunima Bhattacharya
\newpage
\
\newpage
~
\vskip 1.2cm
\centerline{{\bf{\large{DECLARATION}}}}
\vskip 1.2cm
I, hereby declare that the investigation presented in the thesis has been
carried out by me. The work is original and has not been submitted
earlier as a whole or in part for a degree / diploma at this or any
other Institution / University.
\vskip 2.0cm
\rightline{Signature \hspace{1.5cm}}

\vspace{5 mm}

\rightline{Arunima Bhattacharya \hspace{0.9cm}}
\newpage
\
\newpage
\centerline{{\bf {\large List of Publications arising from the thesis \footnote{As it is standard in the High Energy Physics Phenomenology (hep-ph) community the names of the authors on any paper appear in their alphabetical order.}}}}
{\bf {\large Journals}}
\begin{enumerate}
  \item Two loop QCD amplitudes for di-pseudo scalar production in gluon fusion.
  
  \textbf{Arunima Bhattacharya}, Maguni Mahakhud, Prakash Mathews and V. Ravindran.
  
  In: JHEP 02 (2020) 121. arXiv: 1909.08993 [hep-ph].

 \item Next to SV resummed prediction for pseudoscalar Higgs boson production at NNLO$+\overline{\text{NNLL}}$.
 
 \textbf{Arunima Bhattacharya}, M. C. Kumar, Prakash Mathews and V. Ravindran.

 In: Phys.Rev.D 105 (2022) 11, 116015. arXiv: 2112.02341 [hep-ph].
\end{enumerate}
\vskip 1.2cm
\centerline{{\bf {\large List of other publications (not included in the thesis).}}} 
\begin{enumerate}
  \item Two-Loop QCD Amplitudes for Di-pseudo Scalar Production in Gluon Fusion.

  \textbf{Arunima Bhattacharya}, Maguni Mahakhud, Prakash Mathews and V. Ravindran.

  In: Springer Proc.Phys. 277 (2022) 49-53, 24th DAE-BRNS High Energy Physics Symposium.
\end{enumerate}

%\newpage
%
\centerline{\bf {\large Conferences/Workshops attended}}
\begin{enumerate}
 \item{\textbf{The Myriad Colourful ways of understanding extreme QCD Matter}} -
 {Workshop at ICTS-TIFR, Bangalore, India,
 $1^\text{st} - 7^\text{th}$ April 2019.}
 \item{\textbf{Madgraph School 2019}} -
 {School at The Institute of Mathematical Sciences (IMSc), Chennai, India, $18^\text{th} - 22^\text{th}$ November 2019.}
 \item{\textbf{XXXIII SERB Main School}} -
 {School on Theoretical High Energy Physics at SGTB Khalsa College, New Delhi, India,
 $7^\text{th} - 26^\text{th}$ December 2019.}
 \item{\textbf{XXIV DAE-BRNS Symposium on High Energy Physics}} -
  {Virtual symposium organized by the National Institute of Science Education and Research, Jatni, Odisha,
  $14^\text{th}$ - $18^\text{th}$ December 2020.}
  \item{\textbf{$8^\text{th}$ International Workshop on High Precision for Hard Processes (HP2 2022)}} -
  {Conference at the Discovery Museum in Newcastle, UK, organized by IPPP, University of Durham,
  $20^\text{th} - 22^\text{nd}$ September 2022.}
  \item{\textbf{Baryons 2022 - International Conference on the Structure of Baryons}} -
  {Virtual conference at Seville, Spain,
  $7^\text{th} - 11^\text{th}$ November 2022.}
  \item {\textbf{Computer Algebra with FORM for Developers 2023}} -
  {Workshop at IFT - UAM, Madrid, Spain, $12^{\text{th}} - 14^{\text{th}}$ April 2023.}
  \item {\textbf{7th RED LHC workshop 2023}} -
  {Workshop at IFT - UAM, Madrid, Spain, $10^{\text{th}} - 12^{\text{th}}$ May 2023.}
\end{enumerate}
%
%\newpage
%
\centerline{\bf {\large Presentation/Poster}}
\begin{enumerate}
 \item{\textbf{XXIV DAE-BRNS Symposium on High Energy Physics} - }
  {"Two-loop QCD amplitudes for di-pseudo scalar production in gluon fusion",
  virtual symposium organized by the National Institute of Science Education and Research, Jatni, Odisha,
  $14^\text{th} - 18^\text{th}$ December 2020.}

  \item{\textbf{$8^\text{th}$ International Workshop on High Precision for Hard Processes (HP2 2022)}} -
  {"Next-to-soft-virtual resummed prediction for pseudoscalar Higgs boson production at NNLO$+\overline{\text{NNLL}}$",
  Discovery Museum in Newcastle, UK, organized by IPPP, University of Durham,
  $20^\text{th} - 22^\text{nd}$ September 2022.}

  \item{\textbf{Baryons 2022 - International Conference on the Structure of Baryons}} -
  {"Next to SV resummed prediction for pseudoscalar Higgs boson production at NNLO$+\overline{\text{NNLL}}$",
  virtually conference at Seville, Spain, $7^\text{th} - 11^\text{th}$ November  2022.}
\end{enumerate}
\vfill
%
%\rightline{Signature}

\rightline{Arunima Bhattacharya}
\clearpage
%
%\
%
%\newpage
%
\hspace{0pt}
\vfill
{\Fontskrivan\bfseries {\large {Dedicated to my beloved Ma \& Baba, and my husband,}}}
\vskip 3.0cm
\begin{flushright}
{\Fontskrivan\bfseries {\large {For their endless love, support and encouragement.}}}
\end{flushright}
\hspace{0pt}
\vfill
\newpage
\
\newpage
\vskip 1.0cm
\centerline{{\bf{\large ACKNOWLEDGEMENTS}}}
\vskip 0.5cm
\centerline{}
\vspace{-1.0cm}

Undertaking this PhD has been a life-changing experience for me, and it would not have been possible without the support and guidance I received from many people. My experience at SINP was similar to several other PhD students in the country, but I still call it unique.

I want to say a huge thank you to my supervisor Prof.\ Prakash Mathews for all the support and encouragement he gave me. I could not have achieved this milestone without his constant guidance and feedback. He has been supportive and encouraging since the day I approached him, despite my asking for his help with the slightest problems. I would also thank my DC committee members, Prof.\ Munshi Golam Mustafa, Prof.\ Gautam Bhattacharyya, Prof.\ Kaushik Datta, Prof.\ Dilip Kumar Ghosh and Prof.\ Debasish Banerjee, who always pointed out my shortcomings and encouraged me to improve.

This thesis will be incomplete without the constant mentoring, collaboration, and support of Prof.\ V.\ Ravindran. He is one of the pillars on which my research stands.
I also want to show my earnest gratitude to Prof.\ M.\ C.\ Kumar for fruitful collaborations, discussions, and much help for academic and non-academic purposes.
He has been my guide for all the numerical analyses done in the thesis and has been extremely helpful.
Without his constant counselling, this thesis wouldn't have been what it is now.
One of my first collaborators, Prof.\ Maguni Mahakhud, was the one who encouraged me to be an independent and curious researcher.

Thanks to Prof.\ Taushif Ahmed for his personal and professional guidance and advice. The list wouldn't be complete without my friends and colleagues from IMSc --- Pooja, Ajjath, Surabhi and Aparna. I can never thank Pooja enough for addressing my constant questions throughout the analytical computations of the works included in the thesis. I would also like to thank Surabhi for providing the Higgs resummed data for comparisons. And my bonding with Aparna began during the final years of my PhD and is still growing.

Thanks to Saha Institute of Nuclear Physics, affiliated with the Homi Bhabha National Institute, Department of Atomic Energy, for providing me with the financial means to complete this project. I am grateful to my professors at SINP for the exciting courses and discussions during my PhD tenure. Among the non-research staff of the Theory Division at SINP, it is a pleasure to thank Prodyut da, Dola di, Arun da, Sangita di, Nilanjan da and Pradip da for their cooperation in every technical and administrative matter.

Every journey in life requires the support and guidance of elders. I would first like to thank my father, Mr Mayasankar Bhattacharya, who was the first to encourage his daughter from her school days towards this prestigious future. He has been the best father and the most incredible guide. He is the pillar who stands by me in every decision and corrects me whenever I go haywire. My mother, Mrs Srabani Bhattacharya, has always been the backbone of the family and provided me with every possible emotional support. She was my first teacher and is my best friend. It was also at SINP that I met my husband, Ayan Kumar Patra, who has been my booster when I lost confidence in myself during this journey.

Family, friends, and colleagues also play an essential role in this journey. I would start by thanking my colleagues at SINP --- Avik da Senior, Avik Da Junior, Bithika di, Aranya da, Supriyo, Pritam, Ritesh and Udit da. The relaxing adda's in the theory room during the long tiring days, the chit-chats over our evening tea and the fun treats are something I'll cherish my entire life. With passing days,  many old faces were replaced by new ones. But the new faces, namely, Sabyasachi, Suman, Adil, Sandip, Gourav and Pabitra, compensated sufficiently.
I would like to thank my batch mates Astik, Madhurima, Promita, Saikat, Arindam,    Shubhi, Rashika, Subham, Suchanda, Rezwana, Dipali, Upala and Tanmoy.

%Thanks to the taxpayers of this country for their indirect support.
Finally, gratitude and a big thanks to everyone who has been a part of my life, directly or indirectly, since I was born and to everyone I haven't explicitly mentioned in this thesis.

\vfill
\hfill Arunima Bhattacharya

%
%\newpage
%

\tableofcontents
\let\cleardoublepage\clearpage
\listoffigures
\addcontentsline{toc}{chapter}{List of Figures}
\listoftables
\addcontentsline{toc}{chapter}{List of Tables}
\let
\cleardoublepage
\clearpage
\mbox{~}

\chapter*{List of Abbreviations
{\Large{\footnote{Arranged Alphabetically}}}}
\addcontentsline{toc}{chapter}{List of Abbreviations}
\chaptermark{List of Abbreviations}

\begin{tabular}{ |p{3cm}|p{10cm}| }
\hline
Abbreviations &Meaning \\
\hline
\textbf{BSM}      &Beyond Standard Model \\
\textbf{COM}      &Centre of Mass \\
\textbf{DIS}      &Deep Inelastic Scattering \\
\textbf{DOF}      &Degrees of Freedom \\
\textbf{DY}       &Drell-Yan \\
\textbf{EFT}      &Effective Field Theory \\
\textbf{FO}       &Fixed Order \\
\textbf{IR}       &Infrared \\
\textbf{LHC}      &Large Hadron Collider \\
\textbf{$\overline{\text{LL}}$}       &NSV Leading logarithm \\
\textbf{LL}       &SV Leading logarithm \\
\textbf{LO}       &Leading Order \\
\textbf{MI}       &Master Integrals \\
\textbf{$\overline{\text{MS}}$}      &Modified Minimal Subtraction  \\
\textbf{MSSM}     &Minimal Supersymmetric Standard Model \\
\hline
\end{tabular}

\newpage

\begin{tabular}{ |p{3cm}|p{10cm}| }
\hline
Abbreviations     &Meaning \\
\hline
\textbf{$\overline{\text{NLL}}$}      &NSV Next-to-leading logarithm \\
\textbf{$\overline{\text{NNLL}}$}     &NSV Next-to-next-to-leading logarithm  \\
\textbf{NLL}      &SV Next-to-leading logarithm \\
\textbf{NLO}      &Next-to-leading Order \\
\textbf{$\overline{\text{N$^3$LL}}$}  &NSV Next-to-next-to-next-to-leading logarithm \\
\textbf{NNLL}     &SV Next-to-next-to-leading logarithm  \\
\textbf{NNLO}     &Next-to-next-to-leading Order \\
\textbf{N$^3$LO}  &Next-to-next-to-next-to-leading Order \\
\textbf{N$^3$LL}  &SV Next-to-next-to-next-to-leading logarithm \\
\textbf{NSV}      &Next-to Soft-virtual \\
\textbf{PDF}      &Parton Distribution Function \\
\textbf{pQCD}     &Perturbative Quantum Chromodynamics \\
\textbf{QCD}      &Quantum Chromodynamics \\
\textbf{QFT}      &Quantum Field Theory \\
\textbf{SHG}      &Showering and Hadronization Generator \\
\textbf{SM}       &Standard Model \\
\textbf{SSB}      &Spontaneous Symmetry Breaking \\
\textbf{SV}       &Soft-virtual \\
\textbf{UV}       &Ultraviolet \\
\textbf{\textit{vev}}      &Vacuum Expectation Value \\
\textbf{2HDM}     &Two Higgs Doublet Model \\
\hline
\end{tabular}

\chapter*{Summary}
\addcontentsline{toc}{chapter}{Summary}
\chaptermark{Summary}

%The discovery of the long-awaited "Higgs boson" in 2012 was a crowning moment for the Standard Model (SM) of particle physics.
A relevant question following the discovery of the Higgs boson in 2012 is whether this scalar is the only one of its type as predicted by the SM or the first to have been discovered in a family of more such species from an underlying extended scalar sector.
Attempts are being made to relate the available Higgs data with BSM Higgs physics by extending the SM.
The MSSM results from one such attempt.
%Current experimental uncertainties suggest that the discovered Higgs is compatible with the SM predictions and the Higgs sector in the MSSM.
However, venturing into these areas requires precise predictions of relevant observables for scalar and pseudoscalar Higgs bosons.
Compared to the precision up to which the SM Higgs boson production results have been made available, precision computations for the pseudoscalar Higgs boson production need improvements.
This thesis attempts to contribute to the study of the pseudoscalar Higgs boson production.
The central part of this thesis aims to calculate perturbative corrections to the production cross-section of the pseudoscalar Higgs boson in the framework of EFT.
%This requires resumming the whole perturbative series for large logarithmic contributions near the threshold, in addition to computing the FO corrections.

%%%%%%%%%%%%%%%%%%%%%%%%%%%%%%%%
%\section{Two loop QCD amplitudes for di-pseudo scalar production in gluon fusion}

%Pair production of the pseudoscalar Higgs boson is essential to understand the nature of the extended Higgs sector along with that of a single pseudoscalar.
Understanding this extended scalar sector will benefit from the knowledge of the production of a single pseudoscalar Higgs boson and its pair production.
I began by computing radiative corrections to the four-point amplitude,  $g+g \rightarrow A+A$, in massless QCD up to order $a_s^4$ in pQCD \cite{Bhattacharya_2020}.
However, these FO QCD predictions become unreliable in the threshold region of the kinematic phase space because of some large logarithmic corrections which need to be resummed to all orders in perturbation theory.
Subsequently, I focused on the resummation of Next-to-Soft Virtual (NSV) logarithmic terms for the threshold production of a pseudoscalar Higgs boson through gluon fusion at the LHC to $\overline{\text{NNLL}}$ accuracy \cite{Bhattacharya:2021hae}.
We have used the EFT framework in both these works and computed the results for the gluon fusion channel only.
%We chose the EFT framework because the number of loops increases computational complexities. So, such a choice enables physicists to go beyond NLO accuracy with the available tools and make precise and stable predictions with respect to the unphysical scales.

In \cite{Bhattacharya_2020}, we computed the amplitude contributing to the pure virtual part for di-pseudoscalar production up to two loops.
We used the standard approach of the projector method \cite{Banerjee:2018lfq}.
%Consequently, the NNLO computations became doable as they are usually pretty cumbersome.
We have used dimensional regularisation and performed UV renormalisation with a careful treatment of the chiral quantities in dimensional regularisation.
Finally, the UV finite results were found to be consistent with the universal IR  structure of QCD amplitudes \cite{Catani:1998bh}.
%We obtained the finite amplitude by performing loop integrals, and simplifications using packages like QGRAF, FORM, LiteRed, Reduze and Mathematica \cite{Nogueira:1991ex,Vermaseren:2000nd,vonManteuffel:2012np}.
The IR finite part of these amplitudes constitutes a vital component of any NNLO corrections to observables involving a pair of pseudoscalars at the LHC.

In \cite{Bhattacharya:2021hae}, we used the formalism described in \cite{ajjath2020soft} to obtain the resummed analytical results till $\mathcal{O}(a_s^2)$ for a pseudoscalar production.
We observed that these NSV logarithms are significant compared to the conventional SV logarithms.
For a clearer picture, we studied theory uncertainties and observed that the renormalisation scale uncertainties decrease with the inclusion of NSV corrections.
This improvement in theory uncertainties on including the NSV terms contrasted with the observations from the seven-point and the factorisation scale uncertainties, where the uncertainties increased on including the NSV corrections.
These hinted at the significance of including beyond NSV corrections for the $gg$ channel and corrections from other partonic channels.
%We also studied the effects of the other partonic channels on the resummed corrections till NNLO+$\overline{\text{NNLL}}$, compared to the $gg$ channel that we focus on in our work.
%This also helped us to predict a source for the increase in uncertainties in the seven-point scale variation and factorisation scale variation results.
We also studied the mixing between scalar and pseudoscalar Higgs bosons and observed that precise theoretical estimates are needed to study these effects, as the corrections only affected the mixing parameter by a few per cent.
%However, this mixing was studied only for the $2\rightarrow1$ subprocess of pseudoscalar production \textit{via} gluon fusion and as per studies, the simple rescaling we have assumed in our studies will be modified if decay channels of the scalar/pseudoscalar Higgs is considered.
We also studied the PDF uncertainties for five different choices of PDF sets normalised w.r.t the MMHT 2014 PDF set and presented the results at NNLO+$\overline{\text{NNLL}}$ accuracy.
We observed that the CT14 PDF set showed the least uncertainty.

%%%%%%%%%%%%%%%%%%%%%%%%%%%%%%%%%

%To conclude, this thesis deals with higher-order QCD corrections to the observables associated with the pseudoscalar Higgs boson production.
%I have performed the computations using our in-house codes and state-of-the-art techniques.
\textit{
This thesis attempts to contribute to the computation of multi-loops and multi-legs, which are central to making precise theoretical predictions.
Although this thesis includes only the contribution of the gluon channel, we plan to consider contributions from the other parton channels in future works.
}
%This is important because no matter how small, contributions from sub-dominant channels are also necessary along with the dominant ones to reduce dependency on the unphysical scales and make reliable predictions.
%This is also observed in our uncertainty plots in the second project.

%Notable developments in the techniques and methodologies to handle the complexities that arise with the increase in the number of loops and/or external particles have made these computations a reality.
%The main tasks of the particle physics community can be primarily categorised into two parts now: testing the SM with unprecedented accuracy and searching for BSM physics.
%We contribute to this task by means of a simple model in the effective field theory framework where the CP-odd Higgs boson is considered to be just another particle in addition to the CP-even Higgs boson in a CP-conserving model.
%Herein lies the significance of this thesis.

%\pagenumbering{arabic}
\chapter{Introduction}
\pagenumbering{arabic}
\label{chap:Intro}
%%%%%%%%%%%%%&&&%%%%%%%%%

\begin{quote}
 \textit{“Is the purpose of theoretical physics to be no more than a cataloging of all the things that can happen when particles interact with each other and separate? Or is it to be an understanding at a deeper level in which there are things that are not directly observable (as the underlying quantised fields are) but in terms of which we shall have a more fundamental understanding?”}
 
\hfill{\textit{- Julian Schwinger}}
\end{quote}

\begin{figure}[!htb]
\centering
    \includegraphics[scale=0.35]{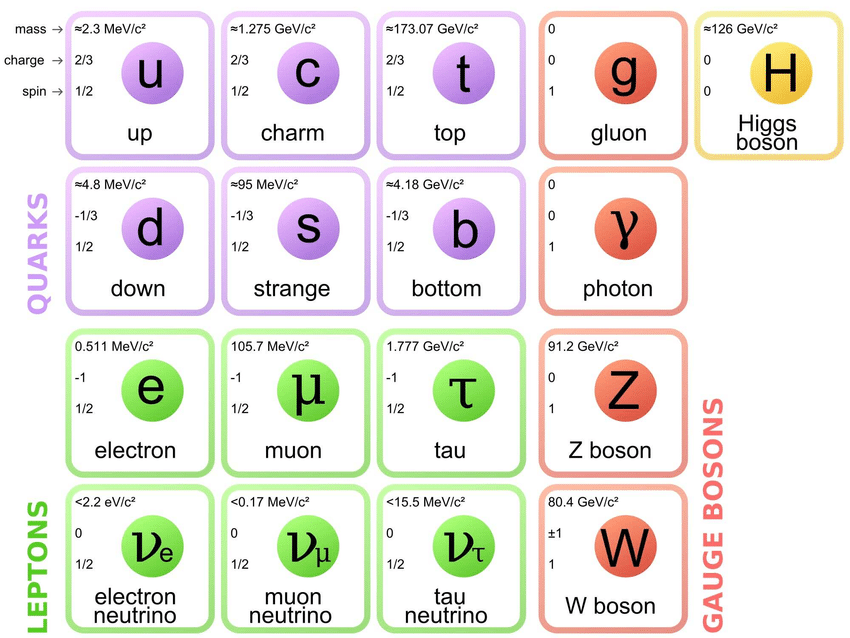}
\label{fig:SM1}
\caption[The Standard Model of Particle Physics]{The Standard Model of Particle Physics \cite{phdthesis:SM}.}
\label{fig:SM}
\end{figure}

Nature stems from only a handful of components, namely the fundamental particles.
The natural laws of the sub-nuclear region ($10^{-13}$ cm and smaller) can be primarily understood by analyzing the consequences of the high-energy collisions between these fundamental particles.
The discovery of the atom in 1803 marked the beginning of an era that would change the world's perception of the universe's evolution and its underlying structure.
Since 1897, which marked the discovery of the electron, followed by the proton discovery in 1898, the nucleus discovery in 1911 and the neutron discovery in 1932, elementary particle physics has come a long way.
These discoveries needed a theory for a satisfactory explanation and, eventually, gain insight into the fundamental structure of matter.
\thr{
Attempts for such a theory finally bore fruit in the early 1970s with the discovery of a robust one now known as the SM of particle physics (see fig.\ \ref{fig:SM}) \cite{salam,PhysRevLett.19.1264,Glashow:1970gm,Weinberg:1973un,Pati:1973uk,Georgi:1974sy,Pati:1974yy,Glashow:1976nt}.
}
The SM did a commendable job of describing and mapping known physics.
It describes the elementary particles and their interactions with one another and with the fundamental forces of nature.
It also proved handy in describing the fundamental particle properties, such as mass, charge and spin.
These fundamental particles can be divided into spin-$1/2$ fermions and integer-spin bosons.
\begin{enumerate}
 \item \underline{Fermions}: These are spin-$1/2$ particles and can be elementary (such as the leptons) or composite (such as the proton or neutron).
 \thr{For instance, quarks are also elementary fermions with three generations and six flavours (up, down, charm, strange, top and bottom)}.

 \item \asr{\underline{Spin-$1$ (vector) Bosons}: These are force carriers which comprise the following:}
    \begin{itemize}
        \item \asr{Photon - mediator of the electromagnetic force,}
        \item \asr{Gluons - mediator of the strong force,}
        \item \asr{Z and W$^\pm$ - mediator of the weak force.}
    \end{itemize}

  \item \thrasr{\underline{Spin-$0$ Boson}: This is the scalar Higgs boson with “even” parity. It can be described as a quantum excitation of the Higgs field. I will explain the Higgs mechanism in section \ref{sec:HiggsMech} in more detail.}
\end{enumerate}
%
%\vspace{-0.478cm}
%

%\newpage

\section{Physics in the LHC Era}
\label{sec:LHCPhysics}

The LHC at CERN is the largest existing particle accelerator. Its successful collaborations lay the stepping stone towards discovering the Higgs boson.
The $4^{\text{th}}$ of July $2012$ marked the historic day when the ATLAS \cite{Aad:2012tfa}, and CMS \cite{Chatrchyan:2012ufa} experiments of the LHC, both announced that a new particle, consistent with the SM Higgs boson, in the mass region around $125$ GeV, has been observed.
It was as early as 1964 when the concept of the Higgs was introduced, and it took 48 long years to observe it finally.
None of the $16$ other particles in the SM was as elusive as this "God Particle".
With its discovery, the SM of particle physics was put on a firm footing.
%In fig.\ \ref{fig:SMtimeline}, it can be seen that the muon and tau were discovered before anyone even predicted their existence.
%\begin{figure}[!htb]
%\begin{center}
% \includegraphics[scale=0.6,keepaspectratio=true]{Extra_Figures/SMtimeline.png}
 % SM.png: 1150x765 px, 72dpi, 40.57x26.99 cm, bb=0 0 1150 765
% \caption[Discovery timeline of the SM of Particle Physics]{Discovery timeline of the SM of Particle Physics \cite{TheEconomist:SMtimeline}}
%\label{fig:SMtimeline}
%\end{center}
%\end{figure}
%
It was evidence of the Brout-Englert-Higgs mechanism \cite{Englert:1964et,Guralnik:1964eu,Higgs:1964pj,Higgs:1964ia,Higgs:1966ev,Kibble:1967sv} which was introduced to explain the occurrence of mass in the SM, gauge invariantly \cite{salam,PhysRevLett.19.1264}.
As the SM is a gauge theory that results in the Lagrangian density being invariant under local $\text{SU}(3)_\text{C} \otimes \text{SU}(2)_\text{L} \otimes \text{U}(1)_\text{Y}$ gauge transformations, the introduction of the covariant derivative becomes essential to achieve local gauge invariance as the derivative acquires a specific form for each symmetry group.
This covariant derivative generates all the interactions described by the SM and introduces the vector boson(s) mediating each interaction.
%
%
%Ian Sample, a science correspondent at \textit{The Guardian}, started live-blogging on this celebrated day from Geneva and ended his report stating the distant implications of this crowning discovery as,
%\begin{quote}
%    \textit{"Months and years of analysis lie ahead to confirm that the particle is the elusive Higgs boson. If so, physicists want to know whether it is the simplest kind of particle put forward in physicists' theories or something more unusual – and more exciting."}
%\end{quote}
Consequently, efforts began to determine whether or not it is the Higgs boson predicted by the SM.
This involves investigating the properties and interactions of the discovered Higgs with the other SM particles \cite{Higgs:1964ia,PhysRevLett.13.508,PhysRev.145.1156,PhysRevLett.13.321,PhysRevLett.13.585,ATLAS:2013sla,ATLAS:2019nkf,2013,2015}.
However, before understanding these developments since the Higgs boson's discovery, one must know what brought the theory to its current state.
The following few sections will give an overview of this.
%
%However, experiments failed to observe this \textit{"God Particle"} for several years, thereby failing to confirm the theory. 
%Hence, the confirmation was a landmark achievement towards explaining the mechanism of how some of the fundamental SM particles acquire mass.
%
%
%The SM, which is often visualized as a table, didn’t begin as a table. As the grand theory of almost everything begins, the SM also began with a collection of several mathematical models that proved to be timeless interpretations of the laws of physics. So if we try to interpret the SM technically, then it can be written in the form of a Lagrangian which is one of the easiest and compact ways of presenting the theory. 
%So, here it is, the SM Lagrangian - or formula - in all its glory, written up by Italian mathematician and physicist Matilde Marcolli :
%\begin{figure}
%\centering
%  \includegraphics[width=.9\linewidth]{SMLagrangian.pdf}
%% SMLagrangian.pdf: 0x0 px, 300dpi, 0.00x0.00 cm, bb=
%\label{fig:SMLag}
%\end{figure}

\section{Higgs and the Higgs Mechanism}
\label{sec:HiggsMech}

The Higgs mechanism enables the SM particles to acquire mass by implementing the process of spontaneous symmetry breaking (SSB) \cite{PhysRev.139.B1006,PhysRevLett.13.321,Higgs:1964pj,Higgs:1964ia,PhysRevLett.19.1264,salam,tHooft:1971qjg,tHooft:1972tcz}.
The process of SSB is fascinating because when a system is spontaneously broken, the ground state solutions of the system fail to exhibit the same symmetry as the initial Lagrangian.
For instance, in a ferromagnet, the ground state does not possess rotational symmetry, unlike its Hamiltonian, since the spins align themselves in a definite but arbitrary direction. Thus, we have infinitely many ground states \textit{i.e.} vacua.
%When SSB occurs, the original symmetry of the system hides itself giving rise to the non-symmetric states.
The point which sets off the symmetry-breaking process corresponds to a critical point in each system which determines if SSB will occur, leading to the degenerate vacua states.
%
%\begin{quote}
%    \textit{"We do not have to look far for examples of spontaneous symmetry breaking. A chair is a case of SSB because the equations governing the atoms of the chair exhibit rotational symmetry. However, a solution to these equations has a definite orientation in space. By the same token, we may say that the macroscopic world is full of finite objects that break spatial isotropy."}

%    \hfill{\textit{- Steven Weinberg (1996)}}
%\end{quote}
%

SSB was first observed in superconductivity but was brought into particle physics by Nambu in 1960 \cite{Nambu:1960xd}.
In 1961, Goldstone successfully applied this method of symmetry breaking to a "simple model" in an attempt to break global U(1) symmetry \cite{Goldstone:1961eq}.
Eventually, Goldstone, Salam and Weinberg proved in the same year that if a global continuous symmetry is spontaneously broken, it generates at least one Nambu-Goldstone boson \cite{Goldstone:1962es}.
So, a direct consequence of SSB in field theory was the generation of this spinless massless particle which was completely unexpected at that time and led to disappointment.

As the case with gauge symmetry was different, a possible extension of the existing theory to the relativistic case was needed.
Englert and Brout \cite{PhysRevLett.13.321}, and Higgs \cite{Higgs:1964ia,Higgs:1964pj} successfully did this extension in 1964 when SSB was first applied to a local symmetry instead of a global one.
These works exhibited how an exact local symmetry can be eliminated from two of the unwanted massless bosons, like the Nambu-Goldstone boson and a gauge boson, by generating a single massive vector boson from their combination in a completely relativistic theory.
%
%\begin{center}
%    \textit{These speculations spanning several years of hard work and brain-storming sessions got closure with the Higgs discovery in 2012.}
%\end{center}
%\begin{quote}
%    \textit{"There is no reason why an invariance of the Hamiltonian of a quantum-mechanical system should also be an invariance of the system's ground state. Hence, symmetry breaking of this type is not only common but also theoretically unremarkable;"}
%
%    \hfill{- Sidney Richard Coleman (1975)}
%\end{quote}
%
%\begin{center}
%    \textit{But what exactly does spontaneous symmetry breaking imply? }
%
%    \textit{Is the symmetry broken, or does it get hidden? }
%\end{center}
%
%

As an example, let us try to introduce mass terms for the W and Z bosons, which require the SU$(2)_{\text{L}} \otimes \text{U}(1)_{\text{Y}}$ symmetry to be spontaneously broken.
For this purpose, consider a scalar field, $\phi$, as depicted below:
\begin{equation}
 \phi = \left(\begin{array}{c}
\phi_{\alpha}\\
\phi_{\beta}
\end{array}\right)= \sqrt{1/2}\left(\begin{array}{c}
\phi_{1}+i\phi_{2}\\
\phi_{3}+i\phi_{4}
\end{array}\right),
\label{eqn:scalarDoublet}
\end{equation}
where $\phi$ is an {SU}(2)$_{\text{L}}$ doublet of complex scalar fields.

\asr{
The scalar field Lagrangian is given by
\begin{equation}
 \mathcal{L} = (\partial_{\mu}\phi)^{\dagger}(\partial^{\mu}\phi) + \mu^{2}\phi^{\dagger}\phi - \lambda(\phi^{\dagger}\phi)^{2},
 \label{eq:scalarfieldlag}
\end{equation}
where $\mu$ and $\lambda$ are the corresponding coefficients.
%is a constant called coupling constant and $\mu$ is often equated to the mass of the field $\phi$.
The term ${\mu}^2 |\phi|^2 + \lambda |\phi|^4$ can be considered as a potential energy term, $V$.
As the potential $V$ must be bounded from below, the $\lambda$ coupling must be positive.
On the other hand, the $\mu^2$ parameter may be positive or negative.
For $\mu^2 > 0$, this theory has a unique ground state with zero
expectation value of the field $\phi$.
%It is therefore invariant under the phase symmetry.
But for $\mu^2 <0$, $V$ has a local maximum and a continuous ring of degenerate minima
%at Φ = v × ei(any phase). None of these minima is
which are not invariant under the U$(1)_{\text{Y}}$ phase symmetry.
%This means that the theory does not have a unique physical vacuum but rather a continuous family of exactly degenerate vacua related to each other by the phase symmetry.
This phenomenon is called SSB.
To achieve local SU$(2)_\text{L} \otimes \text{U}(1)_\text{Y}$ symmetry of the Lagrangian in this scenario, we introduce the covariant derivative
}
\begin{equation}
 \text{D}_{\mu}= \partial_{\mu}+ ig\frac{\tau_{a}}{2}W_{\mu}^{a}+ ig^{\prime}B_{\mu}\frac{Y}{2},
\end{equation}
where $g$ and $g^{\prime}$ represent the coupling constants for SU(2)$_\text{L}$ and U(1)$_\text{Y}$ symmetries, respectively, $\tau_i$ represent the SU(2)$_\text{L}$ spin matrices, and Y represents the hypercharge.
$W_{\mu}^{a}$ and $B_{\mu}$ represent the SU(2)$_\text{L}$ and U(1)$_\text{Y}$ gauge fields, respectively.
The \textit{vev} of the Higgs field can be chosen to be
\begin{equation}
 \phi_{0}= \sqrt{\frac{1}{2}}\left(\begin{array}{c}
0\\
v
\end{array}\right).
\label{eq:vevHiggs}
\end{equation}
The masses of the gauge bosons are determined by using the \textit{vev} of the scalar field $\phi$ appearing in the Lagrangian in eqn.\ \ref{eq:scalarfieldlag}.
The relevant term in the Lagrangian is
\begin{align}
 \left| (-ig\frac{1}{2}\tau.W_{\mu}-i\frac{g^{\prime}}{2}B_{\mu})\phi \right| ^{2}
 &=
 \frac{g^{2}}{8}\left|
 \left(
 \begin{array}{cc} gW_{\mu}^{3}+g^{\prime} B_{\mu} & W_{\mu}^{1}-iW_{\mu}^{2}
 \\
 W_{\mu}^{1}+iW_{\mu}^{2} & -gW_{\mu}^{3}+g^{\prime} B_{\mu}
 \end{array}\right)\left(
 \begin{array}{c}
 0\\
 v
 \end{array}
 \right)
 \right|^{2}
\nonumber \\
&= \frac{g^{2}v^{2}}{8}\left[(W_{\mu}^{1})^{2}+\left(W_{\mu}^{2}\right)^{2}\right] + \frac{v^{2}}{8}(g^{\prime} B_{\mu}-gW_{\mu}^{3})(g^{\prime} B_{\mu}-gW^{3\mu})
\nonumber \\
&= \left(\frac{vg}{2}\right)^{2}W_{\mu}^{+}W^{-\mu} + \left(\frac{v^{2}}{8}\right)\left(W^3_{\mu}B_{\mu}\right)\left(\begin{array}{cc}
g^{2} & -gg^{\prime} \\
-gg^{\prime} & {g^{\prime}}^{2}
\end{array}\right)\left(\begin{array}{c}
W^{3\mu}\\
B^{\mu}
\end{array}\right),
\label{eq:Lag1}
\end{align}
where $W^{\pm} = (W^{1}\mp iW^{2})/\sqrt{2}$.
\asr{
The first term in eqn.\ \ref{eq:Lag1} corresponds to the charged vector bosons and the second term to the neutral vector bosons.
For the charged vector bosons, $W^{\pm}$, we can write the bosonic mass term as
\begin{align}
 \left(\frac{vg}{2}\right)^{2}W_{\mu}^{+}W^{-\mu} &= m_W^2 W_{\mu}^{+} W^{- \mu},
\end{align}
such that $m_W=\frac{1}{2}gv$.
}
%Comparing this with the bosonic mass term \textit{i.e.} $\frac{1}{2}m^{2}B_{\mu}^{2}$, we get masses of the gauge bosons to be $m_W=\frac{1}{2}gv$.
%
The remaining off-diagonal terms of the matrix are
\begin{align}
 \frac{v^{2}}{8}\left[g^{2}\left(W_{\mu}^{3}\right)^{2}-2gg^{\prime} W_{\mu}^{3}B^{\mu}+{g^{\prime}}^{2}B_{\mu}^{2}\right] =
 \frac{v^{2}}{8}\left[gW_{\mu}^{3}-g^{\prime} B_{\mu}\right]^{2}+ 0\left[g^{\prime} W_{\mu}^{3}+gB_{\mu}\right]^{2},
 \label{eqn:Lag2}
\end{align}
where we have introduced the fields as an orthogonal combination of each other, and the zero introduced is an eigenvalue of the $2\times2$ matrix in eqn.\ \ref{eq:Lag1}.
To identify this with the mass form of $\frac{1}{2}m_{Z}^{2}Z_{\mu}^{2}$ and $\frac{1}{2}m_{A}^{2}A_{\mu}^{2}$, we have to normalize the mass terms in the eqn.\ \ref{eqn:Lag2} such that
\begin{align}
A_{\mu} &= \frac{g^{\prime} W_{\mu}^{3}+gB_{\mu}}{\sqrt{g^{2}+{g^{\prime}}^{2}}} \text{ with m$_{A} = 0$},
\nonumber \\
Z_{\mu} &= \frac{gW_{\mu}^{3}-g^\prime B_{\mu}}{\sqrt{g^{2}+{g^\prime}^{2}}} \text{ with m$_{z}= \frac{v}{2}\sqrt{g^{2}+g\prime^{2}}$}.
\end{align}
So, employing the Higgs mechanism, which is a consequence of SSB, we generate masses for the vector bosons.
%Details on spontaneous symmetry breaking can be found in several textbooks like \cite{Georgi:1999wka,Peskin:1995ev,Leader:2011gpt,Das:2008zze}.
%An important thing to note here is that the mass terms for the charged gauge bosons, $W^{\pm}$, does not have the factor of $1/2$ such that $m_W = \dfrac{1}{2} gv$.
%
\asr{
Since all left-handed fermions transform in the fundamental representation of SU$(2)$ and all right-handed ones transform as singlets, the complex SU$(2)$ doublet scalar field $\phi$ in eqn.\ \ref{eqn:scalarDoublet}, that led to the generation of gauge boson masses can also lead to fermion mass generation.
Therefore, fermion masses can be generated in principle with a single Higgs-doublet by making use of $\phi$ and $\tilde{\phi} \equiv \phi^{*}$ as a consequence of SSB.
}

\section{Shortcomings of the SM and moving Beyond}
\label{sec:SMshortcomings}

\asr{
Despite the phenomenal success of the SM, it is widely known that the SM fails to explain certain natural phenomena such as the baryon asymmetry in the universe, the existence of dark matter, tiny non-zero mass of neutrinos, fermion mass hierarchy and the size of CP violation \cite{Martin:1997ns,Virdee:2016mzw}.
}
Most of these problems can be solved by introducing more complex models, including more particles and Higgs fields \cite{Ellis:2002wba}.
The high-energy physics community expects new physics to manifest as tiny deviations from the SM predictions by improving existing precision studies, probing the top and Higgs sectors, and looking for exotic signatures or even a new Higgs boson!
\asr{
The study for new physics requires exhaustive experimental progresses to be accompanied by strong theoretical predictions to a very high accuracy within the SM and beyond because, until now, whatever information has been gained about the scalar Higgs boson from experiments is compatible with that of the SM \cite{CMS:2022dwd}.
}

Extension of the SM scalar sector is a common practice in constructing Beyond Standard Model (BSM) physics because a relevant question following the discovery of the Higgs is whether this scalar is the only one of its type as predicted by the SM or it is the first to have been discovered in a family of more such species arising from an underlying extended scalar sector.
Hence, dedicated efforts have been going on to determine the CP property of the discovered Higgs boson and to identify it
with that of the SM, although there are already indications that it is a scalar with even parity
\cite{2013,2015,CMS:2012vby,ATLAS:2015bcp,ATLAS:2015zhl,CMS:2017len,ATLAS:2017azn}.
The possibility of the observed Higgs boson of $125$ GeV mass being an admixture of scalar and pseudoscalar states cannot also be ruled out. Hence, such possibilities must also be looked into \cite{Artoisenet:2013puc,Gao:2010qx,Maltoni:2013sma}.

To relate the available Higgs data with BSM Higgs physics, an extension of the SM has been obtained by adding another singlet field or a Higgs doublet.
However, while extending to BSM physics, we must be careful that the experimental and theoretical constraints are not violated.
Two of the most important constraints are:
\begin{enumerate}
    \item The $\rho$ parameter, which is defined as the ratio of the neutral current to charged current coupling strengths,
    \item Limits on the existence of flavour-changing neutral currents (FCNC) - Experiments have severely constrained the existence of FCNCs.
    In the SM, tree-level FCNCs are absent automatically, as the mass matrix diagonalises the Higgs-fermion couplings.
    This is not true at higher orders, and this constraint needs to be considered theoretically and phenomenologically.
\end{enumerate}

%The scalar sector of the SM of electroweak interactions is not so well probed as the gauge boson and fermion sectors.
%The simplest possible scalar structure in the SM is that of one SU(2) doublet; the fermion structure with more than one family and mixing between the families is not that simple.

In an attempt to extend the SM, one risks altering the tree-level value of the electroweak $\rho$ parameter.
The $\rho$ parameter contains vital knowledge about the little-known scalar structure.
Consider an SU$(2)_{\text{L}} \otimes $U$(1)_{\text{Y}}$ gauge theory with $n$ scalar multiplets $\varphi$$_{i}$, weak isospin I$_{i}$, weak hypercharge Y$_{i}$ and \textit{vev} of the neutral components v$_{i}$.
With these parameters at the tree level, $\rho$ is given by
\begin{equation}
\rho=\dfrac{\sum_{i=1}^{n}\left[I_{i}\left(I_{i}+1\right)-\tfrac{1}{4}Y_{i}^{2}\right]v_{i}}{\sum_{i=1}^{n}\tfrac{1}{2}Y_{i}^{2}v_{i}}.
\label{eq:rho parameter}
\end{equation}
Experimentally $\rho$ is very close to one. According to eqn.\ \ref{eq:rho parameter}, $\rho=1$ for the following cases:
\begin{enumerate}
\item SU$(2)_{\text{L}}$singlets with Y=0,
\item SU$(2)_{\text{L}}$doublets with Y=$\pm$1.
\end{enumerate}
%where Y is the hypercharge.

A solution to the limitations of the SM is the production of new exotic particles as a consequence of a new symmetry --- Supersymmetry.
The MSSM is one of the simplest forms of supersymmetric theories.
It has five physical Higgs bosons, out of which two are neutral scalars $(h, H)$, one is a pseudoscalar $(A)$, and the remaining two are charged scalars $(H^\pm)$.
The pseudoscalar Higgs boson, which is CP odd, could be as light as the discovered Higgs boson.
Current experimental uncertainties suggest that the discovered Higgs is compatible not only with the SM predictions \cite{Glashow:1961tr,PhysRevLett.19.1264,salam} but also with the Higgs sector in the MSSM \cite{Nilles:1983ge,Haber:1984rc,Barbieri1988,Djouadi:2005gj,Heinemeyer:2004ms,Heinemeyer:2004gx,Degrassi:2002fi,Heinemeyer:2011aa,Benbrik:2012rm,Bottino:2011xv,MSSM2013,Drees:2012fb}.
Considering the far-reaching effects of the Minimal Supersymmetric Extension of the Standard Model, the following section aims at giving a small overview of the MSSM where the hypercharge Y$=\pm1$.
%
%In the case of the SM Higgs boson, the production cross section has been computed in perturbative Quantum Chromodynamics (QCD) to unprecedented accuracy. This is possible thanks to the fact that the top quark degrees of freedom can be integrated out. This results in an effective field theory (EFT) where the scalar Higgs boson couples directly to the gluons even at leading order (LO).

%\section{The SM Lagrangian}

%\section{Quantum Chromodynamics}

%\subsection{The QCD Lagrangian and Feynman Rules}

\subsection[MSSM]{Minimal Supersymmetric Extension of the Standard Model}
\label{subsec:MSSM}

The MSSM is a supersymmetric version of the SM in which the concerned Lagrangian is invariant under the SM gauge group and also supersymmetric invariant.
There are other scalar sectors which are much larger and more complex than the MSSM, for example, models with triplets and models with custodial SU(2) global symmetry.
%But the SM's most straightforward extension is adding scalar doublets and singlets.
However, the Minimal Supersymmetric Extension of the Standard Model is one of the most straightforward extensions where the word "Minimal" refers to a minimal choice of the particle spectrum necessary to make it work.
%
%Like any other supersymmetric theory, this model also involves a generator, $\mathcal{X}$, which is supposed to transform a fermion into a boson and vice-versa.
%\begin{equation}
% \mathcal{X}|Fermion> = |Boson>,\hspace{1mm}\mathcal{X}|Boson> = |Fermion>.
%\end{equation}

In a supersymmetric theory, ordinary quantum fields reform to superfields and the scalar partners of fermions are called "sfermions" ( quarks $\rightarrow$ squarks, leptons $\rightarrow$ sleptons) \cite{Haber:1984rc,Drees:2004jm}.
The MSSM contains two Higgs doublets - one with Y$=+1~(\phi_u)$ and the other with Y$=-1~(\phi_d)$. $\phi_u$ is needed to generate mass for the up-type quarks, and $\phi_d$ to generate mass for the down-type quarks \cite{Inoue:1982ej,Gunion:1986nh,FAYET197414}.
Just as the SM of particle physics, this model must be renormalisable and anomaly free; the corresponding opposite Y doublet, \textit{i.e.} Y$=-1$ for $\phi_u$ and Y$=+1$ for $\phi_d$, is needed for anomaly cancellation in theory.

%The Higgs potential for the MSSM is given by
%
%\begin{align}
%V_{Higgs} & =\left(m_{H_{1}}^{2}+|\mu|^{2}\right)|H_{1}|^{2}+\left(m_{H_{2}}^{2}+|\mu|^{2}\right)|H_{2}|^{2}-B\mu\varepsilon_{ij}\left(H_{1}^{i}H_{2}^{j}+h.c.\right)\\
% & =\dfrac{g^{2}+g'^{2}}{8}\left[|H_{1}|^{2}-|H_{2}|^{2}\right]^{2}+\dfrac{g^{2}}{2}|H_{1}^{\dagger}H_{2}|^{2},
%\end{align}
%where H$_{1}$ and H$_{2}$ represent the two complex Higgs doublets,
%introduced so that the up- and down-type quarks acquire mass, and this also leads to an anomaly-free theory, $g$ is the SU$(2)_L$ gauge coupling, $g'$ is the U$(1)_Y$ gauge coupling, $m_{H_i}~(i=1,2)$ represents the corresponding masses of the doublets, and $\mu$ represents the so-called higgsino mass parameter.
%

Let H$_{1}$ and H$_{2}$ represent the two complex Higgs doublets.
If the MSSM Higgs potential is minimised, then the neutral Higgs fields acquire a vacuum expectation value of $v_1/\sqrt{2}$ or $v_2/\sqrt{2}$, corresponding to the Higgs doublet $(H_1~\text{or }H_2)$ it belongs to \cite{Djouadi:2005gj}.
Consequently, we can define a parameter $\tan\beta$ as a ratio of the \textit{vev} of the Higgs doublets.
In this model, the mass matrix of the CP-even Higgs bosons (h, H) shows dependence on the mass of the pseudoscalar $(m_A)$, the Z boson $(m_Z)$, and $\tan\beta$ on squaring.
The mass of the charged scalars (H$^\pm$) shows dependence on the mass of the pseudoscalar $(m_A)$ and W boson $(m_W)$ on squaring.
%This consequently leads to the conclusion that at the tree level, the boson masses can be expressed in terms of just two parameters in MSSM: the mass of the neutral scalar $(m_A)$ and $\tan\beta$.
These mass matrices result in a significant difference between the SM and the MSSM.

Moreover, in the SM, the Higgs mass is known to be proportional to the Higgs self-coupling $(\lambda)$, which is a free parameter in the theory. In contrast, all the Higgs self-coupling parameters depend on the electroweak gauge couplings squared in MSSM.
Herein lies one of the successes of MSSM --- a solution to the \textit{hierarchy problem}.
The renormalisability of both theories indicates that all higher-order corrections generate finite results in spite of an infinite virtual momentum in loop integrals.
%The Higgs self-coupling $\lambda$ appears in the 4-boson self-interactions, and at one loop, this results in the introduction of an energy scale parameter, $\Lambda$, in the massive constant, $\mu$, of the 2-boson vertex.
%It is at this scale that BSM physics is expected to be observed, and consequently, the one-loop correction becomes much larger than $m_W$.
%MSSM controls the radiative corrections resulting in a small value of the massive constant $\mu^2$, and eventually small values of $m_H^2$ and $m_W^2$.
%The divergences from the quadratic-boson vertex now cancel, leading to a finite result.
%This protects the electroweak scale from large radiative corrections and hence, poses a solution to the hierarchy problem \cite{Inoue:1982pi,Inoue:1983pp}.

%%%%%%%%%%%%%%%%%%%%%%%%%%%%%%%%%%%%%%%%%%%%%%%%%%%%%%%%%%%%%%%%%%%%%%%%%%%%%%%%%%%%%%%%%%%%%%%%%%%%%%%%%%%%%%%55

\section{Outline}
\label{sec:Outline}

\begin{figure}[!htbp]
  \begin{center}
    \includegraphics[scale=0.5,keepaspectratio=true]{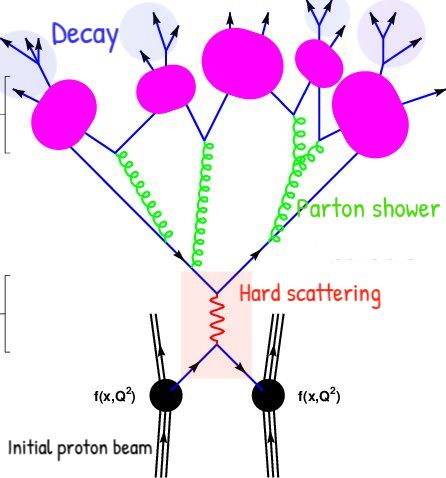}
    % QCD_Illustration.jpg: 446x478 px, 216dpi, 5.25x5.62 cm, bb=0 0 149 159
  \end{center}
  \caption[Schematic diagram of a collision event from an SHG]{Schematic diagram of a collision event from a showering and hadronisation generator (SHG) \cite{article:MonteCarlo}.
  The pink blobs represent hadronisation and $f(x,Q^2)$ the corresponding PDFs.  }
\label{fig:QCDIllustration}
\end{figure}

While works aiming towards building the SM were afoot, simultaneous attempts were being made to develop an understanding of symmetries.
These studies showed that many known symmetries were not exact but approximate. % like the SU$(3)_C$.
For example, the axial vector current was observed to be partially conserved in Beta decays, indicating the breaking of chiral symmetry.
It was actually in the context of strong interactions that SSB was first introduced.
The theory of strong interactions is known as Quantum Chromodynamics (QCD), and QCD is governed by the gauge group SU$(3)_C$.
The SM leptons, Higgs and electroweak gauge fields are SU$(3)_C$ singlets.
QCD is responsible for the production rates of any particle in hadron colliders exhibiting large invariant mass and/or large transverse momentum because of its dominance over the parton luminosity, along with the background signals in the colliders.
Several codes in platforms like PYTHIA, HERWIG and ISAJET are now being made available to generate schematic structures of high-energy collision events at the colliders, starting with a LO hard sub-process.
These are showering and hadronisation generators (SHGs), a general-purpose simulation tool \cite{Dobbs:2001ck,article:MonteCarlo}.
Fig.\ \ref{fig:QCDIllustration} depicts the general structure of the final state of an SHG generated event and shows how much happens in a high-energy collision.

The central idea of this thesis revolves around perturbative QCD (pQCD), where attempts are made to improve our understanding of hard scattering processes in strong interactions.
So after giving an overview of particle physics in chapter \ref{chap:Intro}, I proceed to do the same for QCD in chapter \ref{sec:QCD}. %, which eventually contributes to pQCD.
In chapter \ref{sec:Framework}, I will review EFT and certain other concepts involved in precision calculations which are the building blocks of this thesis.
The first three chapters in this thesis are not an exhaustive review of the respective topics but aim to provide a general idea of the global scenarios that inspired these developments.
For details, refer to \cite{Ellis:318585,Mangano:1998fk,Salam:2010zt,Nason:1997zu} or any standard text.
The following chapters contain our original published works and the ingredients and tools required to comprehend these works.

\clearpage
\newpage
\mbox{~}

\chapter{Quantum Chromodynamics}
\label{sec:QCD}

\begin{quote}
 \textit{“The study of physics is also an adventure. You will find it challenging, sometimes frustrating, occasionally painful, and often richly rewarding.”}

\hfill\textit{- Hugh D. Young.}
\end{quote}

\begin{figure}[!htb]
 \begin{center}
 \includegraphics[scale=0.18,keepaspectratio=true]{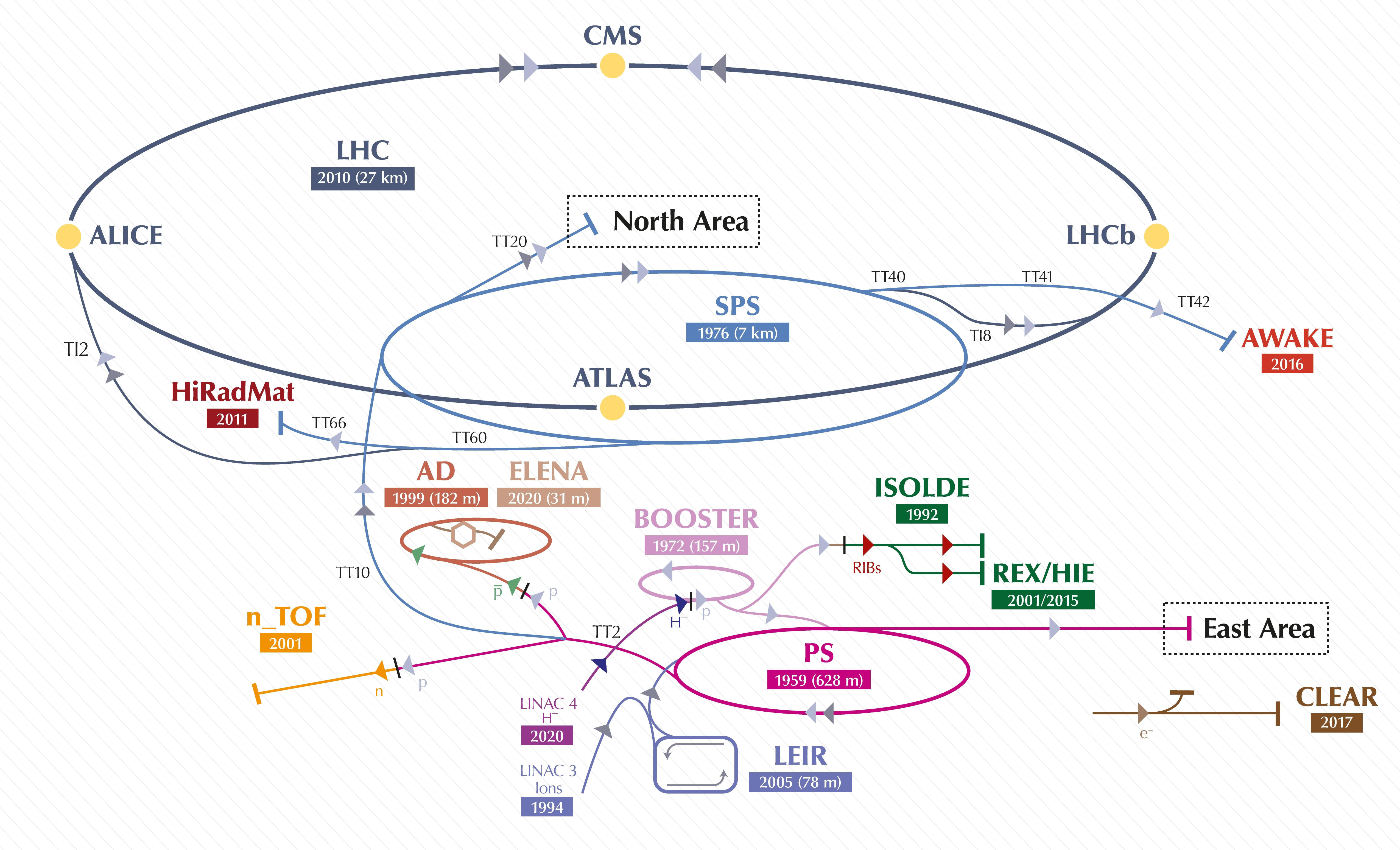}
 % HiggsMechanism.jpg: 980x596 px, 72dpi, 34.57x21.03 cm, bb=0 0 980 596
 \caption[The CERN accelerator complex]{The CERN accelerator complex \cite{VandenBroeck:2693837}}
 \label{fig:CERNAcceleratorComplex}
\end{center}
\end{figure}

The latter half of the 20th century, specifically the 1950s, witnessed the discovery of many new particles and the advent of increasingly powerful particle accelerators.
However, establishing the bedrock of QCD can be traced back to early nuclear physics, which contributed to ideas of strong interactions.
For instance, the discovery of hadrons, like the mesons, began in the Lawrence Berkeley National Laboratory in the USA with the observation of the $\Updelta$ resonances (m$_\Updelta \approx 1230$ MeV) in pion–nucleon collisions.
Eventually, the hyperons and K-mesons were discovered in cosmic-ray experiments.
It was the 1930s that witnessed the discovery of the first particle accelerator with an aim to study the nuclear structure.
Since then, these particle accelerators' development has seen significant improvements such that it did not even take a century to discover the Higgs.
\thr{
%Although time has seen the origin and dissolution of a handful of colliders, the collaborations at CERN are the most impactful ones in recent times.
Although time has witnessed the construction of new colliders and the phasing out of old ones, the collaborations at CERN have emerged as the most impactful colliders in recent times.
}
Fig.\ \ref{fig:CERNAcceleratorComplex} was an official poster released by CERN showing its detectors.
It was only through rigorous experiments, corresponding theories and competent precision phenomenological works that QCD could finally reach where it is now.
%The history of high energy physics in the second half of the twentieth century was driven by a sequence of increasingly more powerful particle accelerators, which allowed matter to be probed at ever smaller distances.
%\begin{center}
\textit{This chapter addresses the experimental discoveries at those machines briefly, which eventually built this challenging and adventurous theory of QCD.}
%\end{center}
%

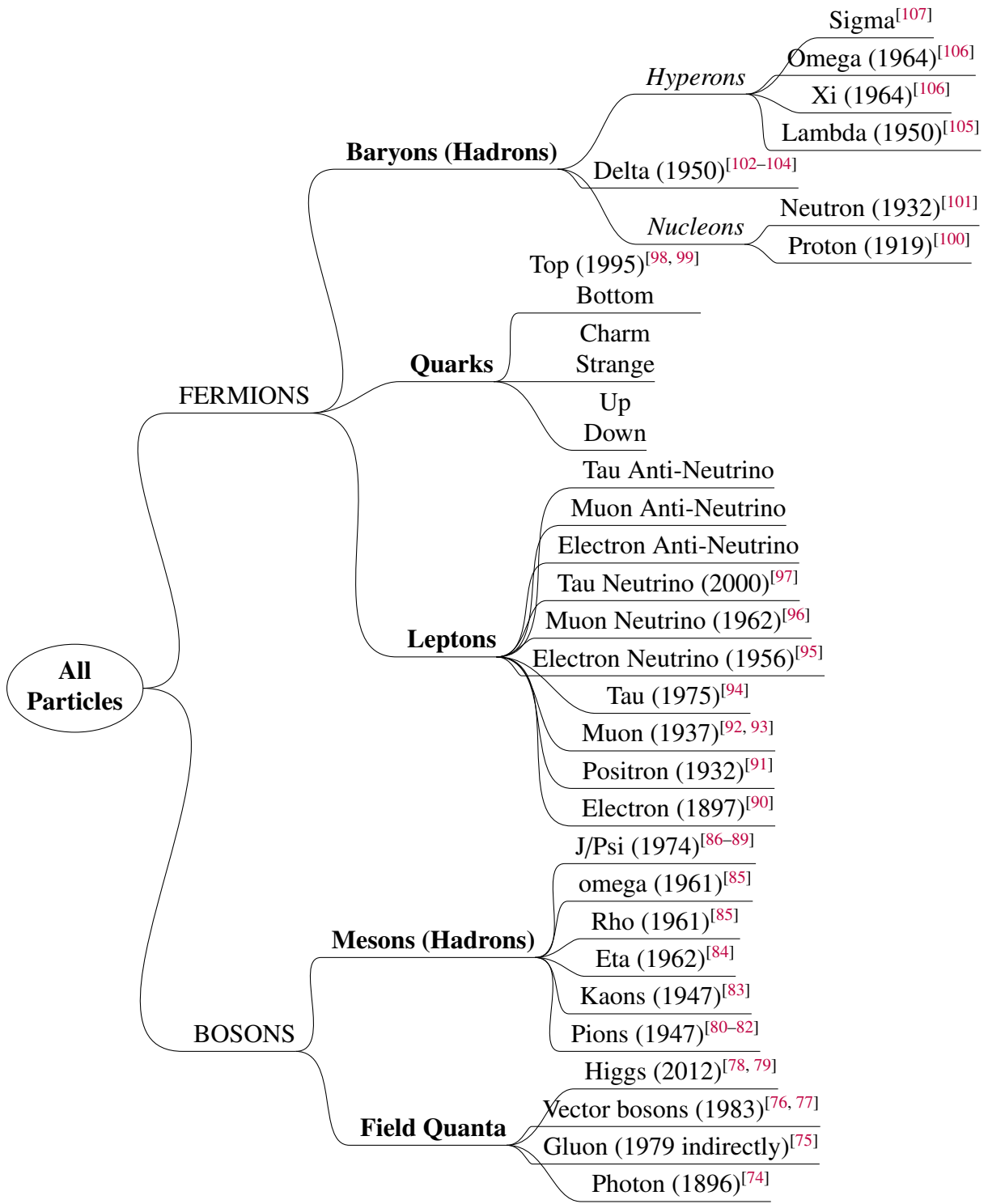
\begin{figure}[!htb]
\centering
 \begin{forest}
    for tree={
      grow=east,
      parent anchor=south east,
      child anchor=south west,
      anchor=south,
      align=center,
      l sep+=2.5pt,
      s sep+=-5pt,
      inner sep=0pt,
      outer sep=0pt,
      % path is based on Gonzalo Medina's answer
      % ref: https://tex.stackexchange.com/questions/176591/typesetting-genealogical-trees/176649#176649
      edge path={
        \noexpand\path [draw, rounded corners=5pt, \forestoption{edge}] (!u.parent anchor) [out=0, in=180] to (.child anchor)\forestoption{edge label} -- (.south east);
      },
      for root={
        ellipse,
        draw,
        parent anchor=east,
      },
    }
    [\textbf{All}\\{\textbf{Particles}}
        [\uppercase{Bosons}
            [\textbf{Field~Quanta}
                [Photon~(1896)\textsuperscript{\cite{Photon:french}}]
                [Gluon~(1979~indirectly)\textsuperscript{\cite{PhysRevLett.43.830}}]
                [Vector~bosons~(1983)\textsuperscript{\cite{EuropeanMuon:1983wih,UA1:1983mne}}]
                [Higgs~(2012)\textsuperscript{\cite{CMS:2012qbp,ATLAS:2012yve}}]
 %               [Graviton]
            ]
            [\textbf{Mesons~(Hadrons)}
                [Pions~(1947)\textsuperscript{\cite{Yukawa:1935xg,Perkins:1947mf,Lattes:1947mx}}]
                [Kaons~(1947)\textsuperscript{\cite{Rochester:1947mi}}]
                [Eta~(1962)\textsuperscript{\cite{Pevsner:1961pa}}]
                [Rho~(1961)\textsuperscript{\cite{Maglich:1976cpa}}]
                [omega~(1961)\textsuperscript{\cite{Maglich:1976cpa}}]
                [J/Psi~(1974)\textsuperscript{\cite{E598:1974sol,Aubert:1975rt,Becker:1975hh,Aubert:1976oaa}}]
            ]
        ]
        [\uppercase{Fermions}
            [\textbf{Leptons}
                [Electron~(1897)\textsuperscript{\cite{Thomson:1897cm}}]
                [Positron~(1932)\textsuperscript{\cite{Anderson:1932zz}}]
                [Muon~(1937)\textsuperscript{\cite{PhysRev.51.884,PhysRev.71.209}}]
                [Tau~(1975)\textsuperscript{\cite{PhysRevLett.35.1489}}]
                [Electron~Neutrino~(1956)\textsuperscript{\cite{Reines:1956rs}}]
                [Muon~Neutrino~(1962)\textsuperscript{\cite{PhysRevLett.9.36}}]
                [Tau~Neutrino~(2000)\textsuperscript{\cite{DONUT:2000fbd}}]
                [Electron~Anti-Neutrino]
                [Muon~Anti-Neutrino]
                [Tau~Anti-Neutrino]
            ]
            [\textbf{Quarks}
                [Up\\Down]
                [Charm\\Strange]
                [Top~(1995)\textsuperscript{\cite{PhysRevLett.74.2626,PhysRevLett.74.2632}}\\Bottom]
            ]
            [\textbf{Baryons~(Hadrons)}
                [\textit{Nucleons}
                    [Proton~(1919)\textsuperscript{\cite{Rutherford:1919fnt}}]
                    [Neutron~(1932)\textsuperscript{\cite{Chadwick:1932ma}}]
                ]
                [Delta~(1950)\textsuperscript{\cite{PhysRev.85.936,PhysRev.85.934,PhysRev.101.1149,}}]
                [\textit{Hyperons}
                    [Lambda~(1950)\textsuperscript{\cite{PhysRev.80.1099}}]
                    [Xi~(1964)\textsuperscript{\cite{PhysRevLett.12.204}}]
                    [Omega~(1964)\textsuperscript{\cite{PhysRevLett.12.204}}]
                    [Sigma\textsuperscript{\cite{ParticleDataGroup:2020ssz}}]
                ]
            ]
        ]
    ]
 \end{forest}
\caption[SM particles schematic classification]{{A schematic classification of the SM particles and their years of discovery}.}
\label{fig:ParticleScheme}
\end{figure}

\thr{
There is a well-established connection between elementary particle physics and representation theory as first noted in the 1930s by Wigner \cite{Wigner:1939cj}.
An outstanding example is the Poincare group, whose irreducible unitary representations provide the quantum mechanical states of various elementary particles.
%As first noted in the 1930s by Wigner \cite{Wigner:1939cj}, the properties of particles can be linked to the structure of Lie groups and Lie algebras.
The classification of SM particles is also done based on irreducible unitary representations of the Poincare group.
They can be indexed by two parameters according to the eigenvalues of the Casimir operators of the Poincare algebra: a continuous parameter $m$ or the “mass" of the particle and a discrete parameter $j$ or the “spin” of a particle.
}
%
%Particles can be classified based on spin or internal structure.
%
Based on the spin, particles can be classified into fermions and bosons, as already discussed in chapter \ref{chap:Intro}.
Based on internal structure, these particles can be classified as:
\begin{enumerate}
 \item \underline{Hadrons} - These are massive and have an internal structure \textit{i.e.} are composite particles. These can be further classified into
    \begin{itemize}
     \item \textbf{Mesons} which are composed of a quark and an anti-quark.
     This makes the mesons possess integral spin (boson).
     \item \textbf{Baryons}, which are composed of three quarks. This makes the baryons exhibit half-integral spin (fermion).
    \end{itemize}
 \item \thrasr{
 \underline{Leptons} - These have half-integral spin and are fundamental particles.
 They experience the weak and electromagnetic forces but not strong force.
 }
 \item \thrasr{
 \underline{Quarks} - These also have half-integral spin and are fundamental particles.
 Unlike leptons, quarks experience strong, weak and electromagnetic interactions.}
 \end{enumerate}
%
%
%Mesons are not the only type of bosons we have.
\thrasr{
In addition to the hadrons, leptons and quarks, there are gauge bosons like the eight gluons, photon, intermediate vector bosons $(W^{\pm}, Z)$, and Higgs.
The intermediate vector bosons and the weak force has a unique impact on the quarks and leptons --- they enable quarks and leptons to change their flavour.
}
Unlike the other gauge bosons which act as force carriers (gluons for strong interactions, photon for electromagnetic interactions, massive vector bosons for weak interactions), the Higgs is a non-gauge boson which makes its presence felt when a non-abelian gauge symmetry is broken spontaneously.
Fig.\ \ref{fig:ParticleScheme} gives a schematic diagram depicting a rough outline of the known SM particle structure and details about the discovery of several SM particles.
%The structural flow of particles can be described by the schematic diagram in fig.\ \ref{fig:ParticleScheme}
%

As depicted in fig.\ \ref{fig:ParticleScheme}, the current particle structure involves years of research and hard work.
The fact that most of an atom's mass is located at the centre of its nuclei, with protons occupying the central core, was known by 1920.
Eventually, radiations from beryllium were studied after bombarding it with alpha particles.
Frédéric and Irène Joliot-Curie observed these radiations as highly penetrating and initially assumed them to consist of high-energy photons.
However, James Chadwick recognised some odd features of this radiation and, in 1932, proposed it to be the neutron.
This changed the entire research scenario because soon, this newly discovered, fairly massive elementary uncharged particle was being used to probe other nuclei.
Hence began, the discovery of a plethora of particles.
This sudden surge of discoveries inspired the need to devise some symmetry in particle physics and find a way to explain the essential properties of strongly interacting particles.
Throughout the 60s, attempts were made to account for this ever-growing number of subatomic particles.
%defining a way to predict the existence of any other particle that can be discovered.
%
The early 1950s witnessed high-energy particles in cosmic ray showers and experiments at the Berkeley and Brookhaven laboratories.
Their fast production in pairs but a comparatively slower decay rate was a puzzle then.
With behaviour like elementary particles during the production time scale and an eventual decay unlike the elementary electron-positron pair creation, these new particles exhibited surprisingly unknown properties.
Finally, in 1956, Gell-Mann, and T. Nakano, with K. Nishijima, independently introduced the "strange" quantum number, which is conserved in strong and electromagnetic (EM) interactions.
Particles with this "strangeness" behaved like stable ones during production (through EM and strong interactions) but decayed slowly like the time scale of weak interactions \cite{PhysRev.92.833,10.1143/PTP.10.581,Gell-Mann:1956iqa,Gell-Mann:1982nvp,Jain:2019ezz}.

\begin{figure}
  \begin{center}
 \includegraphics[scale=0.35,keepaspectratio=true]{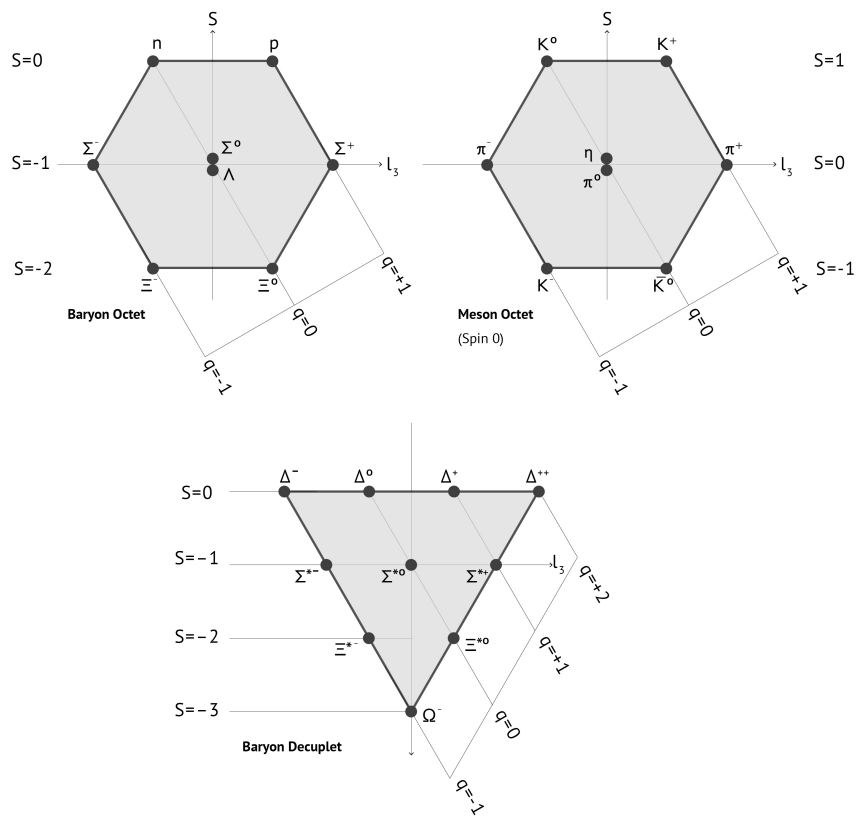}
 % EightFold_Way.png: 864x838 px, 72dpi, 30.48x29.56 cm, bb=0 0 864 838
\end{center}
 \caption[The Eightfold Way]{The Eightfold Way description \cite{Jain:2019ezz}}
 \label{fig:eightfoldway}
\end{figure}

Eventually, in 1961, Murray Gell-Mann \cite{osti_4008239}, and Yuval Ne'eman devised a classification scheme for elementary particles
%based on the SU$(3)$ gauge group
\cite{Gell-Mann:1962yej,Neeman:1961jhl,osti_4008239} into simple groups of eight, based on electric charge, spin, and its other characteristics --- the eightfold way (see fig.\ \ref{fig:eightfoldway}).
\asr{
In 1963, Nicola Cabibbo applied this concept of the eightfold classification to certain baryon decays, which proved that in addition to describing the existence and masses of particles, the $SU(3)$ symmetry could also describe their interactions.}
\thrasr{
Physicist George Zweig proposed in 1964 that "Mesons and baryons are constructed from a set of fundamental particles called aces" which can bind together and exhibit physically observable mass differences \cite{Zweig:1964ruk,Zweig:1964jf}.
In the same year, Gell-Mann independently proposed to use unitary
triplets as fundamental objects to explain the constituents of baryons and mesons.
The unitary triplet, $t$, in Gell-Mann's proposal consisted of an isotopic singlet and an isotopic doublet with electric charges.
The anti-triplet had the opposite signs of the charges.
He then referred to the members of the triplet as quarks and members of the anti-triplet as anti-quarks  \cite{Feynman:1964fk,Gell-Mann:1964ewy}.
%In 1964, the concept of quarks was introduced by Gell-Mann and George Zweig independently as a successful consequence of the eightfold classification scheme.
%
Therefore, Gell-Mann and Zweig independently proposed that the discovered baryons (bound state of three quarks) and mesons (bound state of a quark and an anti-quark) could be composed of quarks and anti-quarks, leading to a further simplification of the theory.
}

%\begin{quote}
% \textit{The $\Omega^-$ is a bound state of three strange quarks. Since this is the ground state, the space wave function should be symmetrical. The three spins of the quarks are aligned to give the spin of the $\Omega^-$. Thus, the $\Omega^-$ wave function does not change if two quarks are interchanged. However, the wave function must be antisymmetric according to the Pauli principle. This was a great problem for the quark model} \cite{Fritsch:2012sw}.
%\end{quote}
%
However, the proposal needed help to solve a significant puzzle --- how the  $\Omega^{-}$ existed with three strange quarks in the same state by contradicting the Pauli exclusion principle.
Oscar Greenberg in 1964 \cite{Greenberg:1964pe}, and Moo-Young Han with Yoichiro Nambu in 1965 \cite{PhysRev.139.B1006} independently proposed that quarks possess an additional SU(3) gauge DOF.
Greenberg introduced this additional DOF as a three-valued charge - a "parastatistics of rank three".
However, instead of three, Han-Nambu considered nine quarks and a symmetry group of the form SU$(3) \times $SU(3), which was assumed to be strongly broken.
Finally, in 1971, Gell-Mann and Harald Fritzsch proposed a different solution by assuming that the three quarks of the same type possess an additional conserved quantum number, "colour",
with the hadronic wave-function being colour singlet and the baryon wave-function being antisymmetric in the colour indices.
%\textbf{Color came to be known as the source of a "strong field" in the theory of QCD by physicists.}
%
Eventually, in 1972, Gell-Mann and Fritzsch started interpreting this colour group as a gauge group, with the colour symmetry being an exact symmetry under the SU(3) gauge group.
The result was a gauge theory similar to quantum electrodynamics (QED).
\begin{quote}
 \textit{"The interaction of the quarks is generated by an octet of massless colour gauge bosons, which we called gluons \cite{Fritzsch:1972jv}. We later introduced the name \textbf{QCD}. We published details of this theory one year later together with Heinrich Leutwyler \cite{Fritzsch:1973pi,}."}

\hfill{\textit{- Harald Fritzsch}}
\end{quote}
The significance of Murray Gell-Mann in the field of particle physics can be summarised by this one fact that he won the Nobel Prize in Physics in 1969 \textit{"for his contributions and discoveries concerning the classification of elementary particles and their interactions"} \cite{Nobel:gellmann}.
1969 proved significant yet again when results of experiments at the Stanford Linear Accelerator Center (SLAC) led by Friedman, Kendall and Taylor probed the proton structure successfully by smashing high-energy electrons into hydrogen atoms.
The results supported the idea that the nucleons and other related heavy particles are indeed composed of constituent particles called quarks \cite{PhysRevLett.23.930,PhysRevLett.23.935}.
The experiments that led to this conclusion were honoured with the 1990 Nobel Prize in Physics.

So QCD emerged as the SU$(3)_\text{C}$ gauge theory of strong interactions within the SM \cite{Fritzsch:1972jv,Fritzsch:1973pi,Gross:1973id,Politzer:1973fx,Ali:2010tw} that describes the building blocks of strongly interacting particles - the quarks (spin\ -\ $\dfrac{1}{2} \hbar$) and gluons (spin\ -\ $\hbar$).
%which make up hadrons such as protons, neutrons and pions.
This theory of strong interactions exhibits two salient features:
\begin{itemize}
 \item Although never observed in isolation as all composite particles are "colourless", the high-energy physics community establishes the existence of these strongly interacting elementary particles with complete confidence.
 The non-existence of quarks in isolation is a notable feature of QCD known as "confinement".

 \item Another noteworthy feature observed in QCD is the asymptotic weakening of bonds between quarks$/$gluons with increasing energy or decreasing distance.
 This feature is known as "asymptotic freedom" \cite{Gross:1973id,Gross:1973ju,Politzer:1973fx,Gross:1974cs,Politzer:1974fr}.
 Asymptotic freedom opens the door to pQCD, enabling us to perform QCD calculations using perturbation theory.
 David Gross, Frank Wilczek and David Politzer were awarded the Nobel prize in Physics in 2004 for this discovery.
\end{itemize}

\section{Basics of QCD}
\label{sec:moreQCD}

%
%This thesis deals with perturbative computations for rare short-distance particle interactions, with the goal to develop reliable theory to search for new physics.
%
%Since the LHC deals with proton-proton collisions, the dynamics of strong interactions play a crucial role in those processes.
During the past years, LHC experiments have mainly focused on studying various facets of strong interactions to provide highly precise results.
Perturbative QCD is being extensively used to perform precision computations in collider physics.
%However, another approach to solving the theory of QCD is known as "Lattice QCD", but discussing it is beyond the scope of this thesis.
In this thesis, I will concentrate on the concepts/techniques involved in pQCD.
%%%%%%%%%%%
In the perturbative regime, the strong coupling constant, $\alpha _\text{s}$, underlying the quark and gluon interactions in QCD, is small in high energy or short distance interactions, allowing the application of perturbation theory techniques.
The small value of $\alpha _\text{s}$ in pQCD has led to the development of an exhaustive framework for computing high energy processes in a power series expansion of $a_\text{s} = \text{g}^2/16 \pi^2 = \alpha_\text{s}/4 \pi$.

The demand to increase the precision of these QCD calculations requires the inclusion of several orders of perturbation theory.
However, including higher-order corrections dramatically increases the complexity of multi-loop calculations.
\thr{
The evaluation of higher-order diagrams involves integrals which give rise to divergences that can be appropriately renormalised.
To obtain UV renormalised results, the singularities are first regularised by employing the technique of dimensional regularisation, where the number of space-time dimensions is taken to be $d = 4 + \varepsilon$, $\varepsilon$ being a tiny change in dimension $d=4$.
Consequently, the resulting singularities appear as poles in $1/\varepsilon^k$ where $k$ is an integer.
Once the regularised results are available, the UV divergences are renormalised by suitably redefining the coupling constants and fields.
The method of UV renormalisation involves absorbing the UV divergences into the appropriate renormalisation constants.
}
Each of these parameters involves the introduction of some scale parameters that are not intrinsic to the theory but tell how the renormalisation was done.
This new unphysical scale is called the renormalisation scale, and physics is independent of it.
The renormalization of a given bare quantity $\hat{Q}$ can be represented as:
\begin{equation}
 \hat{Q} = Z_F(\mu_R ) Q(\mu_R ).
 \label{eq:F&Fhat}
\end{equation}
Here $Z_F$ is the renormalisation constant that absorbs all UV divergences of $\hat{Q}$ making the quantity $Q$ UV finite.

\textit{This method of renormalisation leads to the running of the strong coupling constant in QCD discussed in section \ref{subsec:runningofas}.}

To correlate between experiments and theory, establishing correspondence between the observables obtained at the parton and the hadron levels is now of key concern in phenomenological studies.
The correspondence is necessary because partons cannot be directly measured/observed in experiments, and the particles observed include only hadrons and their decay products.
However, the theoretical predictions of the concerned observable should be made free of all singularities for good correspondence.
In addition to UV renormalisation, the observables must also be made IR (soft and collinear) safe.
Only after removing all singularities can finite predictions be obtained at any order in pQCD.

\subsection{Running of the strong coupling constant}
\label{subsec:runningofas}

To renormalise the strong coupling constant, eqn.~\ref{eq:F&Fhat} can be used to write the unrenormalised strong coupling constant, $\hat{a}_s$ as \cite{Ahmed_2015,Ravindran_2006}
\asr{
\begin{equation}
 \dfrac{\hat{a}_\text{s}}{\mu_0^\varepsilon} S_{\varepsilon} =
 \dfrac{1}{\mu_R^\varepsilon} a_\text{s}(\mu_R^2) Z_g^2(\mu_R^2) =
 \dfrac{1}{\mu_R^\varepsilon} a_\text{s}(\mu_R^2) Z_{a_s}(\mu_R^2) ,
 \label{eq:ashattoas1}
\end{equation}
}
with
\[
S_{\varepsilon}=\exp[(\gamma_E-\ln4\pi)\varepsilon/2].
\]
The renormalisation constant, $Z_{a_{s}}$, up to order $a_s^3$ is given by \cite{TARASOV1980429,Ahmed_2015,Ravindran_2006}
\begin{equation}
Z_{a_{s}}=1+a_{s}\left[\frac{2}{\epsilon}\beta_{0}\right]+a_{s}^{2}\left[\frac{4}{\epsilon^{2}}\beta_{0}^{2}+\frac{1}{\epsilon}\beta_{1}\right]+a_{s}^{3}\left[\frac{8}{\epsilon^{3}}\beta_{0}^{3}+\frac{14}{3\epsilon^{2}}\beta_{0}\beta_{1}+\frac{2}{3\epsilon}\beta_{2}\right]\,.
\label{eq:Zas}
\end{equation}
%
%
%The unrenormalized strong coupling constant, $\hat{a}_{s}$, is related to renormalized one, $a_s$, by \cite{Ahmed_2015,Ravindran_2006}
%
%\begin{equation}
%\hat{a}_{s}S_{\varepsilon}=\left(\dfrac{\mu^{2}}{\mu_{R}^{2}} \right)^{\varepsilon/2}Z_{a_{s}}a_{s},
%\label{eq:ashattoas2}
%\end{equation}
%
In eqn.\ \ref{eq:ashattoas1}, $\mu_0$ is the scale introduced to keep the strong coupling constant dimensionless in $d = 4 + \varepsilon$ space-time dimensions.

As can be seen from eqn.~\ref{eq:ashattoas1}, $\hat{a}_\text{s}$ lacks an explicit dependence on the renormalisation scale $(\mu_R)$.
As a consequence, we get the following renormalisation group equation (RGE):
\begin{equation}
 \mu_R^2 \dfrac{d \hat{a}_\text{s}}{d \mu_R^2} = 0
 \label{eq:asRGE}
\end{equation}
Differentiating eqn.~\ref{eq:ashattoas1} \textit{w.r.t.} $\mu_R^2$ and using the RGE in eqn.\ \ref{eq:asRGE}, we get
\begin{equation}
 \mu_R^2 \dfrac{d\ln a_\text{s}(\mu_R^2)}{d\mu_R^2} = \dfrac{\varepsilon}{2} - \mu_R^2 \dfrac{d\ln Z(\mu_R^2)}{d\mu_R^2}
 \label{eq:asrunning}
\end{equation}
In $\overline{\text{MS}}$ scheme, the QCD $\beta$-function is defined as
\begin{equation}
 \mu_R^2 \dfrac{da_\text{s}(\mu_R^2)}{d\mu_R^2}  = \dfrac{\varepsilon}{2} + \dfrac{1}{a_\text{s}(\mu_R^2)} \beta(a_\text{s}(\mu_R^2))
 \label{eq:QCDbeta}
\end{equation}
where
\begin{equation}
 \beta(a_\text{s}(\mu_R^2)) = - \sum_{n=0}^{\infty} \beta_n a_\text{s}^{n+2}(\mu_R^2).
 \label{eq:betansum}
\end{equation}
The $\beta_i$'s are defined as \cite{TARASOV1980429}
\begin{align}
\label{eqn:betais1}
\beta_{0} & =\frac{11}{3}C_{A}-\frac{4}{3}n_{f}T_{F}\,,\\[0.5ex]
\beta_{1} & =\frac{34}{3}C_{A}^{2}-4n_{f}C_{F}T_{F}-\frac{20}{3}n_{f}T_{F}C_{A}\,,\\[0.5ex]
\beta_{2} & =\frac{2857}{54}C_{A}^{3}-\frac{1415}{27}C_{A}^{2}n_{f}T_{F}+\frac{158}{27}C_{A}n_{f}^{2}T_{F}^{2}%\nonumber%
+\frac{44}{9}C_{F}n_{f}^{2}T_{F}^{2}\nonumber \\
 & -\frac{205}{9}C_{F}C_{A}n_{f}T_{F}+2C_{F}^{2}n_{f}T_{F}
 \label{eq:betais}
\end{align}
where $n_{f}$ is the number of active flavors and $T_{F}=1/2$. The
Casimir operators of the SU(N) color group $C_{F}$ and it's adjoint
$C_{A}$ are represented by
\begin{eqnarray}
C_{F}=\frac{N^{2}-1}{2N},\quad\quad C_{A}=N\,.
\label{eq:CN}
\end{eqnarray}
%The $\beta_i$'s are defined in eqn.~\ref{eq:betais}.
The negative sign in eqn.~\ref{eq:betansum} marks the origin of asymptotic freedom. It is, in fact, a consequence of the colour charge of gluons.
An elaborate explanation of this is given in section \ref{subsec:asympfreedom}.

Using only $\beta_0$ and ignoring the fact that the number of quark degrees of freedom, $n_f$, depends on $\mu_R$, we get a simple solution of the form
\begin{equation}
 a_\text{s}(\mu_R^2) = \dfrac{a_\text{s}(\mu_0^2)}{1 + \beta_0 a_\text{s}(\mu_0^2)\ln \frac{\mu_R^2}{\mu_0^2}} = \dfrac{1}{\beta_0\ln\frac{\mu_R^2}{\Lambda^2}},
 \label{eq:asrunningfinal}
\end{equation}
where $\mu_0$ is a reference scale and $\Lambda$ is known as the QCD scale.

Setting $C_A=3$ and $T_F=1/2$ in eqn.~\ref{eq:betais}, we get $\beta_0 = 11 - 2/3 n_f$. For $n_f < 16.5$, $\beta_0$ remains positive and for $n_f > 16.5$, $\beta_0$ becomes negative. So far we have $6$ quarks in total, \textit{i.e.} $n_f = 6$, resulting in $\beta_0 > 0$.
Hence from eqn.~\ref{eq:asrunningfinal}, we can depict that $a_\text{s} (\mu_R^2)$ decreases as $\mu_R^2$ increases, indicating the presence of asymptotic freedom.
An essential consequence of asymptotic freedom is that the high-energy QCD processes can be reliably computed using pQCD.
Conversely, $a_\text{s} (\mu_R^2)$ increases at low energies and finally exits the perturbative regime at some point, indicating that at low energies, QCD exhibits confinement.

\begin{figure}[!htb]
\centering
 \includegraphics[scale=0.45,keepaspectratio=true]{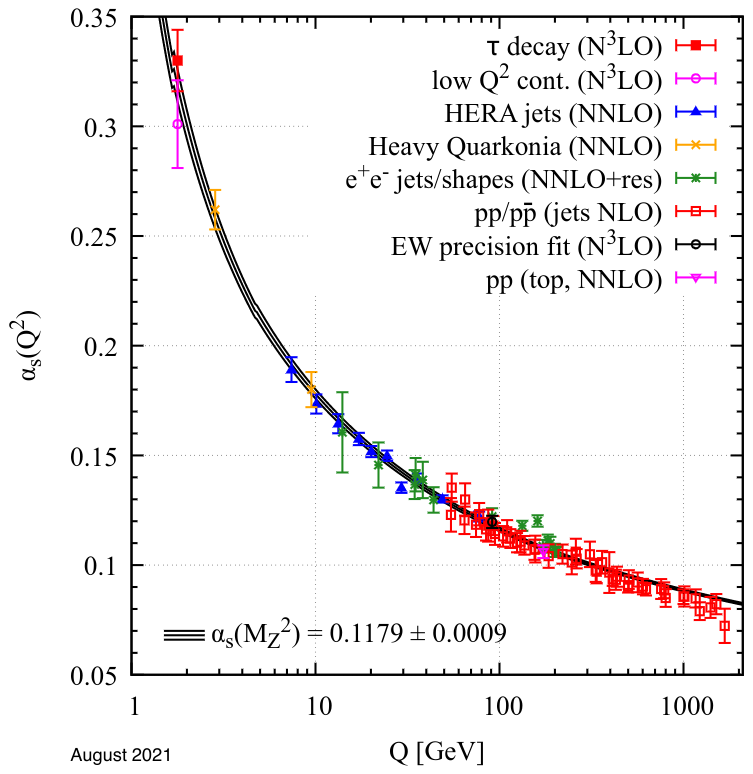}
% alphaSrunning.png: 920x1301 px, 72dpi, 32.46x45.90 cm, bb=0 0 920 1301
 \caption[QCD running coupling constant: experimental data confronting theory]{QCD running coupling constant: experimental data confronting theory \cite{ParticleDataGroup:2022pth}}
 \label{fig:runningcouplingconstant}
\end{figure}

This phenomenon of the running of the strong coupling constant has also been experimentally confirmed as shown in fig.~\ref{fig:runningcouplingconstant}, which depicts a comparison of theoretical and experimental data \cite{Bethke:2004fe,alphasrunning:2008,QCDCoupling:New,ParticleDataGroup:2022pth}.

\subsection{Confinement and Asymptotic freedom}
\label{subsec:asympfreedom}

The 1969 SLAC experiments data showed that the internal structure of a proton is free and pointlike, which eventually raised the question:
\begin{quote}
    "How can the quarks be nearly free inside the proton yet bind so strongly together that they do not exist as free particles?" \cite{Quark:confinement}
\end{quote}

\begin{figure}[!htb]
\centering
 \includegraphics[scale=0.85,keepaspectratio=true]{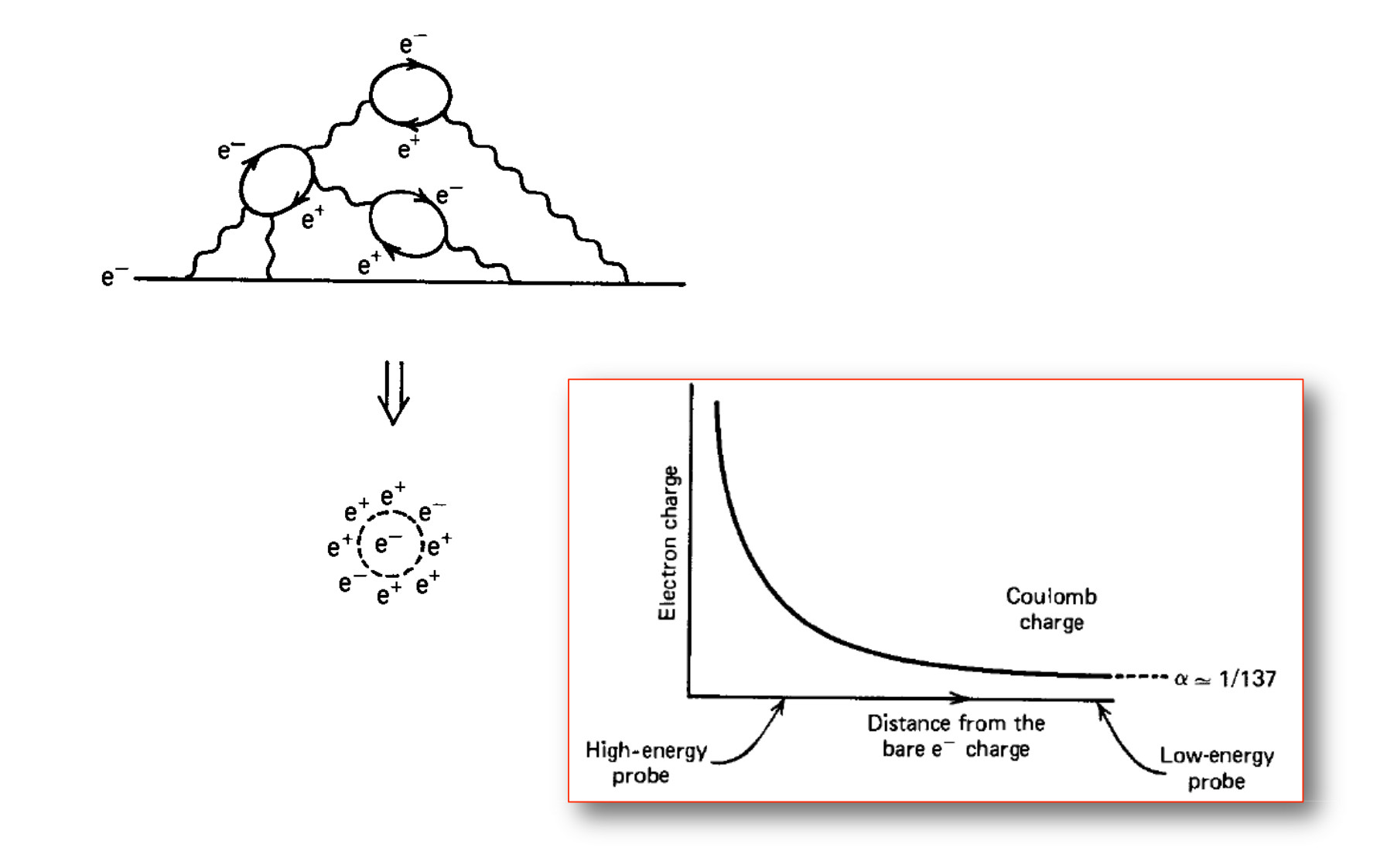}
% alphaSrunning.png: 920x1301 px, 72dpi, 32.46x45.90 cm, bb=0 0 920 1301
 \caption[Screening in QED]{Screening of electric charge from quark loops in QED \cite{screening_antiscreening}.}
 \label{fig:screening}
\end{figure}

\begin{figure}[!htb]
\centering
 \includegraphics[scale=0.3,keepaspectratio=true]{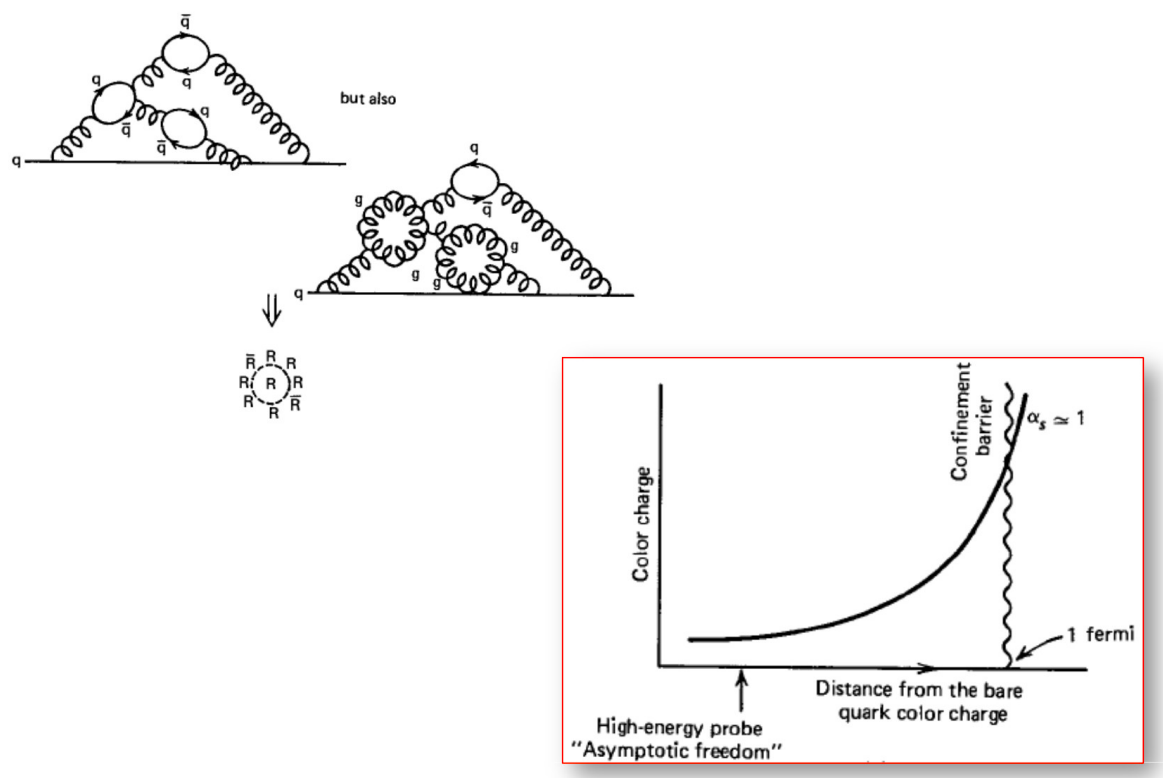}
% alphaSrunning.png: 920x1301 px, 72dpi, 32.46x45.90 cm, bb=0 0 920 1301
 \caption[Antiscreening in QCD]{Screening of colour charges from quark loops and antiscreening from gluon loops in QCD leading to an overall negative sign in the beta function \cite{screening_antiscreening}.}
 \label{fig:antiscreening}
\end{figure}

\textit{In QED, a cloud of virtual photons usually surrounds a charged particle like the electron $(e^-)$, in addition to electron-positron $(e^+ e^-)$ pairs continuously popping in and out of existence.
Since opposite charges attract, these virtual positrons tend to be closer to the electron and screen its charge.
This results in an effective charge, $e(r)$, that becomes smaller with increasing distance leading to the conclusion that the beta function is positive in QED \cite{Halzen:1984mc}.
Fig.\ \ref{fig:screening} pictorially depicts this phenomenon.}
\textit{Similarly, in QCD, virtual $q\overline{q}$ pairs exist in the vacuum.
The charge screening mechanism in QCD should have been the same, resulting in a positive beta function.
However, the gluon self-coupling leads to a surprising result!
The vacuum gets filled with virtual gluon pairs with a colour charge, increasing the effective charge with increasing distance, leading to a negative beta function \cite{Halzen:1984mc}.
This effect is called antiscreening and is diagrammatically expressed in fig.~\ref{fig:antiscreening}.}

This antiscreening effect in QCD causes the quarks inside hadrons to behave as approximate free particles when probed at large enough energies (asymptotic freedom).
Consequently, asymptotic freedom enables us to use perturbation theory and determine quantitative predictions for hard scattering cross-sections in hadronic interactions.
On the other hand, the force holding the internal particles becomes so strong with increasing distance that it becomes impossible to isolate a quark from a hadron (confinement).
%Confinement is a verified mechanism in Lattice QCD. However, its non-perturbative nature makes mathematical proof from first principles almost impossible. Any further discussion on Lattice QCD is also beyond the scope of this thesis.
%

%%%%%%%%%%%%%%%%%%%%%%%%%%%%%%%%%%%%%%
%
\section{A physical picture of hadronic interactions}
\label{sec:physicalpicture}

Correlating theoretical and experimental studies is indirect and requires certain concepts and techniques.
Deep-inelastic scattering (DIS) laid the foundation for these ideas.

\subsection[Deep-inelastic scattering]{Deep-inelastic scattering (DIS)}
\label{sec:DIS}

\begin{center}
 \textit{In the late 1960's SLAC–MIT DIS experiments, it was seen that the rate for electron scattering at large angles is significant, as if protons are made of smaller particles.}
\end{center}
%
%One of the first processes to play a fundamental role in the study of hadronic cross-sections was DIS. 
The DIS of leptons and nucleons is one of our primary sources of information about the internal nucleon structure in terms of the PDFs.
In the simplest case, DIS involves the interaction of an electron (lepton) with a proton (nucleon) by the exchange of a virtual photon as
\begin{equation}
    l(k) + H(P) \rightarrow l^{\prime}(k^{\prime}) + X.
 \label{eq:DIS}
\end{equation}
\begin{figure}[!htb]
 \centering
 \includegraphics[scale=0.34,keepaspectratio=true]{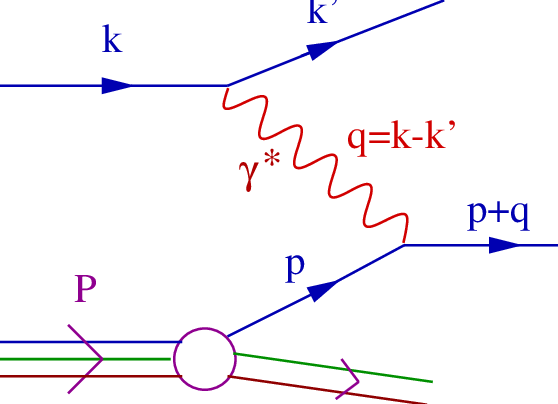}
% DISscattering.jpg: 749x501 px, 72dpi, 26.42x17.67 cm, bb=0 0 749 501
 \caption[Deep-inelastic scattering in QCD]{Deep-inelastic scattering of a lepton from a hadron.
 $k$ and $k'$ are the four-momenta of the incoming and outgoing leptons, $P$ is the four-momentum of the hadron, $q=(k-k^\prime)$ is the four-momentum of the intermediate photon and $p$ is the four-momentum of the scattered parton \cite{article:2008}.}
 \label{fig:DISscattering}
\end{figure}

The process depicted in fig.~\ref{fig:DISscattering} is called deep $(Q^2 =-q^2 \gg M^2)$ inelastic $(W^2  \gg M^2)$ scattering where the invariant squared mass of the subsystem recoiling against the scattered lepton is given by $W^2=(P+q)^2$.

The corresponding cross-section for DIS can be factored into leptonic and hadronic parts as
\begin{equation}
 d{\sigma} \sim L^{\mu\nu}_j ~ W_{\mu\nu}^j
\end{equation}
where $j=\gamma,Z^0,W^\pm$.
Unlike the leptonic tensor, $L^{\mu\nu}_j$, which is calculable in pQCD, determining the hadronic tensor, $W_{\mu\nu}^j$, is difficult because of its non-perturbative nature.
For the hadronic part, the goal is to express it as the most general covariant expression.
Consequently, the $W_{\mu\nu}^j$ is expressed in terms of some structure functions.
This way of expressing the hadronic tensor has been successfully carried out because, even if nothing is known about the structure of the final state, the cross-section can still be parameterized for DIS by the structure functions, $F_{1}(x,Q^{2})$ and $F_{2}(x,Q^{2})$.
The DIS differential cross-section can be written in a general form using the structure functions as
\begin{equation}
 \dfrac{d^2\sigma}{dE^{\prime}d\Omega} = \dfrac{\alpha^{2}}{4E^{2}\sin^{4}({\theta}/{2})}~\dfrac{1}{\nu}
 \left\{ \dfrac{2 \nu}{M} F_{1}(x,Q^{2})\sin^{2}\dfrac{\theta}{2} + F_{2}(x,Q^{2})\right\},
\end{equation}
where $\nu=\dfrac{P.q}{M}$.
For a given CMS energy, the inelasticity $y$ and the Bjorken variable $x$ can be defined as
\begin{align}
    y &= \dfrac{p.k-p.k^{\prime}}{p.k} \equiv 1 - \dfrac{E^{\prime}}{E}, \\
    x &= \dfrac{Q^2}{2~p.q},
\label{eq:BjorkenVariable}
\end{align}
where $E$ and $E^{\prime}$ are the energies of the incoming and outgoing leptons in the laboratory frame of reference.
%
%As is customary, we can slightly rescale the structure functions and define them as below :
%\begin{align}
% F_{1} & \equiv W_{1}~; \nonumber \\
% F_{2} & \equiv \dfrac{\nu}{M^{2}}W_{2}~,
%\end{align}
%where $M$ is the target nucleon mass, $\nu$ is the energy loss between the scattering electrons.

\begin{figure}[!htb]
 \centering
 \includegraphics[scale=0.48,keepaspectratio=true]{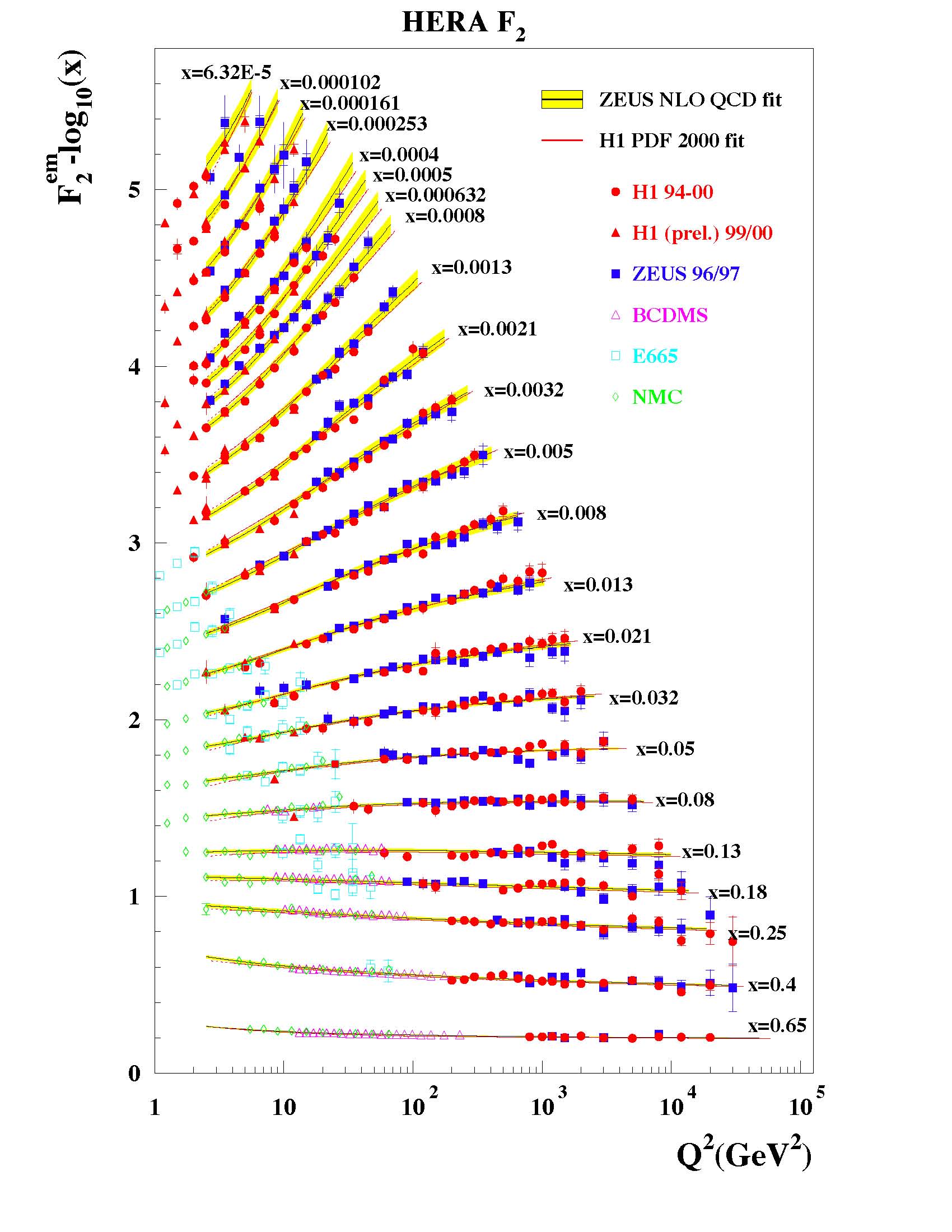}
% BjorkenScaling.png: 582x855 px, 72dpi, 20.53x30.16 cm, bb=0 0 582 855
 \caption[Bjorken Scaling]{The $F_2$ structure function measured by ZEUS, H1 and various other fixed target experiments plotted against $Q^2$. Bjorken scaling is observed at high $x,$ but is seen to be gradually violated towards low $x.$ The line represents a QCD-based fit to the data \cite{BjorkenScaling:Fig}.}
 \label{fig:BjorkenScaling}
\end{figure}

%
As the momentum transfer in scattering increases, the probing wavelength decreases, providing a "deeper" look inside the proton.
Hence, the electron can now be said to scatter on the constituents of the proton --- Partons.
Experimentally, it was observed that the structure functions, $ F_i$'s, must be independent of $Q^2$ for a fixed $x$ \cite{Friedman:1972sy}.
Originally, it was Bjorken who predicted in 1967 that in the limit of large $Q^2$ and fixed $x$, the dimensionless structure functions are depicted by
\begin{align}
 F_{1} & = M~W_{1}(x,Q^2)~; \nonumber \\
 F_{2} & = \nu~W_{2}(x,Q^2)~,
\end{align}
where the Bjorken variable $x$ can be identified as the momentum fraction carried by the struck constituent \cite{Bjorken:1967fb,H1:1993jmo,ZEUS:1993ppj,PhysRev.179.1547}.
The independence of $F_{1,2}$ on $Q^2$ for $x = \text{fixed}$ is called Bjorken scaling.
%
%See fig.\ \ref{fig:BjorkenScaling} for the scaling behaviour of the structure function $F_{2}$.
In fig.\ \ref{fig:BjorkenScaling}, the approximate scaling of the structure-function $F_{2}$ is observed at $x \approx 0.1,$, with scaling violations in the data for lower and higher $x$-regions.
Eventually, in 1969, Callan and Gross put forward the relationship,
\begin{equation}
    2 x F_1(x) = F_2(x).
    \label{eq:Callan-Gross}
\end{equation}
This eqn.\ \ref{eq:Callan-Gross} reflected the fact that the partons are spin-$1/2$ particles inside a proton.

This observation is explained below:
\begin{quotation}
 The parton's (electron's) interaction time with the radiating gauge boson (photon) is much shorter than the actual lifetime of the quantum fluctuation corresponding to the parton.
 So, the corresponding cross-section is proportional to the sum of probabilities to find such partons interacting with the photon.
 Moreover, this will further lead to treating such photon-parton collisions like the collision of two independent quasi-free particles.
 Since the interaction time is shorter than the lifetime of the struck parton, the surrounding partons forming the proton fail to assert their presence to the struck parton.
 The probabilities of finding partons with a momentum fraction $x$ at a scale $Q^2$ inside a proton becomes process-independent in this framework \cite{Campbell:2017hsr}.
\end{quotation}

\vspace{-4mm}
\subsection{Factorisation and the parton model}
\label{sec:Partonmodel}

%\begin{center}
\textit{Scaling behaviour of the structure functions (Bjorken scaling) suggested the hadron to be pictured as an incoherent flux of partons.
Additionally, the space-time picture described in section \ref{sec:DIS} suggests the possibility of separating short- and long-distance physics.
%Hence, we can say that Bjorken scaling and the Callan-Gross formula provided clear evidence of the quark-Parton model.
}
%\end{center}

An essential yet fundamental concept in physics is the existence of fundamental or elementary particles which bind together under the influence of the fundamental forces to constitute a  larger object.
In particle physics, hadrons are the large objects understood as bound states of smaller ones, namely the quarks and gluons.
%In this section, I introduce the Parton model developed to explain DIS.
Over the years, several parton models have been proposed to explain experimental discoveries.
One is the "Naive Parton Model", where the proton was assumed to be an assorted state of loosely bound partons.
% - fermions with electric charges and other neutral species which bind the particles together.
%The observation that partons cannot exchange large momenta through strong interactions was attributed to asymptotic freedom.
%In this "naive" model, a parton is denoted by a fraction of the proton's total momentum that it carries.
%
%This model exhibits the following characteristics:
%\begin{itemize}
% \item $q(\xi)d\xi$ represents the probability that a quark $q$ carries a momentum fraction between $\xi$ and $\xi + d\xi,$ where $0 \le \xi \le 1$,
% \item there is incoherent scattering of the virtual photon off the quark constituents.
%\end{itemize}
%
%This model involves scaling of the structure functions with the probability distribution, $q(x)$, which eventually leads to the Callan-Gross formula given in eqn.\ \ref{eq:Callan-Gross}. This can also be expressed as
%\begin{equation}
%    F_2(x) = 2 x F_1(x) = \sum_{q,\overline{q}} p_k^2 ~x~ q(x),
    \label{eq:Naive_Callan-Gross}
%\end{equation}
%where $p_k$ is the momentum carried by the parton and is expected to carry a fraction of the momentum, $p$, of the proton: $p_k = \xi p$.
%$q(x)$ is better known as the Parton distribution function (PDF).
%
This "Naive" model becomes exceedingly interesting when it shows the scaling of the structure functions, $F(x,Q^2) \rightarrow F(x)$, in the Bjorken limit where $Q^2 \rightarrow \infty$ and $x=\text{fixed}$. Fig.\ \ref{fig:BjorkenScaling} depicts this with the $Q^2$ logarithms in QCD breaking the scaling.
The point to be emphasized here is that the transverse momentum of the parton need not be restricted to the smaller region of phase space only.
A quark can emit a gluon and, in turn, acquire an enormous transverse momentum.
So, a necessity arose to extend/improve the existing parton model.
This was done using QCD by employing the concept of factorisation.
The improved model allowed interactions between the partons \textit{via} gluon exchanges and was called the QCD-improved parton model.

%\begin{center}
\textit{Factorisation plays an essential role in the QCD description of the proton.}
%\end{center}
%
%A function $f_{i}(x)$ denote the probability that a proton contains a Parton of type $i$ and longitudinal fraction $x$.
%The cross-section for a process can now be given by an integral of the rescaled Parton cross-section multiplied by the probability $f_i$ to find a Parton with a fraction $x$ of the incoming hadron's energy \cite{Ellis:1988vi}.
%With the above deductions, the historical Parton model, a direct outcome of Bjorken scaling, was immediately applied to the electron-proton DIS.
%So, when the proton's interior is probed at high energies, the number of Partons that appear to remain unchanged behave almost like free particles - a so-called frozen state.
%
%\subsubsection{Factorisation theorem}
%\label{subsec:factorisation}
%
%\begin{center}
Factorisation leads to the assumption that the differential cross-section for DIS can be written as a convolution of parton densities and the hard scattering process.
%\end{center}
%
\begin{equation}
 d\sigma \left( e(k) + P(P) \rightarrow e(k') + X(p_X) \right) =
 \sum_{i} \int_0^1 d\xi f_{i}\left( \xi \right)
 d{\hat{\sigma}}\left( e(k) + i(\xi P) \rightarrow e(k') + i(p') \right).
 \label{eqn:diffCSDIS}
\end{equation}
Here, the summation is over all possible partonic processes, $i$ represents the type of parton, $\xi P = p$ is the fraction of the hadron's $P$ momentum that the $i$ parton carries out, and $f_i(\xi)$ represents the PDF.
The differential cross-section given in eqn.\ \ref{eqn:diffCSDIS} consists of two components -
one from elastic scattering off the point-like partons given by $d{\hat{\sigma}}$, and the other is the PDF, $f_i(\xi)$, which is universal.
$d{\hat{\sigma}}$ can be determined using pQCD and the $f_i(\xi)$'s are experimentally determined.
Hence, the complete high-energy cross-section can be considered a combination of long- and short-distance effects where the long-distance behaviour is systematically separated from the short-distance behaviour by \textit{factorisation}.
This means that the confining physics does not influence the high-energy scattering, \textit{i.e.} the two are factorized.
Consequently, QCD factorisation theorems prove that the cross-section can be factorized into a process-dependent perturbative piece which can be determined by the methods of pQCD and a non-perturbative part given by the universal PDFs.
%
%\begin{figure}[!htb]
%\centering
% \includegraphics[scale=0.3,keepaspectratio=true]{Extra_Figures/QCDimprovedPM.png}
% QCDimprovedPM.png: 523x486 px, 72dpi, 18.45x17.15 cm, bb=0 0 523 486
% \caption[Factorisation in the QCD Improved Parton Model]{A diagrammatic representation of factorisation in the QCD improved Parton model}
%\label{fig:QCDimprovedPartonModel}
%\end{figure}

%%%%%%%%%%%%%%%%%%%%%%%%%%%%%%%%%%%%%%%%%%%%%%%%%
%
\vspace{-5mm}
\section{QCD computations: Methodologies}
\label{sec:pQCD}

\begin{quotation}
 \textit{"We seem gradually to be groping toward an understanding of the world of subatomic particles, but we do not know how far we have yet to go in this task."}

 \hfill \textit{- Richard Feynman}
\end{quotation}

The need of the hour is to back experimental advancements with sufficient theoretical grounds.
The perturbative picture of QCD has satisfyingly explained experimental outcomes from collider events for a long time.
Hence, strong theoretical grounds can be achieved by including higher and higher-order QCD predictions in the perturbative series.
%An evident and dependable way of improving this situation is by refining our understanding of pQCD.
%
%\newpage
%
In these endless efforts towards improving the precision of cross-section computations, this section summarises two approaches contributing to this direction.
These are also the foundations of my original works included in this thesis.

\subsection{Fixed order computations}
\label{sec:FOcomputation}

The first step towards computing higher order QCD corrections primarily involves the FO expansion of the desired observable, $\sigma$, in powers of the strong coupling constant, $\alpha_\text{s}$, and retaining only a few orders whose computation is possible with the existing tools.
\begin{equation}
 \sigma = \sigma_0 + \alpha_\text{s} \sigma_1 + \alpha_\text{s}^2 \sigma_2 + \cdots.
\label{eq:pSeries}
\end{equation}

As this thesis contributes towards cross-section computations, the hadronic cross-section, $d{\sigma}^X$, can be formulated in the framework of FO calculations as:
\begin{equation}
 d\sigma^X=\sum_{i,j}\int dx_{1}dx_{2}f_{i}^A\left(x_{1}\right)f_{j}^B\left(x_{2}\right)d\hat{\sigma}_{ij}^X \left(s\right),
\label{eq:dsigma1}
\end{equation}
where $X$ represents the final state particle, $s=x_{1}x_{2}S$ is the partonic centre-of-mass (COM) energy, $S$ is the hadronic COM energy, and $f_{i(j)}^{A(B)}\left(x\right)$ are the corresponding PDFs, experimentally determined by the Parton $i~(~j~)$ and the hadron $A~(~B~)$.

The computation of the partonic cross-section $d\hat{\sigma}_{ij}^X \left(s\right)$, introduced in eqn.~\ref{eq:dsigma1} and elaborated in eqn.~\ref{eq:dsigma2}, involves the following elements:
\begin{itemize}
 \item A matrix element, $\mathcal{M}_{ab\rightarrow X}$, constructed from the Feynman diagrams, order-by-order in perturbation theory, contributing to the process $a + b \rightarrow X$.
 \item This matrix element is squared and summed over all incoming and outgoing spins and colours and/or symmetry factors $({\cal{SC}}_{ab})$,
 \item This is further multiplied by the $k$-particle phase space $dPS^{\left(k\right)}$,
 \item And finally divided by the flux factor ${\cal{F}}$.
\end{itemize}
\begin{equation}
 d\sigma_{ij}={\mathcal{SC}_{ab}}\dfrac{1}{\mathcal{F}}\mid{\mathcal{\overline{M}}_{ab\rightarrow X}}\mid^{2}dPS^{\left(k\right)}
 \label{eq:dsigma2}
\end{equation}
At LO, contributions from the square of the born matrix element,
$\mid{\mathcal{\overline{M}}^{(0)}_{ab\rightarrow X}}\mid^{2}$, are required.
%, and the one particle phase space.
However, computations beyond the LO are necessary for precise and stable theoretical predictions.
With the highest contribution from LO results, the corrections gradually become smaller with each increasing order.
Consequently, the higher-order corrections can be disregarded after including the first few orders for a good approximation of the complete result.
This method of computing corrections order-by-order and truncating the perturbative series after a few orders of computation is known as FO approximation.

The perturbative expansion is supposedly well-behaved.
Thus, adding more and more terms in the series brings the approximate result closer and closer to the actual value.
Since $\alpha_\text{s} < 1$, the contributions coming from $\mathcal{O}(\alpha_\text{s}^{n+1})$ can be expected to be smaller than those coming from $\mathcal{O}(\alpha_\text{s}^n)$ and hence, can be ignored.
When retained to the $k$-th order, this order-by-order expansion is called the N$^k$LO correction to the \textit{Born}, with the \textit{Born} being considered the first non-zero term in the expansion.
For instance, in the gluon-initiated Higgs production cross-section, it has been observed that the first contributing term is $\sigma_2$.
So, the first non-zero term starts at $\mathcal{O}(\alpha_\text{s}^{2})$, making it the Born contribution \cite{Harlander_20092,Maltoni:2018dar}.

To begin with precision calculations beyond LO, computing NLO corrections is the first step.
This order involves the appearance of an additional real parton, represented by ${\hat{\sigma}}^R_{ab}$, and one-loop virtual contributions, given by ${\hat{\sigma}}^V_{ab}$.
The complications further increase at the NNLO level, where double real (RR), real-virtual (RV) and double virtual (VV) corrections start contributing.
So beyond LO computations involve the evaluation of virtual (loop) and/or real emission diagrams.
If each of these types of diagrams is considered individually,
% in phenomenological studies,
they exhibit divergences/singularities --- certain ill-defined quantities.
Virtual diagrams generate UV divergences in addition to soft and collinear divergences, which are together called IR divergences.
The singularities from virtual diagrams are resolved with the help of renormalisation and mass-factorisation techniques to obtain a finite pure-virtual partonic coefficient function finally.
\textit{A detailed study on evaluating these virtual diagrams and dealing with the divergences is done in my original work elaborated in chapter \ref{chap:ggAA} for the di-pseudoscalar Higgs boson production process in gluon fusion to two-loops in QCD.}

\thrasr{
Multi-loop Feynman diagrams can be generated using QGRAF/FeynCalc \cite{Nogueira:1991ex}.
This output can then be converted using FORM \cite{Vermaseren:2000nd} to include appropriate Feynman rules and to perform trace of Dirac matrices, contraction of Lorentz indices and colour indices.
At this stage, a considerable number of Feynman integrals containing a set of propagator denominators and a combination of scalar products between loop momenta and independent external momenta appear.
As these large number of Feynman integrals are not all independent, some of their properties are used to establish relations between the integrals, reducing the number of loop integrals.
Exploiting the fact that the total derivative \textit{w.r.t.} any loop momenta of these integrals, evaluates to a surface term, which vanishes, leads to IBP identities \cite{Tkachov:1981wb,Chetyrkin:1981qh}.
Integration-by-parts (IBP) reduction is extensively used for loop amplitude computations.
The goal of IBP reduction is to express “complicated” Feynman integrals as a linear combination of “simple” ones \cite{Chetyrkin:1981qh}.
For instance, consider the integral
\begin{equation}
 F (a) = \int \dfrac{d^d k}{(k^2 - m^2)^a}
 \label{eqn:IBP1}
\end{equation}
where we need to evaluate the integral for integer $a>0$.
Exploiting the fact that the total derivative \textit{w.r.t.} any loop momenta of these integrals, evaluates to a surface term, which vanishes, we get
\begin{equation}
 \int d^d k \dfrac{\partial}{{\partial}k} .
 \left( k \dfrac{1}{(k^2 - m^2)^a} \right) = 0.
 \label{eqn:IBP2}
\end{equation}
Taking the derivatives $\dfrac{\partial}{{\partial}k} . k =
\dfrac{\partial}{{\partial}k_{\mu}} k_{\mu} = d$, we can express the integrand in eqn. \ref{eqn:IBP1} as
\begin{align}
  k . \dfrac{\partial}{{\partial}k} \dfrac{1}{(k^2 - m^2)^a} = &
  - a \dfrac{2 k^2}{(k^2 - m^2)^{a+1}}
  \\ = &
  - 2 a \left[ \dfrac{1}{(k^2 - m^2)^a} + \dfrac{m^2}{(k^2 - m^2)^{a+1}} \right].
  \label{eqn:IBP3}
\end{align}
Consequently, we obtain an iterative IBP relation of the form
\begin{equation}
 (d - 2a) F(a) - 2 a m^2 F(a+1) = 0,
\end{equation}
such that the solution of the integration $F(a)$ is obtained to be
\begin{equation}
 F(a) = \dfrac{d - 2a + 2}{2(a-1)m^2}F(a-1).
\end{equation}
So, Feynman integrals with integer $a > 1$ can be expressed recursively in terms of one integral $F(1) \equiv I_1$, which is now a Master Integral (MI).

The idea of using differential equations to evaluate the MIs dates back to 1991 \cite{Kotikov:1990kg} and has been very successful in recent years due to the ideas presented in \cite{Henn:2013pwa,Henn:2014qga}.
To apply this method to a given integral family, setting up the IBP reduction and identifying the MIs in this family is necessary.
The IBP reduction (imposing boundary conditions on the finitely many IBP identities and then reducing the chosen set of integrals to a fixed number of basis integrals) is made with the aid of various computer programs like the Laporta algorithm \cite{Laporta:2000dsw}, which is one of the most successful ones.
As the algorithm is easy to implement and use and allows for many programming improvements, many modern most powerful reduction programs heavily rely on this algorithm, like AIR \cite{Anastasiou:2004vj}, FIRE \cite{Smirnov:2008iw}, Reduze \cite{Studerus:2009ye}, and Reduze 2 \cite{vonManteuffel:2012np}.
The Laporta algorithm, intrinsically a brute-force search,  is time- and memory-consuming.
Another approach to the reduction is the derivation of symbolic reduction rules.
As nothing is solved in this reduction process, the reduction is swift.
Furthermore, symbolic rules are small and can be easily saved for future calculations.
Recently, another IBP reduction package, LiteRed, has been used, which employs an entirely different approach to the reduction.
First, it tries to find symbolic reduction rules using heuristics and then applies them to the specific reduction.
As the found rules are very lightweight, they can be easily stored for reuse.
}

To conclude this section on the computation of FO predictions, the contributions at any order should satisfy $\sigma^n \gg \sigma^{(n+1)}$ to apply perturbation theory.
If the contributions remain comparable between the $n \text{ and } (n+1)$-th orders, truncating the series at the $n$-th order will result in unreliable theoretical predictions.

\subsection{Threshold corrections and Resummation}
\label{sec:thresholdres}

The QCD perturbative computations of physical quantities in hard hadronic processes are taken a toll by the presence of large logarithmic corrections near the phase space boundary \cite{Catani:1989ne,Curci:1979am,Parisi:1979xd}, mainly due to soft and/or collinear gluon emissions.
%This happens when the hadronic ratio, $\tau=M^2/s$, goes to the limit $\tau\rightarrow1$.
%The processes occurring at the LHC exhibit $\tau$ as very small, \textit{i.e.} far from the threshold region.
%The convolutions of the hard Partonic cross-sections with the PDFs are expected to make resummation effects significant even far from the threshold limit \cite{Appell:1988ie}.
These large logarithmic corrections must be resummed to all orders in perturbation theory to reduce theory uncertainties.
The relevance of resummation can be understood by the following argument \cite{Catani:1996rb,Catani:1997vp}:
\begin{quote}
 \asr{
 Higher-order perturbative corrections from virtual gluons are IR divergent, and the divergences are precisely cancelled by radiations from real gluons, making the perturbative cross-sections finite.
 %However, the cancellation does not necessarily take place order-by-order in perturbation theory.
 In particular kinematic configurations, the real and virtual contributions can be highly unbalanced, spoiling this cancellation mechanism.
 As a result, soft-gluon contribution to QCD cross-sections become significant or singular.

 When $z \rightarrow 1$, the tagged final state must carry most of the total available energy of the process.
 Consequently, the radiative tail of real emission gets strongly suppressed, resulting in a loss of balance with the virtual contribution.
 Then the cancellation of the IR divergences leaves behind finite higher-order contributions of the type $C_{nm} \alpha_S^n \ln^m(1-z),~m \le 2n$, that become large near $z \rightarrow 1$, even in the perturbative regime of $a_s \ll 1$.
 As can be seen from the expression, the size of these logarithmic contributions in cross-section calculations depends on the coefficients $C_{nm}$ and the $z$-shape of the PDFs.
 So, if the relevant PDFs peak in the small-$z$ region, like for gluons, then the mean partonic COM energy becomes small.
 Thus, soft-gluon effects become substantial also before reaching the kinematic boundary.

 In perturbative QCD, it is the hard partonic cross-section which convolutes with the parton luminosity that is resummed.
 This partonic cross-section depends on the partonic COM energy and the dimensionless ratio of the latter to the final state invariant mass.
 %Therefore, resummation is relevant when it is the partonic subprocess that is close to threshold.
 %
 However, as the partonic COM energy can take any value from threshold up to the hadronic COM energy, and its mean value is determined by the shape of the PDFs, one expects threshold resummation to be more important if the average partonic COM energy is small, \textit{i.e.}  if the relevant PDFs peak at small $z$.
 Hence, threshold resummation becomes relevant for certain hadronic processes, like the scalar or pseudoscalar Higgs boson production from gluon fusion.
 }
\end{quote}
%
%This makes the resummation method noteworthy as it involves resumming these large logarithms of the perturbative series to all orders.

%The hard Partonic cross-section is IR-safe and can be calculated using perturbation theory if the perturbative expansion coefficients are small. To add to our problems, this is only sometimes true. At the kinematic edges of phase space, these coefficients are usually enhanced.
The enhanced logarithms make FO predictions unreliable near the threshold region.
The reason behind this disturbing behaviour is explainable.
%The large logarithms arise from three sources: UV, Collinear and Soft origin.
%The reason for partonic sub-processes to approach the threshold regime is explainable:
IR divergences cancel between virtual and real corrections, with the virtual corrections involving integrations that span over the UV, soft and hard energy scales.
After removing these UV divergences by renormalisation, we are left with the region between soft and hard scales.
Despite the cancellation of these soft divergences with those of virtual origins, significant contributions from the soft gluon effects can exist in those kinematic regions where the balance between real and virtual corrections gets disturbed.
This imbalance gives rise to large logarithmic terms that eventually spoil the perturbativity of FO predictions as the convergence of the FO series becomes questionable in this region of phase space.
These large logarithms, arising from the soft gluons, need to be resummed to all orders in perturbation theory.
These corrections, generally known as soft corrections, when supplemented with virtual contributions, comprise the soft-virtual (SV) or threshold or leading-power (LP) corrections.
\textit{To define, resummation is the method of reorganising the perturbative series by an all-order summation of the enhanced logarithms arising from the soft gluons.
Such an approach of resumming a class of large logarithms supplemented with FO results can almost cover the entire kinematic region of the phase space.}

To give an idea, let me first write the partonic cross-section which needs to be resummed in the form as below:
\begin{equation}
 \hat{\sigma}_{ab} (z,q^2,\mu_R^2) =
 \left\{ Z_{ab} (\hat{a}_s,\mu_R^2,\mu^2) \right\}^2
 \left| \hat{F}_{ab} (\hat{a}_s,Q^2,\mu^2) \right|^2 \delta(1-z) \circledast~
 \mathcal{C} \exp[2 \Phi_{ab} (\hat{a}_s,q^2,\mu^2,z) ] \,
 \label{eqn:SVCross}
\end{equation}
$\text{where } a,b = q,\overline{q},g$, $Z_{ab}$ represents the overall UV renormalisation constant, $\hat{F}_{ab}$ is the Form Factor and $\Phi_{ab}$ represents the soft-collinear distribution function.
The $\mathcal{C}$ ordered exponential in eqn.\ \ref{eqn:SVCross} has the following structure:
\begin{equation}
 \mathcal{C}\exp[ f(z) ] = \delta(1-z) + \dfrac{1}{1!} f(z) + \dfrac{1}{2!} f(z) \circledast f(z) + \dfrac{1}{3!} f(z) \circledast f(z) \circledast f(z) + \cdots
\end{equation}
where the function $f(z)$ is a distribution of the kind $\delta(1-z)$, $\mathcal{D}_i  = \left[ \dfrac{\ln^i(1-z)}{(1-z)} \right]_+$ and $\log^i(1-z)$ with $i=0,1,\cdots$.
The symbol $\circledast$ represents Mellin convolution.
For the timing, we will focus on only the SV and the sub-leading or next-to-soft virtual contributions (NSV) to the partonic coefficient function, $\Delta_{ab}$, such that
\begin{equation}
 \dfrac{1}{z} \hat{\sigma}_{ab} (q^2,z,\varepsilon) = \sigma_0 (\mu_R^2)
 \sum_{a^{\prime} b^{\prime}} \Gamma_{a a^{\prime}}^T (z,\mu_F^2,\varepsilon) \circledast
 \left( \dfrac{1}{z} \Delta_{a^{\prime} b^{\prime}} (q^2, \mu_R^2, \mu_F^2, z, \varepsilon) \right)
 \circledast \Gamma_{b^{\prime} b} (z,\mu_F^2,\varepsilon),
 \label{eqn:DeltaAB}
\end{equation}
where $\Gamma_{ab}$ are collinear singular Altarelli-Parisi (AP) kernels \cite{Altarelli:1977zs} and $\sigma_0$ is the Born cross-section.
For the works included in this thesis, we require only those terms that are proportional to distributions $\delta(1-z)$, $\mathcal{D}_i(z)$ and NSV terms, $\log^i(1-z)$ with $i = 0, 1, \cdots$.
We drop any other term resulting from the convolutions when $z \rightarrow 1$.
Substituting eqn.\ \ref{eqn:SVCross} in eqn.\ \ref{eqn:DeltaAB}, the form of $\Delta_{ab}$ needs to be extracted to obtain the master formula depicted in eqns.\ \ref{eq:deltaNSV} and \ref{eqn:PsiA}.
We use this master formula for our computation of the resummed corrections for a pseudoscalar Higgs boson production \textit{via} gluon fusion in chapter \ref{chap:ggA}.
In this thesis, I will represent SV resummed corrections by LL, NLL, NNLL, $\cdots$ for leading logarithmic, next-to-leading logarithmic, next-to-next-to-leading logarithmic corrections, respectively.
For SV+NSV corrections, I will represent the leading logarithmic corrections by $\overline{\text{LL}}$, next-to-leading logarithmic corrections by $\overline{\text{NLL}}$ and next-to-next-to-leading logarithmic corrections by $\overline{\text{NNLL}}$.
As is evident from eqn.\ \ref{eqn:SVCross}, the $z$ space results involve convolutions of the distributions $\delta(1-z)$, $\mathcal{D}_i(z)$ and $\log^i(1-z)$ with $i = 0, 1, \cdots$.
Therefore, we shift to the Mellin space approach using the conjugate variable $N$ for resummation.
For instance, in the Mellin space, the large logarithms of the kind $\mathcal{D}_i(z)$ become functions of $\log^{j+1}(N),~j \le i$ with $\mathcal{O}(1/N)$ suppressed terms.
%in the corresponding $N$ space threshold limit of $N \rightarrow \infty$.
Threshold resummation allows one to resum $\omega = 2 a_s (\mu^2_R) \beta_0 \log(N)$ terms to all orders in $\omega$ and then to organise the resulting perturbative result in powers of the coupling constant $a_s(\mu^2_R) = \dfrac{g_s^2(\mu^2_R)}{16 \pi^2}$, where $g_s$ is the strong coupling constant.
Here, $\beta_0$ is the leading coefficient of QCD beta function.
The convenience of working in the Mellin space is that the convolutions transform to simple products.

Next-to-leading-logarithm (NLL) resummations and their expansions were developed long back and gradually used for different applications like in the computation of double-differential cross-sections using partonic threshold \cite{Kidonakis:1996aq,Kidonakis:1997gm,Laenen:1998qw,Kidonakis:1998ei,Kidonakis:1999ze,Kidonakis:2000ui,Kidonakis:2001nj,Kidonakis:2003tx} and for total cross-sections \cite{Bonciani:1998vc,Bonciani:2003nt}.
Corrections beyond NLL were calculated in \cite{Kidonakis:2001nj,Kidonakis:2003tx,Kidonakis:2003qe,Kidonakis:2004yr,Kidonakis:2006,Moch:2008qy,Kidonakis:2008mu}.
Next-to-next-to-leading-logarithm (NNLL) resummations and expansions were developed in \cite{Kidonakis:2009ev,Mitov:2009sv,Becher:2009kw,Beneke:2009rj,Czakon:2009zw,Ferroglia:2009ep,Ferroglia:2009,Ahrens:2010,Beneke:2010da,Kidonakis:2010dk,Ahrens:2011,Kidonakis:2011zn}.
However certain inclusive \cite{Anastasiou_2015,Duhr:2019kwi,Mistlberger:2018etf,Duhr:2020seh,Anastasiou:2015ema} and differential \cite{PhysRevD.99.034004} observables showed considerable contributions from the sub-leading threshold logarithms at every order in the perturbation theory.
These sub-leading terms are called \textit{next-to-threshold} or \textit{next-to-leading-power (NLP)} or \textit{next-to-soft virtual (NSV) corrections}.

For a better insight, let us look into a few available results comparing these leading and next-to-leading-power logarithmic corrections.
For Higgs production, the scale uncertainty reduces to about $15\%$ at NNLO and further to $10\%$ on including the NNLL resummed predictions \cite{Catani:2003zt,Bolzoni:2006xjf}.
Moving beyond SV resummation, the NSV corrections contribute $25.83 \%$ of the born in the production of the scalar Higgs \textit{via} gluon fusion at $a_{s}^{3}$ order while the SV terms at the same order contribute a mere $-2.28\%$ \cite{Anastasiou:2015ema}.
In DY, the NSV logarithms contribute $1.49\%$ of the born at N$^{3}$LO accuracy, while the SV terms contribute only $0.02\%$ \cite{ajjath2020soft}.
\textit{
Computation of the NSV corrections has been done in my original work \cite{Bhattacharya:2021hae} described in chapter \ref{chap:ggA}, for di-pseudoscalar production via the gluon fusion channel, along with a study of the implications of these sub-leading corrections.
}
%
%This technique of resummation and, consequently, enhancing the precision of computations has been studied. We have applied it to the process of pseudoscalar Higgs boson production \textit{via} gluon fusion to NNLO$+\overline{\text{NNLL}}$ accuracy in chapter \ref{chap:ggA} which is based on the original work \cite{Bhattacharya:2021hae}.

%%%%%%%%%%%%%%%%%%%%%%%%%%%%%%%%%%%%%%%%%%%%%%%%%%
%\clearpage
%\newpage
%\mbox{~}

\chapter{Effective Field Theory and pseudoscalar Higgs boson production}
\label{sec:Framework}

%\begin{center}
% \textit{EFT is a type of approximation for an intrinsic physical theory.}
%\end{center}
%
%
Time has repeatedly made evident that contrasting energy scales dominate our world.
If all physical phenomena had taken place at the same scale, it would have been impossible to understand/discriminate them.
This is because developing one single theory pertaining to all possible/estimated energy scales for all possible physical processes is highly challenging.
Instead, developing an appropriate physical theory for a chosen scale is tractable.
Hence, in this world of multi-scale problems, we can set all scales much larger (smaller) than the energy scale of the physical process under consideration to be infinitely large (small).
Physics, at any scale other than the chosen one, is ignored because their effect(s) can be re-introduced order-by-order if and when needed.
%This basic strategy is used to develop a precise, quantitative and convenient framework.
%
\begin{quotation}
\centering
 \textit{"Wilson's great legacy is that we now regard nearly every quantum field theory as an effective field theory."}

 \hfill \textit{- John Preskill}
\end{quotation}

\section{Effective Field Theory}
\label{sec:EFT}

The EFT framework focuses only on the relevant DOF depending on the scale, and hence, they drastically simplify calculations.
%Consequently, new symmetries may manifest themselves, which otherwise would have remained obscure.
%
With the advent of scale separation as a natural part of physics, EFTs have found a significant base.
As the complexities of QCD calculations increase with increasing precision, EFTs play an essential role by enabling
calculations within the specific limits of the theory.
Depending on the required level of precision for an experimental measurement, whether the mass of a heavier quark (compared to the scale under consideration) can be a subject of concern is decided.
%So, only those degrees of freedom (DOF) are considered, which are supposed to be relevant to the problem in hand for the calculations.
So, we can work in a model where the computations are performed with only those DOF  which can predict the observables under consideration to a good approximation.
In the language of QFT, while including those operators corresponding to the light DOF, the heavier massive particles associated with scales well above the problem at hand are effectively eliminated.
The effects of these assumptions are visible in multi-loop computations, as will be discussed throughout this thesis.

In QCD, the top quark, being the heaviest $(m_\text{t}=172 \text{ GeV})$ in the SM, plays a significant role in the construction of an EFT.
It is also because of this large mass that it can play an essential role in BSM physics by strongly coupling to new physics responsible for mass generation.
EFTs express the amplitude of a decay process in terms of effective operators.
The amplitude can then be defined to be $\propto \text{C}_\text{i} \text{O}_\text{i}$, where $\text{O}_\text{i}$ represent the operators and $\text{C}_\text{i}$ represent the Wilson coefficients.
Transition amplitudes are calculated in the full and effective theories to the desired order in $\alpha_\text{s}$ and matched at a chosen energy scale to determine the values of these Wilson coefficients.
%Summarizing, EFT techniques and their applications prove functional when it is possible to separate the existing scales of the problem definitively.

The SM is known to be a nice, renormalizable ﬁeld theory until now, with reasons to believe it is not a complete theory.
For instance, the candidates for dark matter particles and a convincing description of baryogenesis are not included in the known SM.
A possible solution proposed and being constantly tested is that the SM is a LO approximation to some more fundamental theory in the EFT expansion.
Such a solution would be truly unique because no other effective theory considered so far had a LO term being represented by a perturbatively renormalizable QFT.
%In other words, the leading order term has no encoded information about the scale at which it stops being a meaningful representation of a complete theory.
%The SM remains trustworthy if we move up the energy-scale ladder to the Planck scale, where quantum gravity effects become important.
Aesthetic requirements like gauge coupling uniﬁcation or the absence of ﬁne-tuning of the Higgs mass had initially motivated most theoretical expeditions into the world of BSM physics.
The existence of neutrino oscillations indicates the need for massive neutrinos that are missing in the SM and, consequently, the existence of BSM physics.
There is a possibility that new physics DOF are lying just above the energy scales being dealt with so far, in which case, they can be appropriately and explicitly included in an effective theory.
Many possible extensions of the SM, like the MSSM, described in section \ref{subsec:MSSM}, can be used to build different effective theories.

\textit{This thesis deals with QCD corrections beyond LO using a QCD Effective Lagrangian for the pseudoscalar Higgs boson by integrating out the top quark DOF.
We work in the EFT framework assuming the simple addition of a CP-odd Higgs boson to the existing SM.}
In a linear combination, local composite operators and the Wilson coefficients lead to the effective Lagrangian in eqn.\ \ref{eq:effectiveLag} with five massless quark flavours.
The top quark mass, $m_\text{t}$, dependence is contained in the Wilson coefficients.
These are discussed in the following two sections in detail. %\ref{sec:pseudoscalar}.

\subsection{Top quark mass effects}
\label{sec:topmasseffect}

%EFT techniques have played a pivotal role in interpreting several available LHC results, with the top quark occupying a unique position in most BSM physics scenarios.
%This has also inspired the scrutinization of top quark mass effects in EFT phenomenology.
%The strong coupling of the top sector to Higgs physics, owing to the large top quark Yukawa coupling, adds to this notability.
%The beginning of the LHC Run II era is expected to lead to a unique opportunity for new physics searches.
%This thesis also contributes to this direction.

According to the SM of particle physics, the world of strong interactions involves gluons and six flavours of quarks, with one of the quarks, namely the top quark, being much heavier than the others.
%Consider energy $E$ such that $E<<m_t$, \textit{i.e.} the energy scale is much below the known top quark mass.
In this case, the physics can be described "effectively" by a field theory without the top quark field but containing all higher-dimensional operators built from the fermionic field, $\psi_f$, and the gluon field constrained by the SU$(3)$ gauge symmetry.
This is visible in the pseudoscalar Higgs boson effective Lagrangian given in eqn.\ \ref{eq:effectiveLag}.
Since the full theory is known and weakly coupled at the lower $m_t$ scale, the corresponding effective Lagrangian can be determined order-by-order by a procedure called "matching" \cite{Beneke:EFT}.

An effective QFT integrates out the heavy particles in the theory, leaving non-local interactions from heavy-particle exchanges occurring virtually.
%In an EFT, these non-local interactions are then replaced with a set of local interactions such that the low-energy physics remains unaffected.
%These interactions are suppressed by the inverse power of the masses of the heavy particles.
%So, effectively, only the high energy behaviour of the theory is modified, with its physical validity being limited at energies below the masses of the integrated out heavy particles.
EFT is based on the principle of scale separation, manifested in several other fields of physics also.
Powers of a ratio between the large energy scale and the scale in the problem suppress its effects.
Eliminating heavy particles in the theory with the help of an EFT leads to enormous simplifications.
However, the involvement of all scales in the integration over loop momenta poses a potential problem but is appropriately dealt with using a convenient regularisation scheme, like we use dimensional regularization.
This decoupling of large energy scales is evident in any renormalisable QFT, irrespective of the usage of an EFT \cite{Zhang:2012muc}.

As mentioned, EFTs reduce the complexities of multi-loop computations significantly. Hence, comparisons are made about the errors in evaluating phenomenological results with and without the exact top quark mass effects.
For example, in the case of scalar Higgs boson production, the EFT approach became extremely successful as the difference between the exact and EFT predictions at NNLO accuracy were within 1\% \cite{Ravindran:2003um,Harlander_20091,Harlander_20092,Pak_2010,2021_Harlander}.
Suppose this effective theory approach is rescaled with the exact LO results.
In that case, it provides a reasonably good approximation even at masses outside the region of formal validity of the EFT.
In a similar attempt for pseudoscalar Higgs boson production, the difference between the exact and the effective theory results at NLO reaches $\approx10\%$ for $m_A=500$ GeV with not much change for further increase in the pseudoscalar mass, $m_A$ \cite{Anastasiou_2003,Ravindran:2003um,2016}.

\section{Studying the Pseudoscalar Higgs boson}
\label{sec:pseudoscalar}

Gluon fusion sub-processes, $gg$, and $b\overline{b}$ associated production has the dominant contribution to a neutral Higgs boson, $\Phi$, production at the LHC \cite{Georgi:1977gs,Barger:1991ed,Gunion:1991er,Gunion:1991cw,Kunszt:1991qe} in the large $\tan\beta$ limit, where $\tan\beta$ is the ratio of the \textit{vev} of the Higgs doublets.
This neutral Higgs boson, $\Phi$, can be a CP-even light Higgs, h, or a CP-even heavy Higgs, H, or a CP-odd Higgs boson, A.
Depending on the $\tan\beta$ value and mass of the Higgs boson, couplings are theoretically predicted to be mediated by bottom-quark loops, squark loops or virtual top quarks \cite{Wilczek:1977zn,Dawson:1996xz}.
%The loop-induced ggq~ couplings [4] are mainly mediated by virtual top quarks unless tan/3 is very large, in which case the bottom-quark loops take over. The ggh and g g H couplings also receive contributions from squark loops, which are insignificant for squark masses in excess of about 500 GeV [5].
%
%The pseudoscalar Higgs boson, A, does not couple to squarks at the LO; hence, such contributions remain unobserved for a $ggA$ coupling.
For the production of any neutral Higgs boson \textit{via} gluon fusion, the inclusive cross-section shows a typical increase of 50-70\% under LHC conditions, after including the LO QCD corrections for small $\tan\beta$ values, up to two-loop accuracy \cite{Dawson:1990zj,Djouadi:1991tka,Graudenz:1992pv,Spira:1993bb,Dawson:1993qf,Spira:1995rr}.
However, the top-quark loops regain dominance when $\tan\beta \rightarrow 1$.
In this case, if $M_{\Phi}<<2 m_t$, then constructing an effective Lagrangian by integrating the top quark DOF proves helpful \cite{Chetyrkin:1998mw}.

%Since dedicated efforts are going on to determine the CP property of the one discovered Higgs boson to identify it with that of the SM, even though there are already indications that it is a scalar with even parity \cite{2013,2015},
With attempts to move beyond the shortcomings of the SM as elaborated in section \ref{sec:SMshortcomings}, the requirement of precise predictions for relevant observables of both scalar and pseudoscalar ones has become necessary.
In pQCD, the production cross-section has been computed to unprecedented accuracy for the SM Higgs boson with the help of EFT.
In this case, the scalar Higgs boson couples directly to the gluons at LO.
The EFT approach in the case of scalar Higgs boson production  \cite{Anastasiou_2002,Harlander_2002h,Ravindran_2003} became extremely successful as the difference between the exact and EFT results at NNLO level was found to be within 1\% \cite{Harlander_20091,Harlander_20092,Pak_2010}.

%In the context of light Higgs boson in the MSSM, unlike the SM one, the mass is calculable.
%In \cite{Martin:2007pg,Harlander:2008ju,Kant:2010tf}, higher-order radiative corrections to the mass are obtained with very good accuracy.
%

\subsection{The Pseudoscalar Effective Lagrangian}
\label{subsec:PseudoLag}

Efforts are being made to achieve precision in the predictions for production cross-sections of the pseudoscalar Higgs boson.
For pseudoscalar Higgs boson production at the hadron colliders, NNLO predictions are already available \cite{Ravindran_2003,Harlander_2002a,Anastasiou_2003}.
In \cite{Plehn:1996wb}, the first results on the production rate for the pseudoscalar Higgs boson at the NLO level in QCD appeared with non-zero top quark mass.
In \cite{Dawson:1998py}, the EFT framework was set up by integrating out top quark fields.
Consequently, this opened up the possibility of obtaining observables beyond NLO accuracy, as there is a significant reduction in the number of loops and scales compared to those in the full theory.
\textit{
This thesis deals with the pseudoscalar Higgs boson production through the gluon fusion channel.
Hence, I present below the standard theoretical framework behind all my projects.
}

The effective Lagrangian that describes the interaction of the pseudoscalar field $\Phi^A (x)$ with the gauge field $G^{a \mu\nu}$ and the fermion $\psi$ is given by:
\begin{eqnarray}
\label{eq:EL}
        {\cal L}^A_{eff}  = \Phi^A (x)
        \left[ -\frac{1}{8} C_G O_G (x)
        - \frac{1}{2} C_J O_J (x) \right]
%        + \Phi^A (x) \Phi^A (x) \left[ - \frac{1}{4} C_{G} G^{a \mu \nu} G^a_{\mu \nu} \right]
        \,.
\label{eq:effectiveLag}
\end{eqnarray}
%
%$C_{GG} = C_G/4$
%
This Lagrangian was derived long back in \cite{Chetyrkin:1998mw,Dawson:1998py} where the starting point of the effective theory was the bare Yukawa Lagrangian. %for the interactions of the CP odd Higgs boson with the quarks in the full n$_f$ flavour theory.
\ab{In the limit of large top mass, $m_t\rightarrow\infty$, and keeping only the LO terms, the full theory was re-written as a linear combination of pseudoscalar composite operators, with mass dimension four acting in the effective theory.}

The pseudoscalar gluonic operator, $O_G (x)$, is defined as
\begin{eqnarray}
O_G(x) &=& G^{a\mu\nu}\tilde{G}_{\mu\nu}^a=\epsilon_{\mu\nu\rho\sigma}
G^{a\mu\nu}G^{a\rho\sigma},
\end{eqnarray}
where the gauge field, $G^{a\mu\nu}$, is given by
\begin{eqnarray}
G^{a\mu\nu} &=& \partial^{\mu}G^{a\nu}-\partial^{\nu}G^{a\mu}+g_s f^{abc}
G_{b}^{\mu}G_{c}^{\nu}.
\end{eqnarray}
$f^{abc}$ is the SU(3) structure constant and
$\epsilon_{\mu\nu\rho\sigma}$ is the Levi-Civita tensor.
The pseudoscalar fermionic operator, $O_J (x)$, is defined as the derivative of the flavour singlet axial vector current
\begin{equation}
O_{J}(x)=\partial_{\mu}\left(\bar{\psi}\gamma^{\mu}\gamma^{5}\psi\right) \,.
\label{eq:Fields EL}
\end{equation}
The effective Lagrangian is obtained after integrating out the top quark fields in the limit $m_t\rightarrow\infty$.
Hence, the corresponding Wilson coefficients $C_{G}$ and $C_{J}$ depend on the mass of the top quark, $m_t$ and on the bare parameters of the full theory.
As a result of the Adler-Bardeen theorem \cite{Adler:1969gk}, there is no QCD correction to $C_{G}$ beyond the one-loop level.
On the other hand, $C_{J}$ begins only at second-order in the strong coupling
constant $a_{s}\equiv g_{s}^{2}/16\pi^{2}=\alpha_{s}/4\pi$.
The Wilson coefficients are given by \cite{Ravindran:2003um}
\begin{align}
C_{G}\left(a_{s}\right) & =-a_{s}2^{\tfrac{5}{4}}G_{F}^{\tfrac{1}{2}}\cot\beta \,,
\label{WCG}
\\
C_{J}\left(a_{s}\right) & =-\left[a_{s}C_{F}\left(\dfrac{3}{2}-3\ln\dfrac{\mu_{R}^{2}}{m_{t}^{2}}\right)
+a_{s}^{2}C_{J}^{\left(2\right)}+...\right]C_{G}\,,
\label{WCJ}
\end{align}
where $G_F$ is the Fermi constant,
$\cot \beta$ is the ratio of the two Higgs doublets' vacuum expectation values in a generic two-Higgs doublet model,
$C_F$ is the quadratic Casimir invariant in the fundamental representation
of QCD and $\mu_R$ is the renormalisation scale at which $a_{s}$ is
renormalized.

\newpage
\subsection{Feynman Rules}
\label{subsec:feynmanrules}

The Feynman rules for the various pseudoscalar vertices were derived from the effective Lagrangian given in eqn.\ \ref{eq:effectiveLag} and are presented below:

\tikzset{arrow/.style={->, blue, text = black, style=thick}}
\begin{itemize}
 \item The pseudoscalar propagator is represented as
\begin{figure}[h]
\begin{centering}
\begin{tikzpicture}[line width=0.5 pt, scale=0.55]
\draw[scalarnoarrow] (-2.4,0)--(2.5,0);
%\node at (-2.7,0) {{\color{red}{$a$}}};
%\node at (2.7,0) {{\color{red}{$b$}}};
\node at (0.0,-0.5) {{\color{blue}{$\overleftarrow q$}}};
\node at (0.0,+0.5) {{\color{red}{$A$}}};
\node at (10.0,0.0) {$%- \delta_{ab}
\dfrac{i}{q^2 - m_A^2}$};
%im * (-d_(li'i', li'j')) * Pr(4, q, 0)
\end{tikzpicture}
\end{centering}
\end{figure}
%
%
% \newpage
 \item The relevant pseudoscalar interacting vertices are given below:
\begin{figure}[H]%[htbp!]
\begin{centering}
\begin{tikzpicture}[line width=0.5 pt, scale=0.7]
\draw[gluon] (-2.5,2.0) -- (0,0);
\draw[arrow] (-2.0,2.0) -- (-1.0,1.2); 
\node at (-1.5,2.0) {\color{red}{{$p_1$}}};
\draw[gluon] (-2.5,-2.0) -- (0,0);
\draw[arrow] (-2.0,-2.0) -- (-1.0,-1.2); 
\node at (-1.5,-2.0) {\color{red}{{$p_2$}}};
\draw[fill=blue] (0.11,0) circle (.15cm);
\draw[scalarnoarrow] (0.2,0)--(3.0,0.1);
\node at (1.0,0.4) {{\color{blue}{$\overleftarrow p_3$}}};
\node at (-2.8,2.2)  {{\color{red}{{$g$}}}};
\node at (-2.9,1.5)  {{\color{teal}{\footnotesize{$a,\mu$}}}};
\node at (-2.8,-2.2) {{\color{red}{{$g$}}}};
\node at (-2.9,-1.5) {{\color{teal}{{\footnotesize{$b,\nu$}}}}};
\node at (+3.15,0.0) {{\color{red}{{$A$}}}};
%\node at (+2.9,0.25) {{\color{teal}{\footnotesize{$c$}}}};
%
\node at (10.0,0.0) {$-i ~\delta_{ab} ~\delta^{(4)}(p_1+p_2+p_3) ~\varepsilon_{\mu\nu\rho\sigma} ~p_1^{\rho} ~p_2^{\sigma}$};
%-im * CG * adelta(ae2,ae3) * levieps(le2,le3,p2,p3)
\end{tikzpicture}
 \end{centering}
\end{figure}
%
%\vspace{5cm}
%
\begin{figure}[H]
\begin{centering}
\begin{tikzpicture}[line width=0.5 pt, scale=0.7]
\draw[fermionnoarrow] (-2.5,2.0) -- (0,0);
\draw[arrow] (-2.0,2.0) -- (-1.0,1.2); 
\node at (-1.5,2.0) {\color{red}{{$p_1$}}};
\draw[fermionnoarrow] (-2.5,-2.0) -- (0,0);
\draw[arrow] (-2.0,-2.0) -- (-1.0,-1.2); 
\node at (-1.5,-2.0) {\color{red}{{$p_2$}}};
%
%\draw[fill=blue] (1,-0.9) rectangle (1.25,-1.1);
\draw[fill=blue] (0.0,0.215) rectangle (0.31,-0.22);
\draw[scalarnoarrow] (0.3,0)--(3.2,0.0);
\node at (1.0,0.4) {{\color{blue}{$\overleftarrow p_3$}}};
\node at (-2.8,2.2)  {\color{red}{$q$}};
\node at (-2.9,1.5)  {\color{teal}{\footnotesize{$i,\alpha$}}};
\node at (-2.9,-2.2) {\color{red}{$\overline{q}$}};
\node at (-2.8,-1.5) {\color{teal}{\footnotesize{$j,\beta$}}};
\node at (+3.4,0.0)  {\color{red}{$A$}};
\node at (10.0,0.0) {$i ~\delta_{ij} ~\delta^{(4)}(p_1+p_2+p_3) ~\varepsilon_{\alpha\beta\rho\sigma} ~\gamma^{\alpha} ~\gamma^{\beta} ~\gamma^{\rho} ~(p_{1}+p_{2})^\sigma$ };
%im * CJ * fdelta(ae2,ae3)*( levieps(p2+p3,gle1,gle2,gle3)*gam(le3,ne1,gle1)*gam(ne1,ne2,gle2)*gam(ne2,le2,gle3) )
 \end{tikzpicture}
\end{centering}
\end{figure}

\item All other interacting vertices, not involving the pseudoscalar, are the usual SM vertices and follow the same Feynman rules.
Hence, they are not repeated here.

\end{itemize}

\clearpage
\newpage
\mbox{~}
\chapter{Two loop QCD amplitudes for di-pseudoscalar production in gluon fusion}
\label{chap:ggAA}

Section \ref{sec:pQCD} emphasized that the FO computation of the required observable should be the primary focus in the precision studies of a hadronic process.
Section \ref{sec:SMshortcomings} elaborated on the importance of studying the pseudoscalar Higgs production to the same accuracy as the scalar Higgs boson.
Hence, I began my research on precision studies with the FO computation of di-pseudoscalar production amplitudes \textit{via} gluon fusion up to two loops.
\textbf{\textit{This chapter is based on these FO computation results of the original research work done in collaboration with Maguni Mahakhud, Prakash Mathews and V. Ravindran and elaborates the published article \cite{Bhattacharya_2020}.}}

\section{Prologue}

%%%%%%%%%%%%%&&&%%%%%%%%%

%The CP-odd Higgs bears a certain phenomenological resemblance to the known scalar Higgs.
%So, the following few lines are dedicated to the Higgs research status.
%Eventually, I will discuss the developments in the case of pseudoscalar production.

In \cite{Martin:2007pg,Harlander:2008ju,Kant:2010tf}, higher-order radiative corrections to the light Higgs boson mass in MSSM are obtained to very good accuracy.
Recently, there has been a surge of interest in studying the production of a pair of Higgs bosons to determine Higgs self-coupling, whose strength is a prediction of the SM.
Measurement of this coupling will provide an independent test on the nature of the Higgs boson.
The gluon fusion subprocess producing a pair of Higgs bosons through a heavy quark loop \cite{Glover:1987nx,Plehn:1996wb} is the dominant one at the LHC. However, the cross-section being only a few tens of $fb$ is very difficult to observe.
Higher order QCD corrections not only contribute to the cross-section but also stabilise the predictions against renormalisation $\mu_R$, and factorisation $\mu_F$  scales.
NLO QCD corrections \cite{Dawson:1998py} and later the top quark mass effects are systematically taken into account in \cite{Grigo:2013rya,Frederix:2014hta,Maltoni:2014eza,Degrassi:2016vss,Borowka:2016ehy, Borowka:2016ypz}.
Beyond NLO, an EFT where the top quark degrees of freedom are integrated out is generally used.
At present, the scalar Higgs boson pair production is known to  N$^3$LO level \cite{Chen:2019lzz,deFlorian:2013uza,Grigo:2015dia,deFlorian:2013jea,
Grigo:2013rya,Frederix:2014hta,Maltoni:2014eza,Degrassi:2016vss, Borowka:2016ehy, Borowka:2016ypz} in EFT.
Two-loop virtual amplitudes for the $g+g\to H+H$ process required for the N$^3$LO cross-section computation of the di-Higgs production were obtained in \cite{Banerjee:2018lfq}.
The production of di-Higgs bosons through bottom quark annihilation was obtained up to NNLO level in \cite{H:2018hqz}.
In \cite{Li:2013flc,Maierhofer:2013sha,deFlorian:2015moa}, the fully differential results at the NNLO level are presented.

While there has been a flurry of activities in the context of the scalar Higgs boson, very little is known about the production of pseudoscalar Higgs bosons at the LHC.
Section \ref{sec:pseudoscalar} elaborated on the development of pseudoscalar results starting from NLO computations \cite{Plehn:1996wb} with non-zero top mass to incorporating the EFT framework \cite{Dawson:1998py} convincingly.
%
%In \cite{Plehn:1996wb}, the first results on the production rate at NLO level in QCD for the pseudoscalar at the hadron collider appeared. This was done by keeping non-zero top quark mass. Eventually, \cite{Dawson:1998py} gave the LO contribution keeping finite top mass and the NLO contributions using the EFT framework where the top quark degrees of freedom are integrated out, leading to a reduction in the number of loops compared to those in the full theory.
%
Unlike the case of the CP-even Higgs boson, the inclusive cross-section for the production of pseudoscalar Higgs is known \cite{Harlander:2002vv,Anastasiou:2002wq,Ravindran:2003um} only up to NNLO order in pQCD.
For N$^3$LO predictions, one requires three loop virtual amplitudes and real emission contributions.
The computation of virtual corrections is technically challenging \cite{Ahmed:2015qpa}
as a pseudoscalar Higgs boson couples to the SM fields through
two composite operators that mix under renormalisation. In addition, these operators involve Levi-Civita
tensor and $\gamma_5$, which are hard to define in dimensional regularisation.

Like the production of a single pseudoscalar Higgs boson, pair production is also important to understand the nature of the extended Higgs sector.
In order to reduce the theoretical uncertainties, it is crucial to have QCD radiative corrections under control.
Due to EFT, it is now possible to go beyond NLO with the available tools to make precise as well as stable predictions with respect to the unphysical scales.
At the NNLO level, we require two-loop virtual, one-loop single real emission and double real emission amplitudes.
In this article, as a first step towards going beyond NLO QCD corrections, we compute all the one and two-loop amplitudes that can contribute to the pure virtual part of the cross-section for di-pseudoscalar production in gluon fusion using dimensional regularisation and performing UV renormalisation to obtain UV finite results.

The chapter is organised as follows:  in section \ref{sec:framework}, we describe the mathematical ingredients needed to compute the two-loop virtual amplitudes like the relevant kinematics, the projectors needed to obtain the scalar parts of the amplitudes, the necessary Feynman diagrams, the subtleties involved in defining the Levi-Civita tensor and $\gamma_5$ in dimensional regularisation, and a method to choose the relevant Lorentz tensors $\mathcal{T}_i^{\mu \nu}$.
In section \ref{sec:HAcomparision}, we present a comparative study between the amplitudes of the $g+g \rightarrow H+H$ and $g+g \rightarrow A+A$ subprocesses.
In section \ref{sec:calcdetails}, we give the mathematical details of how we perform UV and operator renormalisations, and eventually, the IR factorisation using the method predicted by Catani in \cite{Catani:1998bh}.
In section \ref{sec:computation}, we include all possible computational details - how to obtain the finite amplitude by performing loop integrals using packages like FORM, LiteRed, Reduze and Mathematica.
We present our final results in section \ref{sec:result}.
We finally summarise our results and conclude this chapter in section \ref{sec:ggAAConc}.

\vspace{-5mm}
\section{Theoretical framework}
\label{sec:framework}

We consider the pair production of pseudoscalar Higgs boson $(A)$ of mass $m_A$ up to the two-loop level in pQCD.
The effective Lagrangian required to obtain the amplitudes for this process is given in eqn.\ \ref{eq:EL}.
We restrict ourselves to the dominant gluon fusion subprocess:
\begin{eqnarray}
\label{eq:amp}
	g(p_1)+g(p_2) \rightarrow A(p_3)+A(p_4)\,,
	\label{eq:ggAAprocess}
\end{eqnarray}
where $p_1$ and $p_2$ are the momenta of the incoming gluons, $p_{1,2}^2=0$ and $p_3$ and $p_4$ are the momenta of the outgoing pseudoscalar Higgs bosons, $p_{3,4}^2=m_A^2$. The momenta can be related as $p_{4} = p_{1} + p_{2} - p_{3}$.

\subsection{Kinematics}
\label{subsec:kinematics}

To encode physical variables like energy and momentum of the scattering process mentioned above in a Lorentz-invariant fashion, we define the Mandelstam variables as
%The Mandelstam variables for the above process are given by
\begin{eqnarray}
s = (p_1+p_2)^2, \quad t = (p_1-p_3)^2, \quad u=(p_2-p_3)^2\,,
\end{eqnarray}
which satisfy $s+t+u=2 m_A^2$.
It is convenient to express these amplitudes in terms of the
dimensionless variables $x$, $y$ and $z$ as
\begin{eqnarray}
	s = m_A^2 {(1+x)^2 \over x},\quad t = -m_A^2 y,\quad u= -m_A^2 z \,,
\end{eqnarray}
which lead to the constraint $x^{-1} + x = y +z$.

\subsection{Projectors}
\label{subsec:projector}

As in the case of di-Higgs production amplitude {\it via} gluon fusion \cite{Glover:1987nx}, the di-pseudoscalar production amplitude can also be decomposed in terms of two second rank Lorentz tensors, ${\cal T}_i^{\mu \nu} \text{ where } i=1,2$, as follows:
\begin{eqnarray}
	{\cal M}^{\mu \nu}_{ab} \varepsilon_\mu (p_1) \varepsilon_\nu (p_2)= \delta_{ab} \left({\cal T}_1^{\mu\nu}~{\cal M}_1
			    + {\cal T}_2^{\mu\nu}~{\cal M}_2\right)
\varepsilon_\mu (p_1) \varepsilon_\nu (p_2)\,
\label{eq:TensorBreak}
\end{eqnarray}
where $\varepsilon_\mu (p_i)$ are the polarisation vectors of the initial state
gluons.
The Lorentz scalar functions, ${\cal M}_i$ where $i=1,2$, are independently gauge invariant.
$\delta_{ab}$ indicates that there is no colour flow from initial
to the final state. The second rank tensors are given by
\begin{eqnarray}
\!\!\!\! {\cal T}_1^{\mu\nu} & \!=\! & g^{\mu \nu} - { p_1^\nu p_2^\mu
                                \over p_1\cdot p_2}\,,
	\\
	\! {\cal T}_2^{\mu\nu} & \!=\! & g^{\mu \nu} + {1 \over p_1\cdot p_2~ p_T^2} \Big(
	m_A^2~ p_2^\mu p_1^\nu - 2 p_1 \cdot p_3~ p_2^\mu p_3^\nu -2 p_2\cdot p_3~ p_3^\mu p_1^\nu
	+2 p_1\cdot p_2~ p_3^\mu p_3^\nu \Big)\,,
\nonumber
\\
\end{eqnarray}
with $p_T^2 = (t u - m_A^4)/s$ being \thr{
the transverse momentum square of a single pseudoscalar Higgs boson expressed}
in terms of the Mandelstam variables.
The tensor ${\cal T}_1^{\mu\nu}$ depends only on the initial state momenta $p_{1,2}$.
Using momentum conservation, it can be seen that ${\cal T}_2^{\mu\nu}$ is symmetric under the interchange of the two pseudoscalar Higgs momenta.
The scalar functions ${\cal M}_{1,2}$ can be obtained from ${\cal M}^{\mu \nu}_{ab}$, by using appropriate $d$-dimensional projectors, $P_{i,ab}^{\mu \nu}$, with $i=1,2$, respectively and the projectors are given by:
\begin{eqnarray}
\nonumber
	P_{1,ab}^{\mu \nu} &=& \frac{\delta_{ab}}{N^2-1} \left(\frac{1}{4}\frac{d-2}{d-3} {\cal T}_1^{\mu\nu} -\frac{1}{4}\frac{d-4}{d-3}{\cal T}_2^{\mu\nu}\right) \,,
%	\nonumber \\
    \\
    P_{2,ab}^{\mu \nu} &=& \frac{\delta_{ab}}{N^2-1} \left(-\frac{1}{4}\frac{d-4}{d-3}{\cal T}_1^{\mu\nu}+\frac{1}{4}\frac{d-2}{d-3}{\cal T}_2^{\mu\nu}\right) \,,
\label{eq:proj}
\end{eqnarray}
where $N$ corresponds to the $\text{SU}(N)$ colour group.

\subsubsection{Obtaining the second rank Lorentz tensors}
\label{subsubsec:LOTensors}

The initial LO amplitude comprises the following nine tensors:
\begin{equation}
g^{\mu\nu},\thinspace p_{1}^{\nu}p_{2}^{\mu},\thinspace p_{1}^{\mu}p_{1}^{\nu},\thinspace p_{1}^{\nu}p_{3}^{\mu},\thinspace p_{1}^{\mu}p_{3}^{\nu},\thinspace p_{2}^{\mu}p_{3}^{\nu},\thinspace p_{3}^{\mu}p_{3}^{\nu},\thinspace p_{2}^{\mu}p_{2}^{\nu},\thinspace p_{2}^{\nu}p_{3}^{\mu}.
\label{eq:All Tensors}
\end{equation}
\asr{
Among these, the tensors $p_{1}^{\mu}p_{1}^{\nu},\thinspace p_{2}^{\mu}p_{2}^{\nu},\thinspace p_{2}^{\nu}p_{3}^{\mu} \text{ and } \thinspace p_{1}^{\mu}p_{3}^{\nu}$ have zero contribution to the amplitude because they independently yield zero on being contracted with the gluon polarisation vectors.
We first computed the LO matrix element and extracted all possible tensor structures.
Then the matrix element was contracted with the gluon polarisation vectors as shown in the left hand side of eqn.\ \ref{eq:TensorBreak}.
This results in the four tensors which have zero contribution to vanish and we are finally left with the non-zero tensors.
These remaining tensors are used to constitute the two independent second rank Lorentz tensors, ${\cal T}_1$ and ${\cal T}_2$, as depicted in the above subsection \ref{subsec:projector}.
}

\subsection{Feynman Diagrams}
\label{subsec:feynmanDiags}

Here we briefly discuss the type of Feynman diagrams contributing up to order ${\cal O} (a_s^4)$ in QCD.

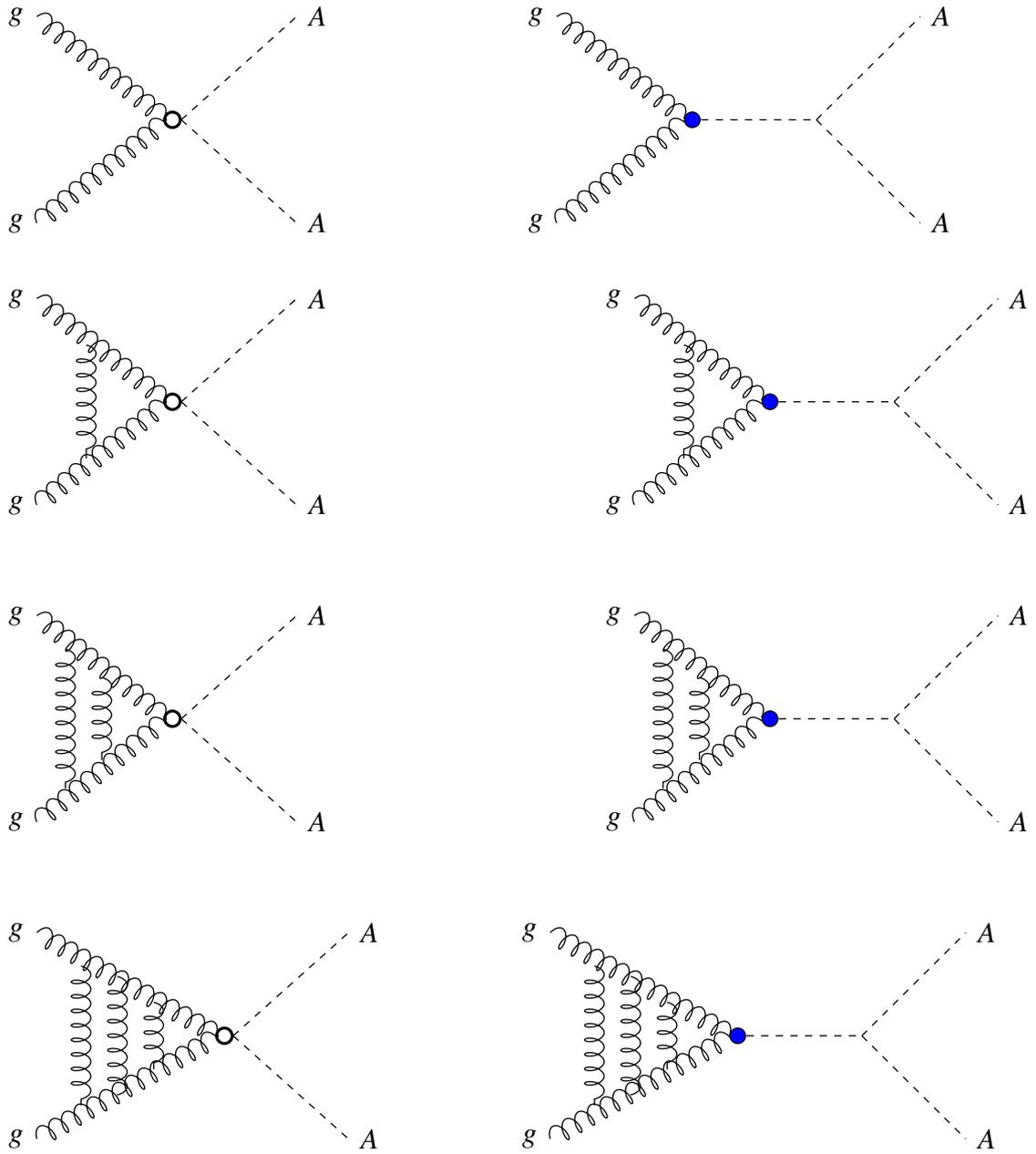
\begin{figure}[htb!]
\begin{centering}
\begin{tikzpicture}[line width=0.5 pt, scale=0.75]
\draw[gluon] (-2.5,2.0) -- (0,0);
\draw[gluon] (-2.5,-2.0) -- (0,0);
\draw[very thick] (0.11,0) circle (.15cm);
\draw[scalarnoarrow] (0.28,0)--(2.5,2.0);
\draw[scalarnoarrow] (0.28,0)--(2.5,-2.0);
\node at (-2.9,2.0) {$g$};
\node at (-2.9,-2.0) {$g$};
\node at (2.9,2.0) {$A$};
\node at (2.9,-2.0) {$A$};
 \end{tikzpicture}
\quad \quad \qquad \qquad 
\begin{tikzpicture}[line width=0.5 pt, scale=0.75]
\draw[gluon] (-2.5,2.0) -- (0,0);
\draw[gluon] (-2.5,-2.0) -- (0,0);
\draw[fill=blue] (0.11,0) circle (.15cm);
\draw[scalarnoarrow] (0.28,0)--(2.5,0);
\draw[scalarnoarrow] (2.5,0)--(4.5,2.0);
\draw[scalarnoarrow] (2.5,0)--(4.5,-2.0);
\node at (-2.9,2.0) {$g$};
\node at (-2.9,-2.0) {$g$};
\node at (4.9,2.0) {$A$};
\node at (4.9,-2.0) {$A$};
%node at (0.75,-3.2) {$ b$};
 \end{tikzpicture}
 \end{centering}
%\end{figure}
%
%
\\
\\
% 1 LOOP
%
%
\begin{centering}
\begin{tikzpicture}[line width=0.5 pt, scale=0.75]
\draw[gluon] (-2.5,2.0) -- (0,0);
\draw[gluon] (-2.5,-2.0) -- (0,0);
\draw[gluon] (-1.55,1.1) -- (-1.55,-1.1);
\draw[very thick] (0.11,0) circle (.15cm);
\draw[scalarnoarrow] (0.28,0)--(2.5,2.0);
\draw[scalarnoarrow] (0.28,0)--(2.5,-2.0);
\node at (-2.9,2.0) {$g$};
\node at (-2.9,-2.0) {$g$};
\node at (2.9,2.0) {$A$};
\node at (2.9,-2.0) {$A$};
 \end{tikzpicture}
\quad \quad \qquad \qquad 
\begin{tikzpicture}[line width=0.5 pt, scale=0.75]
\draw[gluon] (-2.5,2.0) -- (0,0);
\draw[gluon] (-2.5,-2.0) -- (0,0);
\draw[gluon] (-1.55,1.1) -- (-1.55,-1.1);
\draw[fill=blue] (0.11,0) circle (.15cm);
\draw[scalarnoarrow] (0.28,0)--(2.5,0);
\draw[scalarnoarrow] (2.5,0)--(4.5,2.0);
\draw[scalarnoarrow] (2.5,0)--(4.5,-2.0);
\node at (-2.9,2.0) {$g$};
\node at (-2.9,-2.0) {$g$};
\node at (4.9,2.0) {$A$};
\node at (4.9,-2.0) {$A$};
 \end{tikzpicture}
 \end{centering}
\\
\\
% 2 LOOP
%
\begin{centering}
\begin{tikzpicture}[line width=0.5 pt, scale=0.75]
\draw[gluon] (-2.5,2.0) -- (0,0);
\draw[gluon] (-2.5,-2.0) -- (0,0);
\draw[gluon] (-1.25,0.80) -- (-1.25,-0.80);
\draw[gluon] (-1.95,1.35) -- (-1.95,-1.35);
\draw[very thick] (0.11,0) circle (.15cm);
\draw[scalarnoarrow] (0.28,0)--(2.5,2.0);
\draw[scalarnoarrow] (0.28,0)--(2.5,-2.0);
\node at (-2.9,2.0) {$g$};
\node at (-2.9,-2.0) {$g$};
\node at (2.9,2.0) {$A$};
\node at (2.9,-2.0) {$A$};
 \end{tikzpicture}
\quad \quad  \qquad \qquad 
\begin{tikzpicture}[line width=0.5 pt, scale=0.75]
\draw[gluon] (-2.5,2.0) -- (0,0);
\draw[gluon] (-2.5,-2.0) -- (0,0);
\draw[gluon] (-1.25,0.80) -- (-1.25,-0.80);
\draw[gluon] (-1.95,1.35) -- (-1.95,-1.35);
\draw[fill=blue] (0.11,0) circle (.15cm);
\draw[scalarnoarrow] (0.28,0)--(2.5,0);
\draw[scalarnoarrow] (2.5,0)--(4.5,2.0);
\draw[scalarnoarrow] (2.5,0)--(4.5,-2.0);
\node at (-2.9,2.0) {$g$};
\node at (-2.9,-2.0) {$g$};
\node at (4.9,2.0) {$A$};
\node at (4.9,-2.0) {$A$};
 \end{tikzpicture}
 \end{centering}
%\end{figure}
%
%
\\
\\
% 3 LOOP
%
\begin{centering}
\begin{tikzpicture}[line width=0.5 pt, scale=0.75]
\draw[gluon] (-3.5,2.0) -- (0,0);
\draw[gluon] (-3.5,-2.0) -- (0,0);
\draw[gluon] (-1.25,0.65) -- (-1.25,-0.65);
\draw[gluon] (-1.95,1.15) -- (-1.95,-1.15);
\draw[gluon] (-2.65,1.35) -- (-2.65,-1.35);
\draw[very thick] (0.11,0) circle (.15cm);
\draw[scalarnoarrow] (0.28,0)--(2.5,2.0);
\draw[scalarnoarrow] (0.28,0)--(2.5,-2.0);
\node at (-3.9,2.0) {$g$};
\node at (-3.9,-2.0) {$g$};
\node at (2.9,2.0) {$A$};
\node at (2.9,-2.0) {$A$};
 \end{tikzpicture}
\quad \quad \qquad
\begin{tikzpicture}[line width=0.5 pt, scale=0.75]
\draw[gluon] (-3.5,2.0) -- (0,0);
\draw[gluon] (-3.5,-2.0) -- (0,0);
\draw[gluon] (-1.25,0.65) -- (-1.25,-0.65);
\draw[gluon] (-1.95,1.15) -- (-1.95,-1.15);
\draw[gluon] (-2.65,1.35) -- (-2.65,-1.35);
\draw[fill=blue] (0.11,0) circle (.15cm);
\draw[scalarnoarrow] (0.28,0)--(2.5,0);
\draw[scalarnoarrow] (2.5,0)--(4.5,2.0);
\draw[scalarnoarrow] (2.5,0)--(4.5,-2.0);
\node at (-3.9,2.0) {$g$};
\node at (-3.9,-2.0) {$g$};
\node at (4.9,2.0) {$A$};
\node at (4.9,-2.0) {$A$};
\end{tikzpicture}
\end{centering}
\caption[Form factor type sample Feynman diagrams up to 3-loop for $g+g \rightarrow A+A$]{Type-Ia (left panel) corresponds to the  $AAgg$ effective vertex (denoted by circle) which are form factor up to 3-loop and the right panel (Type-Ib) is related to the effective vertex $Agg$ (denoted by shaded circle)
form factor to 3-loop order.}
\label{TypeI}
\end{figure}

To evaluate the 4-point amplitude $g + g \to A +A $ to any order in $a_s$, one needs to calculate the contributing diagrams to that particular order and evaluate the scalar functions ${\cal M}_{1,2}$, using the projectors $P_{i,ab}^{\mu \nu}$, $i=1,2$.
Using the effective Lagrangian eqn.\ \ref{eq:EL}, the higher order corrections to $g + g \to A +A $ amplitude are calculated in massless QCD.
There are two types of diagrams that contribute to this process.
We classify them as type-I and type-II.
The form-factor type diagrams, where a pair of gluons annihilate to a single pseudoscalar which further branches into a pair of pseudoscalars belong to type-I and type-II, contain t and u channel diagrams where each pseudoscalar is coupled to a pair of gluons or quarks.
In type-I, we have two classes of diagrams: type-Ia (left panel of fig.\ \ref{TypeI}), which contains only four-point $AAgg$ effective vertices and type-Ib (right panel of fig.\ \ref{TypeI}), containing both $AAg$ and $AAA$ vertices.
These diagrams contribute at LO, i.e. ${\cal O} (a_s)$, and we need to calculate them to ${\cal O} (a_s^4)$, \textit{i.e.}, up to 3-loop order.
Since these diagrams are related to form factors of $O_G$ between gluons states and $O_J$ between quark and gluon states, we can readily obtain them from \cite{Baikov:2009bg,Gehrmann:2010ue,Ahmed:2015qpa}.

The type-II diagrams consist of (a) two $Agg$ effective vertex (fig.\ \ref{TypeIIa}) and (b) one $Agg$ effective vertex and one $Aq \bar q$ effective vertex as shown in fig.\ \ref{TypeIIb}.
%There is another class in type-II diagrams (type-IIc) involving two $Aq \bar q$ effective vertices, but here do not give their pictorial representation.
Due to the axial anomaly, the pseudoscalar operator for the gluonic field strength mixes with the divergence of the singlet axial vector current.
The $Agg$ effective vertex is proportional to the $C_G$ Wilson coefficient (eqn.\ \ref{WCG}), which is constrained to order ${\cal O} (a_s)$ due to the Adler-Bardeen theorem.
The LO diagram in type-IIa (fig.\ \ref{TypeIIa}) starts at order ${\cal O} (a_s^2)$, and each higher loop order adds an order ${\cal O} (a_s)$.
The $A q \bar q$ effective vertex is proportional to $C_J$, the Wilson coefficient (eqn.\ \ref{WCJ}), which starts at order ${\cal O} (a_s^2)$.
The type-IIb diagrams (fig.\ \ref{TypeIIb}), which consist of one $Agg$ effective vertex and one $Aq \bar q$ effective vertex starting at one loop level at ${\cal O} (a_s^4)$.

\asr{
Since type-I diagrams are known to the required order in $a_s$, the results presented in this paper will mainly include the type-II amplitudes up to two loops in massless pQCD, {\em i.e.}, order ${\cal O} (a_s^4)$.
}
We use dimensional regularisation ($d=4+\varepsilon$) to regularise both UV and IR singularities which appear as poles in $\varepsilon$ in the UV, soft and collinear regions.
Since we will have to deal with the Levi-Civita tensor in the $O_G$ operator and $\gamma_5$ in the $O_J$ operator, both of which are constructed inherently in 4-dimensions, a consistent method to deal with them in $4+\varepsilon$ dimensions is essential.
We discuss the details of a consistent and practical prescription to go over to $4+\varepsilon$ and its implications in the following subsection \ref{subsec:gamma5}.
Hence, the scalar amplitudes ${\cal M}_i$ can be written as a sum of amplitudes resulting from types-I and II diagrams as
\begin{eqnarray}
        {\cal M}_i = {\cal M}_i^{\rm I} + {\cal M}_i^{\rm {II}},\quad \quad \quad i=1,2\,
\end{eqnarray}
and in the following we concentrate only on ${\cal M}_i^{\rm{II}}$.

%

%TREE

\begin{figure}[htb!]
\begin{centering}
\begin{tikzpicture}[line width=0.5 pt, scale=0.75]
\draw[gluon] (-2.5,1.6) -- (0,1.6);
\draw[gluon] (-2.5,-1.6) -- (0,-1.6);
\draw[gluon] (0,1.6) -- (0,-1.6);
\draw[fill=blue] (0,1.6) circle (.15cm);
\draw[fill=blue] (0,-1.6) circle (.15cm);
\draw[scalarnoarrow] (0,1.6)--(2.5,1.6);
\draw[scalarnoarrow] (0,-1.6)--(2.5,-1.6);
\node at (-2.9,1.6) {$g$};
\node at (-2.9,-1.6) {$g$};
\node at (2.9,1.6) {$A$};
\node at (2.9,-1.6) {$A$};
 \end{tikzpicture}
\quad \quad \qquad \qquad 
\begin{tikzpicture}[line width=0.5 pt, scale=0.75]
\draw[gluon] (-2.5,1.6) -- (0,1.6);
\draw[gluon] (-2.5,-1.6) -- (0,-1.6);
\draw[gluon] (0,1.6) -- (0,-1.6);
\draw[fill=blue] (0,1.6) circle (.15cm);
\draw[fill=blue] (0,-1.6) circle (.15cm);
\draw[scalarnoarrow] (0,1.6)--(2.5,-1.6);
\draw[scalarnoarrow] (0,-1.6)--(2.5,1.6);
\node at (-2.9,1.6) {$g$};
\node at (-2.9,-1.6) {$g$};
\node at (2.9,1.6) {$A$};
\node at (2.9,-1.6) {$A$};
\end{tikzpicture}
\end{centering}
\\
\\
%1-Loop
%
\begin{centering}
\begin{tikzpicture}[line width=0.5 pt, scale=0.75]
\draw[gluon] (-2.5,1.6) -- (0,1.6);
\draw[gluon] (-2.5,-1.6) -- (0,-1.6);
\draw[gluon] (0,1.6) -- (0,-1.6);
\draw[gluon] (-1.40,1.6) -- (-1.40,-1.6);
\draw[fill=blue] (0,1.6) circle (.15cm);
\draw[fill=blue] (0,-1.6) circle (.15cm);
\draw[scalarnoarrow] (0,1.6)--(2.5,1.6);
\draw[scalarnoarrow] (0,-1.6)--(2.5,-1.6);
\node at (-2.9,1.6) {$g$};
\node at (-2.9,-1.6) {$g$};
\node at (2.9,1.6) {$A$};
\node at (2.9,-1.6) {$A$};
 \end{tikzpicture}
\quad \quad  \qquad \qquad 
\begin{tikzpicture}[line width=0.5 pt, scale=0.75]
\draw[gluon] (-2.5,1.6) -- (0,1.6);
\draw[gluon] (-2.5,-1.6) -- (0,-1.6);
\draw[gluon] (0,1.6) -- (0,-1.6);
\draw[gluon] (-1.40,1.6) -- (-1.40,-1.6);
\draw[fill=blue] (0,1.6) circle (.15cm);
\draw[fill=blue] (0,-1.6) circle (.15cm);
\draw[scalarnoarrow] (0,1.6)--(2.5,-1.6);
\draw[scalarnoarrow] (0,-1.6)--(2.5,1.6);
\node at (-2.9,1.6) {$g$};
\node at (-2.9,-1.6) {$g$};
\node at (2.9,1.6) {$A$};
\node at (2.9,-1.6) {$A$};
\end{tikzpicture}
\end{centering}
\\
\\
%
%2-Loop
%
\begin{centering}
\begin{tikzpicture}[line width=0.5 pt, scale=0.75]
\draw[gluon] (-2.5,1.6) -- (0,1.6);
\draw[gluon] (-2.5,-1.6) -- (0,-1.6);
\draw[gluon] (0,1.6) -- (0,-1.6);
\draw[gluon] (-0.9,1.6) -- (-0.9,-1.6);
\draw[gluon] (-1.8,1.6) -- (-1.8,-1.6);
\draw[fill=blue] (0,1.6) circle (.15cm);
\draw[fill=blue] (0,-1.6) circle (.15cm);
\draw[scalarnoarrow] (0,1.6)--(2.5,1.6);
\draw[scalarnoarrow] (0,-1.6)--(2.5,-1.6);
\node at (-2.9,1.6) {$g$};
\node at (-2.9,-1.6) {$g$};
\node at (2.9,1.6) {$A$};
\node at (2.9,-1.6) {$A$};
 \end{tikzpicture}
\quad \quad \qquad \qquad 
\begin{tikzpicture}[line width=0.5 pt, scale=0.75]
\draw[gluon] (-2.5,1.6) -- (0,1.6);
\draw[gluon] (-2.5,-1.6) -- (0,-1.6);
\draw[gluon] (0,1.6) -- (0,-1.6);
\draw[gluon] (-0.9,1.6) -- (-0.9,-1.6);
\draw[gluon] (-1.8,1.6) -- (-1.8,-1.6);
\draw[fill=blue] (0,1.6) circle (.15cm);
\draw[fill=blue] (0,-1.6) circle (.15cm);
\draw[scalarnoarrow] (0,1.6)--(2.5,-1.6);
\draw[scalarnoarrow] (0,-1.6)--(2.5,1.6);
\node at (-2.9,1.6) {$g$};
\node at (-2.9,-1.6) {$g$};
\node at (2.9,1.6) {$A$};
\node at (2.9,-1.6) {$A$};
\end{tikzpicture}
\end{centering}
\caption[t- and u- channel sample Feynman diagrams up to 3-loop for $g+g \rightarrow A+A$ with each pseudoscalar coupling to a pair of gluons]{Type-IIa: Sample diagrams of amplitudes up to two-loop involving two
$Agg$ effective vertex.}
\label{TypeIIa}
\end{figure}
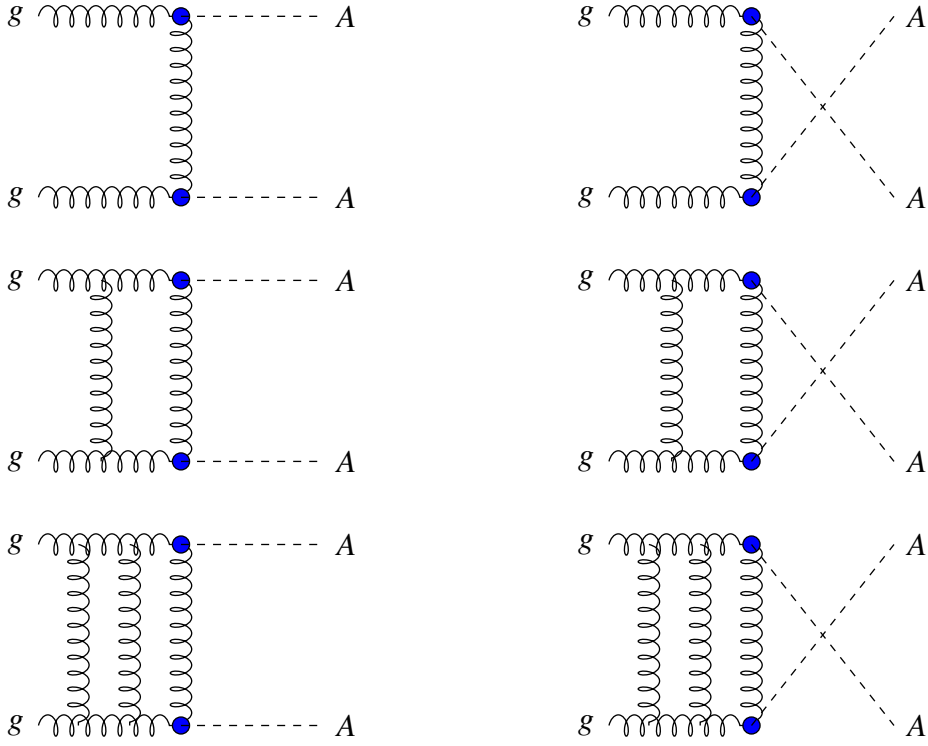
\begin{figure}[htb!]
\begin{centering}
\begin{tikzpicture}[line width=0.5 pt, scale=0.75]
\draw[gluon] (-2.5,1.6) -- (0,1.6);
\draw[gluon] (-2.5,-1.6) -- (0,-1.6);
\draw[gluon] (0,1.6) -- (0,-0.5);
\draw[fermionnoarrow] (0,-0.5) -- (0,-1.6);
\draw[fermionnoarrow] (0,-0.5) -- (1,-1);
\draw[fermionnoarrow] (1,-1) -- (0,-1.6);
\draw[fill=blue] (0,1.6) circle (.15cm);
\draw[fill=blue] (1,-0.9) rectangle (1.25,-1.1);
\draw[scalarnoarrow] (0,1.6)--(2.5,1.6);
\draw[scalarnoarrow] (1,-1)--(2.5,-1);
\node at (-2.9,1.6) {$g$};
\node at (-2.9,-1.6) {$g$};
\node at (2.9,1.6) {$A$};
\node at (2.9,-1.6) {$A$};
 \end{tikzpicture}
\end{centering}
\caption[t- and u- channel sample Feynman diagrams up to 3-loop for $g+g \rightarrow A+A$ with one pseudoscalar coupling to a pair of gluons and the other coupling to a quark-antiquark pair]{Type-IIb Diagram involving mixing of the effective vertices $Agg$ and $A q \bar q$ (denoted by shaded rectangle) which contribute at ${\cal O} (a_s^4)$}
\label{TypeIIb}
\end{figure}
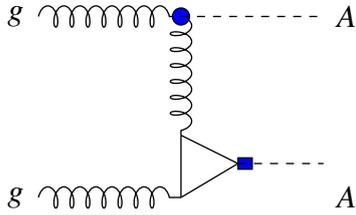

\subsection{$\gamma_5$ within dimensional regularisation}
\label{subsec:gamma5}

Due to axial anomaly, the pseudoscalar gluonic operator
$O_G=\varepsilon_{\mu\nu\rho\sigma} G^{a\mu\nu} G^{a\rho\sigma} $
is related to the divergence of the axial vector current $O_J= \partial_\mu(\bar \psi \gamma^\mu \gamma_5 \psi)$.
Computation of higher-order corrections with chiral quantities involves inherently
$d=4$ dimensional objects like $\gamma_5$ and the Levi-Civita tensor
$\varepsilon_{\mu\nu\rho\sigma}$,
and this warrants a prescription in going away from 4-dimension
{\it i.e.}~$d= 4 +\varepsilon$.  In multi-loop
computations that use dimensional regularisation, the most consistent
 prescription for $\gamma_5$ was proposed by 't~Hooft and Veltman
\cite{tHooft:1972tcz}.  In this prescription, one defines $\gamma_5$ as
\begin{eqnarray}
\gamma_5=\frac{i}{4!} \varepsilon_{\mu_1\mu_2\mu_3\mu_4}
\gamma^{\mu_1}
\gamma^{\mu_2}
\gamma^{\mu_3}
\gamma^{\mu_4} \,,
 \label{g5}
\end{eqnarray}
where the Levi-Civita tensor is purely 4-dimensional, while the Lorentz
indices on the $\gamma^{\mu_i}$ are in $d=4+\varepsilon$ dimensions. To
maintain the anti-commuting nature of $\gamma_5$ with $d$-dimensional
$\gamma^{\mu_i}$, the symmetrical form of the axial current has to
be used
\begin{eqnarray}
J^5_\mu =
\frac{1}{2}
\bar \psi (\gamma_\mu \gamma_5 - \gamma_5 \gamma_\mu) \psi \,,
\end{eqnarray}
which is in concurrence with the above definition of $\gamma_5$ in eqn.\ \ref{g5},
and will lead to
\begin{eqnarray}
J^5_\mu =
\frac{i}{3!} \varepsilon_{\mu \nu_1\nu_2\nu_3}
\bar \psi \gamma^{\nu_1} \gamma^{\nu_2} \gamma^{\nu_3} \psi \,.
\end{eqnarray}
The $O_G$ and  $O_J$ operators now take the form
\begin{equation}
O_G(x)=G^{a\mu\nu}\tilde{G}_{\mu\nu}^a=\varepsilon_{\mu\nu\rho\sigma}
G^{a\mu\nu}G^{a\rho\sigma},
\qquad
O_{J}(x)=\frac{i}{3!} \varepsilon_{\mu \nu_1\nu_2\nu_3}
\partial^{\mu}\left(\bar{\psi}  \gamma^{\nu_1} \gamma^{\nu_2}
 \gamma^{\nu_3} \psi\right) \,.
\label{eq:Fields EL}
\end{equation}
Contraction of two Levi-Civita tensors that result from either $O_G$
operator or the mixing of $O_G$ and $O_J$ operators is given
by \cite{Zijlstra:1992kj}
\noindent
\begin{equation}
\varepsilon_{\mu_{1}\nu_{1}\rho_{1}\sigma_{1}}\varepsilon^{\mu_{2}\nu_{2}\rho_{2}\sigma_{2}}=\left|\begin{array}{cccc}
\delta_{\mu_{1}}^{\mu_{2}} & \delta_{\mu_{1}}^{\nu_{2}} & \delta_{\mu_{1}}^{\rho_{2}} & \delta_{\mu_{1}}^{\sigma_{2}}\\
\delta_{\nu_{1}}^{\mu_{2}} & \delta_{\nu_{1}}^{\nu_{2}} & \delta_{\nu_{1}}^{\rho_{2}} & \delta_{\nu_{1}}^{\sigma_{2}}\\
\delta_{\rho_{1}}^{\mu_{2}} & \delta_{\rho_{1}}^{\nu_{2}} & \delta_{\rho_{1}}^{\rho_{2}} & \delta_{\rho_{1}}^{\sigma_{2}}\\
\delta_{\sigma_{1}}^{\mu_{2}} & \delta_{\sigma_{1}}^{\nu_{2}} & \delta_{\sigma_{1}}^{\rho_{2}} & \delta_{\sigma_{1}}^{\sigma_{2}}
\end{array}\right|\,,
\end{equation}
the Lorentz indices in this determinant could now be considered as
$d$-dimensional, and the consequence would be the addition of only the
inessential ${\cal O} (\varepsilon)$ terms to the renormalized quantity
\cite{Larin:1993tq}.
Though this prescription is not without consequence, a finite renormalisation of the axial vector current \cite{Larin:1991tj} is required to fulfil the chiral Ward identities and the Adler-Bardeen theorem.
This will be discussed further in the section \ref{sec:ren}.

\subsection{A Comparison : $g+g\rightarrow H+H,\ g+g\rightarrow A+A$}
\label{sec:HAcomparision}

The amplitude for di-Higgs production in gluon fusion also has the tensors specified in eqn.\ \ref{eq:All Tensors}.
Hence, the tensorial structure of the amplitude for both the processes, $g+g\rightarrow H+H \text{ and } g+g\rightarrow A+A$, are the same except for some additional terms included in the g$^{\mu\nu}$ tensor.
The tensors, $\mathcal{T}_{1}^{\mu\nu}$ and $\mathcal{T}_{2}^{\mu\nu}$ along with the scalar $\mathcal{M}_{1}$ are exactly same for both the cases.
However, the overall sign for the scalar $\mathcal{M}_{2}$ and the remaining tensors in the two amplitudes is opposite.
Table \ref{tab:;Higgs&PseudoComp} tabulates these results for the two concerned subprocesses.
This table shows that the tensorial structures contributing to the amplitudes for scalar and pseudoscalar Higgs boson productions are the same as the $d$-dimensional projectors.
Also, the $R^{\mu\nu}$ and $B^{\mu\nu}$ tensors in table \ref{tab:;Higgs&PseudoComp}, are defined as
\begin{align}
R^{\mu\nu} = & m_{a}^{2} p_{_{1}}^{\nu}p_{_{2}}^{\mu} - 2\thinspace p_{_{2}}.p_{_{3}}\thinspace p_{_{1}}^{\nu}p_{_{3}}^{\mu} - 2\thinspace p_{_{1}}.p_{_{3}}\thinspace p_{_{2}}^{\mu} p_{_{3}}^{\nu} + 2\thinspace p_{_{1}}.p_{_{2}}\thinspace p_{3}^{\mu}p_{3}^{\nu},\\
B^{\mu\nu} = & \dfrac{p_{1}^{\mu}p_{1}^{\nu}p_{_{2}}.p_{_{3}}}{t} + \dfrac{p_{2}^{\mu}p_{2}^{\nu}p_{_{1}}.p_{_{3}}}{u} - \dfrac{p_{2}^{\nu}p_{3}^{\mu}p_{_{1}}.p_{_{2}}}{u}-\dfrac{p_{1}^{\mu}p_{3}^{\nu}p_{_{1}}.p_{_{2}}}{t},
\end{align}
\asr{
where the constituent tensors of $B^{\mu\nu}$ are those that have zero contribution to the amplitude as discussed in the subsection \ref{subsubsec:LOTensors}.
}

While factorising the matrix element into gauge-invariant tensors, we also found other ways to factorise it.
However, the other methods lead to a structure of the scalars and tensors, $\mathcal{M}_{i}$ and $\mathcal{T}_{i}$ with $i=1,2$, respectively, that do not match with that given in \cite{Banerjee:2018lfq}.
Hence, for convenience, we choose this structure.
The process we are considering involves the gluon fusion channel producing two pseudoscalars/scalars.
A gluon has two polarisation states: $\pm1$ while a pseudoscalar/scalar has $0$.
So, there are only two possible combinations of the states from which we conclude that two gauge-invariant independent tensors are available.

\begin{table}[!htbp]
\renewcommand{\arraystretch}{2.5}
\centering{}%
\begin{tabular}{|c|c|c|}
\hline
 & \textcolor{black}{g+g$\rightarrow$H+H} & \textcolor{black}{g+g$\rightarrow$A+A}\tabularnewline
\hline
\hline
\textcolor{black}{$\mathcal{T}_{1}$} & \textcolor{black}{$g^{\mu\nu}-\dfrac{1}{p_{_{1}}.p_{_{2}}}\left(p_{_{1}}^{\nu}p_{_{2}}^{\mu}\right)$} & \textcolor{black}{same}\tabularnewline
\hline
\textcolor{black}{$\mathcal{M}_{1}$} & \textcolor{black}{$-p_{_{1}}.p_{_{2}}$} & \textcolor{black}{same}\tabularnewline
\hline
\textcolor{black}{$\mathcal{T}_{2}$} & \textcolor{black}{$g^{\mu\nu}+\dfrac{1}{\left(p_{_{1}}.p_{_{2}}\right)p_{T}^{2}}R^{\mu\nu}$} & \textcolor{black}{same}\tabularnewline
\hline
\textcolor{black}{$\mathcal{M}_{2}$} & \textcolor{black}{$-p_{T}^{2}\left(\dfrac{u+t}{2ut}\right)\left(p_{_{1}}.p_{_{2}}\right)$} & \textcolor{black}{$p_{T}^{2}\left(\dfrac{u+t}{2ut}\right)\left(p_{_{1}}.p_{_{2}}\right)$}\tabularnewline
\hline
\textcolor{black}{Remaining tensors} & \textcolor{black}{-$B^{\mu\nu}$} & \textcolor{black}{$B^{\mu\nu}$}
\tabularnewline  \hline
\end{tabular}
\caption{Comparison of the amplitudes for the subprocesses $g+g\rightarrow H+H$ and $g+g\rightarrow A+A$.}
\label{tab:;Higgs&PseudoComp}
\end{table}

%\vspace{-7mm}
\section{Calculation of the amplitude}
\label{sec:calcdetails}

Till now, I have provided all the theoretical tools to compute the amplitude of di-pseudoscalar production \textit{via} gluon fusion.
However, with the vast number of Feynman diagrams, the increase in complications with each order, and us considering only virtual diagrams, the computed amplitude is expected to show irregular behaviour (divergences).
In other words, the bare scattering amplitude of this process given in eqn.\ \ref{eq:ggAAprocess} contains poles in the dimensional regulator $\varepsilon$ of UV and IR origins beyond LO.
The final result must always be finite.
Hence, coupling constant and operator renormalisation is necessary to obtain such UV-finite expressions.
UV renormalisation of the amplitude depicted in eqn.\ \ref{eq:TensorBreak} is done in the $\overline{\text{MS}}$ scheme.
A detailed treatment is shown in the following subsection \ref{sec:ren}.
This UV-finite amplitude is still not finite and exhibits divergences arising from the loop momenta's soft and collinear (IR) configurations.
These IR singularities cancel when real radiation diagrams yielding the same observable in the final state are considered order-by-order while calculating physical quantities and mass factorisation.
These pole structures are handled by using the universal behaviour of IR divergences explained in subsection \ref{subsec:IR}.

\subsection{UV renormalisation, operator renormalisation and mixing}
\label{sec:ren}

For type-I diagrams which begin to contribute at LO, the renormalisation constant $Z_{a_s}$ (eqn.\ \ref{eq:Zas}) up to order ${\cal O}(a_{s}^{3})$ will be needed while for type-II diagrams, one order lower is sufficient.

Apart from the renormalisation of strong coupling in the massless QCD,
the amplitudes require the renormalisation of vertices resulting from the  composite
operators $O_G$ and $O_J$ of the effective Lagrangian eqn.\ (\ref{eq:EL}).
The renormalized operators are denoted by $[~]$ parenthesis, while the
bare quantities without the parenthesis.

The renormalisation of $O_J$ is related to the renormalisation of the
singlet axial vector current $J^\mu_5$ which needs the standard overall
UV renormalisation constant $Z^s_{\overline{MS}}$ and a finite
renormalisation constant $Z^s_5$.  The later is
necessary in dimensional
regularisation in order to ensure the nature of the operator
relation resulting from axial anomaly \cite{Adler:1969er}
\begin{align}
  \label{eq:Anomaly}
  \left[ \partial_{\mu}J^{\mu}_{5} \right] = a_{s} \frac{n_{f}}{2} \left[ G\tilde{G} \right]\,,
\qquad
\qquad
  \text{i.e.}~~~ \left[ O_{J} \right] = a_{s} \frac{n_{f}}{2} \left[ O_{G} \right]\,,
\end{align}
which is true in Pauli-Villars, a 4-dimensional regularisation.
To preserve eqn.\ \ref{eq:Anomaly} in $4+\varepsilon$ dimensions, the multiplicative finite renormalisation
constant $Z^s_5$ is required.
The bare operator, $O_J$, is renormalised
multiplicatively, exactly in the same way as the singlet axial
vector current $J^\mu_5$, through
\begin{align}
  \label{OJRen}
  \left[ O_J \right] = Z^s_5 ~Z^s_{\overline{MS}} ~O_J\,,
\end{align}
whereas the bare pseudoscalar gluon operator $O_G$
mixes with fermionic operator $O_J$ under the renormalisation through
\begin{align}
  \left[ O_G \right] = Z_{GG} ~O_G + Z_{GJ} ~O_J \,,
  \label{OGRen}
\end{align}
with the corresponding renormalisation constants $Z_{GG}$ and
$Z_{GJ}$.
Combining the above two equations in a matrix form, we have
\begin{equation}
  \label{eq:OpMat}
  \left[ O_i \right] = Z_{ij} ~O_j  \,,
\qquad \qquad {\rm where} \quad i,j = \{G, J\}\,,
\end{equation}
\begin{align}
  O \equiv
  \begin{pmatrix}
    O_G\\
    O_J
  \end{pmatrix}
  \qquad\quad &\text{and}  \qquad\quad
                Z \equiv
                \begin{pmatrix}
                  Z_{GG} & Z_{GJ}\\
                  Z_{JG} & Z_{JJ}
                \end{pmatrix}\,,
\end{align}
where $Z_{JG} = 0$ to all orders in perturbation theory and
$Z_{JJ} \equiv Z^s_5 Z^s_{\overline{MS}}$.  The renormalisation constants
required for the above equation are available up to
${\cal O} (a_s^3)$ \cite{Larin:1993tq},
\cite{Zoller:2013ixa}, which was computed using OPE.
For earlier works on this, see \cite{Kataev:1981aw,Kataev:1981gr}.
The same quantities were calculated by some of us using a completely different method \cite{Ahmed:2015qpa} and found to be in full agreement.
%%%%%%%%%%%%
The UV renormalisation constant of the singlet axial vector current $J^\mu_5$ in the
$\overline{MS}$ scheme is
\begin{align}
Z_{\overline{MS}}^s = & 1+a_{s}^{2}\left[C_{A}C_{F}\left\{
-\dfrac{44}{3\varepsilon}\right\} +C_{F}n_{f}\left\{
-\dfrac{10}{3\varepsilon}\right\} \right]\nonumber \\
 & +a_{s}^{3}\left[C_{A}^{2}C_{F}\left\{
-\dfrac{1936}{27\varepsilon^{2}}-\dfrac{7156}{81\varepsilon}\right\}
+C_{F}^{2}n_{f}\left\{ \dfrac{44}{9\varepsilon}\right\} +C_{F}n_{f}^{2}\left\{
\dfrac{80}{27\varepsilon^{2}}-\dfrac{52}{81\varepsilon}\right\} \right]\nonumber \\
 & +a_{s}^{3}\left[C_{A}C_{F}^{2}\left\{ \dfrac{616}{9\varepsilon}\right\}
+C_{A}C_{F}n_{f}\left\{
-\dfrac{88}{27\varepsilon^{2}}-\dfrac{298}{81\varepsilon}\right\} \right] \,,
\end{align}
and the finite renormalisation constant $Z_5^s$ is
\begin{equation}
Z_5^s=1+a_s\left\{ -4C_{F}\right\} +a_s^2 \left\{
22 C_F^2-\dfrac{107}{9}C_A C_F+\dfrac{31}{18}C_F n_f \right\} \,.
\end{equation}
The renormalisation constants for $O_G$ and $O_J$ operators up to two loops
are given by
\begin{align}
  \label{ZGGtZGJ}
  Z_{GG} &= 1 +  a_s \Bigg[ \frac{22}{3\varepsilon}
           C_{A}  -
           \frac{4}{3\varepsilon} n_{f} \Bigg]
           +
           a_s^2 \Bigg[ \frac{1}{\varepsilon^2}
           \Bigg\{ \frac{484}{9} C_{A}^2 - \frac{176}{9} C_{A}
           n_{f} + \frac{16}{9} n_{f}^2 \Bigg\}
           \nonumber\\&
           + \frac{1}{\varepsilon} \Bigg\{ \frac{34}{3} C_{A}^2
         -\frac{10}{3} C_{A} n_{f}  - 2 C_{F} n_{f} \Bigg\} \Bigg],
         \nonumber\\
  Z_{GJ} &=  a_s \Bigg[ - \frac{24}{\varepsilon} C_{F} \Bigg]
                    +
                    a_s^2 \Bigg[ \frac{1}{\varepsilon^2}
                    \Bigg\{ - 176 C_{A} C_{F} + 32 C_{F} n_{f} \Bigg\}
                    \nonumber\\&
                    + \frac{1}{\varepsilon} \Bigg\{ - \frac{284}{3} C_{A} C_{F} +
                    84 C_{F}^2 + \frac{8}{3} C_{F} n_{f} \Bigg\}  \Bigg],
                    \nonumber\\
   Z_{JJ} &= 1 + a_s\left[-4 C_F\right] + a_s^2\Bigg[
-\frac{44}{3\varepsilon}C_A C_F - \frac{10}{3\varepsilon}C_F n_f
   \nonumber\\&
   + 22C_F^2 -\frac{107}{9}C_A C_F  + \frac{31}{18}C_F n_f \Bigg].
\end{align}
The matrix element that would contribute to the $ g + g \to A + A$
amplitude can be obtained {\it via} the insertion of two renormalized operators
$[O_G]$ (eqn.~\ref{OGRen}) and $[O_J]$ (eqn.\ \ref{OJRen}) for each
$A$, which would involve the following operator insertion between
gluon states:
\begin{eqnarray}
\langle g|[O_G O_G] |g \rangle;  \quad
\langle g|[O_G O_J] |g \rangle \quad
{\rm and} \quad
\langle g|[O_J O_J] |g \rangle \,.
\end{eqnarray}
The above operator renormalisation constants (eqn.\ \ref{ZGGtZGJ})
and the strong coupling renormalisation constant $Z_{a_s}$ (eqn.\ \ref{eq:Zas}) would take care of the UV renormalisation.
The gluonic operator $O_G$ couples to gluons at LO (${\cal O} (a_s)$) and
the fermionic operator $O_J$ couples to quarks at LO (${\cal O} (a_s^2)$).  The
basic matrix elements that have to be evaluated diagrammatically
involve the following bare operator combinations:
$\langle g|O_G^2  |g \rangle$, $\langle g|O_G O_J|g \rangle$
and $\langle g|O_J^2  |g \rangle$.  The $O_G^2$ starts to
contribute at tree level at ${\cal O} (a_s^2)$, $O_G O_J$
begins to contribute at one-loop level and at  ${\cal O} (a_s^4)$
while $O_J^2$ starts to contribute at ${\cal O} (a_s^6)$.  Here
we compute $g +g \to A+A$ amplitude to order ${\cal O} (a_s^4)$ and
hence the contributing terms are from $O_G^2$ calculated up to
two-loops and the $O_G O_J$ combination from one-loop.
We need the following renormalized operator $[O_{G} O_{G} ]$
and $[ O_G O_J ]$
which is given by
\begin{eqnarray}
\label{OGOG}
\left[O_G O_G \right] &=& Z_{GG}^2 ~O_G O_G +2Z_{GG} Z_{GJ} ~O_G O_J + Z_{GJ}^2 ~O_J O_J
\,,
\nonumber
\\
\left[O_G O_J \right] &=& Z_{GG} Z_{JJ} ~O_G O_J + Z_{GJ} Z_{JJ} ~O_J O_J\,.
\end{eqnarray}
Sandwiching $\left[O_G O_G \right]$ and $\left[O_G O_J\right]$  between gluon states and using eqn.\ \ref{OGOG}, we obtain up to two loops:
\noindent
\begin{eqnarray}
\mathcal{M}^{\rm{II}}_{GG,g} & = &
Z_{GG}^{2}\Big(\hat {\mathcal{M}}_{GG,g}^{\rm{II}\left(0\right)}+\hat a_{s}\hat {\mathcal{M}}_{GG,g}^{\rm{II},\left(1\right)}
+\hat a_{s}^{2}\hat {\mathcal{M}}_{GG,g}^{\rm{II}\left(2\right)}\Big)\nonumber
\\
 &&
+2Z_{GG}Z_{GJ}\left(\hat a_{s}\hat {\mathcal{M}}_{GJ,g}^{\rm{II}\left(1\right)}+\hat a_{s}^{2}\hat {\mathcal{M}}_{GJ,g}^{\rm{II}\left(2\right)}\right)\nonumber
\\
 &&
+Z_{GJ}^{2}\Big(\hat a_{s}\hat { \mathcal{M}}_{JJ,g}^{\rm{II}\left(1\right)}
+\hat a_{s}^{2}\hat { \mathcal{M}}_{JJ,g}^{\rm{II}\left(2\right)}\Big)\,,
\nonumber\\
\mathcal{M}^{\rm{II}}_{GJ,g} & = &
Z_{GG} Z_{JJ} \Big( \hat a_s \hat {{\cal M}}^{II(1)}_{GJ,g}
                    + \hat a_s^2 \hat {{\cal M}}^{II(2)}_{GJ,g}
              \Big)
+ Z_{GJ} Z_{JJ} \Big(\hat a_s \hat {{\cal M}}^{II(1)}_{JJ,g} + \hat a_s^2 \hat {{\cal M}}^{II(2)}_{JJ,g} \Big)\,
\label{eq:operatorSandwichAmplitude}
\end{eqnarray}
where $\mathcal{M}^{\rm{II}}_{XY,g} = \langle g|\left[O_X O_Y \right]|g\rangle$ and
$\hat {\mathcal{M}}^{\rm{II}}_{XY,g} = \langle g|O_X O_Y |g\rangle$, with $X,Y=\{G,J\}$
that contribute to the type-II diagrams.
$\hat{\mathcal{M}}_{GJ,g}^{\rm{II}\left(2\right)}$,
$\hat{\mathcal{M}}_{JJ,g}^{\rm{II}\left(1\right)}$
and
$\hat{\mathcal{M}}_{JJ,g}^{\rm{II}\left(2\right)}$
do not contribute in our case as they are of order higher than $a_{s}^{4}$ when combined with
their respective Wilson coefficients. Finally,  $\mathcal{M}^{\rm I I}_{GG,g}$
can be expressed in powers of
renormalized $a_s$ as
%, and using Eq.\ref{Zas}, we obtain
%
\begin{equation}
\mathcal{M}^{\rm I I}_{GG,g}=\mathcal{M}_{GG,g}^{{\rm II}\left(0\right)}+a_{s}\mathcal{M}_{GG,g}^{{\rm II}\left(1\right)}+a_{s}^{2}\mathcal{M}_{GG,g}^{{\rm II} \left(2\right)}+\mathcal{O}\left(a_{s}^{3}\right)\,.
\end{equation}
The coefficients $\mathcal{M}_{GG,g}^{{\rm II}\left(i\right)}$ can be  related to
$\hat {\mathcal{M}}_{GG,g}^{{\rm II}\left(i\right)}$ using eqns.\ \ref{eq:Zas} and \ref{ZGGtZGJ}.
Expanding the renormalisation constants $Z_{KL}$ in eqn.\ \ref{ZGGtZGJ} as
\begin{eqnarray}
Z_{KL} = \delta_{KL} + \displaystyle{\sum_{i=1}^\infty} a_s^i Z_{KL}^{(i)},
\quad \quad K,L = \{G,J\}
\end{eqnarray}
we find
\begin{eqnarray}
{\cal M}^{{\rm II}(0)}_{GG,g} &=& \hat {{\cal M}}^{{\rm II}(0)}_{GG,g}\,,
\nonumber\\
{\cal M}^{{\rm II}(1)}_{GG,g} &=& {1 \over \mu_R^\varepsilon}
                           \hat {{\cal M}}^{{\rm II}(1)}_{GG,g}
                           +2 Z^{(1)}_{GG} \hat {{\cal M}}^{{\rm II}(0)}_{GG,g}\,,
\nonumber\\
{\cal M}^{{\rm II}(2)}_{GG,g} &=& {1 \over \mu_R^{2\varepsilon}}
                                \hat {{\cal M}}^{{\rm II}(2)}_{GG,g}
                            +{1 \over \mu_R^{\varepsilon}}
                            \Bigg(
                             {2 \beta_0 \over \varepsilon} \hat {{\cal M}}^{{\rm II}(1)}_{GG,g}
                            +2 Z^{(1)}_{GJ} \hat {{\cal M}}^{{\rm II}(1)}_{GJ,g}
                               +2 Z^{(1)}_{GG} \hat {{\cal M}}^{{\rm II}(1)}_{GG,g}\Bigg)
\nonumber\\
&&                               +\Big( 2 Z^{(2)}_{GG} + (Z^{(1)}_{GG})^2 \Big)
                                \hat {{\cal M}}^{{\rm II}(0)}_{GG,g} \,.
\end{eqnarray}
Similarly for $\mathcal{M}^{\rm I I}_{GJ,g}$,
we find
\begin{equation}
\mathcal{M}^{\rm I I}_{GJ,g}=a_{s}\mathcal{M}_{GJ,g}^{{\rm II}\left(1\right)}+a_{s}^{2}\mathcal{M}_{GJ,g}^{{\rm II} \left(2\right)}+\mathcal{O}\left(a_{s}^{3}\right) \,,
\end{equation}
where
\begin{eqnarray}
{\cal M}^{{\rm II}(1)}_{GJ,g}  &=& {1 \over \mu_R^\varepsilon}
                                 \hat {{\cal M}}^{{\rm II}(1)}_{GJ,g}\,,
\nonumber\\
{\cal M}^{{\rm II}(2)}_{GJ,g} &=& {1 \over \mu_R^\varepsilon}
                            \Bigg( {2 \beta_0 \over \varepsilon} + Z^{(1)}_{JJ} + Z^{(1)}_{GG}\Bigg)
                              \hat {{\cal M}}^{{\rm II}(1)}_{GJ,g}
                           +{1 \over \mu_R^\varepsilon} Z^{(1)}_{GJ} \hat {{\cal M}}^{{\rm II}(1)}_{JJ,g}
                           +{1 \over \mu_R^{2\varepsilon}}
                              \hat {{\cal M}}^{{\rm II}(2)}_{GJ,g}\,.
\end{eqnarray}

We find that the UV singularities that appear at one-loop, and two-loop levels can be taken care of by the coupling constant renormalisation $Z_{a_s}$ and operator renormalisation $Z_{ij}$.
At this point, we would like to stress that there could be additional contact terms required as a result of the behaviour of the product of operators $O_G O_G$ or $O_G O_J$
at short distances. As shown in \cite{Zoller:2013ixa}, we find no contact terms due to these operator products at short distances.
For earlier works on this, see \cite{Kataev:1981aw,Kataev:1981gr}.

\subsection{Infrared factorisation}
\label{subsec:IR}

The UV finite amplitudes we have computed contain only divergences
of IR origin, which appear as poles in the dimensional regularisation parameter $\varepsilon$.
They are expected to cancel against real emission diagrams for the IR-safe observables.
While these singularities disappear in the physical observables, the
amplitudes beyond the LO show a rich universal IR structure. %region.
In \cite{Catani:1998bh}, Catani predicted the IR poles of two-loop $n$-point UV finite amplitudes in terms of certain universal IR anomalous dimensions.
Later, in \cite{Sterman:2002qn}, factorisation and resummation properties of QCD amplitudes were used
to understand the IR structure, and subsequently the, attempts were made to predict the structure
of IR poles beyond two loops in ~\cite{Becher:2009cu,Gardi:2009qi}.
Following \cite{Catani:1998bh}, we obtain
\begin{eqnarray}
\label{mfin}
\mathcal{M}_i ^{\rm{II},(0)} &=& \mathcal{M}_i ^{\rm{II},(0)} \,,
\nonumber\\
\mathcal{M}_i ^{\rm{II},(1)} &=& 2\mathbf{I}_{g}^{(1)}(\varepsilon)\mathcal{M}_i ^{\rm{II},(0)} + \mathcal{M}_i ^{\rm{II},(1),fin} \,,
\nonumber\\
\mathcal{M}_i ^{\rm{II},(2)} &=& 4\mathbf{I}_{g}^{(2)}(\varepsilon)\mathcal{M}_{i} ^{\rm{II},(0)} + 2\mathbf{I}_{g}^{(1)}(\varepsilon)\mathcal{M}_i ^{\rm{II},(1)}
+ \mathcal{M}_{i} ^{B,(2),fin} \,,
\end{eqnarray}
where $\mathbf{I}_{g}^{(1)}(\varepsilon), \,\mathbf{I}_{g}^{(2)}(\varepsilon) $ are the IR singularity operators given by
\begin{align}
\boldsymbol{I}_{g}^{\left(1\right)}\left(\varepsilon\right) & =-\frac{e^{-\frac{\varepsilon}{2}\gamma_{E}}}{\Gamma\left(1+\frac{\varepsilon}{2}\right)}\left(\frac{4C_{A}}{\varepsilon^{2}}-\frac{\beta_{0}}{\varepsilon}\right)\left(-\frac{s}{\mu_{R}^{2}}\right)^{\frac{\varepsilon}{2}} \,,\\
\boldsymbol{I}_{g}^{\left(2\right)}\left(\varepsilon\right) & =-\frac{1}{2}\boldsymbol{I}_{g}^{\left(1\right)}\left(\varepsilon\right)\left[\boldsymbol{I}_{g}^{\left(1\right)}\left(\varepsilon\right)-\frac{2\beta_{0}}{\varepsilon}\right]+\frac{e^{\frac{\varepsilon}{2}\gamma_{E}}\Gamma\left(1+\varepsilon\right)}{\Gamma\left(1+\frac{\varepsilon}{2}\right)}\left[-\frac{\beta_{0}}{\varepsilon}+K\right]\boldsymbol{I}_{g}^{\left(1\right)}\left(2\varepsilon\right)+2\boldsymbol{H}_{g}^{\left(2\right)}\left(\varepsilon\right) \,
\label{eq: I operators}
\end{align}
with

\noindent
\begin{eqnarray}
K & =&\left(\frac{67}{18}-\frac{\pi^{2}}{6}\right)C_{A}-\frac{5}{9}n_{f}\,,\\
\boldsymbol{H}_{g}^{\left(2\right)}\left(\varepsilon\right) & =&\left(-\frac{s}{\mu_{R}^{2}}\right)^{\varepsilon}
\frac{e^{-\frac{\varepsilon}{2}\gamma_{E}}}{\Gamma\left(1+\frac{\varepsilon}{2}\right)}
\frac{1}{\varepsilon}\left\{ C_{A}^{2}\left(-\frac{5}{24}-\frac{11}{48}\zeta_{2}
-\frac{\zeta_{3}}{4}\right)+C_{A}n_{f}\left(\frac{29}{54}+\frac{\zeta_{2}}{24}\right)
\right.
\nonumber\\
&&\left. -\frac{1}{4}C_{F}n_{f}-\frac{5}{54}n_{f}^{2}\right\} \,.
\label{eq: K and H}
\end{eqnarray}

At one loop level, it is straightforward to show analytically that the IR poles agree with the predictions.
For the two-loop case, a fully analytic comparison was possible only for poles $\varepsilon^{-i}$ with $i=2-4$.
Due to a large file size containing the $\varepsilon^{-1}$ pole terms and the absence of relevant tools for handling such files, we had to make the comparison only at a numerical level.
We found full agreement with the predictions of Catani up to the two-loop level for all the IR poles.
Let us discuss this handling of the $1/\varepsilon$ pole for each of our chosen projectors in more detail.

\textbf{Projector 1 : }
The file containing the $1/ \varepsilon $ pole terms is about 16 MB for projector 1.
For convenience, we divide it into four colour structures - $N n_{f}$, $n_{f}^{2}$, $N^{2}$ and $n_{f}/{N}$.
Any other colour structure is a subset of these four, or the coefficient of that colour structure is zero like that of $1/N^{2}$.
The coefficients of the colour structures $N n_{f}$ , $n_{f}^{2}$ and $n_{f}/{N}$ are analytically reduced to zero.
The file size for the $N^{2}$ coefficient is single-handedly about 15.7 MB.
So, we break down these terms further based on their transcendental weights like LT$_0$, which includes all constant pieces, LT$_1$, which includes all polylogarithms of order one and $\pi '$s, LT$_2$ which includes all polylogarithms of order two, LT$_3$ which includes all polylogarithms of order three and LT$_4$ which includes all polylogarithms of order four and the special function Li$_{22}$.
No LT$_4$ terms are found in this case.
LT$_1$ and LT$_2$ terms are reduced to zero analytically.
The LT$_0$ coefficient has a file size of about 12 MB, which was challenging to simplify analytically.
The LT$_3$ coefficient also could not be analytically reduced to zero.
These two coefficients were calculated to be zero numerically to very high precision.
See table~\ref{tab:Proj1LT} for the numerical values of these coefficients for a few specific choices of the parameters, $x$ and $\cos \theta$.
\begin{table}[!htb]
\centering
\renewcommand{\arraystretch}{2}
\begin{tabu} to \textwidth {| X | X | X[2] |}
\hline
\centering $x$ & \centering $\cos\theta$ & \centering $\varepsilon^{-1}$                              \\ \hline
1/13  & 0.2 & $(-2.90366\times10^{-12}+~i~8.99075\times10^{-14})N^{2}$ \\ \hline
1/5   & 0.2 & $(-1.38217\times10^{-12}-~i~1.16271\times10^{-13})N^{2}$ \\ \hline
8/17  & 0.2 & $(-4.71671\times10^{-13}+~i~1.52335\times10^{-14})N^{2}$ \\ \hline
11/15 & 0.2 & $(-2.17893\times10^{-12}-~i~4.48837\times10^{-13})N^{2}$ \\ \hline
1/13  & 0.8 & $(-1.09321\times10^{-12}-~i~4.62596\times10^{-11})N^{2}$ \\ \hline
1/5   & 0.8 & $(-2.27237\times10^{-12}-~i~5.36670\times10^{-12})N^{2}$ \\ \hline
8/17  & 0.8 & $(-7.50942\times10^{-13}+~i~1.06920\times10^{-14})N^{2}$ \\ \hline
11/14 & 0.8 & $(+8.09060\times10^{-12}+~i~7.25492\times10^{-12})N^{2}$ \\ \hline
\end{tabu}
\caption[Numerical values of the transcendalities for the $1/\varepsilon$ coefficients of Projector 1]{Numerical values of the $1/\varepsilon$ coefficients of Projector 1 corresponding to a few specific choices of $x$ and $\cos \theta$.}
\label{tab:Proj1LT}
\end{table}

\textbf{Projector 2 : }
The file containing the $1/\varepsilon $ pole terms for projector 2 is about 48 MB.
We again divide it into different colour structures as before - $Nn_{f}$, $n_{f}^{2}$, $N^{2}$ and $n_{f}/{N}$.
The coefficients corresponding to the colour structure $n_{f}^{2}$ are analytically reduced to zero.
The coefficients of the $N^{2}$ colour factor constitute a file of approximately 46 MB.
For further simplification, we divided it further based on the transcendental weights of the terms into transcendental weight zero (LT$_0$), transcendental weight one (LT$_1$), transcendental weight two (LT$_2$), transcendental weight three  (LT$_3$) and transcendental weight four (LT$_4$).
The coefficient was devoid of any LT$_4$ terms.
All LT$_2$ terms are reduced to zero analytically.
The LT$_0$ coefficient has a file size of 29 MB, and the LT$_1$ coefficient has a file size of 24 MB.
So, the LT$_0$, LT$_1$ and  LT$_3$ coefficients of $N^{2}$ and the coefficients of the colour factors $N n_{f}$ and $n_{f}/{N}$ could not be analytically reduced to zero.
They were numerically calculated to be zero to very high precision.
Table~\ref{tab:Proj2LT} presents the numerical values of these coefficients for a few specific choices of the parameters, $x$ and $\cos \theta$.
\begin{table}[!htbp]
\centering
\renewcommand{\arraystretch}{2.5} 
\begin{tabu} to \textwidth {| X | X | X[2] |}
\hline
\centering $x$ & \centering $\cos\theta$ & \centering $\varepsilon^{-1}$ \\ \hline  
1/13 & 0.2 & $(1.31697\times10^{-11}-~i~5.26061\times10^{-12})N^{2}
             +(2.33534\times10^{-12}-~i~1.23212\times10^{-12})\dfrac{n_{f}}{N} %\\
             -(2.33534\times10^{-12}-~i~1.23212\times10^{-12})N n_{f}$ \\ \hline
1/5  & 0.2 & $(4.07456\times10^{-12}-~i~5.08822\times10^{-13})N^{2} %\\
+(9.30098\times10^{-13}-~i~2.22464\times10^{-13})\dfrac{n_{f}}{N} %\\
-(9.30098\times10^{-13}-~i~2.22464\times10^{-13})Nn_{f}$ \\ \hline
8/17 & 0.2 & $(-4.13837\times10^{-12}+i~4.66444\times10^{-13})N^{2} %\\
             +(4.87421 \times10^{-13}-~i~1.92849\times10^{-14})\dfrac{n_{f}}{N} %\\
             -(4.87421 \times10^{-13}-~i~1.92849\times10^{-14})Nn_{f}$ \\ \hline
11/15 & 0.2 & $(1.83303\times10^{-10}-~i~2.17199\times10^{-13})N^{2} %\\  
-(5.26839\times10^{-13}+~i~2.45716\times10^{-15})\dfrac{n_{f}}{N} %\\
+(5.26839\times10^{-13}+~i~2.45716\times10^{-15})Nn_{f}$ \\ \hline  
1/13  & 0.8 & $(3.18265\times10^{-10}-~i~7.23266\times10^{-12})N^{2} %\\
+(5.60483\times10^{-11}-~i~5.60643\times10^{-12}\dfrac{n_{f}}{N} %\\
+(5.60483\times10^{-11}-~i~5.60643\times10^{-12})N n_{f}$ \\ \hline 
1/5   & 0.8 & $(3.67564\times10^{-11}+~i~5.39447\times10^{-11})N^{2} %\\
+(1.11612\times10^{-11}+~i~1.37015\times10^{-12})\dfrac{n_{f}}{N} %\\
-(1.11612\times10^{-11}+~i~1.37015\times10^{-12})Nn_{f}$ \\ \hline
8/17  & 0.8 & $(-4.68937\times10^{-11}+i~2.95662\times10^{-11})N^{2} %\\
+(2.92453\times10^{-12}+~i~4.21245\times10^{-13})\dfrac{n_{f}}{N} %\\
-(2.92453\times10^{-12}+~i~4.21245\times10^{-13})Nn_{f}$ \\ \hline
11/14 & 0.8 & $(-3.47623\times10^{-9}+~i~9.65797\times10^{-12})N^{2} %\\
+(3.56378\times10^{-12}+~i~2.05351\times10^{-13})\dfrac{n_{f}}{N} %\\
-(3.56378\times10^{-12}+~i~2.05351\times10^{-13})N n_{f}$ \\ \hline
\end{tabu}
\caption[Numerical values of the $1/\varepsilon$ coefficients for Projector 2]{Numerical values of the $1/\varepsilon$ coefficients for Projector 2 corresponding to a few specific choices of $x$ and $\cos \theta$.}
\label{tab:Proj2LT}
\end{table}

After this evaluation, we could confidently predict the UV renormalised and IR factorised part of the amplitude using the method defined in sections \ref{sec:ren} and \ref{subsec:IR}.
The expressions are too large to be explicitly written down. However, they can be obtained from the authors.
They have also been uploaded to the arXiv/JHEP repository and are available on request.

\section{Computational framework}
\label{sec:computation}

Our task of computing the amplitude $g + g \to A + A$ is reduced to the type-II diagrams up to ${\cal O} (a_s^4)$.
This computation of the type-II diagrams involves diagrams with two $Agg$ effective vertices, up to the two-loop level in QCD (Type-IIa) and diagrams with one $Agg$ effective vertex and one $Aq \bar q$ effective vertex, which involves terms up to one loop in QCD (Type-IIb).
Type-IIa involves an effective coupling between a pair of gluons and pseudoscalar. There are two LO, 35 NLO and 789 NNLO diagrams in this case.
Type-IIb includes those diagrams where one of the two outgoing vertices interacts via a fermion loop and the other via a pair of gluons. In this case, there are no LO diagrams but 8 NLO and 236 NNLO diagrams.
Diagrams involving two $A q \bar q$ effective vertices involve no LO diagrams, 8 NLO diagrams and 138 NNLO diagrams, but they start at ${\cal O}(a_s^5)$ and hence, are not considered here.

%\textit{By applying the projectors, $P_{i,ab}^{\mu \nu}$ where $i=1,2$, on the amplitudes, we extract the scalar coefficients, ${\cal M}_{i}$ with $i=1,2$, at every order in the perturbation.}

All the LO, NLO and NNLO Feynman diagrams in massless QCD are generated using QGRAF \cite{Nogueira:1991ex} where additional vertices resulting from the effective Lagrangian given in eqn.\ \ref{eq:EL} are incorporated.
The raw QGRAF output is converted with the help of in-house codes based on FORM \cite{Vermaseren:2000nd} to include appropriate Feynman rules and to perform trace of Dirac matrices, contraction of Lorentz indices and colour indices.
At this stage, we encounter a huge number of one and scalar 2-loop Feynman integrals, which contain a set of propagator denominators and a combination of scalar products between loop momenta and independent external momenta.
These Feynman integrals can be classified by the propagator denominators they contain.
Hence, identifying the momentum shifts required to express each of these diagrams in terms of a standard set of propagators called auxiliary topology is essential.
We use REDUZE2 package~\cite{vonManteuffel:2012np} to achieve this.
The auxiliary topologies needed for the present case are the same as those found in
vector boson pair production \cite{Gehrmann:2013cxs,Gehrmann:2014bfa} at two loops.

As expected, these large numbers of scalar integrals are not all independent.
Some properties of the Feynman integrals in dimensional regularisation are used to establish the relations.
Exploiting the fact that the total derivative \textit{w.r.t.} any loop momenta of these integrals, evaluates to a surface term, which vanishes, leads to IBP identities \cite{Tkachov:1981wb,Chetyrkin:1981qh}.
In addition, the fact that all integrals are Lorentz scalars gives rise to LI identities \cite{Gehrmann:1999as}.
As a result, these integrals can, in turn, be expressed in terms of a much smaller set of integrals which are irreducible and appropriately called master integrals (MI).
Several automated computer algebra packages are available \cite{Anastasiou:2004vj,
Smirnov:2008iw,Studerus:2009ye,vonManteuffel:2012np,Lee:2013mka} that use
the Laporta algorithm \cite{Laporta:2001dd} to reduce these Feynman integrals
to the MIs. We have used the Mathematica-based package LiteRed \cite{Lee:2013mka}
to perform the reductions of all the integrals to MIs. At one loop, there are 10
MIs, while at two-loop, the number is 149. These two-loop MIs are the same as
two-loop four-point functions with two equal-mass external legs. The analytical
result for each MI in terms of Laurent series expansion in $\varepsilon$ is
given in \cite{Gehrmann:2013cxs,Gehrmann:2014bfa}.

At this stage, renormalising the strong coupling constant and the operators $O_G$ and $O_J$, described in section \ref{sec:ren}, removes all the UV singularities.
The singularities that remain are purely of IR origin, and the subsection \ref{subsec:IR} is devoted to it.
%
%This work requires the evaluation of the LO, NLO and NNLO diagrams of type-II, which has been further classified into type-IIa, type-IIb and type-IIc.
%The overall task involved is the same for all the loops and types.
\textit{As several steps are involved in this computation, the next few sections will discuss the steps required for each loop calculation.}

%\vspace{-5mm}
\subsection{FORM - The first step}

We use FORM to compute the matrix elements and their complex conjugates for each Feynman diagram of the process under consideration after generating them using QGRAF as described in section \ref{sec:computation}.

As a first check before proceeding, we multiply the computed matrix element and its conjugate at LO with the gluon polarisation condition
\begin{equation}
 \sum_{\lambda}\varepsilon_{\mu}\left(\lambda\right)\varepsilon_{\nu}^{*}\left(\lambda\right)=-g_{\mu\nu}+\dfrac{n_{\mu}k_{\nu}+n_{\nu}k_{\mu}}{\left(n.k\right)}-\dfrac{n^{2}k_{\mu}k_{\nu}}{\left(n.k\right)^{2}},
 \label{eq:gluonpol}
\end{equation}
where $k$ is the gluon momentum and $n_{\mu}$ is an arbitrary four-vector satisfying the condition $\varepsilon.n = 0$.
To check the gauge invariance of this result, we look for the absence of the arbitrary vector $n_{\mu}$ from the final polarised amplitude.
We also check for crossing symmetry between the Mandelstam variables $t$ and $u$.
After these two fundamental checks, we concluded that our Feynman rules and evaluation methods are correct.

Then we proceed with the computation of the scalar coefficients, $\mathcal{M}_{i}$ where $i=1,2$, by using the projectors given in eqn.\ \ref{eq:proj}.
We multiply the $P_1$ and $P_2$ projectors to each matrix element along with the gluon polarisation condition given in eqn.\ \ref{eq:gluonpol} for each diagram at each order separately.
The possible combinations are: $P_{1}\mathcal{M}_{1}$, $P_{1}\mathcal{M}_{2}$ and $P_{2}\mathcal{M}_{1}$, $P_{2}\mathcal{M}_{2}$.
We finally have two sets of results constituting the scalar coefficients corresponding to the two projectors.

%At the LO, no LiteRed reduction is required as no loop integrals are present.
%
%\textbf{The next few parts of this thesis will involve explaining the above steps order-by-order.}
%
%\subsubsection{LO Diagrams}
%\vspace{-11mm}

\subsection{LiteRed and Reduze - Reduction to MIs}
\label{subsec:LiteRed}

As already specified, we use LiteRed to reduce the large number of loop integrals to the few Master Integrals (MIs), along with REDUZE2, to get the auxiliary topologies.
We take the loop momenta to be $k_{1}$ and $k_{2}$, and all other kinematic relations are defined in the section \ref{sec:framework}.
Each integral family is constituted so that all possible dot products between the loop momenta and the independent external momenta can be determined from the elements of that family.
The number of elements in a family depends on the number of dot products one needs to evaluate for each set.
The propagators are assumed to be massless here.

\textit{At the LO, no LiteRed reduction is required as no loop integrals are present.
So, we begin with the details of one-loop diagrams.}

\subsubsection{NLO Diagrams}

At NLO, we have three independent external momenta $(p_1,p_2,p_3)$ and one loop momentum $(k_1)$.
So, there will be four elements in each integral family (which is equal to the number of dot products required) and a total of three families as defined below:
\begin{enumerate}
\item \textcolor{black}{$\left\{ k_{1},\ k_{1}-p_{_{1}},\ k_{1}-p_{_{1}}-p_{_{2}},\ k_{1}-p_{_{1}}-p_{_{2}}+p_{_{3}}\right\} $}
\item \textcolor{black}{$\left\{ k_{1},\ k_{1}-p_{_{2}},\ k_{1}-p_{_{2}}+p_{_{3}},\ k_{1}-p_{_{1}}-p_{_{2}}+p_{_{3}}\right\} $}
\item \textcolor{black}{$\left\{ k_{1},\ k_{1}+p_{_{3}},\ k_{1}-p_{_{1}}+p_{_{3}},\ k_{1}-p_{_{1}}-p_{_{2}}+p_{_{3}}\right\} $}
\end{enumerate}
The dot products involved in one-loop are:
$k_{1}.k_{1}$, $k_{1}.p_{_{1}}$, $k_{1}.p_{_{2}}$ and $k_{1}.p_{_{3}}$.
As is visible, all these dot products can be derived separately in each of the families.
Because of this, we call all these families independent of each other.
So, the dot products and scalars required at NLO can be obtained from each family.
The MIs obtained are according to the above-chosen integral families and the auxiliary topologies.

\subsubsection{NNLO Diagrams}

Next is the two-loop computation, which differs from the one-loop computation in terms of the choice of integral families.
There are three independent external momenta $(p_1,p_2,p_3)$ and two loop momenta $(k_1,k_2)$ in this case.
So, there will be nine elements in each family corresponding to the nine dot products that need to be determined and a total of six families as below:
\begin{enumerate}
\item \textcolor{black}{$\{ k_{1},\ k_{2},\ ,k_{1}-k_{2},\ k_{1}-p_{_{1}},\ k_{2}-p_{_{1}},\ k_{1}-p_{_{1}}-p_{_{2}},\ k_{2}-p_{_{1}}-p_{_{2}},\ k_{1}-p_{_{1}}-p_{_{2}}+p_{_{3}},$
\\$k_{2}-p_{_{1}}-p_{_{2}}+p_{_{3}}\} $}
\item \textcolor{black}{$\{ k_{1},\ k_{2},\ ,k_{1}-k_{2},\ k_{1}-p_{_{2}},\ k_{2}-p_{_{2}},\ k_{1}-p_{_{2}}+p_{_{3}},\ k_{2}-p_{_{2}}+p_{_{3}},\ k_{1}-p_{_{1}}-p_{_{2}}+p_{_{3}},$
\\$k_{2}-p_{_{1}}-p_{_{2}}+p_{_{3}}\} $}
\item \textcolor{black}{$\{ k_{1},\ k_{2},\ ,k_{1}-k_{2},\ k_{1}+p_{_{3}},\ k_{2}+p_{_{3}},\ k_{1}-p_{_{1}}+p_{_{3}},\ k_{2}-p_{_{1}}+p_{_{3}},\ k_{1}-p_{_{1}}-p_{_{2}}+p_{_{3}},$
\\$k_{2}-p_{_{1}}-p_{_{2}}+p_{_{3}}\} $}
\item \textcolor{black}{$\{ k_{1},\ k_{2},\ ,k_{1}-k_{2},\ k_{1}-p_{_{1}},\ k_{2}-p_{_{1}},\ k_{1}-p_{_{1}}-p_{_{2}},\ k_{2}-p_{_{1}}-p_{_{2}},\ k_{1}-p_{_{1}}-p_{_{2}}+p_{_{3}},$
\\$k_{1}-k_{2}+p_{_{3}}\} $}
\item \textcolor{black}{$\{ k_{1},\ k_{2},\ ,k_{1}-k_{2},\ k_{1}-p_{_{2}},\ k_{2}-p_{_{2}},\ k_{1}-p_{_{2}}+p_{_{3}},\ k_{2}-p_{_{2}}+p_{_{3}},\ k_{1}-p_{_{1}}-p_{_{2}}+p_{_{3}},$
\\$k_{1}-k_{2}-p_{_{1}}\} $}
\item \textcolor{black}{$\{ k_{1},\ k_{2},\ ,k_{1}-k_{2},\ k_{1}+p_{_{3}},\ k_{2}+p_{_{3}},\ k_{1}-p_{_{1}}+p_{_{3}},\ k_{2}-p_{_{1}}+p_{_{3}},\ k_{1}-p_{_{1}}-p_{_{2}}+p_{_{3}},$
\\$k_{1}-k_{2}-p_{_{2}}\} $}
\end{enumerate}
The dot products in two-loop are: $k_{1}.k_{1}$, $k_{1}.p_{_{1}}$, $k_{1}.p_{_{2}}$, $k_{1}.p_{_{3}}$, $k_{2}.k_{2}$, $k_{2}.p_{_{1}}$, $k_{2}.p_{_{2}}$, $k_{2}.p_{_{3}}$
and $k_{1}.k_{2}$.
Here also, it is possible to derive all these dot products separately from each family.
Finally, the dot products and scalars required at NNLO are defined in terms of the above chosen integral families and the auxiliary topologies so that the MIs obtained are per our choice.

\subsection{Mathematica - Obtaining the finite result}

After evaluating the matrix elements, obtaining the MIs, and expressing the scalar coefficients in terms of these MIs, we obtain the finite result by simplifying them and making them free of all divergences (UV and IR).
This calculation is done with the help of Mathematica and other related packages.

%We have explained what needs to be done at each level separately below.
First, all the scalar coefficients, $\mathcal{M}_{i}$ where $i=1,2$, of all the possible Feynman diagrams, are summed order-by-order for $P_{1}$ and $P_{2}$ separately. %and save them in different files.
This summation is done with the help of in-house Mathematica scripts, which also substitute the Mandelstam variables in terms of the dimensionless variables $(x, y, z)$ and the dimension $d = 4 + \varepsilon$.
After all the diagrams are combined, we perform the UV renormalisation.
For simplicity and reducing the program runtime, we reduce the number of variables involved in computation by expressing the Mandelstam variable $u$ in terms of $s$, $t$ and $m_A$ using the relations in subsection \ref{subsec:kinematics}.
Each loop computation involves a specific set of MIs obtained using the Mathematica-based package LiteRed and depended on the scalar products and chosen basis (as discussed in subsection \ref{subsec:LiteRed}).
We obtain the final analytical expressions for each projector that can be renormalised after substituting the results of these MIs, order-by-order.

However, the $\mathcal{M}_{i} '$s at two-loop are too huge for direct substitutions and hence, we simplify them further.
To achieve this, the MIs obtained at two-loop are mapped to a set of known MI results by Gehrmann \cite{Gehrmann:2013cxs,Gehrmann:2014bfa} sharing the same topology.
To summarise, the summed analytic result at each order for each projector has all the scalar coefficients and MIs involved.
At NNLO, the MIs in the expressions are first mapped and then substituted.

%\section{Calculational details}
%\label{sec:calcDetails}
\subsection{Programmatical Details for renormalising}

Along with all the above details, a few other substitutions and assumptions are made to obtain the finite result numerically.
They are discussed below.

\subsubsection{The Wilson Coefficients}

The first set of equations in eqn.\ \ref{eq:operatorSandwichAmplitude} are from the operator $\left[O_{G}O_{G}\right]$ and the second set corresponds to the operator $\left[O_{G}O_{J}\right]$.
The operator $\left[O_{G} O_{G}\right]$ gives an overall $C_{G}^{2}$ coefficient while the operator $\left[O_{G}O_{J}\right]$ gives an overall $C_{G} C_{J}$ coefficient.
These Wilson coefficients are overall factors to the two sets of equations and can be taken out.
We choose them to be $1$ for convenience.
As a consequence, instead of expanding up to $\mathcal{O}\left(a_{s}^{4}\right)$, we expand the perturbative series up to $\mathcal{O}\left(a_{s}^{2}\right)$.
The corresponding Wilson coefficients are taken out as an overall factor for the $\left[O_{G} O_{G}\right]$ and $\left[O_{G} O_{J}\right]$ vertices appropriately.

\subsubsection{The change of observables to cos $\theta$}

For numerical calculations, it is convenient to express the analytic results using the two observables, $\theta$ and $x$.
$\theta$ is related to the Mandelstam variables and the dimensionless variable $z$ depends on $x$, as shown in eqn.\ \ref{eq:costheta}.
We rewrite the analytical result in terms of $\theta$ and $x$
%If we substitute both $x$ and $z$ simultaneously, there is a loss of consistency involved because $z$ is dependant on the value of $x$.
%$\cos\theta$ and $x$
as they are the only independent observables available.
We use the following substitutions to obtain the result in terms of $\cos\theta$ and $x$:
\begin{align}
y & = \dfrac{(1 + x^2)}{x} - z, \nonumber \\
z & = -\dfrac{u}{m_A^2}, \nonumber \\
u & = 2 m_A^2 - s - t, \nonumber \\
s & = m_A^2 \dfrac{(1+x)^2}{x}, \nonumber \\
t & = m_A^2 - \dfrac{s}{2} + \dfrac{s}{2} \beta \cos \theta, \nonumber \\
\beta & = \sqrt{1-\dfrac{4 m_A^2}{s}}, \nonumber \\
Z & = z + 1, \nonumber \\
Y & = y + 1.
\label{eq:costheta}
\end{align}
The first four equations have already been defined before in subsection \ref{subsec:kinematics}.
The redefinition of the Mandelstam variable $t$ in terms of $\cos\theta$ and $\beta$ is for a two-body system with the same mass of the outgoing particles, which in our case is the di-pseudoscalar Higgs boson.
For redefining the Mandelstam variables, we have used the following sets of equations:
\begin{align}
s &= Q^2, \nonumber \\
t &= m_1^2 - \dfrac{s}{2}\bigg(1+\dfrac{m_1^2}{s}-\dfrac{m_2^2}{s}\bigg) + \dfrac{s}{2} ~ \beta ~ \cos \theta, \nonumber \\
u &= m_1^2 + m_2^2 - s - t,
\end{align} where
\begin{align}
\beta &= \sqrt{1+\dfrac{m_1^4}{s^2}-\dfrac{2~m_1^2~m_2^2}{s^2}+\dfrac{m_2^4}{s^2}-\dfrac{2~m_1^2}{s}-\dfrac{2~m_2^2}{s}},
\end{align}
with $m_1 = m_2 = m_A$ for our case.
Initially, our result was in the dimensionless variables $x$ and $z$, which we change to $\cos\theta$ and $x$ using eqn.\ \ref{eq:costheta} for convenience.

\subsection{The Final Results}
\label{sec:result}

We obtain our finite result, after UV renormalisation and IR factorisation,  as two different coefficients, $\mathcal{M}_i$'s, with $i=1,2$.
Below I briefly describe the humongous computational complications involved in this work.

\textbf{Projector 1: }
The file size of the final result is about 71 MB for projector 1.
For convenience, we divided the result into four different colour structures - $Nn_{f}$, $n_{f}^{2}$, $N^{2}$ and $n_{f}/{N}$.
The file containing the coefficient of $N^{2}$ is about 70.6 MB alone.
To simplify this huge result, we divided it into five different transcendalities, as discussed in section \ref{subsec:IR}.
%The $n_{f}^{2}$ colour coefficient has file size 693 bytes. The $Nn_{f}$ colour coefficient has a file size of 2.9 MB. The $\dfrac{n_{f}}{N}$ colour coefficient has a file size of 3.8 MB.

\textbf{Projector 2: }
The file size of the finite part is approximately 133 MB for projector 2.
We again divided the result into four different colour structures as in section \ref{subsec:IR}.
%- $Nn_{f}$, $n_{f}^{2}$, $N^{2}$ and $\dfrac{n_{f}}{N}$. Any other colour structure is either a subset of these four or the coefficient of that colour structure is zero like that of $\dfrac{1}{N^{2}}$.
The file containing the coefficient of $N^{2}$ is again huge.
So, it had to be further split into the different transcendalities (LT$_0$, LT$_1$, LT$_2$, LT$_3$ and LT$_4$).
%The $n_{f}^{2}$ colour coefficient has file size 1.1 kB. The $Nn_{f}$ colour coefficient has a file size of 12.4 MB. The $\dfrac{n_{f}}{N}$ colour coefficient has a file size of 13.8 MB. Thus, we further divided these into the five transcendalities, which did not create much difference. As a result, we could only check the LT$_2$ pole terms analytically.

\begin{figure}[!htb]
	\centering
 \includegraphics[scale=0.18]{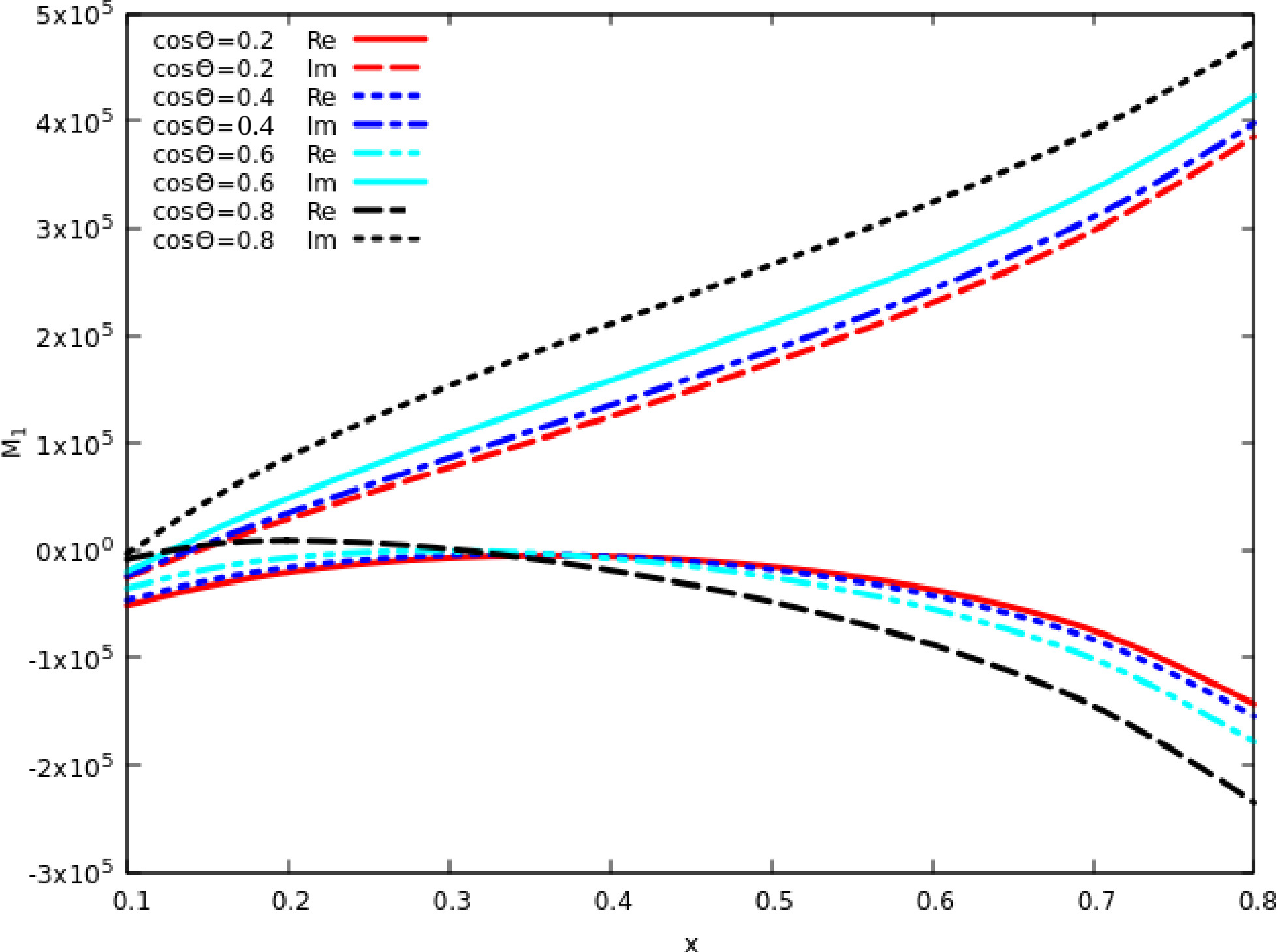}
 % P1_REAL_IMAG_Latex.eps: 1958x1458 px, 300dpi, 16.58x12.34 cm, bb=3 4 473 354
 \caption[Amplitude $\mathcal{M}_1 = \mathcal{M}_{1,GG}^{(2),fin}/\mathcal{M}_{1,GG}^{(0)}$ for Projector 1 with different $\cos{\theta}$ values.]{Amplitude $\mathcal{M}_1 = \mathcal{M}_{1,GG}^{(2),fin}/\mathcal{M}_{1,GG}^{(0)}$ for Projector 1 with different $\cos{\theta}$ values. The upper lines represent the Imaginary (Im) values, and the lower lines the Real (Re) values.}
 \label{fig:Proj1Num}
\end{figure}

\begin{figure}[!htb]
	\centering
 \includegraphics[scale=0.75]{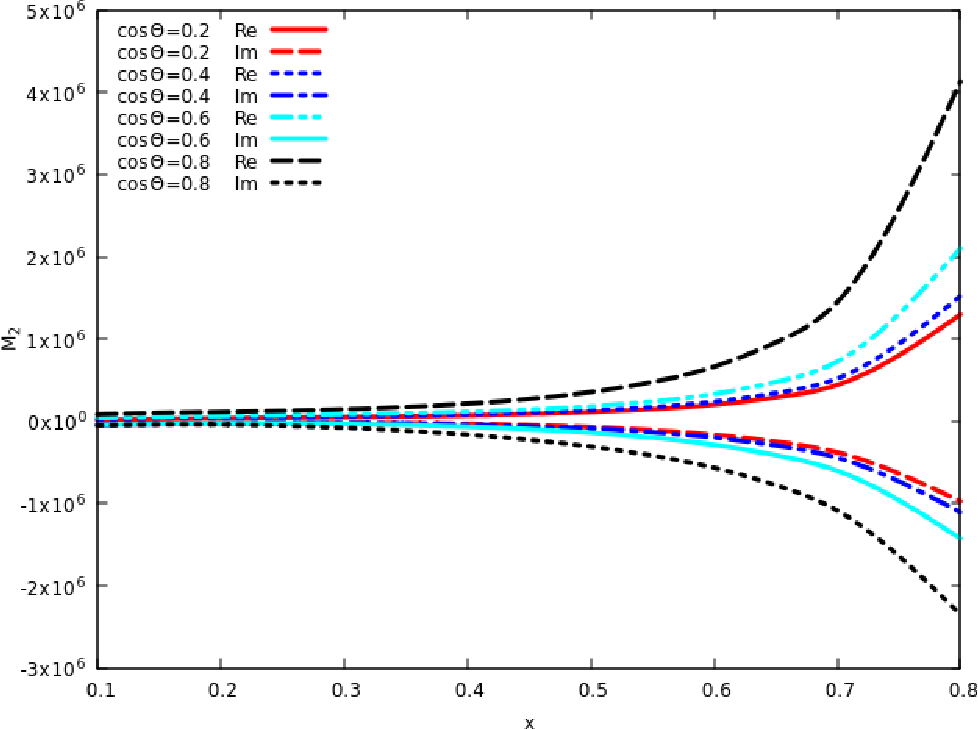}
 % P2_REAL_IMAG_Latex.eps: 1958x1458 px, 300dpi, 16.58x12.34 cm, bb=3 4 473 354
\caption[Amplitude $\mathcal{M}_2 = \mathcal{M}_{2,GG}^{(2),fin}/\mathcal{M}_{2,GG}^{(0)}$ for Projector 2 with different $\cos{\theta}$ values.]{Amplitude $\mathcal{M}_2 = \mathcal{M}_{2,GG}^{(2),fin}/\mathcal{M}_{2,GG}^{(0)}$ for Projector 2 with different $\cos{\theta}$ values. Here the line trend is just opposite to that of the diagram above.}
\label{fig:Proj2Num}
\end{figure}

\textit{After these simplifications, renormalisation and factorisation, the corresponding results are combined to obtain the numerical data depicted in the figs.\ \ref{fig:Proj1Num} and \ref{fig:Proj2Num}.
The LO and NLO analytical expressions of the matrix elements for Type-IIa diagrams are presented in the appendices \ref{appendix:D} and \ref{appendix:E}.
The Born level analytical expressions of the matrix elements for Type-IIb diagrams are presented in the appendices \ref{appendix:F}.}

%\subsubsection{The $\mathcal{M}_{GG}^{II(2)}$ finite part details}

\section{Discussion and Conclusions}
\label{sec:ggAAConc}

In this work, we have presented the two-loop virtual amplitudes relevant for studying the production of a pair of pseudoscalar Higgs bosons \textit{via} the gluon fusion subprocess at the LHC.
This is the dominant sub-process that is sensitive to its self-coupling.
We perform this computation in the EFT framework where the top quark DOF is integrated out.
In our work, the results are obtained in the limit of small $\tan\beta~(\sim 1)$, so that quark Yukawa couplings and masses are related by flavour independent proportionality factor.
Consequently, we treat light quark flavours as massless with vanishing Yukawa couplings and integrate the virtual heavy top quark DOF.
Finally, we are left with interactions of the CP-odd scalar (pseudoscalar) boson, $A$,  with the gluons and light quarks.
We also consistently neglect the Yukawa couplings of light quarks to the pseudoscalar Higgs boson.
For large values of $\tan\beta$, the top Yukawa couplings of the pseudoscalar Higgs boson are suppressed compared to the bottom ones.
So our result is valid for the pseudoscalar Higgs boson of MSSM and 2HDM with small $\tan\beta$ by adjusting the top Yukawa coupling, which appears as an overall factor.

In EFT, the pseudoscalar Higgs boson directly couples to gluons and light quarks through two local composite operators, $O_G$ and $O_J$, respectively, with the strengths proportional to the Wilson coefficients that are calculable in pQCD.
We used dimensional regularisation to regulate both UV and IR divergences.
The composite operators, being CP-odd, contain Levi-Civita tensor and $\gamma_5$.
These inherently four-dimensional objects require careful treatment to deal with them in $d$-dimensions.
Although we followed the prescription advocated by Larin, other procedures may also be used consistently, provided the Ward identities are restored manually with finite counter terms if needed.
%This is also required in Larin's scheme.
This is because, in a purely anti-commuting scheme, it was claimed by Kreimer \textit{et. al.} that a non-cyclic prescription with symmetrisation rules can be used such that no counter terms are needed.
However, since Larin's scheme avoids complicating the loop integrals, it has been used in almost all higher-order calculations of QCD corrections for computational convenience.
This treatment requires additional renormalisation for the singlet axial vector current up to two loops.
In addition, Larin's prescription requires a finite renormalisation constant for singlet axial current, which is also available.

Note that the composite operators mix under UV renormalisation.
The corresponding renormalisation constants are already known. We use them to obtain the UV finite amplitudes up to two-loop.
Unlike the amplitudes involving a pair of Higgs bosons, we do not need any UV contact counter terms here.
The UV finite amplitudes thus obtained contain IR divergences due to massless partons in QCD.
We found that these IR poles agree with Catani's predictions, analytically for poles $\varepsilon^{−i}$ with $i = 2 − 4$, and numerically to high precision for $\varepsilon^{-1}$.
The agreement of the IR pole structure in our pseudoscalar Higgs boson computation results with Catani's universal IR pole structure provided a test on the correctness of our computation.
Our results provide one of the crucial components relevant for studies related to the production of a pair of pseudoscalar Higgs bosons at the LHC up to order ${\cal O}(a_s^4)$.

\label{sec:conc}

\clearpage
\newpage
\mbox{~}

\chapter{Next-to-soft-virtual resummed prediction for pseudoscalar Higgs boson production at NNLO$+\overline{\text{NNLL}}$}
\label{chap:ggA}

\textbf{\textit{This chapter is based on the results of an original research work done in collaboration with M. C. Kumar, Prakash Mathews and V. Ravindran, and elaborates the published article \cite{Bhattacharya:2021hae}.}}

\section{Prologue}
\label{sec:prologue2}

\thr{
The discovery of the $125$ GeV mass Higgs boson in 2012 has put the SM on a firm footing.
As a result, a lot of work has been going on to investigate the properties and interactions of this discovered Higgs boson with the other
SM particles \cite{Higgs:1964ia,PhysRevLett.13.508,PhysRev.145.1156,PhysRevLett.13.321,PhysRevLett.13.585,ATLAS:2013sla,ATLAS:2019nkf,2013,2015}.
Despite such phenomenal successes, it is widely known that the SM fails to explain certain natural phenomena as already discussed in section \ref{sec:SMshortcomings}.
In order to find explanations for these phenomena, one has to go beyond the realm of the SM and supersymmetric theories are expected to address these shortcomings.
%In order to explain these phenomena, one has to go beyond the realm of the SM and supersymmetric theories are expected to provide solutions to the shortcomings of the SM.
One of the most straightforward extensions of the SM is the MSSM which has five Higgs bosons --- two neutral scalars $(h,H)$, one pseudoscalar $(A)$ and two charged scalars $(H^\pm)$.
The pseudoscalar Higgs boson $(A)$ which is CP odd could be as light as the discovered Higgs boson.
Hence, a dedicated effort has been going on to determine the CP property of the discovered Higgs boson, and to identify it with that of the SM, although there are already indications that it is a scalar with even parity \cite{2013,2015,CMS:2012vby,ATLAS:2015bcp,ATLAS:2015zhl,CMS:2017len,ATLAS:2017azn}.
}

Dedicated experimental searches for the pseudoscalar Higgs boson by the CMS and ATLAS collaborations have been carried out from the LHC data for $8$ TeV as well as $13$ TeV proton-proton collisions
\cite{CMS:2014ccx,ATLAS:2014vhc,ATLAS:2017nxi,Touquet:2019kho,Tsuno:2020jpd,ATLAS:2016ivh,CMS:2015grx,ATLAS:2017snw,CMS:2019pzc,ATLAS:2017eiz}.
%\cite{CMS:2020imj,}
%\cite{ATLAS:2015kpj,CMS:2015uzk,}
The searches for heavy scalar resonances decaying into a pair of $\tau$ leptons from the LHC 8 TeV data by CMS \cite{CMS:2014ccx} and ATLAS \cite{ATLAS:2014vhc} have excluded the values of $\tan \beta$ higher than $6.3~(57.6)$ for $m_A=100~(1000)$ GeV at 95\% confidence level (CL).
The experimental searches in the decay channel $A \to \tau \tau$ \cite{ATLAS:2016ivh}
have kept more stringent limits on the pseudoscalar parameter space than those obtained from the $A \to b \bar{b}$ channel \cite{CMS:2015grx}.
The searches for pseudoscalar resonances in the top pair production by the ATLAS collaboration
using data from the $8$ TeV LHC have put upper limits on $\tan \beta$ of the order of unity for $m_A > 500$ GeV \cite{ATLAS:2017snw}.
Similar experimental searches by the CMS collaboration using data from $13$ TeV LHC have probed the pseudoscalar mass from $400$ GeV to $750$ GeV resulting in the exclusion of $\tan \beta$ values below $1.0$ to $1.5$ depending on the value of $m_A$ at 95\% CL \cite{CMS:2019pzc}.
The recent experimental searches for heavy resonances in the di-tau channel from the $13$ TeV LHC data by the ATLAS collaboration have excluded $\tan \beta > 1.0~(48)$ for $m_A = 250$ GeV ($1.5$ TeV) at 95\% CL \cite{ATLAS:2017eiz}.
The higher-order corrections through the NNLO K-factor of about $2.0$ have been used in all these analyses.
The large size of these corrections suggests that higher-order precision calculations through resummation can be useful in the experimental search for heavy scalar resonances.
These studies, which have seen excesses over the background expectation in the Run II data of the LHC, indicate that extensive attempts are going on to discover the BSM Higgs bosons with masses higher than that of the discovered Higgs boson, $m_H=125$ GeV.
With such intense experimental works under process, developing the corresponding theoretical background has become necessary and motivated us for this work.

There is also a possibility that the observed Higgs boson of $125$ GeV mass is an admixture of scalar and pseudoscalar states.
If this is true, then gluon fusion can produce such a mixed scalar in hadron collisions.
Such a possibility has already been explored in \cite{Artoisenet:2013puc,Gao:2010qx,Maltoni:2013sma}.
Identifying a mixed scalar-pseudoscalar state is possible by studying various kinematic distributions of the particles that this mixed state decays into \cite{Artoisenet:2013puc}.
This requires the availability of fully differential distributions, and such a study has been taken up at the NNLO level in \cite{Jaquier:2019bfs}.
The existence of a mixed state indicates possible new physics and hence, the CP violation in the Higgs sector.
Moreover, it can also explain the origin of CP violation in the SM and address the baryogenesis problem \cite{PhysRevD.106.055018,Khoze:2017tjt}.
All these imply the requirement of a detailed study for establishing the spin-parity properties of the discovered scalar boson of $125$ GeV mass.
From the theory side, this necessitates precision studies for the relevant observable corresponding to scalar and pseudoscalar production processes to the same order of precision.

%The last decade has seen phenomenal progress in the experimental field.
%This has initiated an ever-growing need for extremely precise QCD calculations to match theory with experimental data.
Higher-order corrections in perturbative QCD (pQCD) provide a way to achieve the required precision.
The pseudoscalar production cross-sections are available
%involves the study of pseudoscalar particles at the LHC in the large top quark mass limit that has already started in the past
%decade and certain predictions
to NNLO accuracy in QCD \cite{Harlander_2002,Anastasiou_2003,Ravindran_2003}.
The corrections are large and are of the order
of 67$\%$ at NLO and get increased by an additional 15$\%$ at the NNLO level for the renormalisation and factorisation scales
set to $\mu_R=\mu_F=m_A/2$ for the pseudoscalar mass $m_A = 200$ GeV. The large size of these corrections implies that corrections of even higher orders are necessary to achieve precise theoretical results.

A pseudoscalar Higgs production through the gluon fusion channel \textit{via} quark loop and the corresponding LO results for finite quark mass (exact) dependence are already available
\cite{Dawson:1990zj,Djouadi:1991tka,Spira:1995rr,Ravindran:2003um}.
The calculation becomes simpler in the EFT framework where the DOF of the heaviest quark is integrated out.
Consequently, cross-section computations in this framework at higher orders become plausible.
%The effective theory approach is done in the large top quark mass limit, and because of the considerable decrease in the complexity, it provides an opportunity to go beyond NLO.
\thr{
This EFT approach in the case of scalar Higgs boson production  \cite{Anastasiou_2002,Harlander_2002h,Ravindran_2003} became highly successful as the difference between the exact and EFT results at NNLO level was found to be within 1\% \cite{Harlander_20091,Harlander_20092,Pak_2010,2021_Harlander}.
An important point to be mentioned here is that there are indications that this high precision works well for inclusive cross-section results.
However, the uncertainties can be somewhat larger in differential results \cite{Grazzini:2018bsd}.
}
For pseudoscalar production at the hadron colliders, NNLO predictions in the EFT approach are also available \cite{Ravindran_2003,Harlander_2002,Anastasiou_2003}.

In \cite{Anastasiou:2014vaa,Anastasiou:2016cez}, the computation of complete N$^{3}$LO predictions for the scalar Higgs boson production through gluon fusion at the hadron colliders in the effective theory was accomplished.
These third-order corrections increase the cross-section by about $3.1\%$ for the central scale choice of $m_H/2$ while the corresponding scale uncertainty reduces to as small as below $2\%$.
In a recent study for the neutral current Drell-Yan process, the complete N$^3$LO results were calculated for the first time \cite{Duhr:2021vwj}.
The corresponding cross-section is found to be about $0.992$ times that of the NNLO cross-section for the invariant mass region $Q=300$ GeV, indicating a small negative correction from the third order in this kinematic region.
Although the scale uncertainty at the N$^3$LO level is found to be very mild, it has been observed that the scale uncertainty bands for N$^3$LO and NNLO level cross-sections do not overlap.
The next step in the process is the computation of N$^{3}$LO cross-sections for the pseudoscalar production through gluon fusion.
The first task in this direction is to obtain the threshold enhanced cross-section at N$^{3}$LO level, which has already been computed in \cite{Ahmed:2016otz}.
Further, the approximate full N$^3$LO$_A$ results are also available for the pseudoscalar Higgs boson case in \cite{2016}.

Until recently, FO calculations have been done at NLO, NNLO and sometimes even N$^3$LO, depending on the complexity of the process under consideration.
However, these FO QCD predictions have limited applicability because of the presence of large logarithmic contributions that arise in the threshold region of the kinematic phase space.
At the threshold, the emission of soft gluons gives large logarithmic contributions to the cross-section when the partonic COM energy approaches the mass of the outgoing particle.
If these large logarithms arising from the soft gluons can be resummed to all orders in perturbation theory, then the problem of spoiling the reliability of FO perturbative predictions can be solved.
We denote the soft-plus-virtual (SV) resummed results by LL, NLL, etc.
The next-to-next-to-leading logarithmic (NNLL) resummed results \cite{2003,Moch:2005ky,2006,Ravindran:2005vv,Ravindran_2006,2006PRD,2009,2009PRL,2016PRD} give a sizable contribution and reduce scale uncertainties.
These logarithms in the parton level cross-section $d\hat{\sigma}$ computed to order $\alpha_s^k$ will appear as
\begin{equation}
 {\alpha_{s}}^k\left[\dfrac{\ln^{i}\left(1-z\right)}{1-z}\right]_{+}\text{, where }i<2k-1.
 \label{eqn:PlusFunc}
\end{equation}
Here, the subscript $+$ denotes the plus distribution, and $z=m_A^2/s$ with $z\rightarrow1$ limit represents the partonic threshold region.
$m_A$ is the corresponding pseudoscalar mass.
\thr{
Details on the Plus function is given in \ref{appendix:G} along with the generic integrals appearing in the exponents of the threshold resummations.
}
The precise contribution of these parton-level logarithms to the hadron-level cross-section depends on the corresponding parton fluxes in that region.
For the case of the discovered Higgs mass region of $(125$ GeV$)$, the associated gluon flux is large, and hence, these threshold logarithms due to the soft and collinear gluons are also found to be significant.
This way of resumming a set of large logarithms and matching them to the FO results can give robust theoretical predictions.

%These imaginably large logarithmic corrections appear in the coefficient functions $\sigma^{N{^k}LO}$ at every order in perturbation theory.
%When they are dominant, pQCD can only be relied upon if these logarithms are taken into account to all orders, and threshold corrections comprise an important type.
%Resummation of threshold logarithms arising from soft gluon emissions can further improve the results.
%Thus, resummation will lead to a more reliable result of the SM Higgs cross-section.

For scalar Higgs boson production, the FO results are available at N$^3$LO level
\cite{Anastasiou_2015,Anastasiou:2016cez} and the corresponding threshold resummed results have been computed to N$^3$LL accuracy \cite{Bonvini:2014joa,Bonvini_2016_TROLL}.
The computation of the three-loop threshold corrections for pseudoscalar Higgs boson production in the threshold limit has been done in \cite{Ahmed_2015}.
Knowledge of the form factors up to the three-loop level is needed to calculate the threshold corrections for pseudoscalar production to the same level of accuracy.
Further, the UV and IR divergences give rise to most of the logarithms at higher orders in the intermediate stages of the computations.
These logarithms depend on the renormalisation and factorisation scales and are present in the perturbative expansions.
Such logarithms help to estimate the error in theoretical predictions resulting from the truncation of the perturbation theory to a finite order.

%that occur as we truncate the perturbative series because the lesser the dependence on these scales, the more the reliability of the truncated result.
%Logarithms depending on the physical scales or other scaling variables, can sometimes be large enough to ruin the dependability of the truncated perturbative result.
%These logarithms can be systematically summed to all orders as their structure depends on the anomalous dimensions of IR renormalisation group equations. This is called resummation.
%The distributions $\mathcal{D}_i(z)$ that dominate in the threshold region, namely as $z\rightarrow1$, are of the form $\mathcal{D}_i(z)=\left[\dfrac{\ln^{i}\left(1-z\right)}{1-z}\right]_{+}$ and are called threshold logarithms. The terms comprising these distributions and $\delta(1-z)$ are named soft plus virtual (SV) contributions.
%These logarithms convolute with appropriate PDFs, resulting in the hadronic cross-section and giving enormous contributions at every order.
%Because of these large corrections at every order, the dependency of the predictions from the truncated series is somehow ruined.

There are various studies on the computation of SV results in QCD corresponding to a number of observables produced in hadron collisions.
For SV results up to third order, see \cite{Moch:2005ky,Ravindran_2006,de_Florian_2012,Ahmed:2014cla,Catani_2014,Ravindran:2005vv,Ahmed:2014cha,Kumar:2014uwa,PhysRevD.90.053006}.
A series of works in this direction have been carried out for the resummation of the threshold logarithms following the path-breaking works by Sterman \textit{et al.} \cite{STERMAN1987310} and Catani \textit{et al.} \cite{CATANI1989323}.
See \cite{CATANI1996273,Moch:2005ba,Bonvini:2012an,Bonvini:2014joa,Bonvini:2014tea,Bonvini_2016_TROLL} for Higgs production in gluon fusion,
\cite{Bonvini_2016,H:2019dcl} for bottom quark annihilation and \cite{Moch:2005ba,bonvini2010threshold,bonvini2012resummation,h2020resummed,Catani_2014} for Drell-Yan (DY).
In $z$-space, one has to deal with the convolutions of these distributions.
\thr{
So, one moves to the Mellin ($N$-moment) space approach, which uses the conjugate variable $N$ for resummation.
The advantage of working in the Mellin space is that the convolutions become normal products in the $N$-space.
The distributions $\mathcal{D}_i(z)$ become ordinary (or regular) functions of $\text{log}^{i+1}(N)$ where we suppress terms of $\mathcal{O}(1/N)$ in the threshold limit of $N\rightarrow{\infty}$.
}

However, it has also been observed that threshold corrections alone cannot replace the full FO results at higher orders in QCD. It is found that the role of the next-to soft-plus-virtual (NSV) terms, namely $\text{log}^{i}(1-z),~i=0,1,\cdots$, are also important.
These NSV contributions can, in principle, originate from parton channels other than the one corresponding to the Born contribution.
It is also essential to see whether these NSV terms can be resummed systematically to all orders exactly like the leading SV terms.
Several advancements in this direction have been made \cite{Laenen:2008ux,Laenen:2010uz,Bonocore:2014wua,Bonocore:2015esa,Bonocore:2016awd,DelDuca:2017twk,Bahjat-Abbas:2019fqa,Moch:2009hr,Beneke:2018gvs,Beneke:2019mua,Beneke:2019oqx}.
In \cite{ajjath2020soft,Ajjath:2022kyb}, the theories of mass factorisation, renormalisation group invariance and Sudakov K-plus-G equation have been exploited to provide a result in $z$- and $N$-space to predict the NSV terms for DY and Higgs boson production to all orders in perturbation theory.

In \cite{2021}, the NSV resummation has been achieved to Leading Logarithmic ($\overline{\text{LL}}$) accuracy for colour singlet production processes in hadron collisions.
It has been observed that the resummation of NSV logarithms in the diagonal channel gives large contributions to the cross-sections, while those from the $qg$ channel are found to give negative contributions of about $3\%$ in the high mass region.
\thr{
The NLP threshold corrections were seen to have a notable impact on the cross-section results, and therefore, is a motivation to further develop the understanding of NLP corrections.
%It has also been noticed that including NSV resummed results increase scale uncertainties obtained from the SV resummation.
%
However, the uncertainty bands were observed to increase upon including NLP LL effects for both dQCD and SCET in this work on colour-singlet cross-sections and the authors expect this to be reversed once NLP NLL contributions are included.
}

In this article, we computed the NSV corrections for the pseudoscalar Higgs boson production process at N$^3$LO level based on the formalism developed in \cite{Ravindran_2006,Ravindran:2005vv,ajjath2020soft,Ajjath:2022kyb}.
We further study the phenomenological impact of resumming these NSV logarithms to
$\overline{\text{NNLL}}$ accuracy after being systematically matched to the FO NNLO ones.
We represent the resummed results for the production of pseudoscalar Higgs boson at leading logarithmic, next-to-leading logarithmic and next-to-next-to-leading logarithmic accuracy by $\overline{\text{LL}}$, $\overline{\text{NLL}}$ and $\overline{\text{NNLL}}$, when we take into
account both SV and NSV threshold logarithms together.
As a start, in table \ref{tab:SVNSVratio}, I present a comparison of the SV and SV+NSV corrections at $\mathcal{O}(a_s^2)$ for a pseudoscalar Higgs boson production \textit{via} gluon fusion to depict the significance of the SV+NSV corrections.
In table \ref{tab:SVNSVratio}, $m_A$ depicts the pseudoscalar mass, NNLL the SV corrections at $a_s^2$ order, $\overline{\text{NNLL}}$ the corresponding SV+NSV corrections.
NNLO represents the FO contribution to the same level of accuracy.
The magnitude of implications these NSV corrections can have on the total cross-section is evident from the table \ref{tab:SVNSVratio}.
Hence, determining the role of these sub-leading terms, appearing as $\ln^{i}\left(1-z\right),i=0,1,\cdots$, in phenomenological computations is substantial.

The theoretical basis for this work has already been discussed in subsection \ref{subsec:PseudoLag}.
Hence, we structure this chapter as below.
In the next section \ref{sec:TC}, I will discuss the structure of the threshold corrections, and the formalism \cite{ajjath2020soft,Ajjath:2022kyb} we have used for this work in detail.
This involves a detailed explanation of the fundamental building blocks contributing to the cross-section in the threshold region for the pseudoscalar Higgs boson production.
In section \ref{sec:NSVres}, I will present the analytical results after resumming the NSV logarithms to $\overline{N^3LL}$ accuracy and explain how to correlate these results with that of the available scalar Higgs using \cite{2016}.
In section \ref{sec:resNSVformalism}, I will briefly recollect the NSV resummation formalism developed in \cite{ajjath2020soft,Ajjath:2022kyb} in the Mellin space followed by the numerical results for the pseudoscalar Higgs boson production in section \ref{sec:Numerical}.
Then I will conclude this chapter in section \ref{sec:Summary}.

\begin{table}[!htbp]
 \centering
 \renewcommand{\arraystretch}{1.5}
 \begin{tabular}{|l||r||r|}
 \hline
 \textbf{ $m_A$ (in GeV)}
 & \textbf{NNLL/NNLO (in \%)}
 & \textbf{$\overline{\text{NNLL}}/$NNLO (in \%)}
 \\ \hline \hline
 125                       & 11.8189   &   17.0234 \\ \hline
 700                       & 12.8902   &   15.8511 \\ \hline
 1000                      & 13.2377   &   16.2727 \\ \hline
 1500                      & 14.8419   &   18.4658 \\ \hline
 2000                      & 16.5992   &   21.0971 \\ \hline
 2500                      & 18.5168   &   24.1535 \\ \hline
 \end{tabular}
\caption{SV and SV+NSV corrections normalized \textit{w.r.t} FO results at $\mathcal{O}(a_s^2)$ for pseudoscalar Higgs boson production \textit{via} gluon fusion.}
\label{tab:SVNSVratio}
\end{table}

\section{Threshold Corrections}
\label{sec:TC}

The inclusive cross-section for a pseudoscalar Higgs boson production at the hadron colliders can be computed using \cite{Ahmed:2016otz}
\begin{equation}
\sigma^{A}\left(\tau,m_{A}^{2}\right)=\sigma^{A,\left(0\right)}\left(\mu_{R}^{2}\right)\sum_{a,b=q,\bar{q,g}}\int_{\tau}^{1}dy\ \Phi_{ab}\left(y,\mu_{F}^{2}\right)\Delta_{ab}^{A}\left(\dfrac{\tau}{y},m_{A}^{2},\mu_{R}^{2},\mu_{F}^{2}\right),
\label{eq:sigmaA}
\end{equation}
where $\sigma^{A,\left(0\right)}\left(\mu_{R}^{2}\right)$ is the
born cross-section at the parton level with finite top mass dependence and is given by
\begin{equation}
\sigma^{A,\left(0\right)}\left(\mu_{R}^{2}\right) = \frac{\pi\sqrt{2}G_F}{16} a_s^2 \cot^2{\beta} |\tau_A f(\tau_A)|^2.
\end{equation}
In the above equation, $\tau_A=4m_t^2/m_A^2$.
The function $f(\tau_A)$ can be expressed as
\begin{equation}
f(\tau_A) =
     \begin{cases}
       \arcsin^2 \frac{1}{\sqrt{\tau_A}}&\quad\tau_A\geq1;\\
       -\frac{1}{4}\bigg(\ln \frac{1-\sqrt{1-\tau_A}}{1+\sqrt{1-\tau_A}}+i\pi\bigg)^2 &\quad\tau_A<1. \\
     \end{cases}
\end{equation}
The parton flux is given by
\begin{equation}
\Phi_{ab}\left(y,\mu_{F}^{2}\right)=\int_{y}^{1}\dfrac{dx}{x}f_{a}\left(x,\mu_{F}^{2}\right)f_{b}\left(\dfrac{y}{x},\mu_{F}^{2}\right),
\end{equation}
where $f_{a}$ and $f_{b}$ are PDFs with $a$ and $b$ representing the initial state partons, renormalised at the factorisation scale $\mu_{F}$.
Here, $\Delta_{ab}^{A}\left(\tau/y,m_{A}^{2},\mu_{R}^{2},\mu_{F}^{2}\right)$ represents the parton level cross-section for the sub-process initiated by $a$ and $b$ partons.
This is the final result obtained after performing the UV renormalisation at scale $\mu_R$ and the mass factorisation at scale $\mu_F$.

The chief aim of this article is to study how the contribution from soft gluons affects the cross-section for pseudoscalar production at hadron colliders.
We  get the final result that is IR-safe by adding the soft part of the
cross-section to the UV renormalised virtual part.
However, more than this is needed as mass factorisation is needed to be done using appropriate counter terms.
This combination is known as the soft-plus-virtual (SV) cross-section; the remaining part is known as the hard part.
While the resummed results provide reliable predictions that can be compared against the experimental data, it is crucial to find the role of sub-leading terms, namely $\log^{i}\left(1-z\right)$, $i=0,1,\cdots$.
We call them next-to-SV (NSV) contributions.
Thus, we write the partonic cross-section as
\begin{equation}
\Delta_{ab}^{A}\left(z,q^{2},\mu_{R}^{2},\mu_{F}^{2}\right)=\Delta_{ab}^{A,NSV}\left(z,q^{2},\mu_{R}^{2},\mu_{F}^{2}\right)+\Delta_{ab}^{A,hard}\left(z,q^{2},\mu_{R}^{2},\mu_{F}^{2}\right)\label{eq:deltaA}
\end{equation}
with $z\equiv q^{2}/\hat{s}=\tau/\left(x_{1}x_{2}\right)$.
The threshold SV contributions $\Delta_{ab}^{A,SV}\left(z,q^{2},\mu_{R}^{2},\mu_{F}^{2}\right)$ contains distributions of type $\delta\left(1-z\right)$ and $\mathcal{D}_{i}$ where the latter is defined as
\begin{equation}
\mathcal{D}_{i}\equiv\left[\dfrac{\ln^{i}\left(1-z\right)}{1-z}\right]_{+}\label{eq:Distribution}
\end{equation}
On the contrary, the hard part of the cross-section $\Delta_{ab}^{A,hard}\left(z,q^{2},\mu_{R}^{2},\mu_{F}^{2}\right)$ contains all regular terms in $z$.
The NSV cross-section in $z$-space is computed in $d=4+\varepsilon$ dimensions using \cite{ajjath2020soft}.

Since we work only with the gluon fusion channel, from now on, I will present the results corresponding to the $gg$-channel only \textit{i.e.} $a=g,~b=g$.
So, we express the finite partonic coefficient function as
\begin{equation}
\Delta_{gg}^{A,NSV}\left(z,q^{2},\mu_{R}^{2},\mu_{F}^{2}\right)=\mathcal{C}\exp\left\{ \varPsi_{g}^{A}\left(z,q^{2},\mu_{R}^{2},\mu_{F}^{2},\varepsilon\right)\right\} \mid_{\varepsilon=0}
\label{eq:deltaNSV}
\end{equation}
where $\varPsi_{g}^{A}\left(z,q^{2},\mu_{R}^{2},\mu_{F}^{2},\varepsilon\right)$ is a finite distribution with $\mathcal{C}$ being the Mellin convolution.
The convolution $\mathcal{C}$ defined as
\begin{equation}
\mathcal{C}e^{f\left(z\right)}=\delta\left(1-z\right)+\dfrac{1}{1!}f\left(z\right)+\dfrac{1}{2!}f\left(z\right)\otimes f\left(z\right)+...\label{eq:Convolution}
\end{equation}
Here $\otimes$ represents Mellin convolution and $f\left(z\right)$ is a distribution of the kind $\delta\left(1-z\right)$ and $\mathcal{D}_{i}$.
The subscript $g$ signifies the gluon-initiated production of the pseudoscalar Higgs boson.
An equivalent formalism can be done in the Mellin ($N$-moment) space, which replaces the distributions in $z$ by continuous functions of the variable $N$.
In this space, the threshold limit of $z\rightarrow1$ changes to $N\rightarrow\infty$.

The finite distribution $\varPsi_{g}^{A}$ depends on the form factors $\mathcal{F}_{g}^{A}\left(\hat{a}_{s},Q^{2},\mu^{2},\varepsilon\right)$ with $Q^2=-q^2$, the overall operator UV renormalisation constant $Z_{g}^{A}\left(\hat{a_{s}},\mu_{R}^{2},\mu^{2},\varepsilon\right)$, the soft collinear distribution $\varPhi_{g}\left(\hat{a}_{s},q^{2},\mu^{2},z,\varepsilon\right)$ and the mass factorisation kernels $\varGamma_{gg}\left(\hat{a}_{s},\mu_{F}^{2},\mu^{2},z,\varepsilon\right)$.
These quantities acquire the following form as below \cite{Ahmed:2016otz}
\begin{align}
\varPsi_{g}^{A}\left(z,q^{2},\mu_{R}^{2},\mu_{F}^{2},\varepsilon\right)= & \left(\ln\left[Z_{g}^{A}\left(\hat{a_{s}},\mu_{R}^{2},\mu^{2},\varepsilon\right)\right]^{2}+\ln\mid\mathcal{F}_{g}^{A}\left(\hat{a}_{s},Q^{2},\mu^{2},\varepsilon\right)\mid\right)\delta\left(1-z\right)\nonumber \\
 & +\ 2\varPhi_{g}\left(\hat{a}_{s},q^{2},\mu^{2},z,\varepsilon\right)-2\mathcal{C}\ln\varGamma_{gg}\left(\hat{a}_{s},\mu_{F}^{2},\mu^{2},z,\varepsilon\right).
 \label{eqn:PsiA}
\end{align}

In the following sections, I will describe how to get these ingredients to compute the NSV resummed cross-section for the pseudoscalar Higgs boson production at N$^{3}$LO.

\subsection{Operator Renormalisation Constant}
\label{subsec:ORC}

After the strong coupling constant renormalisation through $Z_{a_{s}}$, the form factor $\mathcal{F}_{g}^{A}\left(\hat{a}_{s},Q^{2},\mu^{2},\varepsilon\right)$ still does not become completely UV finite as the $\gamma_{g}^{A}$ present are non-zero.
This additional renormalisation is called the overall operator renormalisation and is done using the constant $Z_{g}^{A}$.
This can be determined by solving the underlying RG equation
\cite{Ahmed:2016otz}:
\begin{equation}
\mu_{R}^{2}\dfrac{d}{d\mu_{R}^{2}}\ln Z_{g}^{A}\left(\hat{a_{s}},\mu_{R}^{2},\mu^{2},\varepsilon\right)=\sum_{i=1}^{\infty}a_{s}^{i}\gamma_{g,i}^{A},
\end{equation}
where $\gamma_{g,i}^{A}$ up to three-loop $(i = 3)$ are obtained as following:
\begin{align}
\gamma_{g,1}^{A} & =\beta_{1},\\
\gamma_{g,2}^{A} & =\beta_{2},\\
\gamma_{g,3}^{A} & =\beta_{3}.\label{eq:gammaAgi}
\end{align}

Using the above RG equation and the solutions of $\gamma_{g,i}^{A}$'s, we obtain the overall renormalisation constant up to the three-loop level:
\begin{align}
Z_{g}^{A}= & 1+a_{s}\left[\dfrac{22}{3\varepsilon}C_{A}-\dfrac{4}{3\varepsilon}n_{f}\right] \nonumber \\
 & +a_{s}^{2}\bigg[\dfrac{1}{\varepsilon^{2}}\left\{ \dfrac{484}{9}C_{A}^{2}-\dfrac{176}{9}C_{A}n_{f}+\dfrac{16}{9}n_{f}^{2}\right\} + \dfrac{1}{\varepsilon}\left\{ \dfrac{34}{3}C_{A}^{2}-\dfrac{10}{3}C_{A}n_{f}-2C_{F}n_{f}\right\} \bigg]\ \nonumber \\
 & + a_{s}^{3}\bigg[\dfrac{1}{\varepsilon^{3}}\left\{ \dfrac{10648}{27}C_{A}^{3}-\dfrac{1936}{9}C_{A}^{2}n_{f}+\dfrac{352}{9}C_{A}n_{f}^{2}-\dfrac{64}{27}n_{f}^{3}\right\} \nonumber \\
 & +\dfrac{1}{\varepsilon^{2}}\left\{ \dfrac{5236}{27}C_{A}^{3}-\dfrac{2492}{27}C_{A}^{2}n_{f}-\dfrac{308}{9}C_{A}C_{F}n_{f}+\dfrac{280}{27}C_{A}n_{f}^{2}+\dfrac{56}{9}C_{F}n_{f}^{2}\right\} \nonumber \\
 & +\dfrac{1}{\varepsilon}\left\{ \dfrac{2857}{81}C_{A}^{3}-\dfrac{1415}{81}C_{A}^{2}n_{f}-\dfrac{205}{27}C_{A}C_{F}n_{f}+\dfrac{2}{3}C_{F}^{2}n_{f}+\dfrac{79}{81}C_{A}n_{f}^{2}+\dfrac{22}{27}C_{F}n_{f}^{2}\right\} \bigg].
 \label{eq:ZAg}
\end{align}
As can be seen, $Z_{g}^{A}=Z_{GG}$ where $Z_{GG}$ is the renormalisation constant for two $O_G$ operators, which have been discussed in detail in \cite{Ahmed_2015,Bhattacharya_2020}.

\subsection{The Form Factor}
\label{subsec:FF}

The unrenormalised form factor $\hat{\mathcal{F}}_{g}^{A,(n)}$, expanded in terms of components, can be depicted as below:
\begin{equation}
  \label{eq:FF3}
  \mathcal{F}_{g}^{A} \equiv \sum_{n=0}^{\infty} \left[ {\hat{a}}_{s}^{n}
  \left( \frac{Q^{2}}{\mu^{2}} \right)^{n\frac{\varepsilon}{2}}
  S_{\varepsilon}^{n}  {\hat{\mathcal{F}}}^{A,(n)}_{g}\right] \,.
\end{equation}
For the choice of scale, $\mu_{R}^{2}=\mu_{F}^{2}=q^{2}$, the unrenormalized results are given in \cite{Ravindran:2004mb} up to two loops and up to three loops in \cite{Ahmed:2016otz}. These are required in computing the NSV resummed cross-section, as discussed below.
The fact that QCD amplitudes obey factorisation property, gauge and renormalisation
group (RG) invariances lead to the consequence that the form factor
$\mathcal{F}_{g}^{A}\left(\hat{a}_{s},Q^{2},\mu^{2},\varepsilon\right)$ satisfies
the following KG-differential equation \cite{Sudakov:1954sw, Mueller:1979ih, Collins:1980ih, Sen:1981sd, Magnea:1990zb, Ahmed:2016otz}:
\begin{equation}
Q^{2}\cfrac{d}{dQ^{2}}\ln\mathcal{F}_{g}^{A}\left(\hat{a}_{s},Q^{2},\mu^{2},\varepsilon\right)=\dfrac{1}{2}\left[K_{g}^{A}\left(\hat{a}_{s},\dfrac{\mu_{R}^{2}}{\mu^{2}},\varepsilon\right)+G_{g}^{A}\left(\hat{a}_{s},\dfrac{Q^{2}}{\mu_{R}^{2}},\dfrac{\mu_{R}^{2}}{\mu^{2}},\varepsilon\right)\right]
\end{equation}
All the poles in $\varepsilon$ are contained in the $Q^{2}$ independent function $K_{g}^{A}$ and those which are finite as $\varepsilon\rightarrow0$ are contained
in $G_{g}^{A}$.
The solution of the above KG equation is given in a desirable form as below \cite{Ravindran:2005vv}:
\begin{equation}
\ln\mathcal{F}_{g}^{A}\left(\hat{a}_{s},Q^{2},\mu^{2},\varepsilon\right)=\sum_{i=1}^{\infty}\hat{a}_{s}^{i}\left(\dfrac{Q^{2}}{\mu^{2}}\right)^{i\dfrac{\varepsilon}{2}}S_{\varepsilon}^{i}\hat{\mathcal{L}}_{g,i}^{A}\left(\varepsilon\right)\label{eq:lnFAg}
\end{equation}
with
\begin{align}
\hat{\mathcal{L}}_{g,1}^{A}\left(\varepsilon\right)= & \dfrac{1}{\varepsilon^{2}}\left\{ -2A_{g,1}^{A}\right\} +\dfrac{1}{\varepsilon}\left\{ G_{g,1}^{A}\left(\varepsilon\right)\right\} ,\\
\hat{\mathcal{L}}_{g,2}^{A}\left(\varepsilon\right)= & \dfrac{1}{\varepsilon^{3}}\left\{ \beta_{0}A_{g,1}^{A}\right\} +\dfrac{1}{\varepsilon^{2}}\left\{ -\dfrac{1}{2}A_{g,2}^{A}-\beta_{0}G_{g,1}^{A}\left(\varepsilon\right)\right\} +\dfrac{1}{\varepsilon}\left\{ \dfrac{1}{2}G_{g,2}^{A}\left(\varepsilon\right)\right\} ,\\
\hat{\mathcal{L}}_{g,3}^{A}\left(\varepsilon\right)= & \dfrac{1}{\varepsilon^{4}}\left\{ -\dfrac{8}{9}\beta_{0}^{2}A_{g,1}^{A}\right\} +\dfrac{1}{\varepsilon^{3}}\left\{ \dfrac{2}{9}\beta_{1}A_{g,1}^{A}+\dfrac{8}{9}\beta_{0}A_{g,2}^{A}+\dfrac{4}{3}\beta_{0}^{2}G_{g,1}^{A}\left(\varepsilon\right)\right\} \nonumber \\
 & +\dfrac{1}{\varepsilon^{2}}\left\{ -\dfrac{2}{9}A_{g,3}^{A}-\dfrac{1}{3}\beta_{1}G_{g,1}^{A}\left(\varepsilon\right)-\dfrac{4}{3}\beta_{0}G_{g,2}^{A}\left(\varepsilon\right)\right\} +\dfrac{1}{\varepsilon}\left\{ \dfrac{1}{3}G_{g,3}^{A}\left(\varepsilon\right)\right\} .
\end{align}
In the above expressions, $X^{A}_{g,i}$ with $X=A,B,f$ and
$\gamma^{A}_{g, i}$ are defined through the series expansion in powers
of $a_{s}$:
\begin{align}
  \label{eq:ABfgmExp}
  X^{A}_{g} &\equiv \sum_{i=1}^{\infty} a_{s}^{i}
              X^{A}_{g,i}\,,
              \qquad \text{and} \qquad
              \gamma^{A}_{g} \equiv \sum_{i=1}^{\infty} a_{s}^{i} \gamma^{A}_{g,i}\,\,.
\end{align}
$A_{g,i}^{A}$'s are known as the cusp anomalous dimensions and are given by \cite{Moch_2005}
\begin{align}
A_{g,1}^{A}= & 4C_{A},\\
A_{g,2}^{A}= & 8C_{A}^{2}\left(\dfrac{67}{18}-\zeta_{2}\right)+8C_{A}n_{f}\left(-\dfrac{5}{9}\right),\\
A_{g,3}^{A}= & 16C_{A}^{3}\left(\dfrac{245}{24}-\dfrac{67}{9}\zeta_{2}+\dfrac{11}{6}\zeta_{3}+\dfrac{11}{5}\zeta_{2}^{2}\right)+16C_{A}C_{F}n_{f}\left(-\dfrac{55}{24}+2\zeta_{3}\right)\nonumber \\
 & +16C_{A}^{2}n_{f}\left(-\dfrac{209}{108}+\dfrac{10}{9}\zeta_{2}-\dfrac{7}{3}\zeta_{3}\right)+16C_{A}n_{f}^{2}\left(-\dfrac{1}{27}\right).
 \label{eqn:Cusp}
\end{align}
The $G_{g,i}^{A}\left(\varepsilon\right)$ are the resummation functions
which obey the following decomposition in terms of collinear $\left(B_{g}^{A}\right)$,
soft $\left(f_{g}^{A}\right)$ and UV $\left(\gamma_{g}^{A}\right)$
anomalous dimensions \cite{Ahmed:2016otz}:
\[
G_{g,i}^{A}\left(\varepsilon\right)=2\left(B_{g,i}^{A}-\gamma_{g,i}^{A}\right)+f_{g,i}^{A}+C_{g,i}^{A}+\sum_{k=1}^{\infty}\varepsilon^{k}g_{g,i}^{A,k},
\]
where the constants $C_{g,i}^{A}$ are given by \cite{Ravindran_2006}
\begin{align}
C_{g,1}^{A} & =0,\\
C_{g,2}^{A} & =-2\beta_{0}g_{g,1}^{A,1},\\
C_{g,3}^{A} & =-2\beta_{1}g_{g,1}^{A,1}-2\beta_{0}\left(g_{g,2}^{A,1}+2\beta_{0}g_{g,1}^{A,2}\right).
\end{align}
The above decomposition is crucial in the sense that the functions
$f_{g,i}^{A}$ are, like the cusp anomalous dimensions $A_{g,i}^{A}$,
universal up to the factor $\dfrac{C_{A}}{C_{F}}$, i.e., $f_{g,i}^{A}=\dfrac{C_{A}}{C_{F}}f_{q,i}^{A}$.
These functions exhibit the same maximally non-Abelian colour structure
as the cusp anomalous dimensions with \cite{Moch_2005}
\begin{align}
f_{q,1}^{A}= & 0,\\
f_{q,2}^{A}= & 2C_{F}\left\{ -\beta_{0}\zeta_{2}-\dfrac{56}{27}n_{f}+C_{A}\left(\dfrac{404}{27}-14\zeta_{3}\right)\right\} ,\\
f_{q,3}^{A}= & C_{F}C_{A}^{2}\left(\dfrac{136781}{729}-\dfrac{12650}{81}\zeta_{2}-\dfrac{1316}{3}\zeta_{3}+\dfrac{352}{5}\zeta_{2}^{2}+\dfrac{176}{3}\zeta_{2}\zeta_{3}+192\zeta_{5}\right)\nonumber \\
 & +C_{A}C_{F}n_{f}\left(-\dfrac{11842}{729}+\dfrac{2828}{81}\zeta_{2}+\dfrac{728}{27}\zeta_{3}-\dfrac{96}{5}\zeta_{2}^{2}\right)+C_{F}^{2}n_{f}\bigg(-\dfrac{1711}{27}\nonumber \\
 & +4\zeta_{2}+\dfrac{304}{9}\zeta_{3}+\dfrac{32}{5}\zeta_{2}^{2}\bigg)+C_{F}n_{f}^{2}\left(-\dfrac{2080}{729}-\dfrac{40}{27}\zeta_{2}+\dfrac{112}{27}\zeta_{3}\right).
 \label{eqn:soft}
\end{align}
The collinear anomalous dimensions are given by
\begin{align}
B_{g,1}^{A}= & \frac{11}{3}C_{A}-\frac{4}{3}n_{f}T_{F}\,,\\
B_{g,2}^{A}= & \dfrac{1}{2}\left\{ C_{A}^{2}\left(\dfrac{64}{3}+24\zeta_{3}\right)-\dfrac{32}{3}n_{f}T_{f}C_{A}-8n_{f}T_{f}C_{f}\right\} ,\\
B_{g,3}^{A}= & 16C_{A}C_{f}n_{f}\left(-\dfrac{241}{288}\right)+16C_{A}n_{f}^{2}\left(\dfrac{29}{288}\right)-16C_{A}^{2}n_{f}\bigg(\dfrac{233}{288}+\dfrac{1}{6}\zeta_{2}+\dfrac{1}{12}\zeta_{2}^{2}\nonumber \\
 & +\dfrac{5}{3}\zeta_{3}\bigg)+16C_{A}^{3}\bigg(\dfrac{79}{32}-\zeta_{2}\zeta_{3}+\dfrac{1}{6}\zeta_{2}+\dfrac{11}{24}\zeta_{2}^{2}+\dfrac{67}{6}\zeta_{3}-5\zeta_{5}\bigg)\nonumber \\
 & +16C_{F}n_{f}^{2}\left(\dfrac{11}{144}\right)+16C_{F}^{2}n_{f}\left(\dfrac{1}{16}\right).
\end{align}
$f_{g}^{A}$'s were introduced for the first time in the
article \cite{Ravindran:2004mb}. The article \cite{Ravindran:2004mb} has shown that
$f_{g}^{A}$'s fulfil the maximally non-Abelian property up to two-loop level whose validity is reconfirmed in \cite{Moch_2005} at the three-loop level.
The quantities $X$ depend only on the partons involved in the initial state of any process.
It is simply because of this reason that we have used the existing results up to three loops for a gluon-gluon initiated process:
\begin{align}
  \label{eq:1}
  X^{A}_{g} = X_{g}\,.
\end{align}
$f_{g}$ can be found in \cite{Ravindran:2004mb,Moch_2005},
$A_{g,i}$ in~\cite{Moch:2004pa,Moch_2005,Vogt:2004mw,Vogt:2000ci}
and $B_{g,i}$ in \cite{Vogt:2004mw,Moch_2005} up to three loop
level. For the NSV resummed cross-section at N$^{3}$LO, we need
$g^{A,1}_{g,3}$ in addition to the quantities arising from one and two
loops. The form factors for the pseudoscalar production up to two loops
can be found in~\cite{Ravindran:2004mb}, and the three-loop one is
calculated very recently in the article~\cite{Ahmed_2015} by some
of us. However, in this computation of NSV resummed cross-section at N$^{3}$LO, the form factor is needed in a definite form, slightly different
from the one presented in the recent article \cite{Ahmed_2015}.
With some effort, the required form can be extracted from the above-mentioned recent work. For the convenience of
our readers, we have presented the form factors $\mathcal{F}^{A}_{g}$ up to three loops at the
beginning of this section. The $g_{g,i}^{A,k}$'s up to three loop level are given below \cite{Ahmed:2016otz}:
\begin{align}
  \label{eq:g31}
  g^{A,1}_{g,1} &= {\dis{C_{A}}} \Bigg\{ 4 + \zeta_2 \Bigg\}\,,
                  \nonumber\\
  g^{A,2}_{g,1} &= {\dis{C_{A}}} \Bigg\{ - 6 - \frac{7}{3} \zeta_3 \Bigg\}\,,
                  \nonumber\\
  g^{A,3}_{g,1} &= {\dis{C_{A}}} \Bigg\{ 7 - \frac{1}{2} \zeta_2 + \frac{47}{80} \zeta_2^2 \Bigg\}\,,
                  \nonumber\\ \nonumber\\
  g^{A,1}_{g,2} &= {\dis{C_{A}^2}} \Bigg\{ \frac{11882}{81} + \frac{67}{3}
                  \zeta_2 - \frac{44}{3} \zeta_3 \Bigg\}
                  +
                  {\dis{C_{A} n_{f}}} \Bigg\{ - \frac{2534}{81} - \frac{10}{3} \zeta_2 -
                  \frac{40}{3} \zeta_3 \Bigg\}
                  \nonumber\\
                & + {\dis{C_{F} n_{f}}} \Bigg\{ - \frac{160}{3}
                  + 12 \ln \left(\frac{\mu_R^2}{m_t^2}\right) + 16 \zeta_3 \Bigg\}\,,
                  \nonumber\\
  g^{A,2}_{g,2} &= {\dis{C_{F} n_{f}}} \Bigg\{ \frac{2827}{18} - 18
                  \ln \left(\frac{\mu_R^2}{m_t^2}\right)  - \frac{19}{3} \zeta_2 - \frac{16}{3}
                  \zeta_2^2  -
                  \frac{128}{3} \zeta_3 \Bigg\}
                  + {\dis{C_{A} n_{f}}} \Bigg\{
                  \frac{21839}{243}  - \frac{17}{9} \zeta_2
                  \nonumber\\
                &+
                  \frac{259}{60} \zeta_2^2  +
                  \frac{766}{27} \zeta_3 \Bigg\}
                  + {\dis{C_{A}^2}} \Bigg\{ -
                  \frac{223861}{486}  + \frac{80}{9} \zeta_2 +
                  \frac{671}{120} \zeta_2^2  +
                  \frac{2111}{27} \zeta_3 + \frac{5}{3} \zeta_2
                  \zeta_3  - 39 \zeta_5 \Bigg\}\,,
                  \nonumber\\ \nonumber\\
  g^{A,1}_{g,3} &=  {\dis{n_{f} C_{J}^{(2)}}}  \Bigg\{ - 6 \Bigg\}
                  + {\dis{C_{F}
                  n_{f}^2}} \Bigg\{ \frac{12395}{27}  -
                  \frac{136}{9} \zeta_2  -
                  \frac{368}{45} \zeta_2^2 - \frac{1520}{9} \zeta_3  -
                  24  \ln \left(\frac{\mu_R^2}{m_t^2}\right)
                  \Bigg\}
                  \nonumber\\
                &+
                  {\dis{C_{F}^2 n_{f}}} \Bigg\{ \frac{457}{2} + 312 \zeta_3 -
                  480 \zeta_5 \Bigg\}
                  +
                  {\dis{C_{A}^2 n_{f}}} \Bigg\{ - \frac{12480497}{4374} -
                  \frac{2075}{243} \zeta_2  - \frac{128}{45} \zeta_2^2
                  \nonumber\\
                &-
                  \frac{12992}{81} \zeta_3 - \frac{88}{9} \zeta_2
                  \zeta_3 + \frac{272}{3} \zeta_5 \Bigg\}
                  +
                  {\dis{C_{A}^3}} \Bigg\{ \frac{62867783}{8748} +
                  \frac{146677}{486} \zeta_2  - \frac{5744}{45}
                  \zeta_2^2  -
                  \frac{12352}{315} \zeta_2^3
                  \nonumber\\
                &- \frac{67766}{27}
                  \zeta_3 - \frac{1496}{9} \zeta_2 \zeta_3  -
                  \frac{104}{3} \zeta_3^2 + \frac{3080}{3} \zeta_5 \Bigg\}
                  +
                  {\dis{C_{A} n_{f}^2}} \Bigg\{ \frac{514997}{2187} -
                  \frac{8}{27} \zeta_2  + \frac{232}{45} \zeta_2^2
                  \nonumber\\
                &+
                  \frac{7640}{81} \zeta_3 \Bigg\}
                  + {\dis{C_{A} C_{F} n_{f}}} \Bigg\{
                  - \frac{1004195}{324}  + \frac{1031}{18} \zeta_2 +
                  \frac{1568}{45} \zeta_2^2 + \frac{25784}{27} \zeta_3
                  + 40 \zeta_2 \zeta_3  + \frac{608}{3} \zeta_5
                  \nonumber\\
                &+ 132 \ln \left(\frac{\mu_R^2}{m_t^2}\right) \Bigg\}\,.
\end{align}

\subsection{Mass Factorisation Kernel}
\label{subsec:MFK}

The partonic cross-section $\Delta_{ab}^{A,NSV}\left(z,q^{2},\mu_{R}^{2},\mu_{F}^{2}\right)$
is UV finite after performing the coupling constant and overall operator
renormalisations using $Z_{a_{s}}$ and $Z_{g}^{A}$. However, it still exhibits collinear divergences and, thus, requires mass factorisation.
This section deals with the issue of collinear divergence and also provides a recipe for removing them.
The $\overline{MS}$ scheme is used to remove the collinear singularities arising in the massless limit of partons using the mass factorisation kernel $\varGamma\left(\hat{a}_{s},\mu^{2},\mu_{F}^{2},z,\varepsilon\right)$.
The kernel satisfies the following RG equation \cite{Ravindran:2005vv,Ahmed:2016otz}:
\begin{equation}
\mu_{F}^{2}\dfrac{d}{d\mu_{F}^{2}}\varGamma\left(z,\mu_{F}^{2},\varepsilon\right)=\dfrac{1}{2}P\left(z,\mu_{F}^{2}\right)\otimes\varGamma\left(z,\mu_{F}^{2},\varepsilon\right),
\end{equation}
where $P\left(z,\mu_{F}^{2}\right)$ are the Altarelli-Parisi splitting
functions (matrix-valued). We can expand $P\left(z,\mu_{F}^{2}\right)$
and $\varGamma\left(z,\mu_{F}^{2},\varepsilon\right)$ in powers of
the strong coupling constant $a_{s}$ as follows,
\begin{equation}
P\left(z,\mu_{F}^{2}\right)=\sum_{i=1}^{\infty}a_{s}^{i}\left(\mu_{F}^{2}\right)P^{\left(i-1\right)}\left(z\right)
\end{equation}
and
\begin{equation}
\varGamma\left(z,\mu_{F}^{2},\varepsilon\right)=\delta\left(1-z\right)+\sum_{i=1}^{\infty}\hat{a}_{s}^{i}\left(\dfrac{\mu_{F}^{2}}{\mu^{2}}\right)S_{\varepsilon}^{i}\varGamma^{\left(i\right)}\left(z,\varepsilon\right).
\end{equation}
We can solve the RGE for the kernel. The solutions in the $\overline{MS}$
scheme contains only the poles in $\varepsilon$ and are given by \cite{Ravindran:2005vv}
\begin{align}
\varGamma_{II}^{\left(1\right)}\left(z,\varepsilon\right)= & \dfrac{1}{\varepsilon}P_{II}^{\left(0\right)}\left(z\right),\\
\varGamma_{II}^{\left(2\right)}\left(z,\varepsilon\right)= & \dfrac{1}{\varepsilon^{2}}\left(\dfrac{1}{2}P_{II}^{\left(0\right)}\left(z\right)\otimes P_{II}^{\left(0\right)}\left(z\right)-\beta_{0}P_{II}^{\left(0\right)}\left(z\right)\right)+\dfrac{1}{\varepsilon}\left(\dfrac{1}{2}P_{II}^{\left(1\right)}\left(z\right)\right)\\
\varGamma_{II}^{\left(3\right)}\left(z,\varepsilon\right)= & \dfrac{1}{\varepsilon^{3}}\bigg(\dfrac{4}{3}\beta_{0}^{2}P_{II}^{\left(0\right)}\left(z\right)-\beta_{0}P_{II}^{\left(0\right)}\left(z\right)\otimes P_{II}^{\left(0\right)}\left(z\right)\nonumber \\
 & +\dfrac{1}{6}P_{II}^{\left(0\right)}\left(z\right)\otimes P_{II}^{\left(0\right)}\left(z\right)\otimes P_{II}^{\left(0\right)}\left(z\right)\bigg)+\dfrac{1}{\varepsilon^{2}}\bigg(\dfrac{1}{2}P_{II}^{\left(0\right)}\left(z\right)\otimes P_{II}^{\left(1\right)}\left(z\right)\nonumber \\
 & -\dfrac{1}{3}\beta_{1}P_{II}^{\left(0\right)}\left(z\right)-\dfrac{4}{3}\beta_{0}P_{II}^{\left(1\right)}\left(z\right)\bigg)+\dfrac{1}{\varepsilon}\left(\dfrac{1}{3}P_{II}^{\left(2\right)}\left(z\right)\right).
\end{align}
The relevant values of $P^{(0)}(z), P^{(1)}(z) ~\text{and}~ P^{(2)}(z)$ are computed in the articles~\cite{Moch:2004pa, Vogt:2004mw}.
Only the diagonal Altarelli-Parisi kernels contribute to our analysis.
So, we expand the corresponding AP splitting functions around $z = 1$, dropping all those terms that do not contribute to NSV.
The AP splitting functions near $z = 1$ take the following form \cite{ajjath2020soft} for the gluon-fusion channel:
\begin{align}
P_{gg,i}\left(z,a_{s}\left(\mu_{F}^{2}\right)\right) &=
2~\bigg[B_{g,i}^{A}\left(a_{s}\left(\mu_{F}^{2}\right)\right) \delta\left(1-z\right) + A_{g,i}^{A}~\mathcal{D}_{0}\left(z\right) +C_{i}^{g}\left(a_{s}\left(\mu_{F}^{2}\right)\right)\log\left(1-z\right)
\nonumber \\
& +D_{i}^{g}\left(a_{s}\left(\mu_{F}^{2}\right)\right)\bigg] + \mathcal{O}\left(1-z\right).
\end{align}
$C_{i}^{g}$ and $D_{i}^{g}$ are constants that can be obtained from the splitting functions $P_{gg,i}$.
Just as the cusp and the collinear anomalous dimensions were expanded in powers of $a_s(\mu_F^2)$, the constants $C^g$ and $D^g$ can also be expanded similarly as below:
\begin{eqnarray}
C^g(a_s(\mu_F^2)) = \sum_{i=1}^\infty a_s^i(\mu_F^2) C_i^g,
\quad \quad
D^g(a_s(\mu_F^2)) = \sum_{i=1}^\infty a_s^i(\mu_F^2) D_i^g \,,
\end{eqnarray}
where $C_i^g$ and $D_i^g$ to third order are available in \cite{Moch:2004pa,Vogt:2004mw,2021_Blumlein}.
They are as follows:
\begin{align}
C_{1}^{g}= & 0,\\
C_{2}^{g}= & 16C_{A}^{2},\\
C_{3}^{g}= & C_{A}^{3}\left(\dfrac{2144}{9}-64\zeta_{2}\right)+n_{f}C_{A}^{2}\left(-\dfrac{320}{9}\right).
\end{align}
and
\begin{align}
D_{1}^{g}= & -4C_{A},\\
D_{2}^{g}= & C_{A}^{2}\left(-\dfrac{268}{9}+8\zeta_{2}\right)+C_{A}n_{f}\left(\dfrac{40}{9}\right),\\
D_{3}^{g}= & C_{A}^{3}\left(-166+\dfrac{56}{3}\zeta_{3}+\dfrac{1072}{9}\zeta_{2}-\dfrac{176}{5}\zeta_{2}^{2}\right)+n_{f}C_{A}^{2}\bigg(\dfrac{548}{27}+\dfrac{112}{3}\zeta_{3}\nonumber \\
 & -\dfrac{160}{9}\zeta_{2}\bigg)+n_{f}C_{A}^{2}T_{f}\left(\dfrac{80}{3}\right)+n_{f}C_{F}C_{A}\left(\dfrac{86}{3}-32\zeta_{3}\right)\nonumber \\
 & +n_{f}C_{F}C_{A}T_{f}\left(16\right)+n_{f}^{2}C_{A}\left(\dfrac{16}{27}\right).
\end{align}

\subsection{Soft Collinear Distribution}
\label{subsec:SCD}

The resulting expression obtained after using the operator renormalisation constant and mass factorisation kernel must still be made free of certain residual divergences.
These residual divergences get cancelled against the contribution arising from soft gluon emissions.
This is why the finiteness of $\Delta_{gg}^{A,NSV}\left(z,q^{2},\mu_{R}^{2},\mu_{F}^{2}\right)$ in the limit $\varepsilon\rightarrow0$ requires the soft-collinear distribution $\varPhi_{g}\left(\hat{a}_{s},q^{2},\mu^{2},z,\varepsilon\right)$ which has pole structure in $\varepsilon$ similar to that of residual divergences.
$\varPhi_{g}\left(\hat{a}_{s},q^{2},\mu^{2},z,\varepsilon\right)$ satisfy the K+G type differential equation \cite{ajjath2020soft}:
\begin{equation}
q^{2}\dfrac{d}{dq^{2}}\varPhi_{g}=\dfrac{1}{2}\left[\overline{K_{g}}\left(\hat{a}_{s},\dfrac{\mu_{R}^{2}}{\mu^{2}},\varepsilon,z\right)+\overline{G_{g}}\left(\hat{a}_{s},\dfrac{q^{2}}{\mu_{R}^{2}},\dfrac{\mu_{R}^{2}}{\mu^{2}},\varepsilon,z\right)\right],
\end{equation}
where $\overline{K_{g}}$ contains all the divergent terms and $\overline{G_{g}}$ contains all finite functions of $\left(z,\varepsilon\right)$.
The solution of $\varPhi_{g}\left(\hat{a}_{s},q^{2},\mu^{2},z,\varepsilon\right)$ is obtained by solving the above equation to be
\begin{equation}
\varPhi_{g}\left(\hat{a}_{s},q^{2},\mu^{2},z,\varepsilon\right)=\sum_{i=1}^{\infty}\hat{a}_{s}^{i}\left\{ \dfrac{q^{2}\left(1-z\right)^{2}}{\mu^{2}z}\right\} ^{i\dfrac{\varepsilon}{2}}S_{\varepsilon}^{i}\left(\dfrac{i\varepsilon}{1-z}\right)\hat{\phi}_{g}^{\left(i\right)}\left(z,\varepsilon\right).
\end{equation}
The term $\left\{ \dfrac{q^{2}\left(1-z\right)^{2}}{\mu^{2}z}\right\}$ in the parenthesis results from two body phase space while $\dfrac{\hat{\phi}_{g}^{\left(i\right)}\left(z,\varepsilon\right)}{1-z}$ comes from the square of the matrix elements for corresponding amplitudes.
The function $\hat \phi^{(i)}_c(z,\varepsilon)$ is regular as $z \rightarrow 0$ but
contains poles in $\varepsilon$.
The $1/(1-z)$ term has been factored out explicitly so that it generates all the distributions ${\cal D}_j$ and $\delta(1-z)$ and
NSV terms $\log^k(1-z), k=0,\cdots$ when combined with the factor $((1-z)^2)^{i \varepsilon/2}$ and
$\hat \phi^{(i)}_c(z,\varepsilon)$ at each order in $\hat a_s$.  Note that the term $z^{-i \varepsilon/2}$ inside
the parenthesis does not give terms like ${\cal D}_j$ and $\delta(1-z)$ but they contribute to NSV terms $\log^j(1-z) ,j=0,1,\cdots$ which is expanded around $z=1$.
A more detailed study has been given in \cite{ajjath2020soft}.
The components of the $z$ space resummed cross-sections get contributions from both the form factor and the soft distribution functions.
The form factor contributes to $\delta\left(1-z\right)$ part, and the soft distributions functions contribute to $\delta\left(1-z\right)$ as well as to the distributions $\mathcal{D}_{j}$.
This is done using the following relation \cite{Ravindran_2006}:
\begin{equation}
\dfrac{1}{1-z}\left[\left(1-z\right)^{2}\right]^{i\tfrac{\varepsilon}{2}}=\dfrac{1}{i\varepsilon}\delta\left(1-z\right)+\left(\dfrac{1}{1-z}\left[\left(1-z\right)^{2}\right]^{i\tfrac{\varepsilon}{2}}\right)_{+}.
\end{equation}

We rewrite $\varPhi_{g}\left(\hat{a}_{s},q^{2},\mu^{2},z,\varepsilon\right)$ in a convenient form which separates SV terms from the NSV.
Hence, we decompose it as
\begin{equation}
\varPhi_{g}=\varPhi_{g}^{SV}+\varPhi_{g}^{NSV},
\end{equation}
in such a way that $\varPhi_{g}^{SV}$ contains only SV terms and the remaining $\varPhi_{g}^{NSV}$ contains next to soft-virtual terms in the limit $z\rightarrow1$.
The form of $\varPhi_{g}^{SV}$ is given by
\begin{equation}
\varPhi_{g}^{SV}\left(\hat{a}_{s},q^{2},\mu^{2},z,\varepsilon\right)=\sum_{i=1}^{\infty}\hat{a}_{s}^{i}\left(\dfrac{q^{2}\left(1-z\right)^{2}}{\mu^{2}}\right)^{i\dfrac{\varepsilon}{2}}S_{\varepsilon}^{i}\left(\dfrac{i\varepsilon}{1-z}\right)\hat{\phi}^{SV,\left(i\right)}_{g}\left(\varepsilon\right),
\label{eq:phicapg}
\end{equation}
where
\begin{equation}
\hat{\phi}^{SV,\left(i\right)}_{g}\left(\varepsilon\right) =
\mathcal{\hat{L}}_{g,\left(i\right)}^{A}\left(\varepsilon\right)
\left\{A^{A}_{g,(i)}\rightarrow-A^{A}_{g,(i)},
G^{A}_{g,(i)}\left(\varepsilon\right) \rightarrow
\overline{\mathcal{G}}^{A}_{g,(i)}\left(\varepsilon\right)\right\}.
\end{equation}
The $z$ independent constants $\overline{\mathcal{G}}_{g,i}^{A}\left(\varepsilon\right)$ in
$\hat{\phi}^{SV,(i)}_{g}\left(\varepsilon\right)$ can be obtained using the form factors, mass factorisation kernels and coefficient functions expanded in powers of $\varepsilon$ to the desired accuracy.
This is achieved by comparing the poles as well as non-pole terms in $\varepsilon$ of
$\hat{\phi}^{SV,(i)}_{g}(\varepsilon)$ with those coming from the form factors, overall renormalisation constants and splitting functions.
The $z$-independent constants $\overline{\mathcal{G}}_{g,i}^{A}\left(\varepsilon\right)$
are defined as \cite{ajjath2020soft,Ahmed:2016otz}
\begin{equation}
\overline{\mathcal{G}}_{g,i}^{A}\left(\varepsilon\right)=-f_{g,i}^{A}+\overline{C}_{g,i}^{A}+\sum_{k=1}^{\infty}\varepsilon^{k}\overline{\mathcal{G}}_{g,i}^{A,k},
\end{equation}
where
\begin{align}
\overline{C}_{g,1}^{A}= & 0,\\
\overline{C}_{g,2}^{A}= & -2\beta_{0}\overline{\mathcal{G}}_{g,1}^{A,1},\\
\overline{C}_{g,3}^{A}= & -2\beta_{1}\overline{\mathcal{G}}_{g,1}^{A,1}-2\beta_{0}\left(\overline{\mathcal{G}}_{g,2}^{A,1}+2\beta_{0}\overline{\mathcal{G}}_{g,1}^{A,2}\right).
\end{align}
The expressions for $\hat{\phi}_{g}^{SV,\left(i\right)}\left(\varepsilon\right)$'s are explicitly given in Appendix \ref{appendix:B}.

As verified by some of us (up to N$^3$LO) in \cite{Ahmed:2016otz}, due to the universality of the soft gluon contribution,
$\varPhi_{g}^{SV}\left(\hat{a}_{s},q^{2},\mu^{2},z,\varepsilon\right)$ for pseudoscalar Higgs boson production must be the same as that of the scalar Higgs boson production \textit{via} gluon fusion:
\begin{align}
  \label{eq:PhiAgPhiHg}
  \varPhi^{A}_{g} &= \varPhi^{H}_{g} = \varPhi_{g},
                 \nonumber\\
  \text{i.e.}~~ {\overline {\cal G}}^{A,k}_{g,i} &=  {\overline {\cal G}}^{H,k}_{g,i} =  {\overline {\cal G}}^{k}_{g,i}\,.
\end{align}
Here, ${\Phi}^{H}_{g}$ and $ {\overline {\cal G}}^{H,k}_{g,i}$ can be used for any gluon fusion process as these are independent of the operator insertion.
The constants, ${\overline {\cal G}}_{g,1}^{H,1}, {\overline {\cal G}}_{g,1}^{H,2},
{\overline {\cal G}}_{g,2}^{H,1}$, were determined from the result of the explicit computations of soft gluon emissions to the Higgs boson production in \cite{Ravindran:2003um} and later, these corrections were further extended to all orders in the dimensional
regularization parameter $\varepsilon$ in \cite{de_Florian_2012},
using which $ {\overline {\cal G}}_{g,1}^{H,3}$ and $ {\overline {\cal G}}_{g,2}^{H,2}$
are extracted in \cite{Ahmed:2016otz}.
A detailed description of these constants, the ${\overline {\cal G}}^{H,k}_{g,i}$'s or the
${\overline {\cal G}}^{A,k}_{g,i}$'s, that are used in the evaluation of
$\varPhi_{g}^{SV}\left(\hat{a}_{s},q^{2},\mu^{2},z,\varepsilon\right)$, are already given in \cite{Ahmed:2016otz}.
Hence, we do not repeat them here but give the final results for $\hat{\phi}_{g}^{SV,\left(i\right)}\left(\varepsilon\right)$ in Appendix \ref{appendix:B} after applying the available relevant expressions.
The third order constant $ {\overline {\cal G}}_{g,3}^{H,1}$ is computed from the result of the SV cross-section for the production of the Higgs boson at N$^{3}$LO \cite{Anastasiou:2014vaa} which was presented in the article~\cite{Ahmed:2014cla}.
%The $\overline{\mathcal{G}}_{g,i}^{A,k}$ required to get the SV cross-sections up to N$^{3}$LO are given in \cite{Ahmed:2016otz}.
The $\overline{\mathcal{G}}_{g,i}^{A,k}$ required to get the SV cross-sections up to N$^{3}$LO are given below \cite{Ahmed:2016otz}:
\begin{align}
  \label{eq:calG}
  {\overline {\cal G}}^{H,1}_{g,1} &= {\dis{C_A}} \Bigg\{ - 3 \zeta_2 \Bigg\} \,,
                                     \nonumber\\
  {\overline {\cal G}}^{H,2}_{g,1} &= {\dis{C_A}} \Bigg\{ \frac{7}{3} \zeta_3 \Bigg\} \,,
                                     \nonumber\\
  {\overline {\cal G}}^{H,3}_{g,1} &=  {\dis{C_A}} \Bigg\{ - \frac{3}{16} {\zeta_2}^2 \Bigg\} \,,
                                     \nonumber\\
  {\overline {\cal G}}^{H,1}_{g,2} &=  {\dis{C_A n_f}}  \Bigg\{ - \frac{328}{81} + \frac{70}{9} \zeta_2 + \frac{32}{3} \zeta_3 \Bigg\}
                                     +  {\dis{C_A^{2}}} \Bigg\{ \frac{2428}{81} - \frac{469}{9} \zeta_2
                                     + 4 {\zeta_2}^2 - \frac{176}{3} \zeta_3 \Bigg\} \,,
                                     \nonumber\\
  {\overline {\cal G}}^{H,2}_{g,2} &=   {\dis{C_A^{2}}} \Bigg\{
                                     \frac{11}{40} {\zeta_2}^2  -
                                     \frac{203}{3} {\zeta_2} {\zeta_3}
                                     + \frac{1414}{27} {\zeta_2} +
                                     \frac{2077}{27} {\zeta_3}  + 43
                                     {\zeta_5}  - \frac{7288}{243}  \Bigg\}
                                     \nonumber\\
                                   & +  {\dis{C_A n_f}} \Bigg\{
                                     -\frac{1}{20} {\zeta_2}^2 -
                                     \frac{196}{27} {\zeta_2} -
                                     \frac{310}{27} {\zeta_3} +  \frac{976}{243} \Bigg\}\, ,
                                     \nonumber\\
  {\overline {\cal G}}^{H,1}_{g,3} &=
                                     {\dis{C_A}^3} \Bigg\{\frac{152}{63}
                                     \;{\zeta_2}^3  + \frac{1964}{9} \;{\zeta_2}^2
                                     + \frac{11000}{9} \;{\zeta_2}
                                     {\zeta_3}  - \frac{765127}{486} \;{\zeta_2}
                                     +\frac{536}{3} \;{\zeta_3}^2 -  \frac{59648}{27} \;{\zeta_3}
                                     \nonumber\\
                                   &- \frac{1430}{3} \;{\zeta_5}
                                     +\frac{7135981}{8748}\Bigg\}
                                     + {\dis{C_A}^{2} {n_f}}
                                     \Bigg\{-\frac{532}{9}
                                     \;{\zeta_2}^2 -   \frac{1208}{9} \;{\zeta_2} {\zeta_3}
                                     +\frac{105059}{243} \;{\zeta_2} +  \frac{45956}{81} \;{\zeta_3}
                                     \nonumber\\
                                   &+\frac{148}{3} \;{\zeta_5} - \frac{716509}{4374} \Bigg\}
                                     +  {\dis{C_{A} {C_F} {n_f}}} \
                                     \Bigg\{\frac{152}{15} \;{\zeta_2}^2
                                     - 88 \;{\zeta_2} {\zeta_3}
                                     +\frac{605}{6} \;{\zeta_2} + \frac{2536}{27} \;{\zeta_3}
                                     +\frac{112}{3} \;{\zeta_5}
                                     \nonumber\\
                                   &- \frac{42727}{324}\Bigg\}
                                     +  {\dis{C_{A} {n_f}^2}} \
                                     \Bigg\{\frac{32}{9} \;{\zeta_2}^2 - \frac{1996}{81} \;{\zeta_2}
                                     -\frac{2720}{81} \;{\zeta_3} + \frac{11584}{2187}\Bigg\}   \,.
\end{align}

Having discussed the computation of the cross-section's SV part, we now calculate the corresponding NSV part.
The form of $\varPhi_{g}^{NSV}$ is given by \cite{ajjath2020soft},
\begin{equation}
\varPhi_{g}^{NSV}\left(\hat{a}_{s},q^{2},\mu^{2},z,\varepsilon\right)=\sum_{i=1}^{\infty}\hat{a}_{s}^{i}\left(\dfrac{q^{2}\left(1-z\right)^{2}}{\mu^{2}}\right)^{i\dfrac{\varepsilon}{2}}S_{\varepsilon}^{i}\varphi_{g}^{NSV,\left(i\right)}\left(z,\varepsilon\right).
\end{equation}
The $\varphi_{g}^{NSV,\left(i\right)}\left(z,\varepsilon\right)$ coefficients can be expressed as a sum of singular and finite part in $\varepsilon$ given by \cite{ajjath2020soft},
\begin{equation}
\varphi_{g}^{NSV,\left(i\right)}\left(z,\varepsilon\right)=\varphi_{s,g}^{NSV,\left(i\right)}\left(z,\varepsilon\right)+\varphi_{f,g}^{NSV,\left(i\right)}\left(z,\varepsilon\right).
\label{eq:phiNSV}
\end{equation}
\thr{
We have given the explicit expressions of the singular coefficients, $\varphi_{s,g}^{NSV,\left(i\right)}\left(z,\varepsilon\right)$, for a pseudoscalar Higgs production \textit{via} gluon fusion in Appendix \ref{appendix:C}.
}
$\varphi_{s,NSV}^{c\left(i\right)}\left(z,\varepsilon\right)$ are the singular coefficients with the following substitution \cite{ajjath2020soft},
\begin{equation}
\varphi_{s,NSV}^{c\left(i\right)}\left(z,\varepsilon\right) =
\overline{K}_{g,i}\left(\varepsilon\right)
\left\{A_{g}^A \rightarrow L_g^A\left(z\right)\right\},
\end{equation}
where $L^{I}\left(a_{s}\left(\mu_{R}^{2}\right),z\right)$, with $I=A$ in this work, can be
expanded in powers of $a_{s}\left(\mu_{R}^{2}\right)$ as
\begin{equation}
L^{I}\left(a_{s}\left(\mu_{R}^{2}\right),z\right)=\sum_{i=1}^{\infty}a_{s}^{i}\left(\mu_{R}^{2}\right)L_{i}^{I}\left(z\right)
\end{equation}
with
\begin{equation}
L_{i}^{I} = C_{g,i}^{A}\log\left(1-z\right)+D_{g,i}^{A}.
\end{equation}
The constants $\overline{K}_{c,i}$ are given in details in \cite{Ravindran_2006} as below:
\begin{align}
\overline{K}_{c,1}\left(\varepsilon\right) = &
\dfrac{1}{\varepsilon} ~ 2 A_{c,1},\\
\overline{K}_{c,2}\left(\varepsilon\right) = &
\dfrac{1}{\varepsilon^{2}}\left(-2\beta_{0} A_{c,1}\right) +
\dfrac{1}{\varepsilon}~A_{c,2},\\
\overline{K}_{c,3}\left(\varepsilon\right) = &
\dfrac{1}{\varepsilon^{3}} \left( \dfrac{8}{3} \beta_{0}^{2} A_{c,1} \right) +
\dfrac{1}{\varepsilon^{2}}\left( -\dfrac{2}{3} \beta_{1} A_{c,1} -
\dfrac{8}{3}\beta_{0} A_{c,2} \right) +
\dfrac{1}{\varepsilon} \left(\dfrac{2}{3}A_{c,3}\right).
\end{align}
However, for our computation, we choose only $c=g$.

The coefficients $\varphi_{f,g}^{NSV,\left(i\right)}\left(z,\varepsilon\right)$
are finite as $\varepsilon\rightarrow0$ and can be written in terms
of the finite coefficients $\mathcal{G}_{L,i}^{g}\left(z,\varepsilon\right)$
as \cite{ajjath2020soft},
\begin{align}
\varphi_{f,g}^{NSV,\left(1\right)}\left(z,\varepsilon\right)= & \dfrac{1}{\varepsilon}\mathcal{G}_{L,1}^{g}\left(z,\varepsilon\right),\\
\varphi_{f,g}^{NSV,\left(2\right)}\left(z,\varepsilon\right)= & \dfrac{1}{\varepsilon^{2}}\left\{ -\beta_{0}\mathcal{G}_{L,1}^{g}\left(z,\varepsilon\right)\right\} +\dfrac{1}{2\varepsilon}\mathcal{G}_{L,2}^{g}\left(z,\varepsilon\right),\\
\varphi_{f,g}^{NSV,\left(3\right)}\left(z,\varepsilon\right)= & \dfrac{1}{\varepsilon^{3}}\left\{ \dfrac{4}{3}\beta_{0}^{2}\mathcal{G}_{L,1}^{g}\left(z,\varepsilon\right)\right\} +\dfrac{1}{\varepsilon^{2}}\left\{ -\dfrac{1}{3}\beta_{1}\mathcal{G}_{L,1}^{g}\left(z,\varepsilon\right)-\dfrac{4}{3}\beta_{0}\mathcal{G}_{L,2}^{g}\left(z,\varepsilon\right)\right\} \nonumber\\
& +\dfrac{1}{3\varepsilon}\mathcal{G}_{L,3}^{g}\left(z,\varepsilon\right),
\end{align}
where
\begin{align}
\mathcal{G}_{L,1}^{g}\left(z,\varepsilon\right)= & \sum_{j=1}^{\infty}\varepsilon^{j}\mathcal{G}_{L,1}^{g,\left(j\right)}\left(z\right),\\
\mathcal{G}_{L,2}^{g}\left(z,\varepsilon\right)= & -2\beta_{0}\mathcal{G}_{L,1}^{g,\left(1\right)}\left(z\right)+\sum_{j=1}^{\infty}\varepsilon^{j}\mathcal{G}_{L,2}^{g,\left(j\right)}\left(z\right),\\
\mathcal{G}_{L,3}^{g}\left(z,\varepsilon\right)= & -2\beta_{1}\mathcal{G}_{L,1}^{g,\left(1\right)}\left(z\right)-2\beta_{0}\left(\mathcal{G}_{L,2}^{g,\left(1\right)}\left(z\right)+2\beta_{0}\mathcal{G}_{L,1}^{g,\left(2\right)}\left(z\right)\right)+\sum_{j=1}^{\infty}\varepsilon^{j}\mathcal{G}_{L,3}^{g,\left(j\right)}\left(z\right).
\label{eq:curlyGg}
\end{align}
The coefficients $\mathcal{G}_{L,i}^{g,\left(j\right)}\left(z\right)$ given
in the above equations are parameterized in terms of $\log^{k}\left(1-z\right),
k = 0,1,\cdots$ while all other terms that vanish as $z\rightarrow1$ are dropped
\begin{equation}
\mathcal{G}_{L,1}^{g,\left(j\right)}\left(z\right)=\sum_{k=0}^{i+j-1}\mathcal{G}_{L,i}^{g,\left(j,k\right)}\left(z\right)\log^{k}\left(1-z\right).
\end{equation}
The highest power of the log$\left(1-z\right)$ terms at every order depends
on the order of the perturbation, \textit{i.e.} the power of $a_{s}$ and also
the power of $\varepsilon$ at each order in $a_{s}$.

The expansion coefficients $\varphi_{g,i}^{\left(k\right)}$ are related to $\mathcal{G}_{L,i}^{g,\left(j,k\right)}$ as below \cite{ajjath2020soft}:

\begin{align}
\label{eq:phigk1}
\varphi_{g,1}^{\left(k\right)}= & \mathcal{G}_{L,1}^{g,\left(1,k\right)},\ \ \ \ \ \ k=0,1\\
\varphi_{g,2}^{\left(k\right)}= & \dfrac{1}{2}\mathcal{G}_{L,2}^{g,\left(1,k\right)}+\beta_{0}\mathcal{G}_{L,1}^{g,\left(2,k\right)},\ \ \ \ \ \ k=0,1,2\\
\varphi_{g,3}^{\left(k\right)}= & \dfrac{1}{3}\mathcal{G}_{L,3}^{g,\left(1,k\right)}+\dfrac{2}{3}\beta_{1}\mathcal{G}_{L,1}^{g,\left(2,k\right)}+\dfrac{2}{3}\beta_{0}\mathcal{G}_{L,2}^{g,\left(2,k\right)}+\dfrac{4}{3}\beta_{0}^{2}\mathcal{G}_{L,1}^{g,\left(3,k\right)},\ \ \ \ \ \ k=0,1,2,3.
\label{eq:phigk3}
\end{align}

\section{Next to SV results}
\label{sec:NSVres}

Using all the above available ingredients, we can calculate the NSV coefficient functions for the pseudoscalar Higgs boson production from gluon fusion in terms of the expansion coefficients, $\varphi_{g,i}^{\left(k\right)}$'s, defined in section \ref{subsec:SCD}.
The SV corrections to the production of pseudoscalar Higgs boson are available up to order $a_{s}^{3}$. In contrast, the corresponding NSV corrections are available to order $a_{s}^{2}$.
On the contrary, the NSV corrections to the production of scalar Higgs boson are available up to order $a_{s}^{3}$.
As given in \cite{2016}, the similarity between pseudoscalar Higgs boson production and scalar Higgs boson production is exploited, which leads to the conclusion that the pseudoscalar result can be approximated from the available scalar Higgs boson result using
\begin{align}
\Delta_{gg}^{A}\left(z,q^{2},\mu_{R}^{2},\mu_{F}^{2}\right)=& \dfrac{g_{0}\left(a_{s}\right)}{g_{0}^{H}\left(a_{s}\right)}\bigg[\Delta_{gg}^{H}\left(z,q^{2},\mu_{R}^{2},\mu_{F}^{2}\right)
+\delta\Delta_{gg}^{A}\left(z,q^{2},\mu_{R}^{2},\mu_{F}^{2}\right)\bigg].
\label{deltaAH}
\end{align}
Eqn.\ \ref{deltaAH} effectively defines $\delta\Delta_{gg}^{A}\left(z,q^{2},\mu_{R}^{2},\mu_{F}^{2}\right)$ as the correction to the scalar Higgs coefficient functions such that the rescaling $g_{0}\left(a_{s}\right)/g_{0}^{H}\left(a_{s}\right)$ converts them to the pseudoscalar coefficients.
Here, $\Delta_{gg}^{A}\left(z,q^{2},\mu_{R}^{2},\mu_{F}^{2}\right)$ represents the coefficient function for pseudoscalar Higgs boson and $\Delta_{gg}^{H}\left(z,q^{2},\mu_{R}^{2},\mu_{F}^{2}\right)$ represents the same for scalar Higgs boson.
Moreover, ${g_{0}\left(a_{s}\right)}$ is the constant function of resummation for pseudoscalar Higgs, and ${g_{0}^{H}\left(a_{s}\right)}$ is the analogous function for scalar Higgs.

\mk{
All the above ingredients are known up to NNLO, which has led to the successful computation of $\delta\Delta_{gg}^{A}\left(z,q^{2},\mu_{R}^{2},\mu_{F}^{2}\right)$ up to two-loop level.
It has been shown in \cite{2016} that the $\delta\Delta_{gg}^{A}\left(z,q^{2},\mu_{R}^{2},\mu_{F}^{2}\right)$ corrections vanish at the one-loop level,
and at the two-loop level, these $\delta\Delta_{gg}^{A}\left(z,q^{2},\mu_{R}^{2},\mu_{F}^{2}\right)$ corrections contain only the next-to-next-to-soft terms.
It is conjectured in \cite{2016} that this can be true for all higher orders.
If that is so, then these $\delta\Delta_{gg}^{A}\left(z,q^{2},\mu_{R}^{2},\mu_{F}^{2}\right)$ corrections do not contain any NSV terms at ${\cal{O}}(a_{s}^{3})$.
Moreover, in \cite{2016}, the authors suggest that to define an approximate $\Delta_{gg}^{A}\left(z,q^{2},\mu_{R}^{2},\mu_{F}^{2}\right)$ at N$^3$LO, the unknown ${\cal{O}}(a_{s}^{3})$ contributions to $\delta\Delta_{gg}^{A}\left(z,q^{2},\mu_{R}^{2},\mu_{F}^{2}\right)$ in eqn.\ \ref{deltaAH} can be set to zero with a sufficiently good approximation.
By setting $\delta\Delta_{gg}^{A}\left(z,q^{2},\mu_{R}^{2},\mu_{F}^{2}\right)$ to zero this way in eqn. \ref{deltaAH}, one can obtain the approximate N$^3$LO cross-sections denoted by N$^3$LO$_A$ \cite{Ball_2013,Bonvini:2014joa,Bonvini:2014tea,Bonvini:2015ira,Bonvini_2016_TROLL,Bonvini_2014_TROLL,2016,Bonvini_2018,Bonvini_2018_1,Bonvini:2018xvt}.
Hence, in our analysis, we simply rescale the Higgs SV$+$NSV coefficient functions to
obtain the corresponding ones of the pseudoscalar as
\begin{align}
\Delta_{gg}^{A,NSV}\left(z,q^{2},\mu_{R}^{2},\mu_{F}^{2}\right)= & \dfrac{g_{0}\left(a_{s}\right)}{g_{0}^{H}\left(a_{s}\right)}\bigg[\Delta_{gg}^{H,NSV}\left(z,q^{2},\mu_{R}^{2},\mu_{F}^{2}\right)
\bigg].
\label{eq:deltaAHnew}
\end{align}
The rescaling components, $g_0(a_s)$ and $g_0^H(a_s)$, are known from resummation
\cite{Bonvini:2014joa,CATANI1989323,Anastasiou:2014vaa,Laenen:2008ux} while the scalar Higgs coefficient function, $\Delta_{gg}^{H,NSV}\left(z,q^{2},\mu_{R}^{2},\mu_{F}^{2}\right)$, are obtained from \cite{Anastasiou:2014lda,Anastasiou:2016cez}.
%However, the pseudoscalar Higgs coefficient function, $\Delta_{gg}^{A,NSV}\left(z,q^{2},\mu_{R}^{2},\mu_{F}^{2}\right)$ (and consequently the correction $\delta\Delta_{gg}^{A,NSV}\left(z,q^{2},\mu_{R}^{2},\mu_{F}^{2}\right)$), are not known at ${\cal{O}}(a_{s}^{3})$.
%
The rescaling ratio $g_{0}\left(a_{s}\right)/g_{0}^{H}\left(a_{s}\right)$ up to $a_{s}^{3}$ order is given below:
\begin{align}
\dfrac{g_{0}\left(a_{s}\right)}{g_{0}^{H}\left(a_{s}\right)}= & 1 + a_s (8~C_A) -\frac{1}{3} a_s^2 \bigg[-215~C_A^2+2 ~C_A ~n_f + 3 ~C_F ~n_f \bigg\{ 31-12~
    \log\bigg(\dfrac{m_t^2}{\mu_R^2}\bigg)\bigg\} \bigg]
    \nonumber \\
     & + \frac{1}{81} a_s^3
\bigg[C_A^3 \bigg(-11880 ~\zeta_2-5616~ \zeta_3+68309\bigg)+C_A^2~
    n_f \bigg(-216 ~\zeta_2-1296 ~\zeta_3+1973\bigg)
    \nonumber \\
     & + C_A~ C_F~ n_f
    \bigg\{7776~ \log\bigg(\dfrac{m_t^2}{\mu_R^2}\bigg) -7128 ~\zeta_2+6048
    ~\zeta_3-67094\bigg\} + n_f \bigg(432 ~\zeta_2-631\bigg)
    \nonumber \\
     & + 9~ C_F^2~ n_f \bigg(96 ~\zeta_3+763\bigg)+8~ C_F~ n_f^2 \bigg(162
    ~\zeta_2+565\bigg)-324~C_J^{(2)}\bigg].
 \label{eq:g0Abg0H}
\end{align}

It has been shown that the $\varphi_{g,i}^{\left(k\right)}$'s given in eqns.\  [\ref{eq:phigk1}-\ref{eq:phigk3}] for the scalar and the pseudoscalar Higgs boson productions in gluon fusion are identical to each other at the two-loop level.
The same is also noticed in the DY process and scalar Higgs production \textit{via} bottom quark annihilation up to the two-loop level.
However, for the quark annihilation process, it is found that such universality breaks down at third order for $k=0,1$.
%In \cite{ajjath2020soft}, it has been shown that the $\varphi_{g,i}^{\left(k\right)}$'s given in Eqns.~\ref{eq:phigk} can be deduced from the scalar Higgs boson production in gluon fusion and are identical to that of the pseudoscalar Higgs boson production {\it{via}} gluon fusion with the universality breaking at third order for $k=0,1$.
This was checked in \cite{ajjath2020soft} using the state-of-art results from \cite{Anastasiou_2015,Duhr:2019kwi,Mistlberger:2018etf,Duhr:2020seh}.
In our present work, when we explicitly compute $\varphi_{g,3}^{\left(0\right)}$ and $\varphi_{g,3}^{\left(1\right)}$ from eqn.\ \ref{eq:deltaAHnew}, we notice that they are identical for both scalar and pseudoscalar Higgs boson productions {\it{via}} gluon fusion.
%This can be attributed to the fact that the method we are using from \cite{2016} is an approximate one for computing the unknown expansion coefficients, $\varphi_{g,i}^{\left(k\right)}$'s.
Hence, the universality of the $\varphi_{g,i}^{\left(k\right)}$'s at third order can be checked only when the explicit N$^3$LO results are available for the pseudoscalar Higgs boson production in gluon fusion.
%at N$^3$LO is possible in the future, then these coefficients could be determined exactly.
%So, we can say that the expressions obtained for $\varphi_{g,3}^{\left(0\right)}$ and $\varphi_{g,3}^{\left(1\right)}$ in this section cannot be completely true.
}

As depicted in Section 3 of \cite{ajjath2020soft}, once we have the $\varphi_{g,i}^{\left(k\right)}$ values up to a particular order, we can predict the coefficients of the highest logarithms in the finite partonic coefficient functions for a few consecutive higher orders.
In our case, we could predict these coefficients for $\Delta_{gg}^{A,NSV}\left(z,q^{2},\mu_{R}^{2},\mu_{F}^{2}\right)$ by virtue of the formalism developed in \cite{Ravindran_2006,Ravindran:2005vv,ajjath2020soft}.
Using the evaluated $\varphi_{g,i}^{\left(k\right)}$ values, we could predict the three highest logarithms of $\Delta_{gg}^{A,NSV}\left(z,q^{2},\mu_{R}^{2},\mu_{F}^{2}\right)$, up to ${\cal{O}}(a_{s}^{7})$ from ${\cal{O}}(a_{s}^{4})$.
\mk{
Even when we use the available $\varphi_{g,i}^{\left(k\right)}$'s for $k=0,1$, we could predict these highest logarithmic terms because they are independent of the unknown variables, $\varphi_{g,3}^{\left(0\right)}$ and $\varphi_{g,3}^{\left(1\right)}$.
}
%These predictions match the result obtained using the relationship developed in \cite{2016}. The NSV coefficient functions calculated from the available Higgs boson results can be matched with that obtained using the developed formalism in \cite{Ravindran_2006,Ravindran_2006_B,ajjath2020soft}. Till ${\cal{O}}(a_{s}^{3})$, the results match exactly using the above defined $\varphi_{g,i}^{\left(k\right)}$ values.
The ingredients needed to match the other higher-order results of the pseudoscalar Higgs boson still need to be made available.
So, we compute the highest logarithms up to ${\cal{O}}(a_{s}^{7})$ from ${\cal{O}}(a_{s}^{4})$ of the $\Delta_{gg}^{A,NSV}\left(z,q^{2},\mu_{R}^{2},\mu_{F}^{2}\right)$ result.
%These do match and provide a consistency check on the results obtained.
%

The explicit results of the SV+NSV cross-section up to N$^3$LO are given in the Appendix \ref{appendix:A} where it can be seen that the general form of the output is depicted as:
\begin{equation}
\Delta_{g,i}^{A,NSV}\left(z,q^{2}\right)=\delta(1-z)[\cdots] + \sum_{j=1/2,1,\cdots}^{i}\mathcal{D}_{(2j-1)}[\cdots] + \sum_{j=1/2,1,\cdots}^{i}\log^{(2j-1)}(1-z)[\cdots].
\end{equation}
where $i = 1,2,3$, and $[\cdots]$ represent coefficients of the corresponding distribution, logarithmic term or delta function.
The constant terms in the above result correspond to the $\log^0(1-z)$ coefficients.
Although we do all the analytical computations in $z$-space, we shift to the Mellin ($N$-moment) space for numerical analysis.
The next section is dedicated to this change from $z$ to $N$-moment space.

\section{Resummation of the NSV results in Mellin Space}
\label{sec:resNSVformalism}

As we shift from the $z$-space to the Mellin space, the Mellin moment of the partonic coefficient function, $\Delta_{gg}$, is given by \cite{ajjath2020soft}
\begin{eqnarray}
\label{DeltaN}
\Delta_{gg,N}(q^2,\mu_R^2,\mu_F^2) = C_0(q^2,\mu_R^2,\mu_F^2) \exp\left(
\Psi_N^g (q^2,\mu_F^2)
\right)\,,
\label{eqn:NspaceExp}
\end{eqnarray}

In order to study the all-order behaviour of the coefficient function, $\Delta_{gg}$, in the $N$-moment space, it is convenient to use the following form of the partonic coefficient function \cite{ajjath2020soft}:
\begin{eqnarray}
\label{resumz}
\Delta_{gg}(q^2,\mu_R^2,\mu_F^2,z)= C^g_0(q^2,\mu_R^2,\mu_F^2)
~~{\cal C} \exp \Bigg(2 \Psi^g_{\cal D} (q^2,\mu_F^2,z) \Bigg)\,,
\end{eqnarray}
where
\begin{eqnarray}
\label{phicint}
\Psi^g_{\cal D} (q^2,\mu_F^2,z) &=& {1 \over 2}
\int_{\mu_F^2}^{q^2 (1-z)^2} {d \lambda^2 \over \lambda^2}
P_{gg} (a_s(\lambda^2),z)  + {\cal Q}^g(a_s(q^2 (1-z)^2),z)\,,
\end{eqnarray}
with
\begin{eqnarray}
\label{calQc}
{\cal Q}^g (a_s(q^2(1-z)^2),z) &=&  \left({1 \over 1-z} \overline G^g_{SV}(a_s(q^2 (1-z)^2))\right)_+ + \varphi_{f,g}(a_s(q^2(1-z)^2),z).
\end{eqnarray}
The coefficient $C_0^g$ is $z$ independent and is expanded in powers of $a_s(\mu_R^2)$ as
\begin{eqnarray}
\label{C0expand}
C_0^g (q^2,\mu_R^2,\mu_F^2) = \sum_{i=0}^\infty a_s^i(\mu_R^2) C_{0i}^g (q^2,\mu_R^2,\mu_F^2)\,,
\end{eqnarray}
where the coefficients $C^g_{0i}$ are calculated in \cite{Ahmed_2015} for pseudoscalar.
Eqn.\ \ref{resumz} gives the $z$-space resummed result.

Now it is easy to compute the Mellin moment of $\Delta_{gg}$.
%This formalism is suitable for obtaining only SV and NSV terms.{\color{black}{If we use this formalism to predict beyond NSV terms, such as those proportional to ${\cal O}((1-z)^n \log^j(1-z));~n,j\ge 0$ in $z$-space and terms like ${\cal O}(1/N^2)$ in $N$-space, they will not be correct.}} Hence, we compute the Mellin moment of Eqn.(\ref{resumz}) in the appropriate limit of $N$ such that the resulting expression in $N$-space correctly predicts all the SV and NSV terms.
The limit $z\rightarrow 1$ translates to $N\rightarrow \infty$ in the $N$-moment space, and to include NSV terms, we need to keep ${\cal O}(1/N)$ corrections in the large $N$ limit.
The Mellin moment of the exponent in eqn. \ref{eqn:NspaceExp} acquires the following form:
\begin{eqnarray}
\Psi_N^g = \Psi_{\rm{SV},N}^g + \Psi_{\rm{NSV},N}^g ,
\end{eqnarray}
where we can split $\Psi_N^g$ in such a way that all those terms that are functions of $\log^j(N),~j=0,1,\cdots$ are
kept in $\Psi_{{\rm SV},N}^g$ and the remaining terms that are proportional to $(1/N) \log^j(N),~j=0,1,\cdots$ are contained
in $\Psi_{\rm{NSV},N}^g$.  Hence,
\begin{eqnarray}
\label{PsiSVN}
	\Psi_{\rm{SV},N}^g = \log(g_0^g(a_s(\mu_R^2))) + g_1^g(\omega)\log(N) + \sum_{i=0}^\infty a_s^i(\mu_R^2) g_{i+2}^g(\omega) \,,
\end{eqnarray}
where the $g^g_i(\omega)$'s, identical to those in \cite{CATANI1989323,Moch:2005ba,H:2019dcl}, are obtained from the resummed formula for SV terms
and $g^g_0(a_s)$ is expanded in powers of $a_s$ as (see \cite{Moch:2005ba})
\begin{eqnarray}
	\log(g_0^g(a_s(\mu_R^2))) = \sum_{i=1}^\infty a_s^i(\mu_R^2) g^g_{0,i}\quad \,.
\end{eqnarray}

These SV coefficients are defined as
\begin{align}
 g_{0,0} = &~ 0, \\
 g_{0,1} = &~ 2 f_1 \gamma_E + 2 A_1 \gamma_E^2 + 2 A_1 \gamma_E \log(\mu_F^2/\mu_R^2) - 2 A_1 \gamma_E \log(q^2/\mu_R^2) + 2 A_1 \zeta_2, \\
 g_1 = &~ \dfrac{2 A_1}{\beta_0}\left[ 1 - \log(1-\omega) + \dfrac{\log(1-\omega)}{\omega} \right],
 \\
 g_2 = &~ \dfrac{A_1 \beta_1}{\beta_0^3} \left[ {\omega} +
 {\log(1-\omega)} + \dfrac{\log^2(1-\omega)}{2} \right] - \dfrac{A_2 \omega}{\beta_0^2} - \dfrac{A_2 \log(1-\omega)}{\beta_0^2} - \dfrac{f_1 \log(1-\omega)}{\beta_0}
 \nonumber \\
 & + \dfrac{A_1 \log\left({\mu_F^2}/{\mu_R^2}\right) \omega}{\beta_0} - \dfrac{2 A_1 \gamma_E \log(1-\omega)}{\beta_0} + \dfrac{A_1 \log\left({q^2}/{\mu_R^2}\right) \log(1-\omega)}{\beta_0}.
 %(bt1*((A1*w)/bt0 + (A1*zlwm1)/bt0 +  (A1*zlwm2)/(2*bt0)))/bt0^2.
 \label{eqn:gis}
\end{align}
where $f_i$'s are the soft anomalous dimensions and $A_i$'s are the cusp anomalous dimensions given in eqns.\ \ref{eqn:soft} and \ref{eqn:Cusp}, respectively.
$\beta_i$'s are the standard QCD beta functions defined in eqns.\ \ref{eqn:betais1} - \ref{eq:betais} and $\omega$ is defined as $\omega = 2 \beta_0 a_s (\mu^2_R ) \log(N)$.
%
%These coefficients are also provided in the ancillary files of \cite{ajjath2020soft}.

The function $\Psi_{\rm{NSV},N}^g$ is given by
\begin{align}
\label{PsiNSVN}
 \Psi_{\rm{NSV},N}^g = {1 \over N} \sum_{i=0}^\infty a_s^i(\mu_R^2) \bigg(
 \bar g_{i+1}^g(\omega)  +  h^g_{i}(\omega,N) \bigg)\,,
\end{align}
with
%  \Bigg)\,,
%\nonumber\\
\begin{align}
	h^g_i(\omega,N) = \sum_{k=0}^{i} h^g_{ik}(\omega)~ \log^k(N).
\label{eq:hcik}
\end{align}
where the resummation constants, $\bar g_{i}^g(\omega) $ and $h^g_i(\omega,N)$, are explicitly given in \cite{ajjath2020soft}.

In Mellin ($N$-moment) space, the compact expression of the exponent in eqn.\ \ref{eqn:NspaceExp} is seen to depend on functions of $\omega$ as a resummed $a_s$ is used to perform the integral.
Additionally, because of this resummed $a_s$, the $N$ space perturbative expansion could be organised so that $\omega$ behaves as order one at every order in $a_s(\mu^2_R)$.
Since the $N$-space result is obtained from the $z$-space computation, they naturally contain the exact same information.
However, to resum the large threshold logarithmic terms to all-order in the perturbative structure, the $N$-space result proves handy because the all-order structure becomes more transparent in the Mellin space.
So, the resummed $N$-space result becomes technically easy to use for phenomenological studies.

Additionally, the master formula in eqn.\ \ref{eq:deltaNSV} displays a predictive power \textit{i.e.}, if each of its components is known up to a definite order, then this formula can predict SV and SV+NSV terms to all-orders in perturbation theory.
For example, in our case of a pseudoscalar Higgs boson production \textit{via} gluon fusion, the components of the exponent was known till NNLO accuracy and we could predict the coefficients of the three highest logarithms up to $\mathcal{O}(a_s^7)$.
These coefficients are given in table \ref{tab:PredLog}.
\ab{
\begin{table}[!htbp]
\begin{center}
\renewcommand{\arraystretch}{1.5}
  \begin{tabular}{ | l | c | c | c | }
    \hline
    $\mathbf{\mathcal{O}(a_s^4)}$ & \textbf{Log$\mathbf{^7(1-z)}$} & \textbf{Log$\mathbf{^6(1-z)}$} & \textbf{Log$\mathbf{^5(1-z)}$}
    \\ \hline
            & \scriptsize{$-\dfrac{4096}{3}C_A^4$} & \scriptsize{$\dfrac{98560}{9}C_A^4 - \dfrac{7168}{9}n_f C_A^3$}  & \scriptsize{$-\dfrac{335104}{9}C_A^4 + \dfrac{174208}{27} n_f C_A^3 - \dfrac{4096}{27} n_f^2 C_A^2 + 23552 \zeta_2 C_A^4 $}
    \\ \hline
    $\mathbf{\mathcal{O}(a_s^5)}$ & \textbf{Log$\mathbf{^9(1-z)}$} & \textbf{Log$\mathbf{^8(1-z)}$} & \textbf{Log$\mathbf{^7(1-z)}$}
    \\ \hline
            & \scriptsize{$-\dfrac{8192}{3}C_A^5$} & \scriptsize{$\dfrac{96256}{3}C_A^5 - \dfrac{8192}{3}n_f C_A^4$}  & \scriptsize{$-\dfrac{131685640}{81}C_A^5 + \dfrac{2569216}{81} n_f C_A^4 - \dfrac{81920}{81} n_f^2 C_A^3 + \dfrac{2262144}{3} \zeta_2 C_A^5$}
    \\ \hline
    $\mathbf{\mathcal{O}(a_s^6)}$ & \textbf{Log$\mathbf{^{11}(1-z)}$} & \textbf{Log$\mathbf{^{10}(1-z)}$} & \textbf{Log$\mathbf{^9(1-z)}$}
    \\ \hline
            & \scriptsize{$-\dfrac{65536}{15}C_A^6$} & \scriptsize{$\dfrac{9490432}{135}C_A^6 - \dfrac{180224}{27} n_f C_A^5$}  & \scriptsize{$-\dfrac{4458496}{9}C_A^6 + \dfrac{8493056}{81} n_f C_A^5 - \dfrac{327689}{81} n_f^2 C_A^4 + \dfrac{671744}{3} \zeta_2 C_A^6$}
    \\ \hline
    $\mathbf{\mathcal{O}(a_s^7)}$ & \textbf{Log$\mathbf{^{13}(1-z)}$} & \textbf{Log$\mathbf{^{12}(1-z)}$} & \textbf{Log$\mathbf{^{11}(1-z)}$}
    \\ \hline
            & \scriptsize{$-\dfrac{262144}{45}C_A^7$} & \scriptsize{$\dfrac{3309568}{27}C_A^7 - \dfrac{1703936}{135} n_f C_A^6$}  & \scriptsize{$-\dfrac{92717056}{81}C_A^7 + \dfrac{115835488}{45} n_f C_A^6 - \dfrac{917504}{81} n_f^2 C_A^5 + \dfrac{1310720}{3} \zeta_2 C_A^7$}
    \\ \hline
  \end{tabular}
\end{center}
\caption[Predictive coefficients of the highest logarithms for higher orders in Mellin space threshold resummation]{Coefficients of the three highest power logarithmic terms from $\mathcal{O}(a_s^4)$ to $\mathcal{O}(a_s^7)$ predicted from the all-order structure of the master formula in Mellin space}
\label{tab:PredLog}
\end{table}
}
Similarities between the pseudoscalar Higgs boson data given in table \ref{tab:PredLog} and the corresponding scalar Higgs data studied in \cite{ajjath2020soft} have been observed.
The two highest power logarithms in the predicted results of each order match precisely with that of the scalar Higgs data (first two columns in table \ref{tab:PredLog}).
Differences arise in the coefficients of the lowest power of the logarithm, say $\log^{k}(1-z)$ for $a_s^i$ where $k=i+1$, that we could predict for the coefficients $C_A^i$.
All other coefficients match exactly with the Higgs data.

\section{Numerical Results and Discussion}
\label{sec:Numerical}

In this section, we will present our numerical results of the NSV corrections at N$^3$LO level in QCD for the production of a pseudoscalar Higgs boson at the LHC.
Our predictions are based on EFT, where the top quarks are integrated out at higher orders.
%However, we retain the top quark mass dependence at LO.
%, and we are only left with light quarks.
The term C$_J^{(2)}$ in the Wilson coefficient $C_J(a_s)$ is taken to be zero in our analysis because it is not available in the literature yet.
We have set cot $\beta=1$ in our numerical analysis for simplicity. Results for other values of cot $\beta$ can be easily obtained by rescaling the cross-sections with cot$^2\beta$.
At LO, we have retained the full top quark mass dependence, while the EFT approach has been used for higher-order corrections.
We use MMHT 2014 PDFs throughout where the LO, NLO and NNLO parton level cross-sections are convoluted with the corresponding order-by-order central PDF sets.
However, for N$^3$LO cross-sections, we use MMHT2014nnlo68cl PDFs.
The strong coupling constant is provided by the respective PDFs from LHAPDF.

To estimate the impact of QCD corrections, we define the K-factors as
\begin{equation}
K_{\left(1\right)}^X=\dfrac{\sigma_\text{NLO}^X}{\sigma_\text{LO}},\ \ \ \ \ K_{\left(2\right)}^X=\dfrac{\sigma_\text{NNLO}^X}{\sigma^\text{LO}},
\label{eq:KfactorFO}
\end{equation}
where X is either SV or SV+NSV or Full which includes all possible sub-processes or Full$(gg)$ which includes only the gluon-gluon$(gg)$ sub-process.
When the Full$(gg)$ case is computed, only the $gg$ sub-process is considered at the $n$-th order, while at the lower orders $(k<n)$, all possible sub-processes are taken together.

\begin{figure}[!htb]
\centering
\subfloat{%
  \includegraphics[clip,scale=0.5]{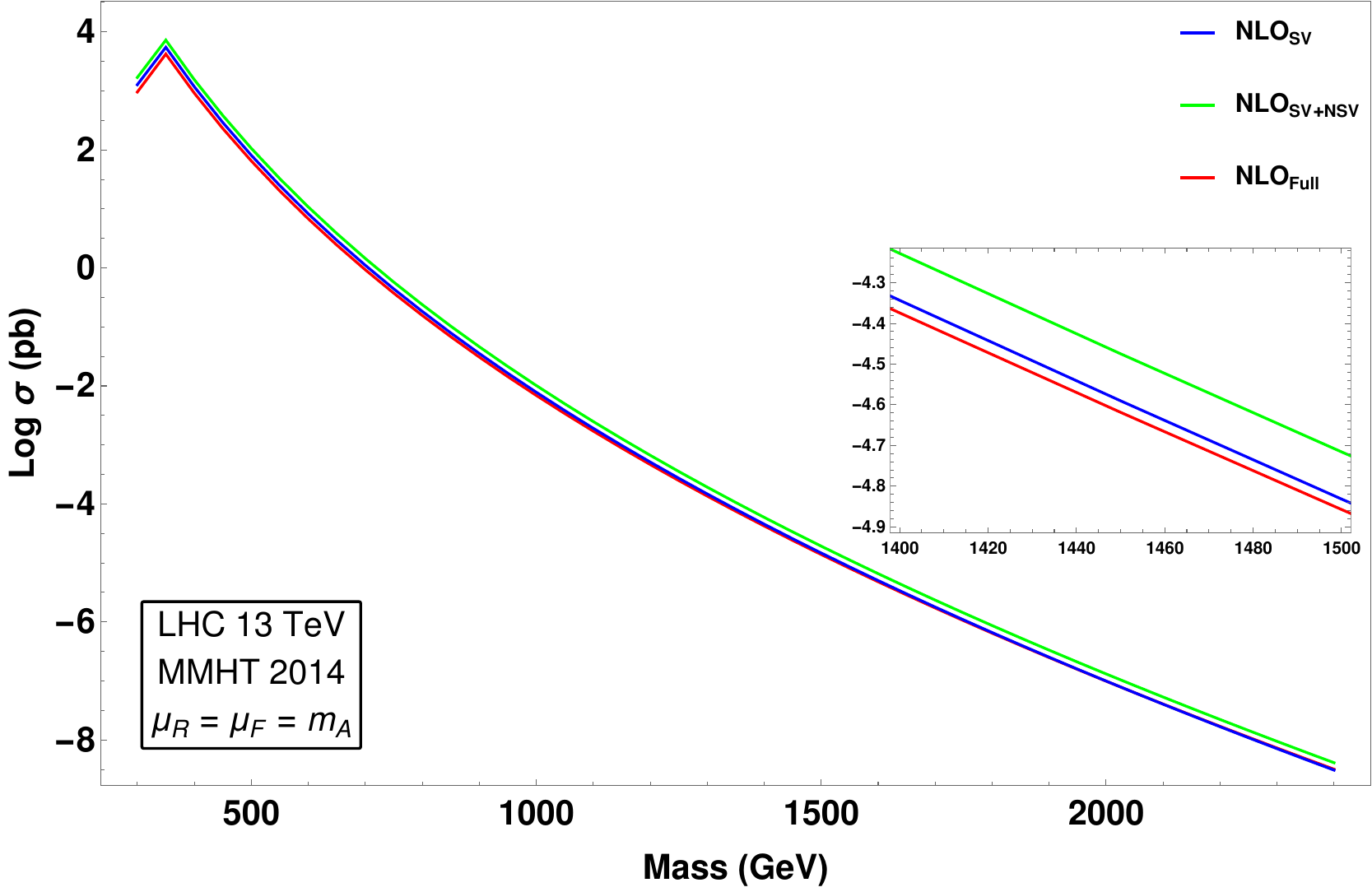}%
}
\vspace{0.5cm}
\subfloat{%
  \includegraphics[clip,scale=0.5]{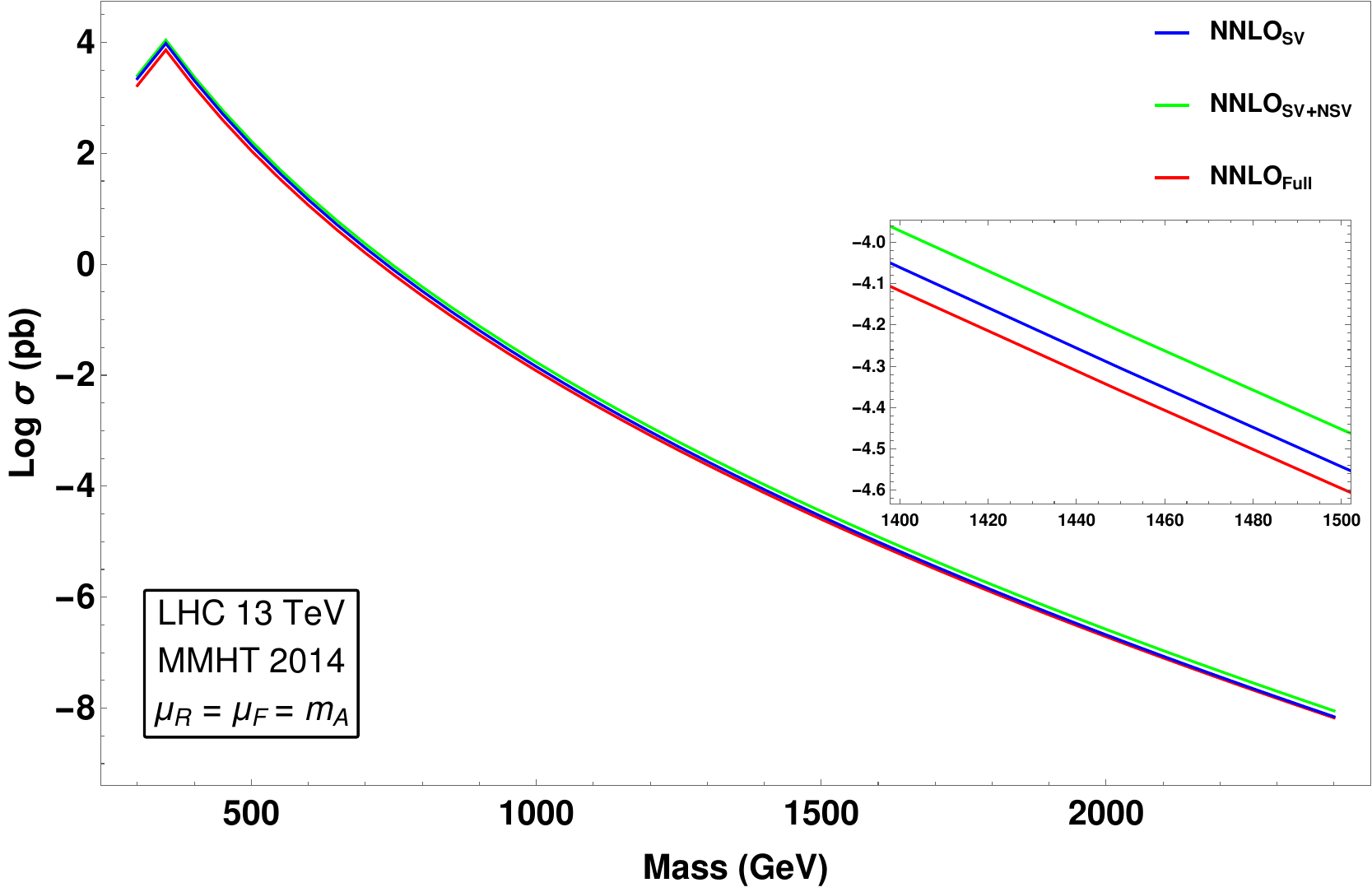}%
}
\caption[Pseudoscalar production cross-section at NLO level and NNLO level]{Pseudoscalar production cross-section at NLO level (left panel) and NNLO level (right panel) with a comparison between fixed order SV, SV+NSV and Full results for 13 TeV LHC}
\label{fig:cross12}
\end{figure}

\begin{figure}[!htbp]
\centering
\subfloat{%
  \includegraphics[clip,scale=0.42]{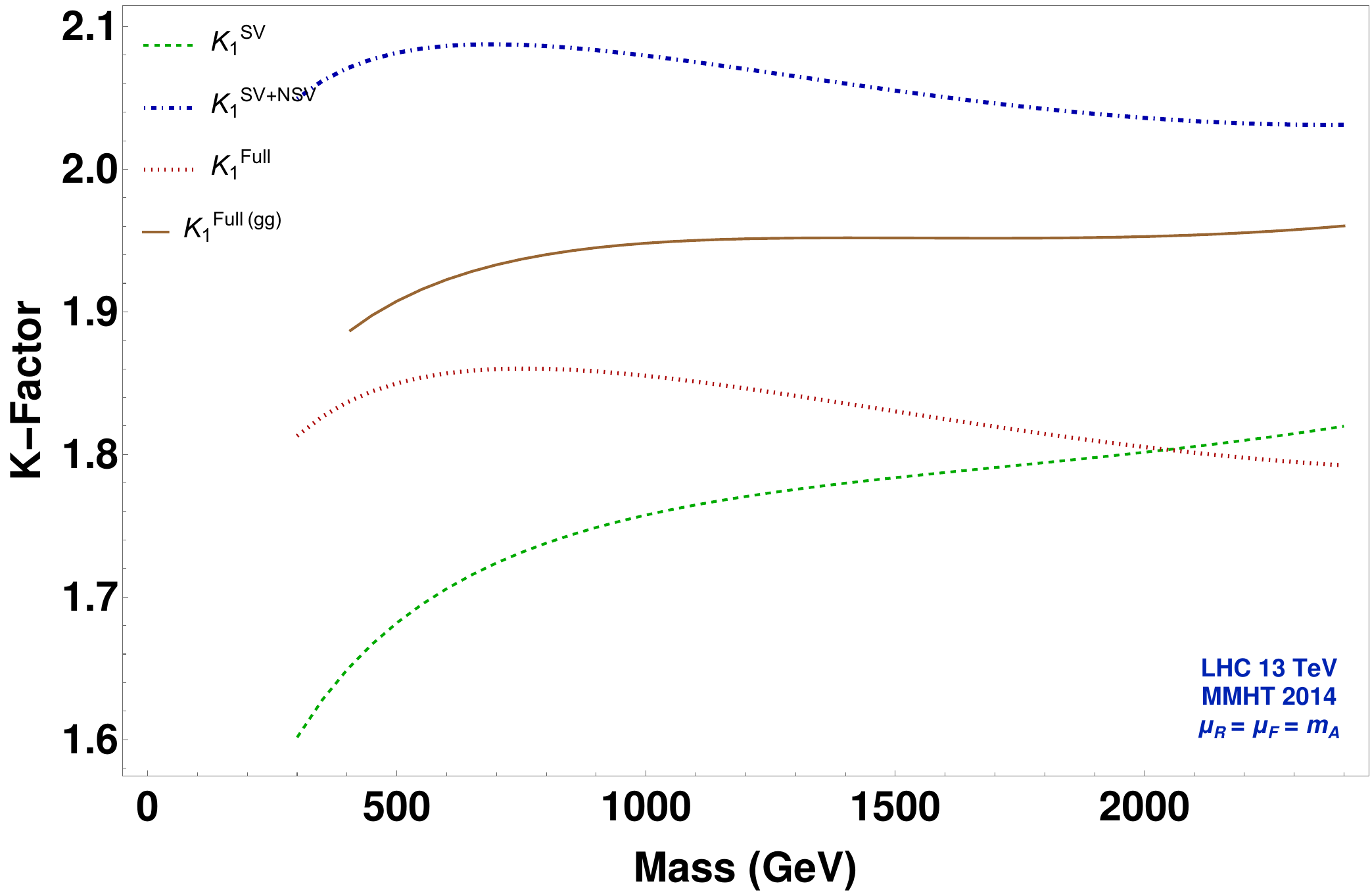}%
}
\vspace{0.5cm}
\subfloat{%
  \includegraphics[clip,scale=0.42]{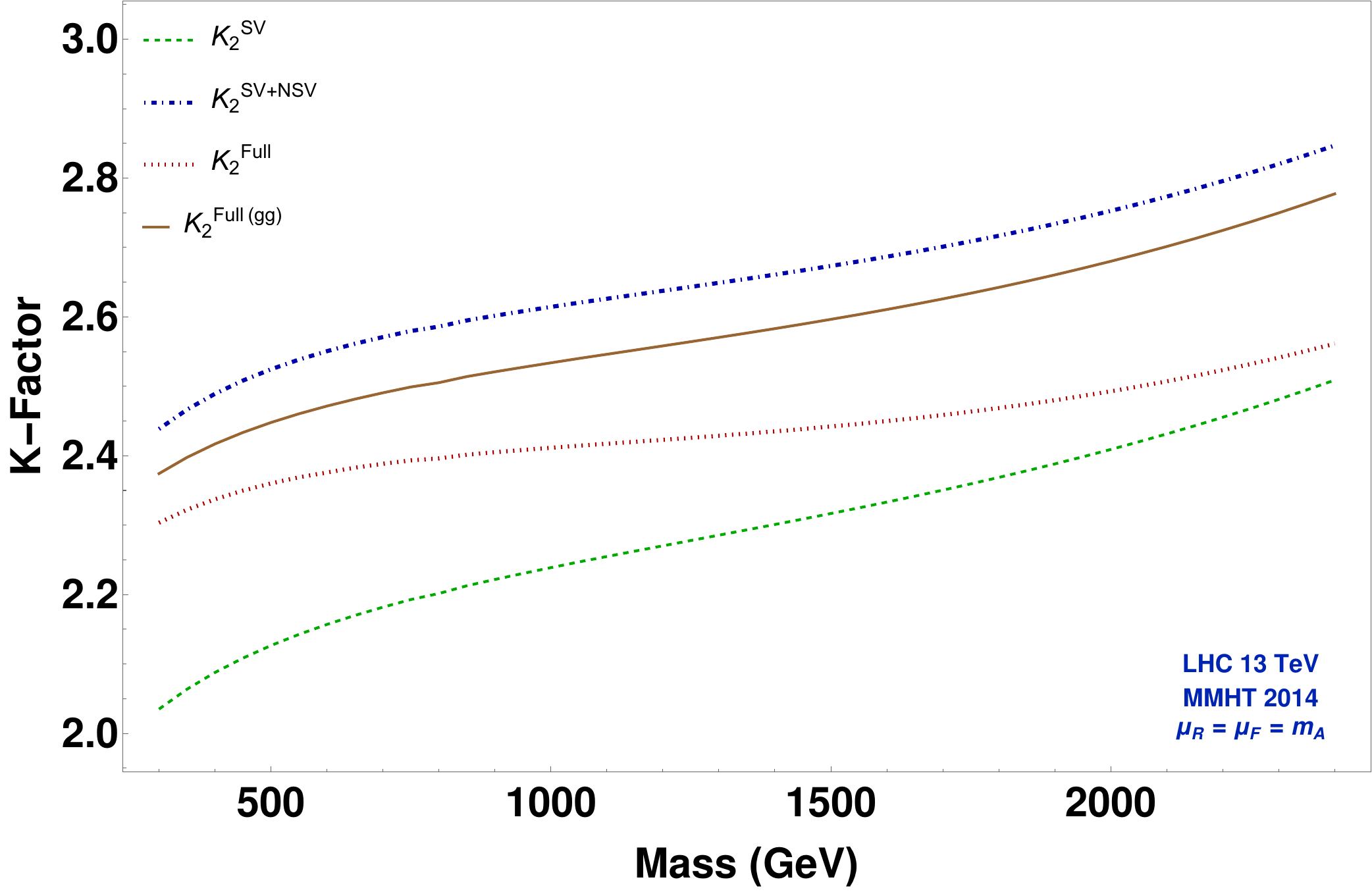}%
}
\caption[K-Factor for pseudoscalar production cross-section at NLO level and NNLO level]{K-Factor for pseudoscalar production cross-section at NLO level (left panel) and NNLO level (right panel) with a comparison between fixed order SV, SV+NSV,  Full (all sub-processes included) and Full (only $gg$ sub-process included at $n$-th order) results for 13 TeV LHC.}
\label{fig:kfactor12}
\end{figure}

In fig.\ \ref{fig:cross12}, we plot the pseudoscalar production cross-section as a function of its mass $m_A$ at NLO(left panel) and NNLO (right panel) by varying $m_A$ from $300$ GeV to $2400$ GeV.
Here the "SV" results contain the SV threshold logarithms and $\delta(1-z)$ contributions, while the "NSV" corrections retain the NSV logarithms in the gluon fusion channel only.
We note that the NSV corrections at higher orders also arise from other partonic channels, a detailed study of which is beyond the scope of the present work and will be
presented elsewhere.
As shown in fig.\ \ref{fig:cross12}, the SV results give a sizable contribution to the FO results; however, they underestimate the latter. Including NSV corrections increases the cross-section substantially, but they overestimate the FO results for the mass range of $m_A$ we have considered.
%
%On the contrary, when we consider the full result, it has a negative impact on the NSV
%results, and hence, the overall result lowers.
%

To better study the impact of these corrections at SV and beyond, we depict in fig.\ \ref{fig:kfactor12} the corresponding K-factors defined in eqn.\ \ref{eq:KfactorFO}.
Since our NSV corrections include only the gluon fusion channel, it is worthwhile to have a comparison with the complete result of the gluon fusion channel, including the pure regular contributions. %{\color{red}{To this end, we also consider the K-factor () by taking into account only the gluon fusion channel contributions at higher orders.}}
From these K-factors, we notice that the SV corrections converge to the FO result (denoted by $K_i^{Full}$) in the high mass region and differ significantly in the small mass region. However, the FO result here does contain contributions from other parton channels. The difference noticed between the full FO result and the FO result for the pure gluon channel (denoted by $K_i^{Full(gg)}$) contribution can be understood from the presence of these other parton channels, which have a negative contribution to the pure gluon contribution.
We also notice that the SV+NSV corrections behave similarly to the total gluon fusion channel contribution, with the former being slightly higher than the latter.
\mk{This indicates that the regular or beyond NSV corrections also have a negative impact but are smaller in magnitude compared to the NSV corrections at higher orders.}

%When we add the NSV results, there is a considerable change. However, the K-factor shows a negative effect when we include all possible sub-processes (denoted by $K_i^{Full}$). On the contrary, when only the gluon-gluon sub-process (denoted by $K_i^{Full(gg)}$) is considered, the K-factor shows an increase from the case where all the sub-processes are considered.

Next, we study the impact of the resummation of these NSV logarithms on the pseudoscalar production cross-section at NLL and NNLL accuracy.
For the resummed cross-section, we do the matching as below:
\begin{equation}
 \sigma^{(\text{matched})}=\sigma^{\text{SV+NSV}}_{\text{resum}}-\sigma^{\text{SV+NSV}}\bigg|_{(\text{FO})}+\sigma^{(\text{FO})}.
\end{equation}

\begin{figure}[!htb]
\centering
  \includegraphics[keepaspectratio,scale=0.4]{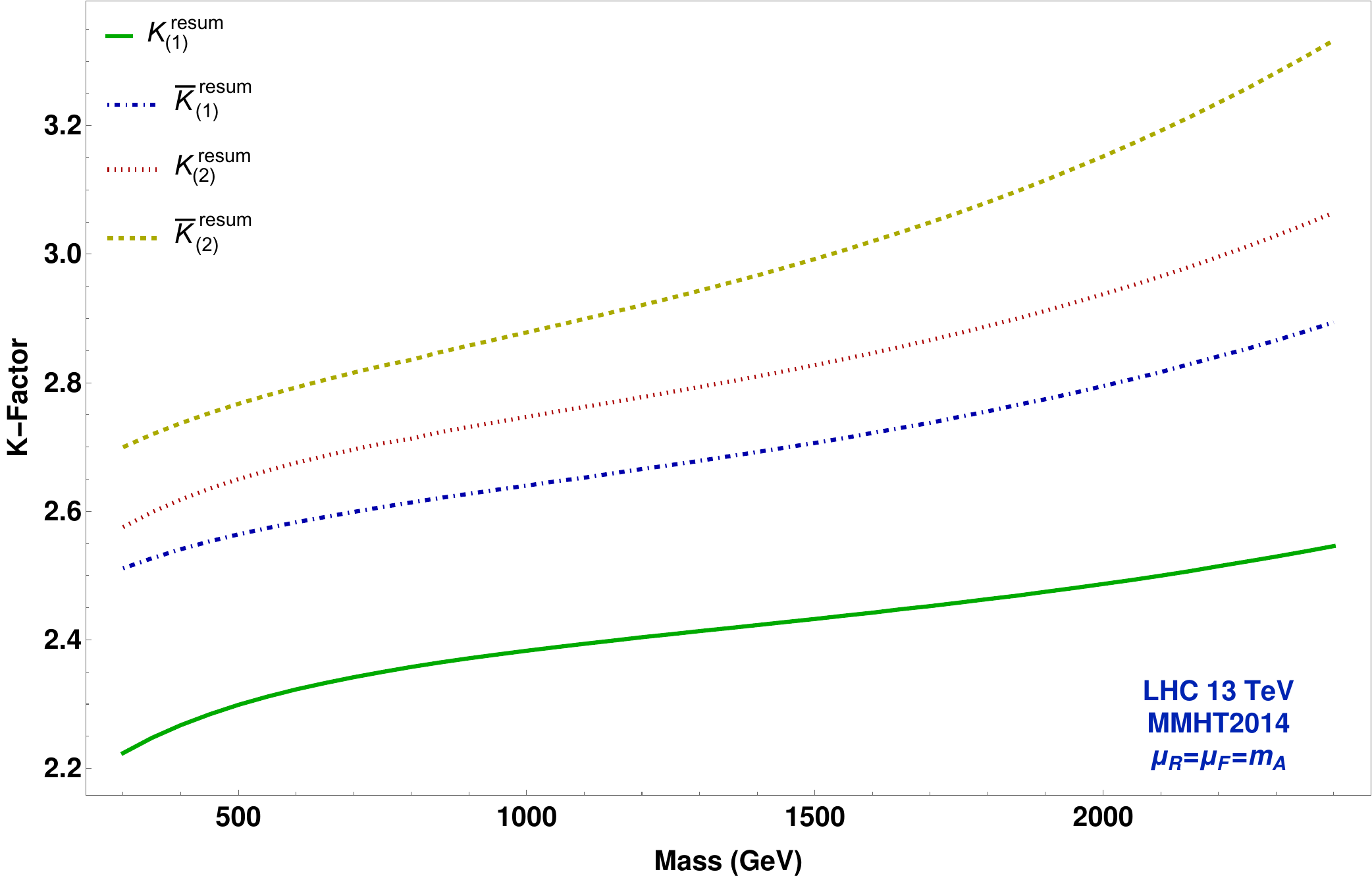}
% Resum_NLO_NNLO.pdf: 0x0 px, 300dpi, 0.00x0.00 cm, bb=
% PDFUncertainty.pdf: 0x0 px, 300dpi, 0.00x0.00 cm, bb=
  \caption[Resummed K-factor plots for 13 TeV LHC]{Resummed K-factor plots for 13 TeV LHC taking MMHT 2014 as the reference PDF and choosing the central scale $\mu_R = \mu_F = m_A$.}
  \label{fig:sub1}
\label{fig:resumKfactor}
\end{figure}

In fig.\ \ref{fig:resumKfactor}, we depict the resummed K-factors at NLO and NNLO orders and contrast them against the corresponding ones due to the SV resummation for a wide range of pseudoscalar mass {\it{i.e.}} $300 < m_A < 2400$ GeV.
We define these K-factors as
\begin{align}
K_{\left(1\right)}^\text{resum}=\dfrac{\sigma_\text{NLO+NLL}}{\sigma_\text{LO}},\ \ \ \ \ K_{\left(2\right)}^\text{resum}=\dfrac{\sigma_\text{NNLO+NNLL}}{\sigma^\text{LO}},
\nonumber \\
\nonumber \\
\overline{K}_{\left(1\right)}^\text{~resum}=\dfrac{\sigma_{\text{NLO}+\overline{\text{NLL}}}}{\sigma_\text{LO}},\ \ \ \ \ \overline{K}_{\left(2\right)}^\text{~resum}=\dfrac{\sigma_{\text{NNLO}+\overline{\text{NNLL}}}}{\sigma^\text{LO}},
\end{align}
The resummed NLO SV K-factor $({K}_{(1)}^{\text{resum}})$ varies from 2.2 (at $m_A=300$ GeV) to about 2.5 (at $m_A=2400$ GeV).
However, the inclusion of NSV logarithms at $\overline{\text{NLL}}$ accuracy increases these results by about 30\% in the low mass region and by about 40\% of LO in the high mass region {\it{i.e.}} $(\overline{K}_{(1)}^{~\text{resum}})$ varies from $2.5$ to about $2.9$.
The resummation of NSV logarithms to $\overline{\text{NNLL}}$ accuracy has similar behaviour and enhances the SV resummed results by about
\mk{10\% (30\%)}
in the low (high) mass region.

%
%the least stability followed by the resummed NLO NSV K-factor (K$_1^{\text{resum,NSV}}$). Then comes the NNLO SV K-factor (K$_2^{\text{resum,SV}}$) followed by the NNLO NSV K-factor (K$_2^{\text{resum,NSV}}$).
%
%

\begin{figure}[!htb]%[H]%
\centering
\subfloat{%
  \includegraphics[clip,scale=0.5]{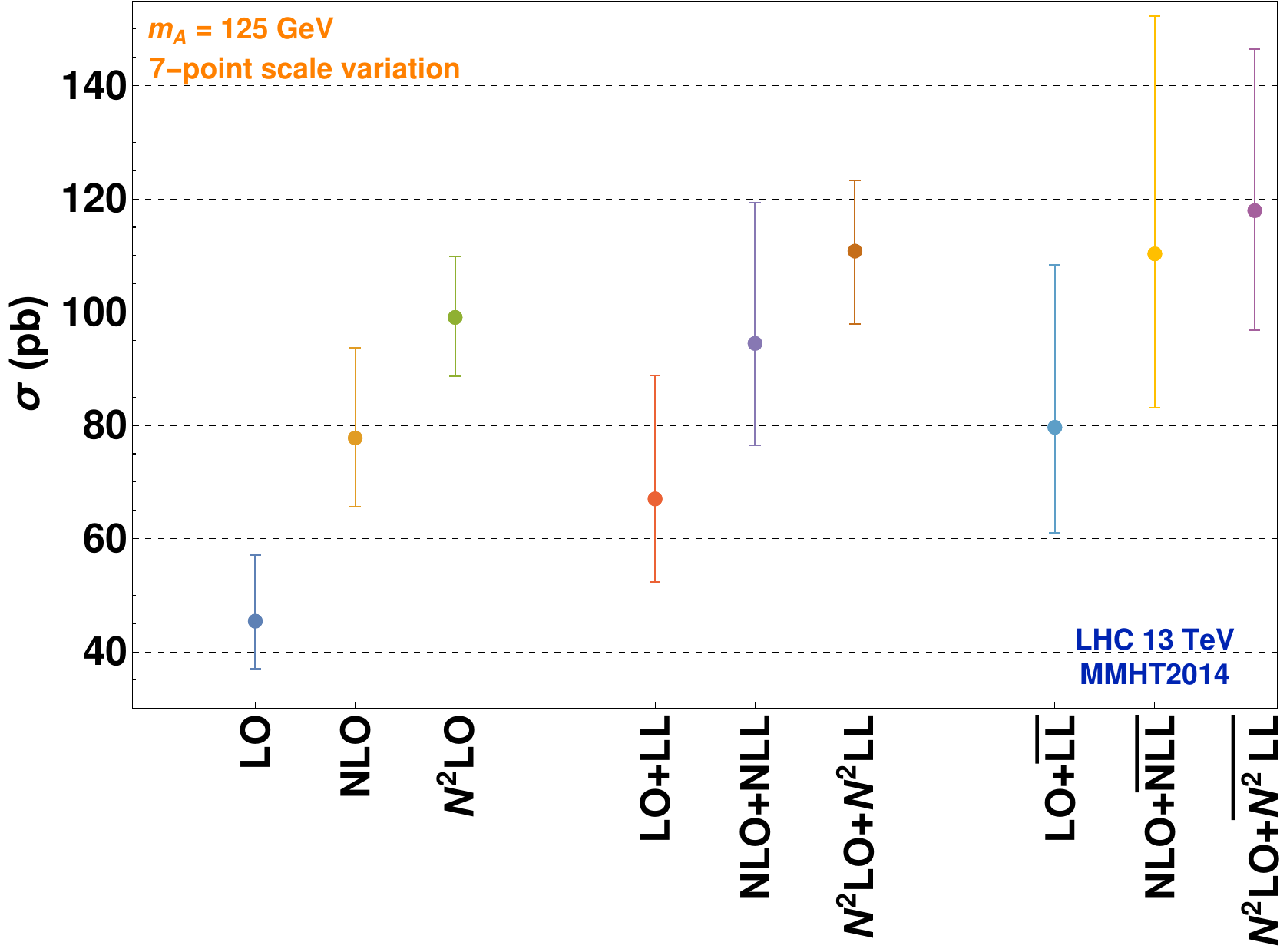}%
}

\subfloat{%
  \includegraphics[clip,scale=0.5]{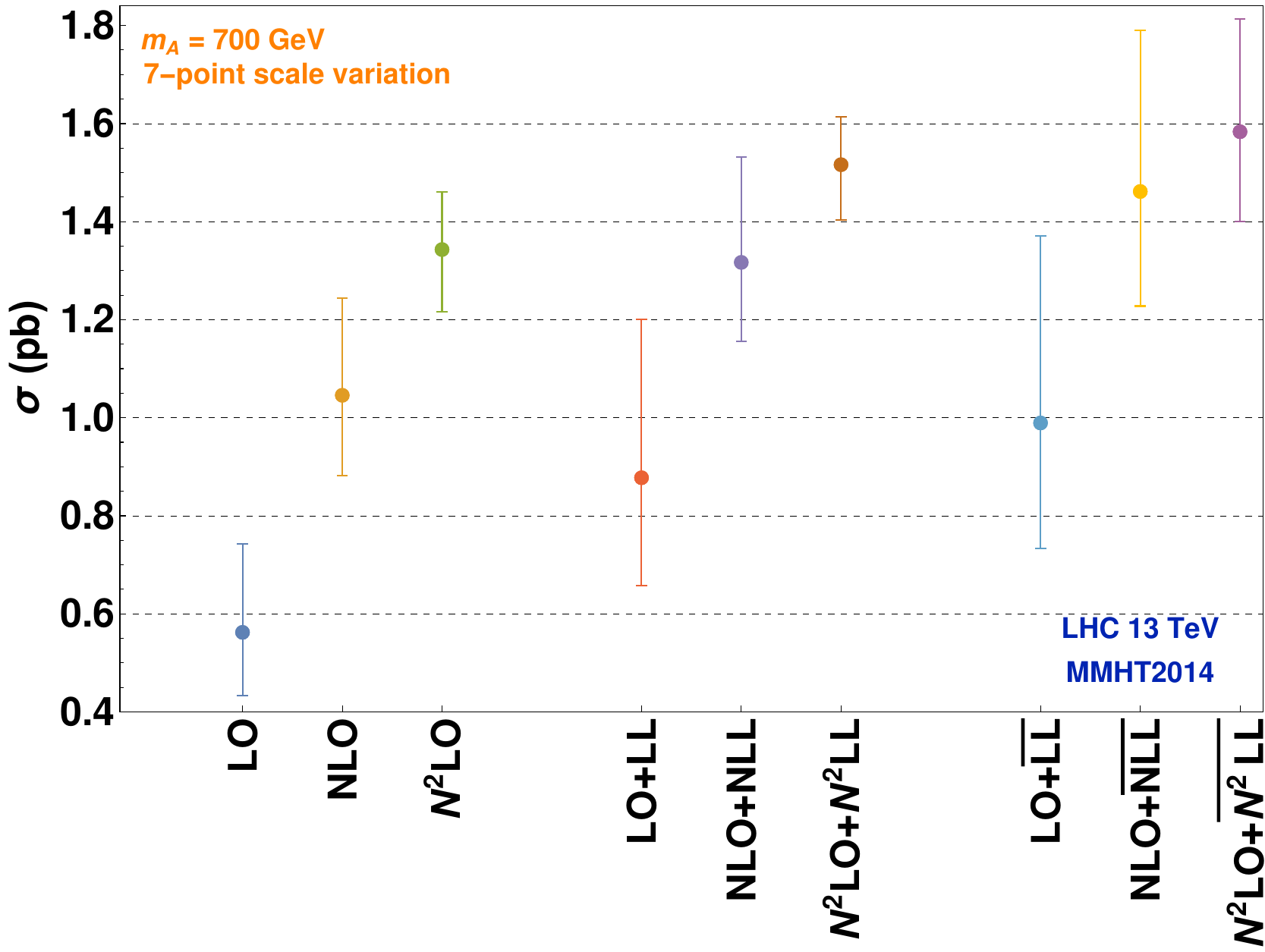}%
}
\caption[Uncertainty plot with 7-point scale uncertainty for $m_A$=125, 700 GeV]
{Uncertainty plot with 7-point scale uncertainty for $m_A$=125 GeV (top figure) and $m_A$=700 GeV (bottom figure) for 13 TeV LHC with MMHT 2014 PDF.}
\label{fig:PS_7pt_ErrorBar}
\end{figure}

\begin{figure}[!htb]
\centering
\subfloat{%
  \includegraphics[clip,scale=0.5]{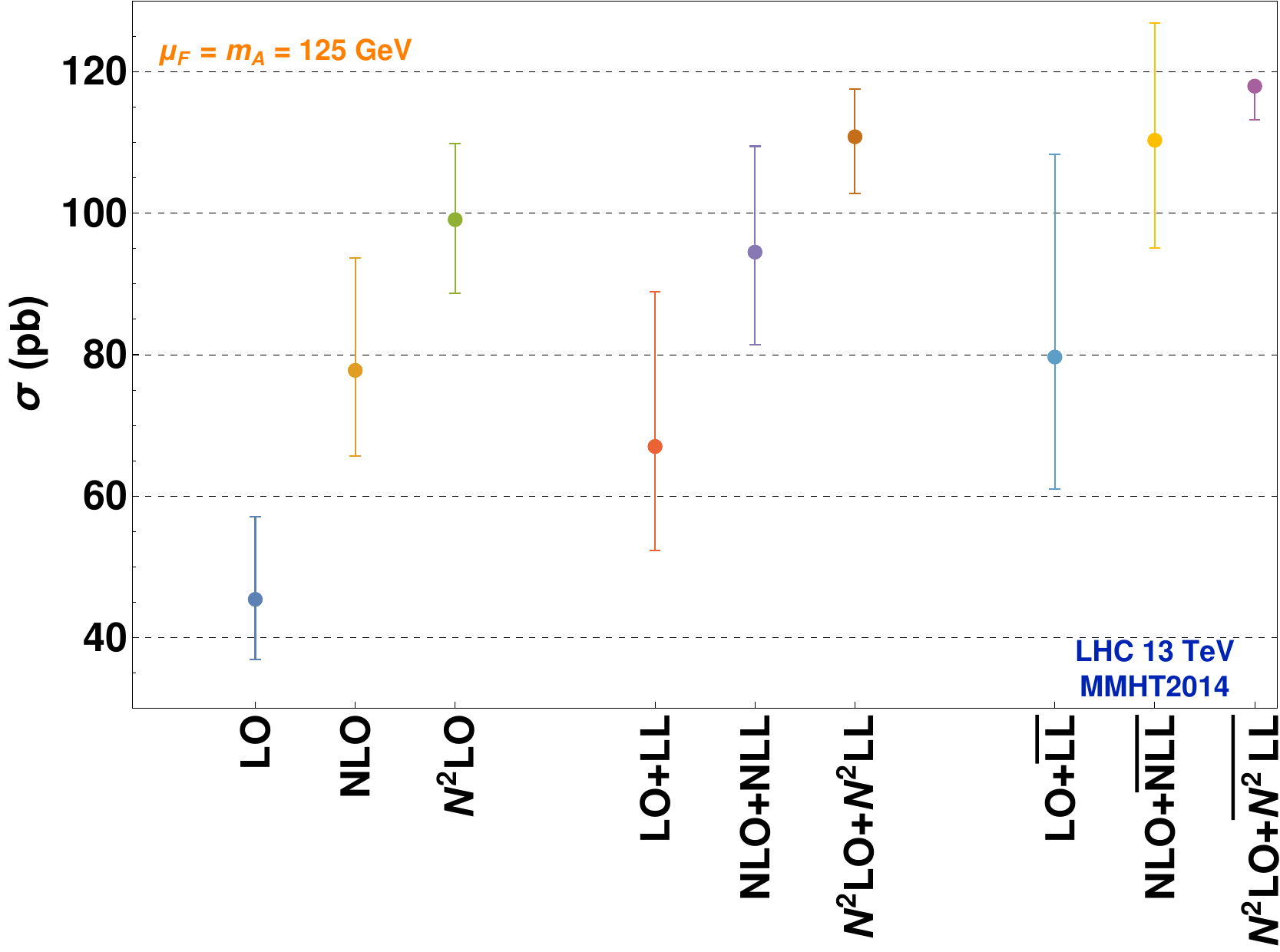}%
}

\subfloat{%
  \includegraphics[clip,scale=0.5]{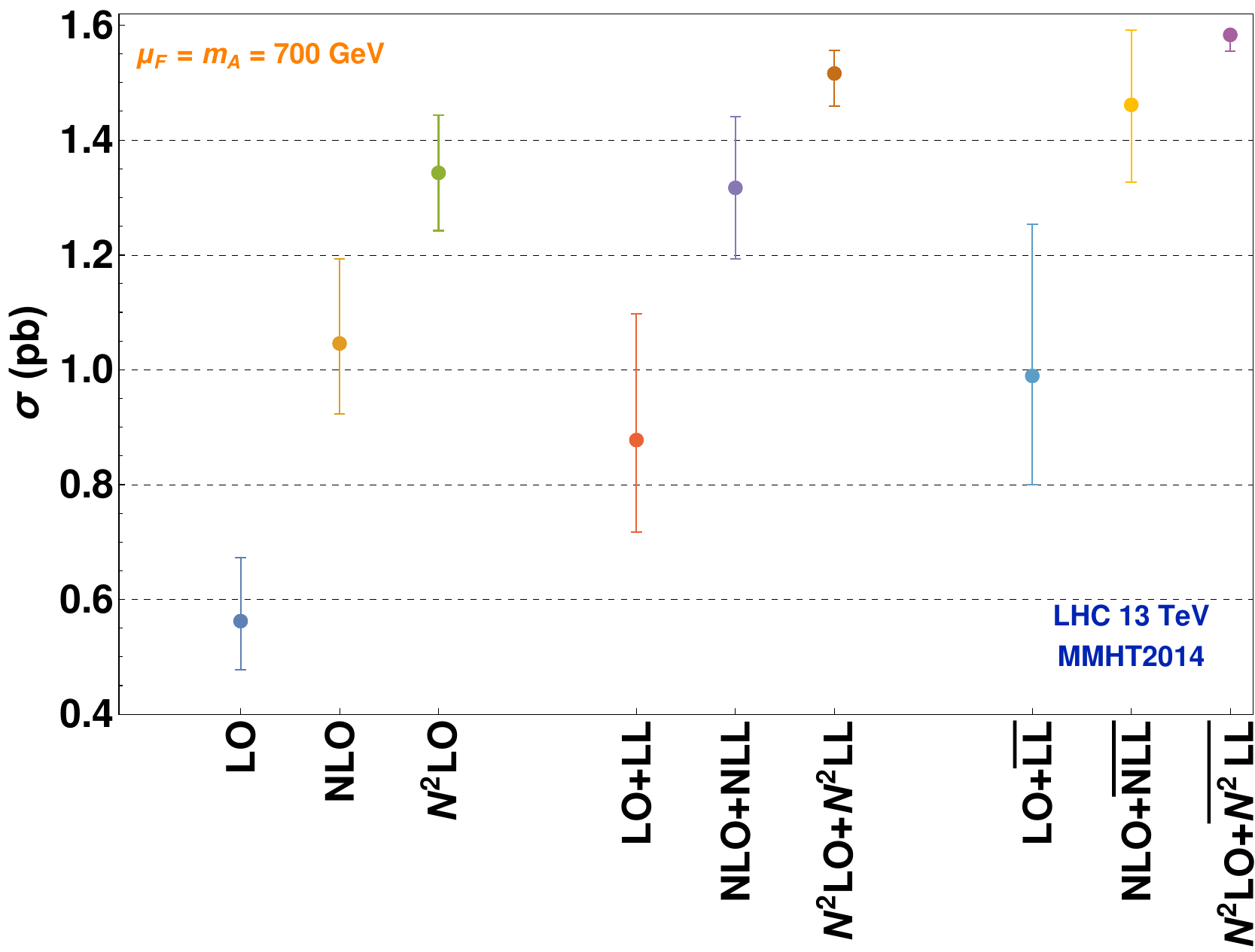}%
}

\caption[Uncertainty plot with $\mu_F$ scale fixed for $m_A$=125, 700 GeV]
{Uncertainty plot with $\mu_F$ scale fixed for $m_A$=125 GeV (top figure) and $m_A$=700 GeV (bottom figure) for 13 TeV LHC with MMHT 2014 PDF.}
\label{fig:PS_125GeV_ErrorBar}
\end{figure}

\begin{figure}[!htb]
\centering
\subfloat{%
  \includegraphics[clip,scale=0.5]{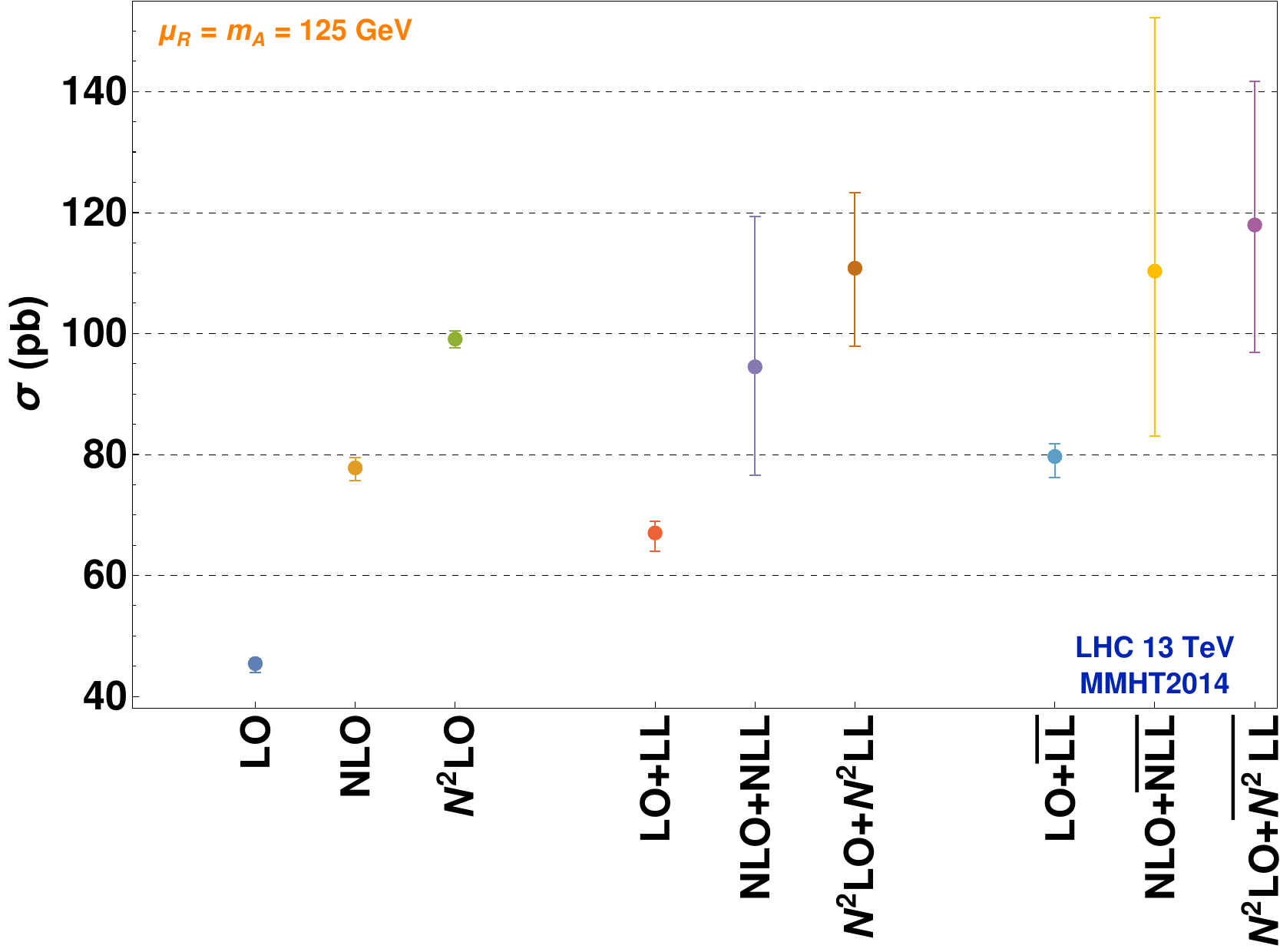}%
}

\subfloat{%
  \includegraphics[clip,scale=0.5]{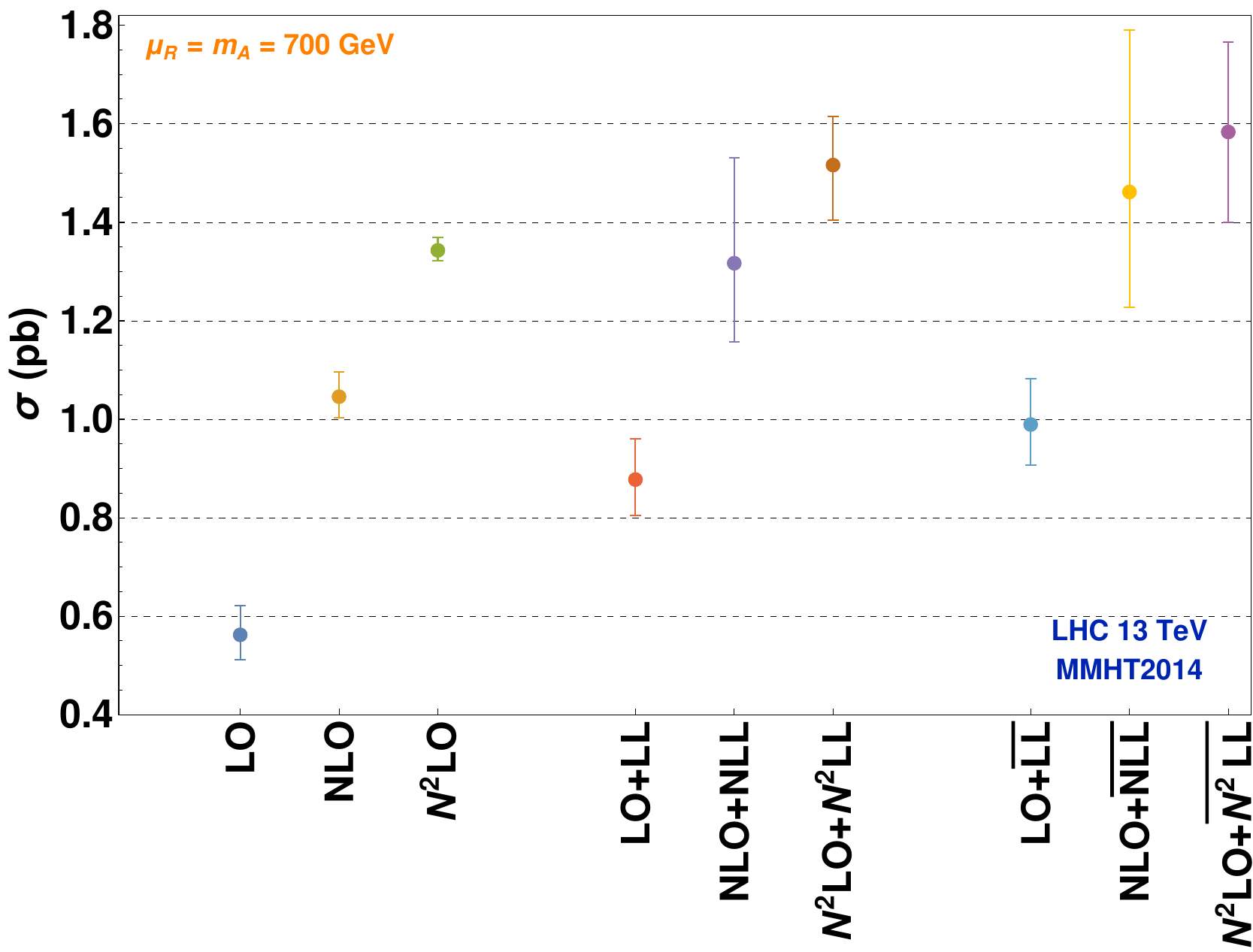}%
}

\caption[Uncertainty plot with $\mu_R$ scale fixed for $m_A$=125, 700 GeV]{Uncertainty plot with $\mu_R$ scale fixed for $m_A$=125 GeV (top figure) and $m_A$=700 GeV (bottom figure) for 13 TeV LHC with MMHT 2014 PDF.}
\label{fig:PS_700GeV_ErrorBar}
\end{figure}

\begin{figure}[!htbp]
\centering
\subfloat{%
  \includegraphics[clip,scale=0.35]{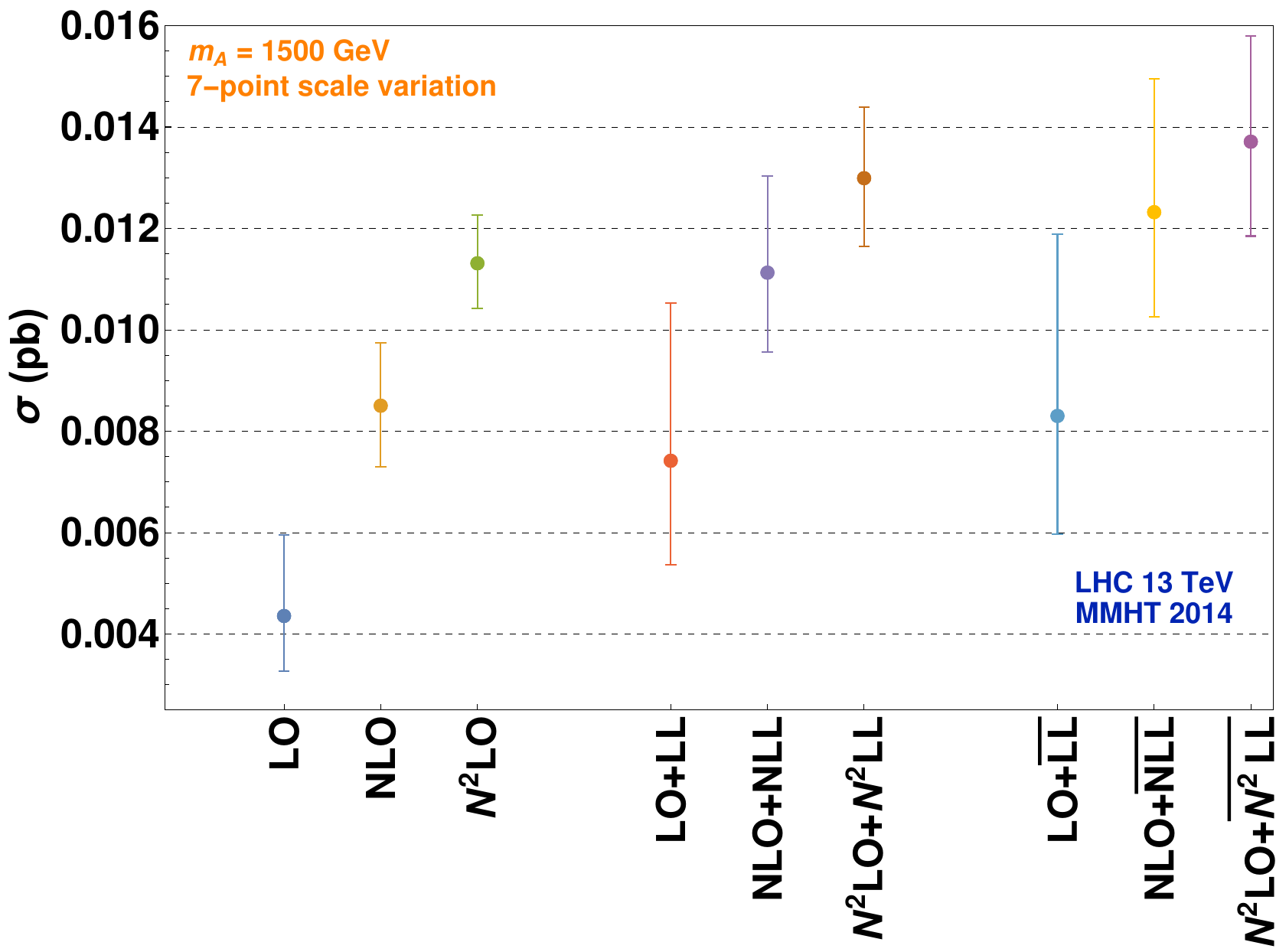}%
}

\subfloat{%
  \includegraphics[clip,scale=0.35]{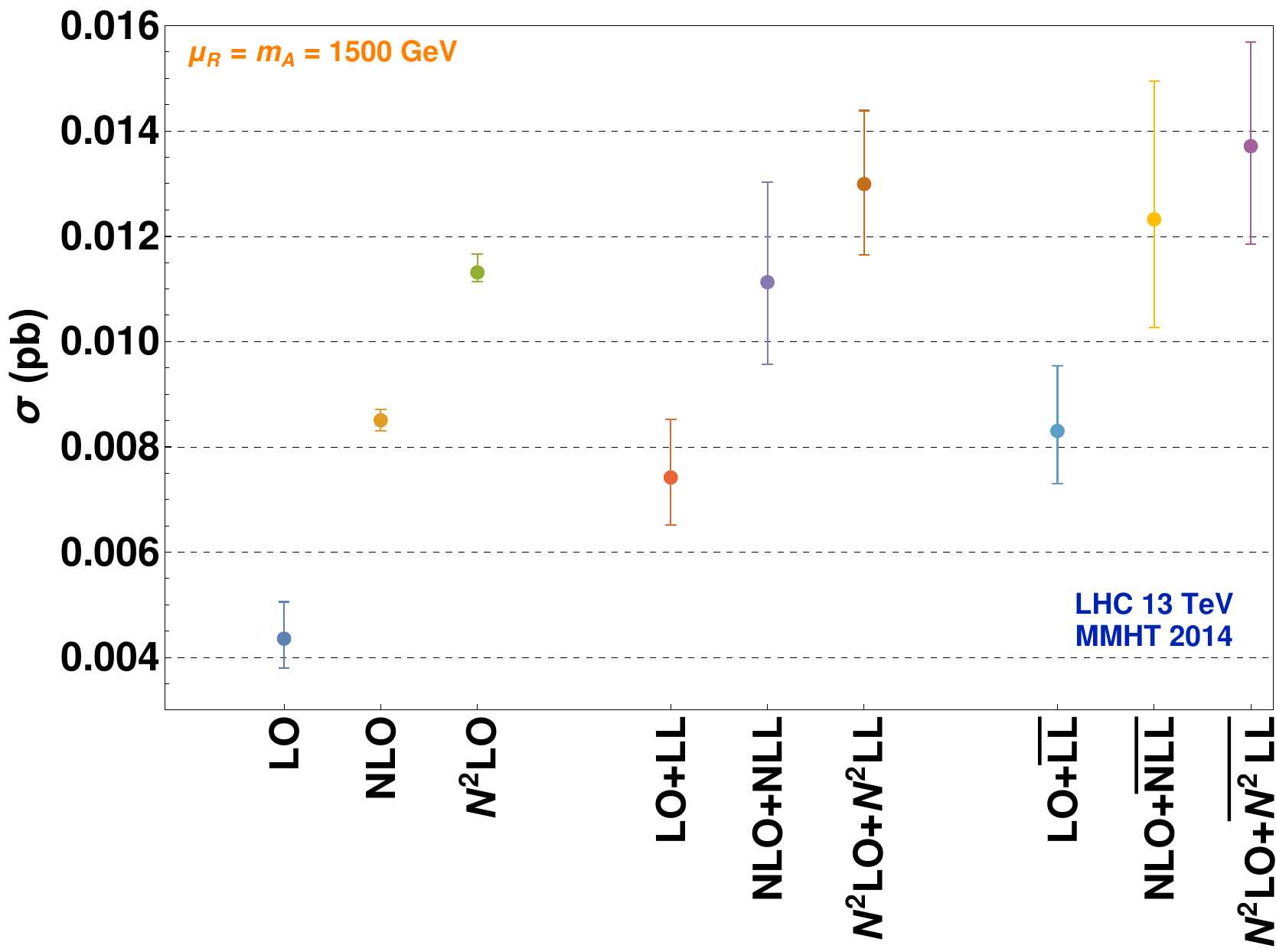}%
}

\subfloat{%
  \includegraphics[clip,scale=0.35]{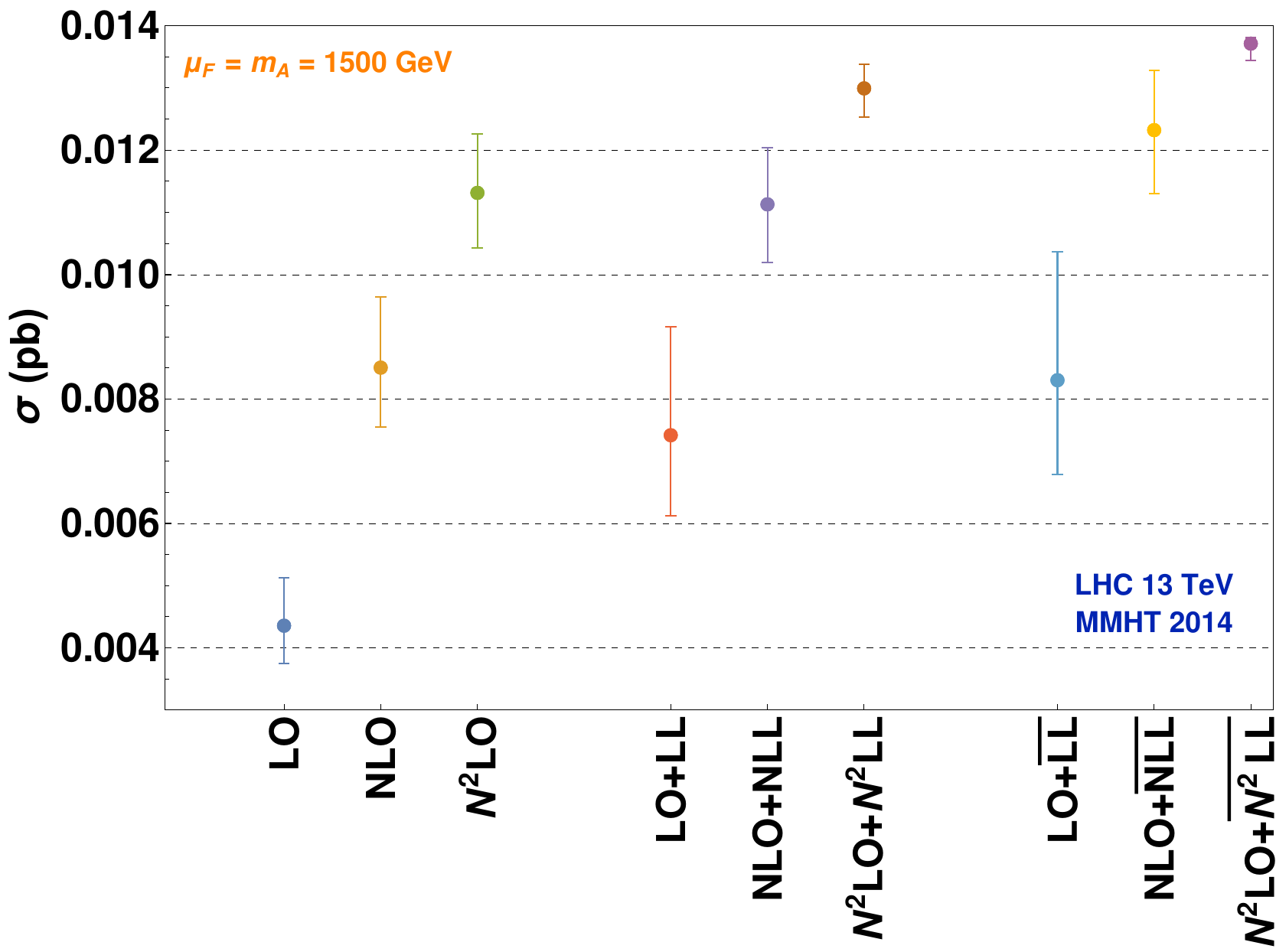}%
}
\caption[Percentage scale uncertainty for $m_A=1500$ GeV]{Uncertainty plots for 13 TeV LHC with MMHT 2014 PDF for seven-point scale variation (top), $\mu_F$ scale variation (middle) and $\mu_R$ scale variation (bottom) for $m_A=1500$ GeV.}
\label{fig:PS_1500GeV_ErrorBar}
\end{figure}

%All the above analyses have been done in the central scale choosing $\mu_R$ = $\mu_F$ = $m_A$.
We will next study our predictions' theoretical uncertainties due to the unphysical scales, $\mu_R$ and $\mu_F$.
We will present the conventional $7$-point scale uncertainties and make the following scale choices
$(\mu_R/m_A, \mu_F/m_A)$: $(0.5,0.5), (0.5,1), (1.0, 0.5), (1.0, 1.0)$,
$(1.0,2.0)$, $(2.0,1.0)$ and $(2.0,2.0)$ for a given value of $m_A$.
\mk{
All these uncertainties are presented to NNLO+NNLL and NNLO+$\overline{\text{NNLL}}$ accuracy.
In all the uncertainty plots, the first three results correspond to FO results, the next three correspond to SV resummation, and the last three represent the NSV resummed ones.
We first plot the $7$-point scale uncertainty involved for the total pseudoscalar production cross-section at each order for $m_A=125$ GeV (top) and $m_A=700$ GeV (bottom) in fig.\ \ref{fig:PS_7pt_ErrorBar}.
We observe in fig.\ \ref{fig:PS_7pt_ErrorBar} that the 7-point scale uncertainties get reduced on going from NLO to NNLO, NLO+NLL to NNLO+NNLL and NNLO+$\overline{\text{NNLL}}$ for both $m_A=125$ and $700$ GeV.
However, these 7-point scale uncertainties are found to increase while going from the SV to NSV resummation.
To better understand this aspect, we study the scale variations
due to $\mu_R$ and $\mu_F$ separately by varying one of them between $[m_A/2,~2m_A]$ and keeping the other fixed at $m_A$, for both low and high mass regions.
In fig.\ \ref{fig:PS_125GeV_ErrorBar}, we present the $\mu_R$ scale uncertainties by keeping $\mu_F$ fixed at $m_A$.
Here we notice that at second order, the NSV resummed results have smaller scale uncertainties than the SV resummed ones, which are smaller than the FO scale uncertainties for both $m_A=125$ and $700$ GeV.
In fig.\ \ref{fig:PS_700GeV_ErrorBar}, we study the $\mu_F$ scale uncertainties by keeping $\mu_R$ fixed at $m_A$.
In this case of $\mu_F$ scale uncertainties, we notice that, contrary to the case of $\mu_R$ scale uncertainties, the NSV resummed results exhibit larger scale uncertainties than the SV resummed ones, which are larger than the corresponding FO ones for both $m_A=125$ and $700$ GeV.
%
%This can be attributed to the missing contributions from other partonic channels while estimating the $\mu_F$ scale uncertainties in our SV and NSV resummed predictions.
}
%
%%%% TO BE UPDATED FROM HERE
%that after including the NSV resummation results, these scale uncertainties get reduced for the $125$ GeV mass region. At the same time, they do not reduce for $m_A=700$ GeV.
It is also to be noted that for the NSV resummation done in this analysis, logarithms involving the factorisation scale get resummed to all orders only for the gluon channel.
However, factorisation scale dependence also enters through other parton channels, which are not included in the present analysis.
Thus, if we include all possible sub-processes, the factorisation scale uncertainties are expected to reduce, otherwise showing an imbalance when the $gg$ channel is considered alone.
This indicates that the observed larger scale uncertainties in the NSV resummed results for the $700$ GeV mass region are due to the missing parton channels in the present analysis.
A similar set of results have been given in fig.\ \ref{fig:PS_1500GeV_ErrorBar} for $m_A=1500$ GeV, and they depict the same behaviour as observed at $m_A=125,~700$ GeV.
In table~\ref{tab:700GeV_table}, we tabulate the percentage errors due to the 7-point scale uncertainty, $\mu_F$ scale uncertainty and $\mu_R$ scale uncertainty cases for the pseudoscalar mass of $700$ GeV.

\begin{table}[!htb]
\centering
\begin{tabularx}{1.0\textwidth}{
  | >{\raggedright\arraybackslash}X
  | >{\centering\arraybackslash}X
  | >{\centering\arraybackslash}X
  | >{\centering\arraybackslash}X
  | >{\centering\arraybackslash}X
  | >{\centering\arraybackslash}X
  | >{\centering\arraybackslash}X
  | >{\centering\arraybackslash}X
  | >{\centering\arraybackslash}X
  | >{\raggedleft\arraybackslash}X | }
 \hline
 \textbf{Order}            & \textbf{Percentage uncertainty (7-point variation)} & \textbf{Percentage uncertainty ($\mu_F$ fixed)}  & \textbf{Percentage uncertainty ($\mu_R$ fixed)} \\ \hline \hline
\rule{0pt}{10pt} LO                            & 54.92 & 34.70 & 19.54 \\ \hline
\rule{0pt}{10pt} NLO                           & 34.61 & 25.81 &  8.91 \\ \hline
\rule{0pt}{10pt} NNLO                          & 18.18 & 14.96 &  3.48 \\ \hline
                                                      \hline
\rule{0pt}{10pt} LO+LL                         & 61.84 & 43.35 & 17.72 \\ \hline
\rule{0pt}{10pt} NLO+NLL                       & 28.45 & 18.80 & 28.45 \\ \hline
\rule{0pt}{10pt} NNLO+NNLL                     & 13.89 &  6.42 & 13.89 \\ \hline
                                                      \hline
\rule{0pt}{10pt} LO+$\overline{\text{LL}}$     & 64.40 & 45.83 & 17.71 \\ \hline
\rule{0pt}{10pt} NLO+$\overline{\text{NLL}}$   & 38.52 & 18.13 & 38.52 \\ \hline
\rule{0pt}{10pt} NNLO+$\overline{\text{NNLL}}$ & 26.09 &  1.82 & 23.13 \\ \hline
\end{tabularx}
\caption[Percentage of uncertainty in the pseudoscalar production cross-sections due to scale variations for $m_A=700$ GeV]{Percentage of uncertainty in the pseudoscalar production cross-sections due to scale variations, for the pseudoscalar mass $m_A=700$ GeV at 13 TeV LHC, using MMHT 2014 PDF. The second column corresponds to $7$-point scale variations,
the third column represents the uncertainty due to the $\mu_R$ scale variations while keeping $\mu_F$ fixed and
the last column represents the uncertainty due to the $\mu_F$ scale variations while keeping $\mu_R$ fixed.}
\label{tab:700GeV_table}
\end{table}
\ab{
While moving from $125$ GeV to $1500$ GeV for the uncertainty studies, we also measured the scale uncertainties for other values of the pseudoscalar mass, $m_A$.
The data for $m_A$ values ranging from 125 GeV to 2000 GeV are presented in a tabular form in table~\ref{tab:7pt_Comp_table} for the 7-point scale uncertainties only.
Here we observe that at the NLO and NNLO levels, the 7-point scale uncertainties decrease as we go from 125 GeV to about 1000 GeV, and then, they slowly vary with further increase in the $m_A$ values.
We notice a similar behaviour for NLO+NLL, NNLO+NNLL, NLO+$\overline{\text{NLL}}$ and NNLO+$\overline{\text{NNLL}}$.
Additionally, we observed that not only for $7$-point scale variation, increasing the mass changes the percentage uncertainties very slowly for the $\mu_F$ and $\mu_R$ scale variation cases too.
%From this, we concluded that the effects of the other parton channels decrease at higher mass regions. Hence, the $gg$-channel results show reliability alone.
%Till $m_A=500$ GeV, changes in the uncertainty data are observed. Beyond this mass range, the percentage uncertainties do not change much compared to that observed in the lower mass region.
Large uncertainties are observed for the $7$-point and $\mu_F$ scale variation results, especially in the lower mass regions.
In the case of $7$-point and $\mu_F$ scale variation, the NSV resummed results show maximum uncertainty, with these uncertainties minimising for the $\mu_R$ scale variation.
The uncertainty predictions always show significant improvement after incorporating the NSV resummed corrections for only the $\mu_R$ scale variation case.
%In the $\mu_R$ scale variation case, the LO$+\overline{\text{LL}}$ uncertainties are maximum followed by NLO$+\overline{\text{NLL}}$ and then NNLO$+\overline{\text{NNLL}}$.
%The NSV corrections in this analysis get resummed to all orders only for the gluon channel; these observations lead to the conclusion that the factorisation scale dependence enters through other parton channels, which are not included in the present analysis.
So, there may be effects of beyond NSV terms in the $ gg $ channel and those from the other parton channels, which we do not include here, that lead to such behaviour.
}
%So, like depicted earlier in the figs.\ \ref{fig:PS_7pt_ErrorBar},\ref{fig:PS_125GeV_ErrorBar},\ref{fig:PS_700GeV_ErrorBar} and \ref{fig:PS_1500GeV_ErrorBar}, the uncertainties show significant improvement for the $\mu_R$ scale variation results over all mass regions.
%
%\begin{figure}[!htbp]
%\centering
%\subfloat{%
%  \includegraphics[clip,scale=0.6]{Plots_New_Chap4/7Point_NNLO.pdf}%
%}
%
%\subfloat{%
%  \includegraphics[clip,scale=0.6]{Plots_New_Chap4/muFVar_NNLO.pdf}%
%}
%
%\subfloat{%
%  \includegraphics[clip,scale=0.6]{Plots_New_Chap4/muRVar_NNLO.pdf}%
%}
%
%\caption[Scale uncertainty at NNLO for $m_A$ ranging from 125 GeV to 1000 GeV]{Uncertainty plots for 13 TeV LHC with MMHT 2014 PDF at NNLO for seven-point scale variation (top), $\mu_F$ scale variation (middle) and $\mu_R$ scale variation (bottom) for $m_A$ ranging from 125 GeV to 1000 GeV.}
%\label{fig:PS_NNLO_ErrorBar}
%\end{figure}
%
%$m_A = 125,~700,~1000,~1500,~2000,~2500,~3500$ GeV.
%
%In fig.\ \ref{fig:PS_LO+NLO+NNLO_ErrorBar}, I present the uncertainties involved in the NSV resummed corrections at the LO, NLO and NNLO for seven-point scale variation, $\mu_F$ scale variation and $\mu_R$ scale variation separately over a wide mass range. The increase in stability in the high-mass region is also evident here, at all orders of precision. In the case of seven-point scale variation, the LO$+\overline{\text{LL}}$ shows the least uncertainty, whereas the NLO$+\overline{\text{NLL}}$ and NNLO$+\overline{\text{NNLL}}$ corrections depict more substantial uncertainties.
%
%
\begin{table}[!htb]
\renewcommand{\arraystretch}{1.5}
\begin{center}
 \begin{tabular}{| L{2.5cm} | C{1.2cm} | C{1.2cm} | C{1.2cm} | C{1.2cm} | C{1.2cm} | C{1.2cm} |}
 \hline
 \textbf{Order} & \multicolumn{5}{c}
 {\textbf{7-point scale uncertainty} for $\mathbf{m_A}$ (in GeV) }
 & \\ \cline{2-7}
 & \textbf{$\mathbf{125}$}
 & \textbf{$\mathbf{350}$}
 & \textbf{$\mathbf{700}$}
 & \textbf{$\mathbf{1000}$}
 & \textbf{$\mathbf{1500}$}
 & \textbf{$\mathbf{2000}$}
 \\ \hline
\hline
\rule{0pt}{10pt} LO                 & 44.46 & 49.16 & 54.92 & 57.91 & 61.67 & 64.84
\\ \hline
\rule{0pt}{10pt} NLO                & 36.02 & 33.39 & 34.61 & 27.59 & 28.70 & 29.80
\\ \hline
\rule{0pt}{10pt} NNLO               & 21.33 & 18.24 & 18.18 & 16.53 & 16.23 & 16.24
\\ \hline
\hline
\rule{0pt}{8pt} LO+LL               & 54.49 & 55.92 & 61.84 & 65.13 & 69.45 & 73.29
\\ \hline
\rule{0pt}{8pt} NLO+NLL             & 45.39 & 33.17 & 28.45 & 32.22 & 31.15 & 30.97
\\ \hline
\rule{0pt}{8pt} N$^2$LO+N$^2$LL     & 22.94 & 16.29 & 13.89 & 20.75 & 21.09 & 21.82
\\ \hline
\hline
\rule{0pt}{8pt} LO+$\overline{\text{LL}}$
& 59.38 & 73.17 & 64.40 & 67.35 & 71.37 & 75.04 \\ \hline
\rule{0pt}{8pt} NLO+$\overline{\text{NLL}}$
& 62.71 & 45.66 & 38.52 & 40.42 & 38.08 & 36.92 \\ \hline
\rule{0pt}{2pt} N$^2$LO+$\overline{\text{N$^2$LL}}$
& 42.11 & 30.50 & 26.09 & 29.31 & 28.78 & 28.94 \\ \hline
\end{tabular}
\caption[Percentage of uncertainty in the pseudoscalar production cross-sections due to $7$-point scale variations for different values of pseudoscalar mass]
{Percentage of uncertainty in the pseudoscalar production cross-sections due to $7$-point scale variations for different values of pseudoscalar mass at 13 TeV LHC.}
\label{tab:7pt_Comp_table}
\end{center}
\end{table}

%\begin{figure}[!htbp]
%\centering
%\subfloat{%
%  \includegraphics[clip,scale=0.6]{Plots_New_Chap4/SevptVar_LO+NLO+NNLO.pdf}%
%}

%\subfloat{%
%  \includegraphics[clip,scale=0.6]{Plots_New_Chap4/muRVar_LO+NLO+NNLO.pdf}%
%}

%\subfloat{%
%  \includegraphics[clip,scale=0.6]{Plots_New_Chap4/muFVar_LO+NLO+NNLO.pdf}%
%}

%\caption[Seven-point scale variation, $\mu_F$ scale variation and $\mu_R$ scale variation results till two-loop for $m_A$ ranging from 125 GeV to 1000 GeV]{Uncertainty plots for 13 TeV LHC with MMHT 2014 PDF for seven-point scale variation (top), $\mu_F$ scale variation (middle) and $\mu_R$ scale variation (bottom) for $m_A$ ranging from 125 GeV to 1000 GeV at LO, NLO and NNLO accuracy.}
%\label{fig:PS_LO+NLO+NNLO_ErrorBar}
%\end{figure}

\begin{figure}[!htbp]
\centering
  \includegraphics[scale=0.35]{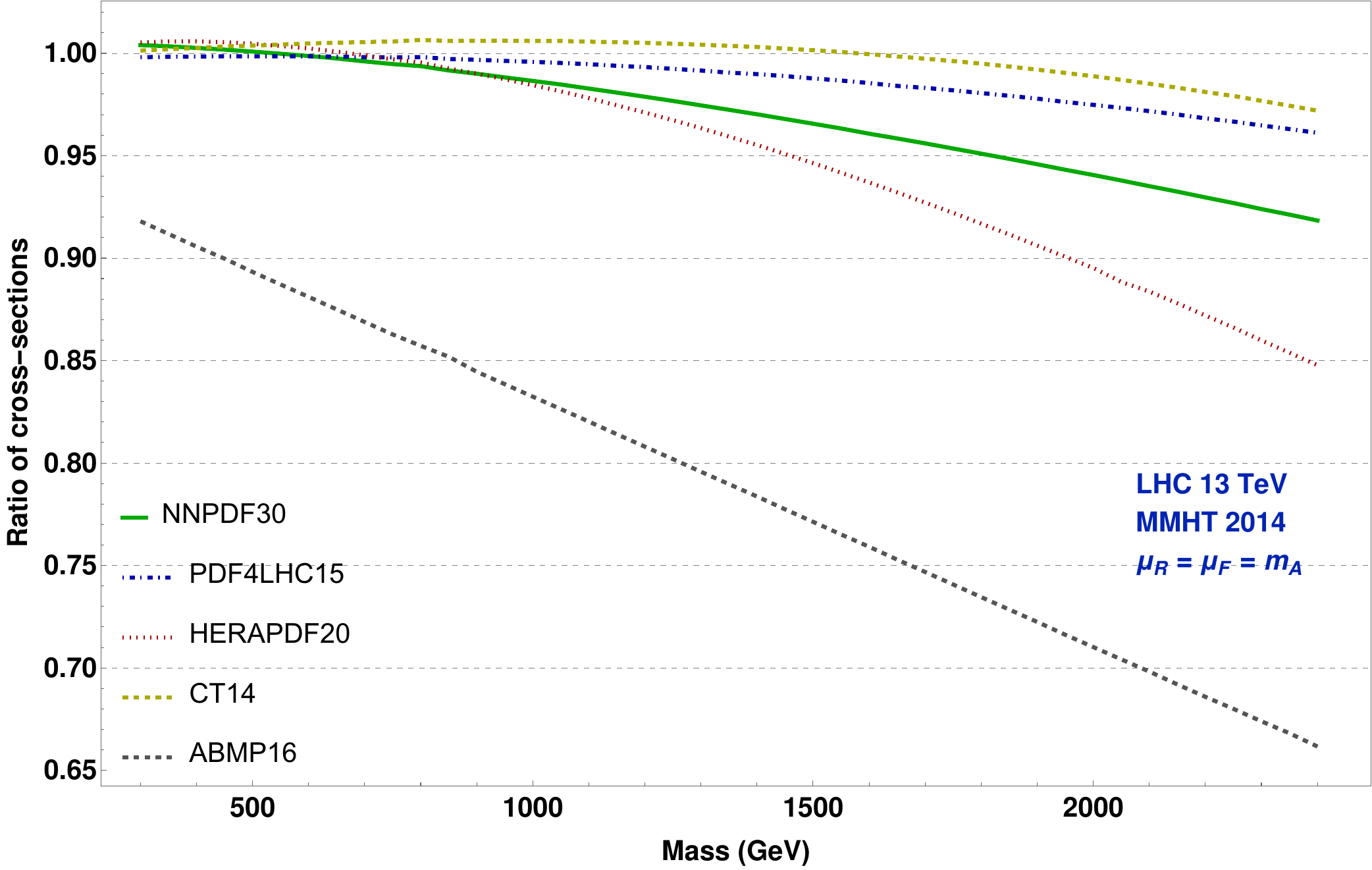}
% PDFUncertainty.pdf: 0x0 px, 300dpi, 0.00x0.00 cm, bb=
  \caption{Pseudoscalar production cross-sections at NNLO$+\overline{\text{NNLL}}$ for different choices of PDF's, normalized w.r.t. those obtained from MMHT2014 PDF's.}
\label{fig:pdfuncertainty}
\end{figure}
We also estimate the uncertainty due to the choice of parton densities in our calculation for completeness.
Fig.~\ref{fig:pdfuncertainty} depicts the PDF uncertainties involved in the calculations due to the choice of different PDF sets.
For this analysis, we choose NNPDF30, PDF4LHC15, HERAPDF20, CT14 and ABMP16 as the reference PDFs and present the results for the production cross-section at NNLO+$\overline{\text{NNLL}}$, normalised with the corresponding results obtained from our default choice of MMHT2014 PDF set.
$\mu_F$ scale uncertainties show similar behaviour.

To further extend this study, I present in table \ref{tab:MSHTvsMMHT} a comparative study of the percentage uncertainties for two different PDF choices --- the MMHT 2014 PDF and a new MSHT20 PDF.
This new MSHT20 set of PDFs has been determined recently and is expected to supersede the MMHT2014 sets \cite{Bailey:2020ooq}.
We check the 7-point scale uncertainties, $\mu_R$ scale variation uncertainties and $\mu_F$ scale variation uncertainties for both the sets at FO and after SV and NSV resummation.
We observe that the stability of the results improves slightly with the new PDF set of MSHT but is not significant.
This shows that all possible parton channels and even beyond NSV corrections are important to improve precision.
NSV corrections for the $gg$ channel may not be sufficient.
\begin{table}[!htb]
\centering
\begin{tabular}{lccccc c ccccc}
\hline
 \textbf{Order}
 & \multicolumn{2}{c}{\textbf{7-point variation}}
 && \multicolumn{2}{c}{\textbf{$\mu_R$ scale variation}}
 && \multicolumn{2}{c}{\textbf{$\mu_F$ scale variation}} \\ \hline
 \cline{2-3} \cline{4-5} \cline{6-7} \\ \hline
 & \textbf{MMHT} & \textbf{MSHT} && \textbf{MMHT} & \textbf{MSWT} && \textbf{MMHT} & \textbf{MSHT} \\ \hline
LO   & 44.46 & 42.81 && 4.92 & 5.48 &&  44.46 & 42.81 \\
NLO  & 36.02 & 35.15 && 4.92 & 4.69 &&  36.02 & 35.15 \\
NNLO & 21.33 & 21.32 && 2.81 & 2.77 &&  21.33 & 21.32 \\ \hline
LO+LL     & 54.49 & 51.92 && 7.29  & 7.77  &&  54.49 & 51.92 \\
NLO+NLL   & 45.39 & 44.13 && 45.39 & 44.13 &&  29.64 & 28.90 \\
NNLO+NNLL & 22.94 & 22.91 && 22.94 & 22.91 &&  13.37 & 13.35 \\ \hline
LO$+\overline{\text{LL}}$     & 59.38 & 56.38 && 6.96   & 7.43 &&  59.38 & 56.38 \\
NLO$+\overline{\text{NLL}}$   & 62.71 & 61.00 && 62.71 & 61.00 &&  28.83 & 28.05 \\
NNLO$+\overline{\text{NNLL}}$ & 42.11 & 42.12 && 37.97 & 37.95 &&  4.42  & 4.40  \\ \hline
\end{tabular}
\caption[Percentage uncertainty with PDFs MMHT2014 and MSHT2020, for 13 TeV LHC and $m_A=125$ GeV]
{Comparison of percentage uncertainty with two different PDF choices, MMHT2014 and MSHT20, for 13 TeV LHC and $m_A=125$ GeV.}
\label{tab:MSHTvsMMHT}
\end{table}

\mk{
The requirement of these high-precision computations is consequential in discovering the pseudoscalar Higgs and crucial in establishing the CP properties of the discovered Higgs boson.
As mentioned in section \ref{sec:prologue2}, the observed Higgs boson at the LHC can be an admixture of scalar-pseudoscalar states. This mixing can be parameterised by a single mixing angle $\alpha$ if any further decay of the produced Higgs is not considered.
In \cite{Jaquier:2019bfs}, the authors concluded that if a single Higgs boson production is considered for any arbitrary value of the mixing angle while neglecting its decay, then the results up to NNLO may be obtained by a simple rescaling of the scalar and pseudoscalar cross-sections as below:
\begin{equation}
 \sigma = \cos^2\alpha \cdot \sigma_H + \sin^2\alpha \cdot \sigma_A
 \label{eqn:mixing}
\end{equation}
where $\sigma$ is the superposed cross-section, $\sigma_H$ is the pure Higgs cross-section and $\sigma_A$ is the pure pseudoscalar Higgs cross-section.
Regarding this mixing angle, a pure scalar state can be represented by $\alpha=0$ and a pure pseudoscalar state by $\alpha=\pi/2$.
Gluon fusion at the LHC can produce a mixed scalar-pseudoscalar Higgs boson.
This possibility has been studied in \cite{Jaquier:2019bfs} through NNLO.
In table \ref{tab:Mixing_table}, we give the production cross-sections at the LHC for different values of this mixing angle $\alpha$ in the $125$ GeV mass range.
\begin{table}[!htb]
\centering
\begin{tabularx}{0.9\textwidth}{
  | >{\raggedright\arraybackslash}X
  | >{\centering\arraybackslash}X
  | >{\centering\arraybackslash}X
  | >{\centering\arraybackslash}X
  | >{\centering\arraybackslash}X
  | >{\centering\arraybackslash}X
  | >{\centering\arraybackslash}X
  | >{\centering\arraybackslash}X
  | >{\centering\arraybackslash}X
  | >{\raggedleft\arraybackslash}X | }
 \hline
 \vspace{0.06pt} \textbf{K-Factor}
 & \vspace{0.06pt}$\alpha=0$
 & \vspace{0.06pt}$\alpha=1$
 & \vspace{0.06pt}$\alpha=\pi/4$
 & \vspace{0.06pt}$\alpha=\pi/6$
                               \\ \hline
                               \hline
\rule{0pt}{10pt} K$_{(1)}$            & 1.6990 & 1.7124 & 1.7083 & 1.7048 \\ \hline
\rule{0pt}{10pt} K$_{(2)}$            & 2.1571 & 2.1814 & 2.1741 & 2.1677 \\ \hline
                                      \hline
\rule{0pt}{10pt} K$_{(1)}^{resum}$    & 2.0033 & 2.0803 & 2.0570 & 2.0368 \\ \hline
\rule{0pt}{10pt} K$_{(2)}^{resum}$    & 2.2785 & 2.4392 & 2.3907 & 2.3485 \\ \hline
                                      \hline
\rule{0pt}{10pt} $\overline{K}_{(1)}^{resum}$  & 2.3425 & 2.4284 & 2.4025 & 2.3799 \\ \hline
\rule{0pt}{10pt} $\overline{K}_{(2)}^{resum}$  & 2.4737 & 2.5966 & 2.5595 & 2.5272 \\ \hline
\end{tabularx}
\caption[K-factors for the production cross-sections of a mixed state of scalar and pseudoscalar Higgs bosons.]
{\mk{K-factors for the production cross-sections of a mixed state of scalar and pseudoscalar Higgs bosons.
The production cross-sections are computed for $m_A=125$ GeV at 13 TeV LHC using MMHT2014 PDFs for the central scale choice
of $\mu_R=\mu_F=m_A$.  Each column represents different values of the mixing angle $\alpha$.
}}
\label{tab:Mixing_table}
\end{table}
Here, we observe that the inclusion of the SV resummed results increases the NLO (NNLO) cross-sections by about $30\%~(12\%)$ of the LO results, while further addition of the NSV resummed results increases the NLO+NLL (NNLO+NNLL) cross-sections by about $33\%~(20\%)$ of LO ones for the pure scalar state.
However, for the pure pseudoscalar state, we observe that the inclusion of the SV resummed results increases the NLO (NNLO) cross-sections by about $37\%~(26\%)$ of the LO ones while further addition of the NSV resummed results, increase the NLO+NLL (NNLO+NNLL) cross-section by approximately $35\%~(16\%)$ of the LO ones.
For the mixed cases considered, the inclusion of the SV resummed results increases the NLO (NNLO) cross-sections by about $35\%~(22\%)$ of the LO ones for $\alpha=\pi/4$ and by approximately $33\%~(18\%)$ of the LO ones for $\alpha=\pi/6$.
Similarly, the addition of the NSV resummed results increases the NLO+NLL (NNLO+NNLL) cross-sections by approximately $35\%~(17\%)$ of the LO ones for $\alpha=\pi/4$ and by about $34\%~(18\%)$ of the LO ones for $\alpha=\pi/6$.
Overall, we observe that the corresponding QCD corrections change only by a few per cent as we change the mixing angle.
In this scenario, if the pseudoscalar Higgs boson production cross-section is made available to a precision comparable to that of the scalar Higgs boson, then it can prove helpful in extracting the mixing angle to a better accuracy.

It is necessary to clarify here that several angular observables corresponding to the Higgs boson's decay products must be studied to establish its properties. When such a Higgs decay is considered, the simple reweighing formula in eqn.\ \ref{eqn:mixing} fails. Consequently, the corresponding K-factors similar to those in table \ref{tab:Mixing_table} get modified slightly. However, such a detailed analysis (including the angular distributions of the decay products) is beyond the scope of this article.
}

Finally, we attempt to give predictions for the pseudoscalar production cross-section after including the resummed results to $\overline{\text{N$^3$LL}}$ accuracy and then matching them to the FO N$^3$LO$_A$ results.
Here, N$^3$LO$_A$ results represent the approximate full FO results at third order in QCD, which we have taken from the public code
\cite{Ball_2013,Bonvini_2014_TROLL,Bonvini_2016_TROLL,Bonvini_2018,Bonvini_2018_1}.
In the left panel of fig.\ \ref{fig:crossK3}, we plot the pseudoscalar production cross-section as a function of its mass $m_A$ at third order and present the results up to N$^3$LO$_A+\overline{\text{N$^3$LL}}$.
In the right panel, we give the corresponding K-factors obtained by normalising with the LO cross-sections.
%We have the SV results (FO SV) available and also the full results (FO FULL) from Bonvini's public code \cite{Ball_2013,Bonvini_2014_TROLL,Bonvini_2016_TROLL,Ahmed_2016,Bonvini_2018,Bonvini_2018_1}. In this work, we calculate the NSV results (FO NSV). We also do the resummation using these SV (Resum SV) and NSV (Resum NSV) results.
\mk{We also give K-factors for the FO case by keeping only the third order SV and SV+NSV results and compare with those of the N$^3$LO$_A$.
\begin{figure}[!htbp]
\centering
\begin{subfigure}{.5\textwidth}
  \centering
  \includegraphics[scale=0.265]{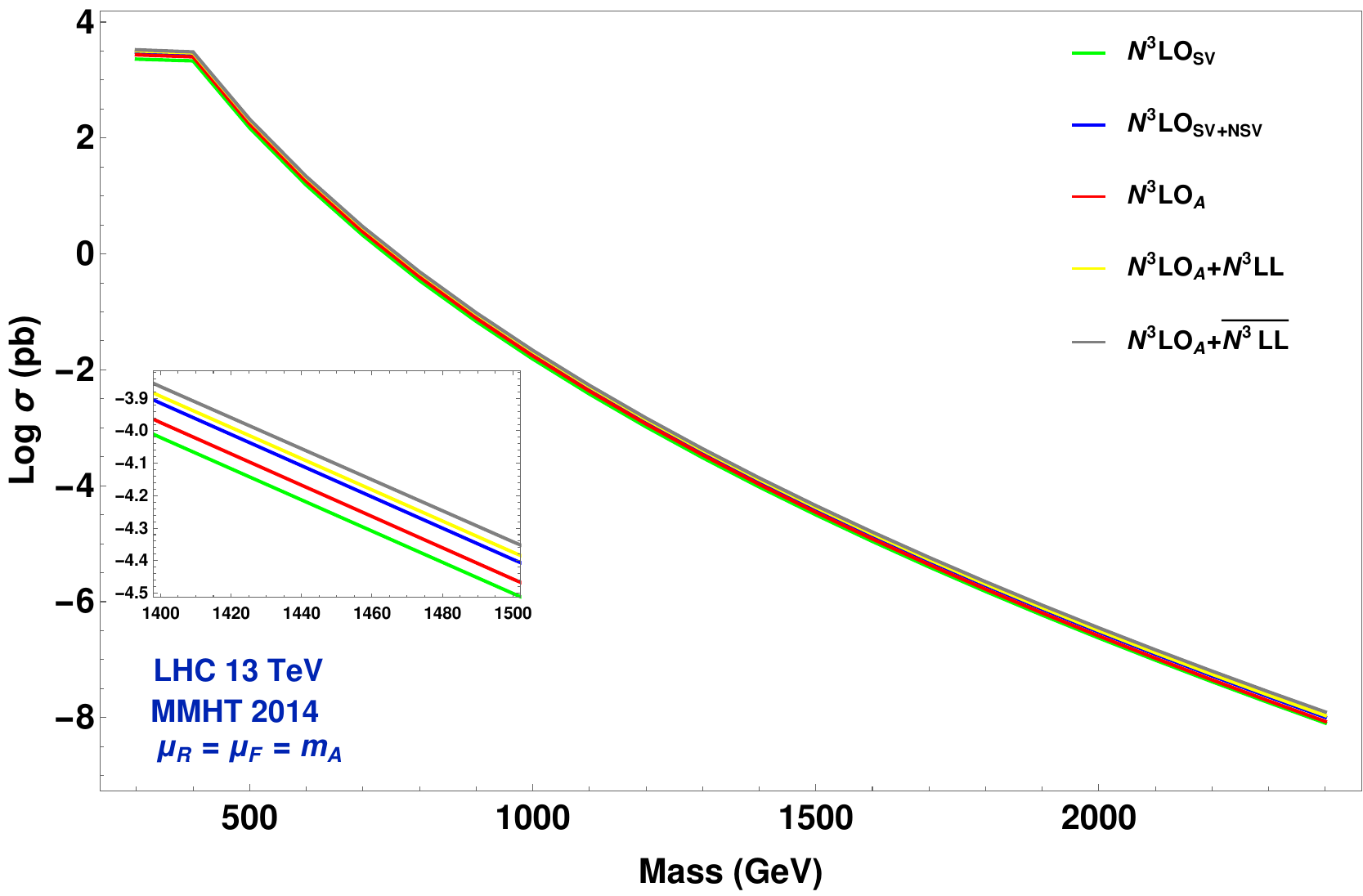}
% N3LO_CROSS_comparison.pdf: 0x0 px, 300dpi, 0.00x0.00 cm, bb=
% NLO.pdf: 0x0 px, 300dpi, 0.00x0.00 cm, bb=
  \label{fig:sub3}
\end{subfigure}%
\begin{subfigure}{.5\textwidth}
  \centering
  \includegraphics[scale=0.265]{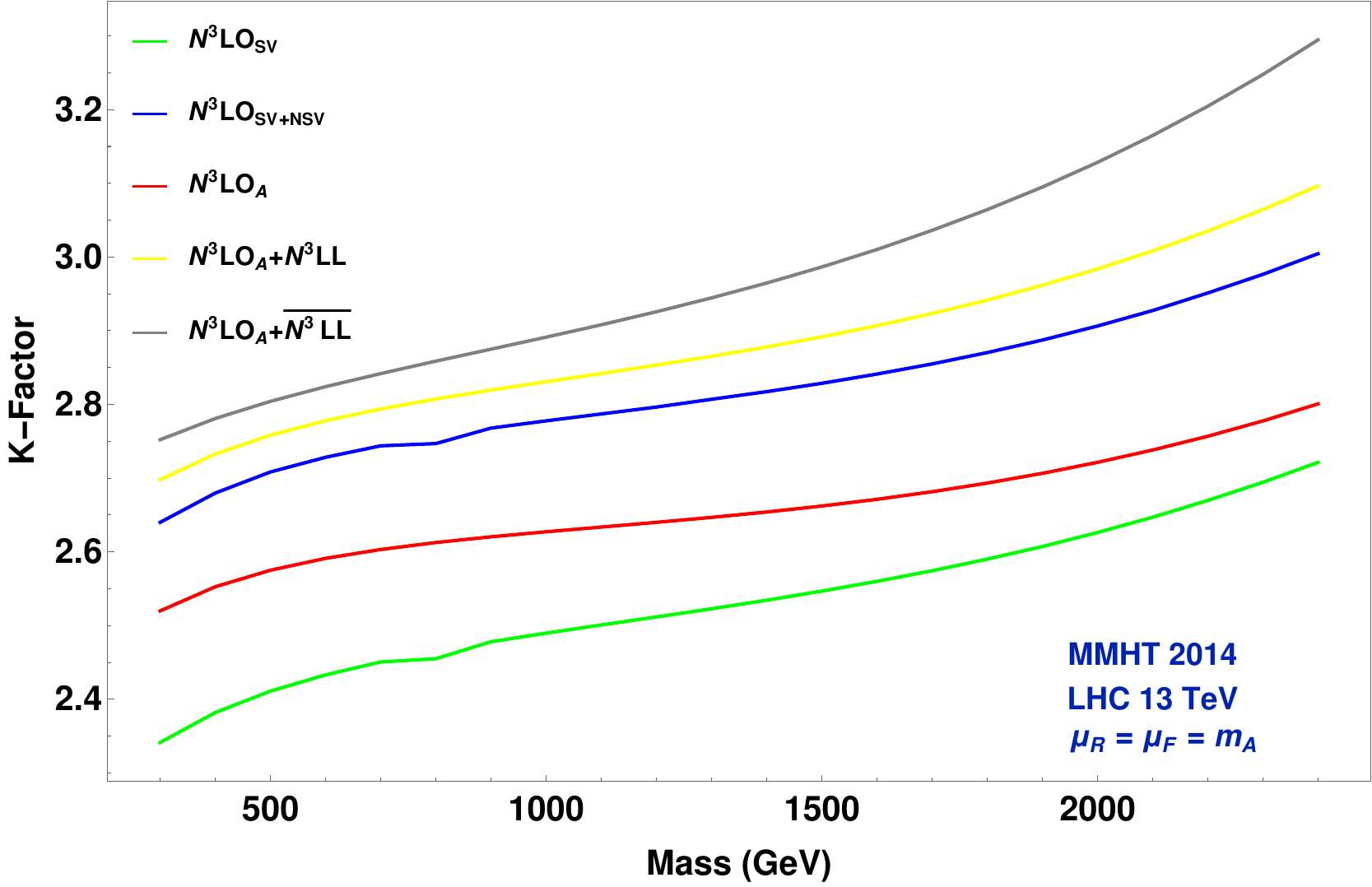}
% N3LO_Kfactor_comparison.pdf: 0x0 px, 300dpi, 0.00x0.00 cm, bb=
  \label{fig:sub4}
\end{subfigure}
%Ratio of cross-sections
\caption[Pseudoscalar production cross-section at third order in QCD and the corresponding K-Factors]
{Pseudoscalar production cross-section at third order in QCD (left panel) and the corresponding K-Factors (right panel). The corresponding resummed results to N$^3$LL accuracy are also given after matching them to the approximate full N$^3$LO$_A$ results.}
\label{fig:crossK3}
\end{figure}
The N$^3$LO$_\text{SV}$ results are closer to the N$^3$LO$_A$ ones in the high mass region, while N$^3$LO$_\text{SV+NSV}$ results get a bit closer to the N$^3$LO$_A$ ones in the low mass region.
We observe that the inclusion of NSV corrections at the FO level or their resummation through NSV substantially increases the cross-sections.
}
%This also suggests the requirement of including NSV corrections from other parton channels.
%
However, a detailed phenomenological study at $\overline{\text{N$^3$LL}}$ level accuracy, including the estimation of theoretical uncertainties, requires additional terms such as the resummation constant $h_{33}^g(\omega)$ that are yet to be determined.
%We see that the FO SV result shows the least stability. The FO Bonvini full results show a reduction from the FO NSV results.
%However, when we do the resummation, there is an increase from the full results to resummed SV results and the resummed
%NSV results show a further increment.
In each of the above cases, the N$^3$LO level PDFs still need to be made available.
However, it can be anticipated that the uncertainty will get further reduced if the full N$^3$LO results with the corresponding PDF's are made available.

\section{Summary}
\label{sec:Summary}

In this work, we have performed the NSV resummation for the pseudoscalar production process to $\overline{\text{NNLL}}$ accuracy in QCD at the LHC.
We further make a detailed phenomenology of the same and present our results for 13 TeV LHC.
We have computed the NSV corrections at both first and second orders in QCD and compared them with the full NLO and NNLO corrections for the gluon fusion sub-process.
We find that these NSV corrections are potentially large and enhance the pseudoscalar production cross-sections much more than the conventional SV or threshold logarithms.
We also give numerical results for the NSV resummed results to $\overline{\text{NNLL}}$ accuracy by systematically matching to the FO NNLO cross-sections.
We find that these NSV resummed predictions contribute substantially to the cross-section compared to the SV resummed results.
We estimate the size of these corrections in terms of the resummed K-factors defined
{\it{w.r.t.}} the LO and find them to be as large as $3$ at NNLO+$\overline{\text{NNLL}}$ accuracy in the high mass region.
We also estimate in our predictions the uncertainties due to the choice of various parton densities and those due to the unknown renormalisation and factorisation scales.
We find that the conventional 7-point scale uncertainties do not get an improvement
after performing the NSV resummation, suggesting the requirement of including NSV contributions from other parton channels as well as beyond NSV contributions in the gluon fusion channel.
Specifically, we notice that the $\mu_F$ scale variations, while keeping $\mu_R$ fixed, lead to large uncertainties in the pseudoscalar production cross-sections in the high mass region.
However, for the pure $\mu_R$ variation, keeping $\mu_F$ fixed, we find that the scale uncertainties get reduced significantly to about $1.8\%$ for $m_A=700$ GeV and are much smaller than those of the SV resummed results $(\sim 6.4\%)$.
We also present the production cross-sections for mixed scalar-pseudoscalar state and study the impact of QCD corrections on them for different values of the mixing angle $\alpha$. We find that these QCD corrections change with the mixing angle by a few per cent. We anticipate that these precision results will be helpful in future analyses aiming to study the CP properties of the discovered Higgs boson.

\clearpage
\newpage
\mbox{~}

\chapter{Conclusion}
\label{chap:conclusion}

According to the current scenario, the fundamental job of the particle physics community can be categorised into two parts: testing the SM with unprecedented accuracy and searching for physics beyond the SM.
%The pseudoscalar Higgs boson we study in this thesis is not a part of the SM particle spectrum.
Although intensive searches for the pseudoscalar Higgs boson have been on for quite some time, conclusive evidence has yet to be found.
Improved precision of the theoretical predictions is necessary for making conclusive remarks about the existence of this BSM particle.
%So, we need to revamp our understanding of this particle.
%
%
%Precise theoretical predictions play a very crucial role towards achieving these goals.
Higher-order corrections to scattering amplitudes largely contribute to precision studies, with the QCD corrections contributing substantially to any typical observable.
This thesis tries to contribute to the precision studies for observables associated with the pseudoscalar Higgs boson production.
We perform our computations in the EFT framework in a CP-conserving model by adding one particle to the existing SM particle structure --- a CP-odd Higgs boson.
%Consequently, the only differences we observed from the studies of the scalar Higgs boson were those because of the Levi-Civita tensors and $\gamma_5$ matrices, which made the results much more complicated. However, the results depicted structural similarities to the scalar Higgs.

The fundamental need in studying these precision results is the computation of the contributing Feynman diagrams - all possible corrections from real and virtual diagrams and those from all possible parton channels.
As a first step towards this, we focused only on the gluon fusion channel in the two original works included in this thesis \cite{Bhattacharya_2020,Bhattacharya:2021hae}.
I began by computing the FO corrections till two-loop for the $2\rightarrow2$ process of di-pseudoscalar Higgs boson production \textit{via} gluon fusion \cite{Bhattacharya_2020}.
In this work, we observed how complex and extensive the pseudoscalar amplitude is.
Consequently, to include the effects of threshold logarithms, we tried to use an approximate method where instead of beginning the study with Feynman diagrams, we employed a known formalism tested for the DY and scalar Higgs boson \cite{ajjath2020soft,Ajjath:2022kyb}.
We extended this study to the pseudoscalar Higgs boson production process since the SV and NSV logarithms' structure is expected to be the same in the simple extension of the SM we are using.
Our resummed results depict predictions similar to those in \cite{Das:2019bxi,ajjath2020soft}.
I will elaborate on my works and their consequences in the following two paragraphs.

In chapter \ref{chap:ggAA}, we have computed the virtual amplitudes for producing a pair of pseudoscalar Higgs bosons \textit{via} the gluon fusion subprocess at the LHC up to NNLO accuracy.
Here the CP-odd scalar (pseudoscalar) Higgs boson, $A$, interacts with gluons and light quarks.
Our results are valid for the pseudoscalar Higgs boson of MSSM and 2HDM with small $\tan\beta$ values by adjusting the top Yukawa coupling, which appears as an overall factor.
In the EFT framework we use, the pseudoscalar Higgs boson directly couples to gluons and light quarks through two local composite operators, $O_G$ and $O_J$, respectively, with strengths proportional to the Wilson coefficients that are calculable in pQCD (see eqn. \ref{eq:effectiveLag}).
We used dimensional regularisation to regulate the existing UV and IR divergences.
%The Levi-Civita tensor and $\gamma_5$ matrices, which are inherently four-dimensional objects, enter the structure of the CP-odd Higgs boson results because of the CP-odd nature of the composite operators, $O_G$ and $O_J$.
The composite operators, being for a CP-odd particle, contain Levi-Civita tensor and $\gamma_5$, which are inherently four-dimensional objects.
Hence, careful treatment was needed to deal with them in $d$-dimensions which we executed by following the prescription advocated by Larin.
This treatment requires additional renormalisation for the singlet axial vector current up to two loops.
In addition, Larin's prescription requires a finite renormalisation constant for singlet axial current, which is also available.
These composite operators exhibit mixing under UV renormalisation.
The corresponding renormalisation constants are known, and we use them to obtain the UV finite amplitudes up to two-loop.
Unlike the amplitudes involving a pair of Higgs bosons, we do not need any UV contact counter terms here.
Finally, we obtain the UV finite amplitudes. However, these still contain IR divergences which arise from the massless partons in QCD.
Since we did not compute the real corrections in this work, we used the predictions of Catani on the universal structure of IR poles of two-loop $n$-point UV finite amplitudes \cite{Catani:1998bh} to factorise the IR singularities.
We found that the IR poles agreed with Catani's predictions.
We could do the evaluation analytically for poles $\varepsilon^{−i}$ with $i = 2 − 4$, and numerically to high precision for $\varepsilon^{-1}$ (see tables \ref{tab:Proj1LT} and \ref{tab:Proj2LT}).
For poles, $\varepsilon^{-1}$, the analytical verification could not be done because of the large size of the results.
This check of the pole structure provided a test on the correctness of our computation.
Our results provide one of the essential components relevant for studies related to the production of a pair of pseudoscalar Higgs bosons at the LHC up to order ${\cal O}(a_s^4)$.

Next, as described in chapter \ref{chap:ggA}, we performed the NSV resummation for a pseudoscalar Higgs boson production \textit{via} gluon fusion to $\overline{\text{NNLL}}$ accuracy in QCD at the LHC \cite{Bhattacharya:2021hae}.
We made a detailed phenomenological study of this process and presented our numerical results for $13$ TeV LHC.
We have computed the NSV corrections at both first and second orders in QCD and compared them with the full NLO and NNLO corrections for the gluon fusion sub-process.
These NSV corrections are observed to be potentially significant and enhance the pseudoscalar production cross-sections much more than the conventional SV logarithms.
We also give numerical results for the NSV resummed results to $\overline{\text{NNLL}}$ accuracy by systematically matching them to the FO NNLO cross-sections.
%These NSV resummed predictions significantly contribute to the cross-sections compared to the SV resummation results.
We estimate the size of the NSV corrections in terms of the resummed K-factors defined {\it{w.r.t.}} the LO and find them significant compared to the SV results.
We also observe that including other parton channels and the regular terms in $z$ negatively contribute to the results from the gluon fusion channel.
We also estimate in our predictions the uncertainties due to the choice of various PDFs and those due to the unknown renormalisation and factorisation scales.
We find that the conventional 7-point scale uncertainties do not improve after performing the NSV resummation.
We also observe that the $\mu_F$ scale variations, while keeping the $\mu_R$ scale fixed, lead to large uncertainties in the NSV resummed results compared to the FO and SV resummed cross-sections.
However, for pure $\mu_R$ variation, keeping the $\mu_F$ scale fixed, we find that the scale uncertainties get reduced significantly and are much smaller than those of the SV resummed results, which is as expected.
These theory uncertainty studies indicate the requirement of including NSV contributions from other parton channels and beyond NSV contributions in the gluon fusion channel for improved stability.
We also evaluated the production cross-sections for mixed scalar-pseudoscalar states and studied the impact of QCD corrections on them for different values of the mixing angle $\alpha$.
We find that these QCD corrections change with the mixing angle by a few per cent, indicating the necessity of improving the known precision results.
Studying this mixing demands that the pseudoscalar observables be made available up to an order comparable to that of the scalar Higgs boson.
%These FO and resummed results for the pseudoscalar Higgs boson are helpful for future analyses aiming to study the CP properties of the discovered Higgs boson.
%
 \textit{We have used state-of-the-art techniques and in-house codes to perform all the computations presented in this thesis.
 The methodologies we have used for calculating the pseudoscalar observables can be generalised to any other sub-process with the same number of incoming and outgoing partons.}

 \asr{
 As this analysis is limited by the lack of necessary tools of its time like N$^3$LO PDFs, high-order MIs and efficient hardware and software for higher-order precision computations, producing an analysis utilising the full Run $2$ data set and the future Run $3$ data for more sensitivity to multiple BSM scenarios requires much improvement.
 To fully take advantage of the increased amount of data, the analysis could be improved, for example, by studying the other parton channels that contribute to $AA$ and $A$ production, the corresponding higher-order corrections and a combination of explicit FO and resummed predictions to the highest order possible.
 In addition to including the effects of all parton channels, this would involve integrations over the entire phase space region.
 This type of computation is expected to consume vast amounts of time to evaluate each piece of data.
 One idea to achieve this goal is to use the slicing method, which can be used to compute a few data points covering the entire phase space region \cite{Harris:2001sx}.
 This data set can then be extrapolated to the entire phase space region.
 Improving precision measurements is essential, as any deviation from the known SM data could hint at new physics \cite{ZurbanoFernandez:2020cco,ILC:2013jhg,Barklow:2015tja,Blondel:2019qlh,Boughezal:2022cbl}.
%Therefore, improving the parameter space in BSM models is crucial for investigating new physics signals.
 These calculations can improve phenomenological results, reduce theoretical uncertainties and impose new stringent constraints on the Higgs self-couplings, which can prove significant for investigating new physics signals.
%, for instance, studies on the mixing of the scalar-pseudoscalar Higgs bosons.
 }

%\clearpage
%\newpage
%\mbox{~}

\begin{appendices}

\chapter{For the process of di-pseudo scalar production in gluon fusion}

\section{Born level Matrix element for Type-IIa Diagrams}
\label{appendix:D}

For projector $1$, $\mathcal{M}_{GG,1}^{II,(0)}$, and projector $2$, $\mathcal{M}_{GG,2}^{II,(0)}$, the amplitudes are given by

\begin{align}
 \mathcal{M}_{GG,1}^{II(0)}= & \dfrac{i}{2} {(1 + \epsilon) ~m_A^2} \dfrac{(1+x)^2}{x}, \\
\mathcal{M}_{GG,2}^{II(0)}= &
\dfrac{i}{4} (1+{\epsilon}) ~m_A^2 \dfrac{(1+x^2) (z ( 1 + x^2) - x~(1 + z^2))}{x~z~(1 + x^2 - x~z)}.
\end{align}
%respectively.

\section{One-loop level Matrix element for Type-IIa Diagrams}
\label{appendix:E}

At one-loop, the renormalised (finite) amplitudes for the two projectors, $\mathcal{M}_{GG,1}^{II,(1),fin}$ for projector $1$, and $\mathcal{M}_{GG,2}^{II,(1),fin}$ for projector $2$, are given by

\begin{center}
 \includegraphics[scale=0.37]{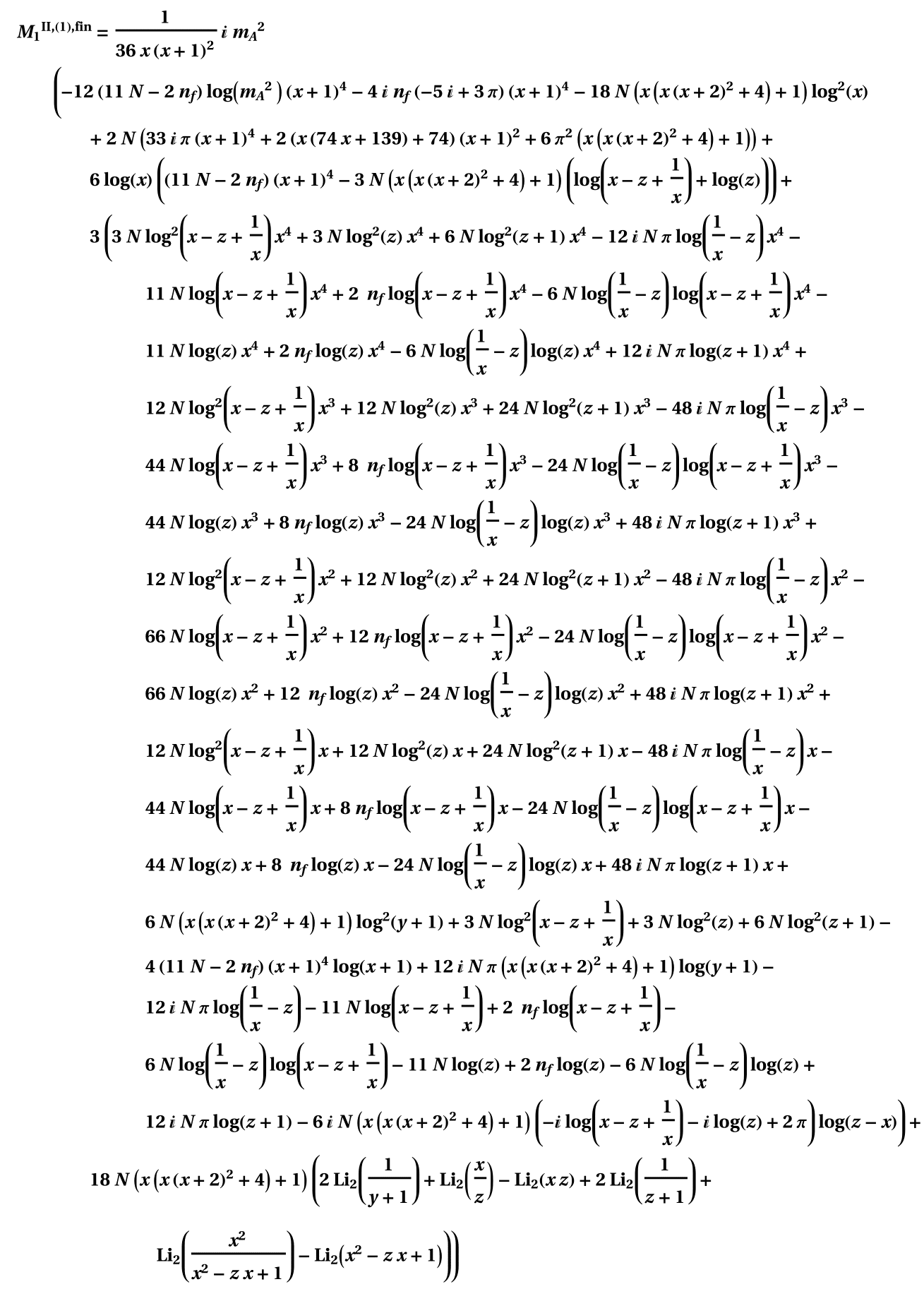}
 % Finite1L11.png: 1700x2200 px, 72dpi, 59.97x77.61 cm, bb=0 0 1700 2200
\end{center}

\begin{center}
 \includegraphics[scale=0.95]{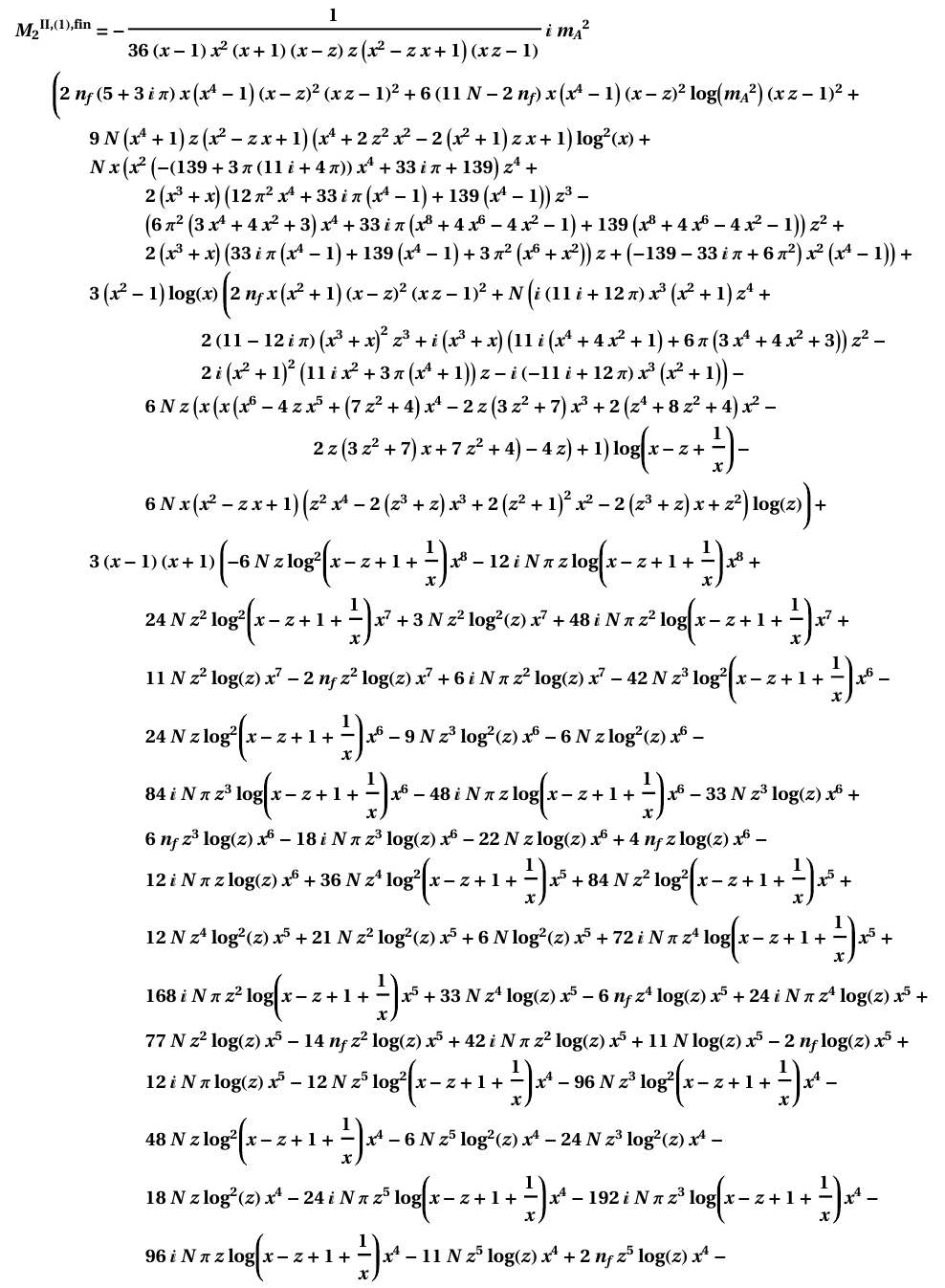}
 % Finite1L11.png: 1700x2200 px, 72dpi, 59.97x77.61 cm, bb=0 0 1700 2200
\end{center}

\begin{center}
 \includegraphics[scale=1]{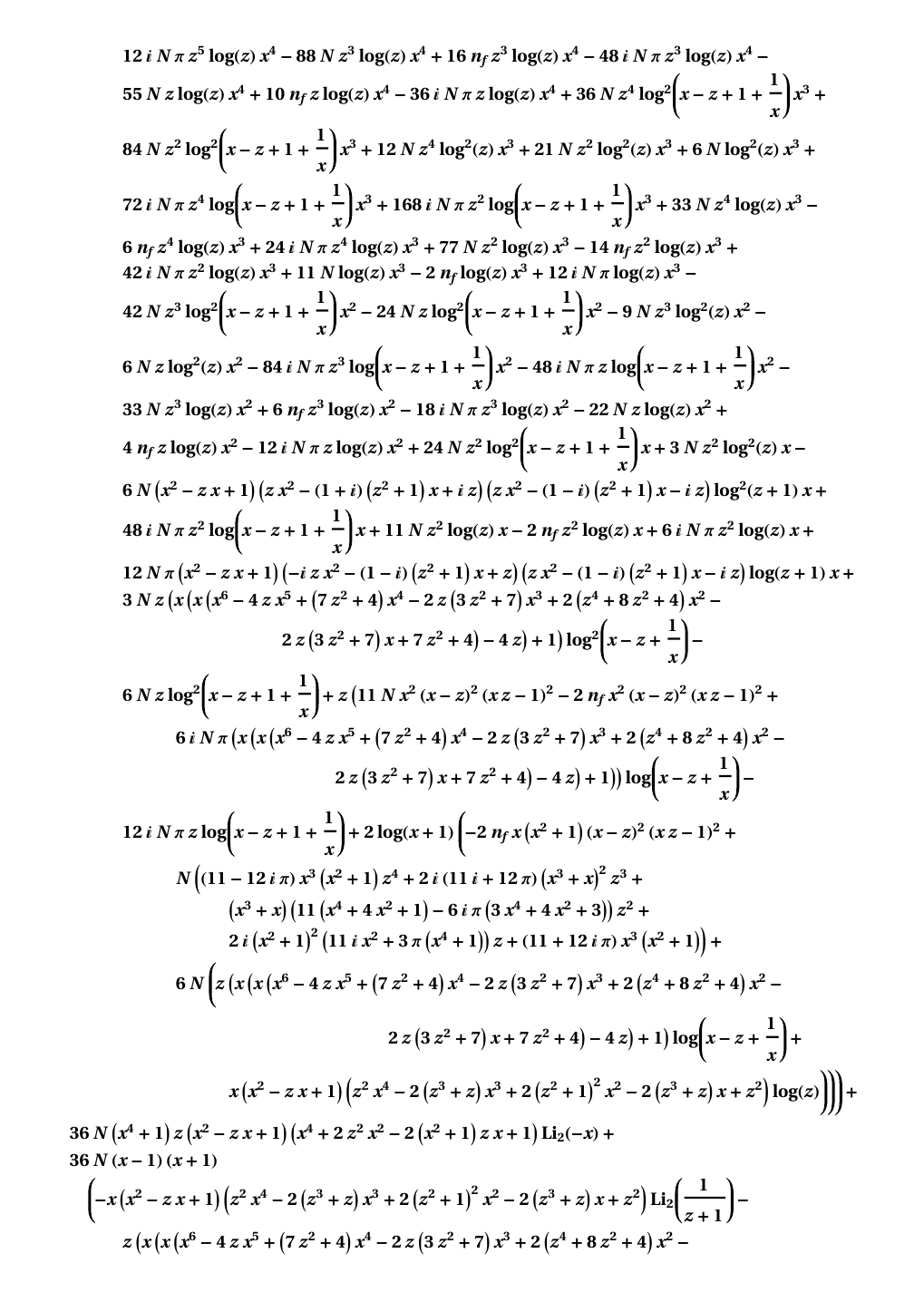}
 % Finite1L11.png: 1700x2200 px, 72dpi, 59.97x77.61 cm, bb=0 0 1700 2200
\end{center}

\begin{center}
 \includegraphics[scale=1]{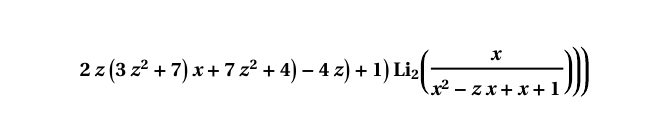}
 % Finite1L11.png: 1700x2200 px, 72dpi, 59.97x77.61 cm, bb=0 0 1700 2200
\end{center}

\section{Born level Matrix element for Type-IIb Diagrams}
\label{appendix:F}
%The $\mathcal{M}_{GJ}^{II(1)}$ finite part details:

The renormalised amplitude is the same as the initial unrenormalised amplitude at the Born level for $\mathcal{M}_{GJ,~i}^{II(1)}$ with $i=1$ for projector $1$ and $i=2$ for projector $2$ respectively.
They are given by
\begin{align}
\mathcal{M}_{GJ,~1}^{II(1)}= &
-\dfrac{24 ~i~ m_A^{2}n_{f}\left(1+x\right)^{2}}{x}, \\
\mathcal{M}_{GJ,~2}^{II(1)}= &
-\dfrac{\left(12~i~\right)m_A^{2}n_{f}\left(1-x^{2}\right)^{2}\left(1+x^{2}\right)\sin^{2}\theta}{\left(x\left(1+x^{2}\right)^{2}\right)-x\left(1-x^{2}\right)^{2}\cos^{2}\theta}.
\end{align}

\clearpage
\newpage
\mbox{~}

\chapter{For the process of Next to Soft Virtual Corrections to Pseudoscalar Higgs boson production from gluon fusion}

\section{Moment integrals to $\mathcal{O}(1/N)$}
\label{appendix:G}

\asr{
Computing resummed predictions beyond the eikonal approximation requires both conceptual and practical tools.
An essential tool is the evaluation of the generic integrals appearing in the exponents of the threshold resummations, to our required accuracy, \textit{i.e.} including all correction of order $1/N$ \cite{Laenen:2008ux}.
To this accuracy, the threshold-resummed partonic cross-sections can be expressed as
\begin{equation}
 \ln \hat{\sigma}(N) =
 \int_0^1 dz \dfrac{z^{N-1} - 1}{1-z} \sum_{p=0}^{\infty} f_1^{(p)} \ln^p(1-z)
 +
 \int_0^1 dz z^{N-1} \sum_{p=0}^{\infty} f_2^{(p)} \ln^p(1-z) + H,
\end{equation}
where $H$ constitutes all N-independent terms and $f_i^{(p)}$ are some coefficients.
Let us substitute
\begin{align}
\label{eqn:Dpn}
 \mathcal{D}_p(N) = & \int_0^1 dz \dfrac{z^{N-1} - 1}{1-z} \ln^p(1-z),
 \\
 \mathcal{J}_{p}(N) = & \int_0^1 dz z^{N-1} \ln^p(1-z).
 \label{eqn:JpN}
\end{align}
The definition of $\mathcal{D}_p(N)$ in eqn. \ref{eqn:Dpn} is the Mellin transform of the plus distribution defined in eqn. \ref{eqn:PlusFunc}.
$\mathcal{J}_{p}(N)$ represents the Mellin transform of the logarithmic terms that contribute to NLP accuracy.
These integrals are solved using generating functions in \cite{Laenen:2008ux} and are finally expressed as
\begin{align}
\label{eqn:Plus}
 \mathcal{D}_p   = & \dfrac{1}{p+1} \sum_{k=0}^{p+1} \Gamma_k(N) \begin{pmatrix}
p+1 \\
k
\end{pmatrix}
 (- \ln N)^{p+1-k} + \mathcal{O}\left( \dfrac{\ln^m N}{N^2} \right),
 \\
 \mathcal{J}_{p} = & \dfrac{1}{N} \sum_{k=0}^p \Gamma^{(k)}(1) \begin{pmatrix}
p \\
k
\end{pmatrix} (- \ln N)^{p-k} + \mathcal{O}\left( \dfrac{\ln^m N}{N^2} \right).
\label{eqn:Logs}
\end{align}
Here $\Gamma^{(k)}$ is the $k$th derivative of the Euler gamma function.

The generic form of a Plus distribution can be written as
\begin{equation}
 \left[ \dfrac{f(z)}{1-z} \right]_+ = \dfrac{f(z)}{1-z} - \delta(1-z) \int_0^z dy \dfrac{f(y)}{1-y},
\end{equation}
where $f$ is some smooth function of $z$ on the integration domain $0 < z < 1$.
While working in the Mellin space, we express this Plus distribution as given in eqn.\ \ref{eqn:Dpn}, defining $f(z) =  \ln^p(1-z)$.
}

\section{NSV Cross-section Results}
\label{appendix:A}

In this section, we present our analytical results of the SV+NSV coefficient functions up to N$^3$LO order. 
Expanding the SV+NSV coefficient function in eqn.~\ref{eq:deltaA}, in powers of $a_s$, we obtain
\begin{equation}
\Delta_{g}^{A,NSV}\left(z,q^{2}\right)=\sum_{i=0}^{\infty}a_{s}^{i}~\Delta_{g,i}^{A,NSV}\left(z,q^{2}\right).
\end{equation}
where
\begin{align}
 \Delta_{g,0}^{A,NSV} &= \delta(1-z),
                            \nonumber\\
 \Delta_{g,1}^{A,NSV} &= \delta(1-z) \bigg[8 C_A \big(1 + \zeta_2 \big) \bigg] +
                            \mathcal{D}_1 \bigg[16 C_A \bigg] - \bigg[16 C_A\bigg] \log(1-z) + 8 C_A,
                            \nonumber\\
 \Delta_{g,2}^{A,NSV} &= \delta(1-z) \bigg[C_A^2\bigg( + \dfrac{494}{3} -
                         \dfrac{220\zeta_3}{3} - \dfrac{4\zeta_2^2}{5} + \dfrac{1112\zeta_2}{9}\bigg) 
                         \nonumber\\
                         &+ C_A n_f \bigg( - \dfrac{82}{3} - \dfrac{80\zeta_2}{9} - \dfrac{8\zeta_3}{3} \bigg) + C_F n_f \bigg( - \dfrac{160}{3} + 12 \log\dfrac{\mu_R^2}{m_t^2} + 16 \zeta_3\bigg)\bigg]
                         \nonumber\\
                         & + \mathcal{D}_0 \bigg[ C_A^2 \bigg( - \dfrac{1616}{27} 
                         + \dfrac{176 \zeta_2}{3} + 312 \zeta_3 \bigg)
                         + C_A n_f \bigg(\dfrac{224}{27}  - 
                         \dfrac{32 \zeta_2}{3} \bigg) \bigg]
                         \nonumber\\
                         &+ \mathcal{D}_1 \bigg[ C_A^2 \bigg( + \dfrac{2224}{9} 
                         - 160 \zeta_2\bigg) - C_A n_f \bigg(\dfrac{160}{9}\bigg) \bigg]
                        \nonumber\\
                         & + \mathcal{D}_2 \bigg[-C_A^2 \bigg(\dfrac{176}{3}\bigg) +C_A n_f\bigg(\dfrac{32}{3}\bigg)\bigg] + \mathcal{D}_3 \bigg[C_A^2\bigg(128\bigg)\bigg] - \bigg[C_A^2\bigg(128\bigg)\bigg] \log^3(1-z)
                         \nonumber\\
                         & + \bigg[C_A^2\bigg(\dfrac{920}{3}\bigg) - C_A n_f \bigg(\dfrac{32}{3}\bigg)\bigg] \log^2(1-z) + \bigg[C_A^2\bigg(-\dfrac{2740}{9} + 160 \zeta_2\bigg) \nonumber\\
                         & + C_A n_f \bigg(\dfrac{244}{9} \bigg)\bigg] \log(1-z)
                          + \bigg[C_A^2 \bigg(\dfrac{4340}{27} - \dfrac{608 \zeta_2}{3} - 312\zeta_3 \bigg) - C_A n_f \bigg( \dfrac{392}{27}  + \dfrac{32 \zeta_2}{3} \bigg) \bigg],
                         \nonumber\\
 \Delta_{g,3}^{A,NSV} &= 512~ C_A^3 \mathcal{D}_5 - 512~ C_A^3 \log^5(1-z) + 
                         \log^4(1-z) \bigg[\dfrac{22592~ C_A^3}{9} - \dfrac{1280~ C_A^2 n_f}{9}\bigg] 
                         \nonumber \\
                         & + \mathcal{D}_4 \bigg[-\dfrac{7040~ C_A^3}{9} + \dfrac{1280~ C_A^2 n_f}{9} \bigg] + \mathcal{D}_3 \bigg[-\dfrac{10496~ C_A^2 n_f}{27} + \dfrac{256~ C_A n_f^2}{27} + C_A^3 \bigg(\dfrac{86848}{27} - 3584~ \zeta_2\bigg)\bigg]  \nonumber \\
                         & + \log^3(1-z) \bigg[\dfrac{6560~ C_A^2 n_f}{9} - \dfrac{256~ C_A n_f^2}{27} + C_A^3 \bigg(-\dfrac{138656}{27} + 3584~ \zeta_2\bigg) \bigg] 
                         \nonumber \\
                         &
                         + C_A \bigg[ n_f^2 \bigg(\dfrac{3136}{29}  
                         - \dfrac{304~ \zeta_2}{9} - \dfrac{320~ \zeta_3}{27} \bigg)
                         + C_F n_f \bigg(-\dfrac{16826}{27}
                         + 96~ \log\dfrac{\mu_R^2}{m_t^2}
                         \nonumber \\
                         &
                         + \dfrac{80~ \zeta_2}{3}
                         + \dfrac{64~ \zeta_2^2}{5} + \dfrac{2384~ \zeta_3}{9} \bigg) \bigg]
                         + C_A^2 n_f \bigg(-\dfrac{479636}{729} + \dfrac{79096~ \zeta_2}{81} + \dfrac{272~ \zeta_2^2}{5} 
                         \nonumber \\
                         & + \dfrac{13120~ \zeta_3}{9} \bigg)
                         + \log^2(1-z) \bigg[C_A \bigg(-32~ C_F n_f 
                         + \dfrac{944~ n_f^2}{27} \bigg) + C_A^2 n_f \bigg( - \dfrac{34984}{27} + \dfrac{2180~ \zeta_2}{3} \bigg) 
                         \nonumber \\
                         & + C_A^3 \bigg( \dfrac{239744}{27} - \dfrac{34352~ \zeta_2}{3} - 11584 ~\zeta_3 \bigg)\bigg]  
                         + \mathcal{D}_2 \bigg[C_A \bigg(32~ C_F n_f - \dfrac{640~ n_f^2}{27} + C_A^2 n_f \bigg(\dfrac{16928}{27} - \dfrac{2176 ~\zeta_2}{3} \bigg)
                         \nonumber \\
                         & + C_A^3 \bigg(-\dfrac{79936}{27} + \dfrac{11968~ \zeta_2}{3} + 11584~ \zeta_3 \bigg) \bigg] 
                        + \log(1-z) \bigg[ C_A \bigg\{ n_f^2 \bigg(-\dfrac{2608}{81} + \dfrac{256~ \zeta_2}{9}\bigg) 
                        \nonumber \\
                         & + C_F n_f \bigg(1040 - 192 \log\dfrac{\mu_R^2}{m_t^2} - \dfrac{16~ \zeta_2}{3} - 384~ \zeta_3 \bigg)\bigg\}  
                         + C_A^2 n_f \bigg( \dfrac{125536}{81} - \dfrac{14480~ \zeta_2}{9} - \dfrac{2992~ \zeta_3}{3} \bigg) 
                         \nonumber \\                         
                         & + C_A^3 \bigg( - \dfrac{221824}{27} + \dfrac{87520~ \zeta_2}{9} + \dfrac{9856~ \zeta_2^2}{5} + 23168~ \zeta_3 \bigg) \bigg] 
                         + \mathcal{D}_1 \bigg[C_A^3 \bigg( \dfrac{414616}{81} - \dfrac{13568~ \zeta_2}{3} - \dfrac{9856~ \zeta_2^2}{5} 
                         \nonumber \\
                         & - \dfrac{22528~ \zeta_3}{3}) + C_A^2 n_f \bigg(-\dfrac{79760}{81} + \dfrac{6016~ \zeta_2}{9} + \dfrac{2944~ \zeta_3}{3} \bigg) + C_A \bigg\{ n_f^2 \bigg( \dfrac{1600}{81} - \dfrac{256~ \zeta_2}{9}\bigg) 
                         \nonumber \\
                         & + C_F n_f \bigg(-1000 + 192~ \log\dfrac{\mu_R^2}{m_t^2} + 384~ \zeta_3 \bigg) \bigg\} \bigg] 
                         \nonumber \\
                         &
                         + C_A^3 \bigg(\dfrac{2650990}{729} - \dfrac{489184~ \zeta_2}{81} - \dfrac{26272~ \zeta_2^2}{15} - \dfrac{330184~ \zeta_3}{27}
                         \nonumber \\
                         & + \dfrac{23200~ \zeta_2~ \zeta_3}{3} - 11904~ \zeta_5 \bigg) 
                         + \mathcal{D}_0 \bigg[ C_A^2 n_f \bigg( \dfrac{173636}{729} - \dfrac{41680~ \zeta_2}{81} - \dfrac{544~ \zeta_2^2}{15} - \dfrac{7600~ \zeta_3}{9} \bigg) 
                         \nonumber \\
                         &
                         + C_A \bigg \{ C_F~ n_f \bigg(\dfrac{3422}{27} 
                         - 32 ~\zeta_2 - \dfrac{64~ \zeta_2^2}{5} - \dfrac{608~ \zeta_3}{9} \bigg)
                         \nonumber\\
                         &
                         + n_f^2 \bigg( -\dfrac{3712}{729} + \dfrac{640~ \zeta_2}{27} + \dfrac{320~ \zeta_3}{27} \bigg) \bigg\}
                         + C_A^3 \bigg( - \dfrac{943114}{729} + \dfrac{175024~ \zeta_2}{81}
                         \nonumber\\                    
                         & + \dfrac{4048~ \zeta_2^2}{15} + \dfrac{210448 \zeta_3}{27} - \dfrac{23200~ \zeta_2~ \zeta_3}{3} + 11904~ \zeta_5 \bigg) \bigg]
                         \nonumber \\
                         & +  \delta(1-z) \bigg[-4 n_f C_J^{(2)}~ \log^2(1-z) 
                         + C_F n_f^2 \bigg( \dfrac{1498}{9} - \dfrac{40~ \zeta_2}{9} - \dfrac{32~ \zeta_2^2}{45} - \dfrac{224~ \zeta_3}{3} \bigg) 
                         \nonumber \\
                         & + C_A^3 \bigg( \dfrac{114568}{27} + \dfrac{266155~ \zeta_2}{162} - \dfrac{4007~ \zeta_2^2}{10} - \dfrac{64096~ \zeta_2^3}{105} - 3932~ \zeta_3 + \dfrac{7832~ \zeta_2 ~\zeta_3}{3} + \dfrac{13216~ \zeta_3^2}{3}
                         \nonumber\\
                         &
                         - \dfrac{30316~ \zeta_5}{9} \bigg)
                         + C_F^2~ n_f \bigg( \dfrac{457}{3} + 208~ \zeta_3 - 320~ \zeta_5 \bigg)
                         + C_A^2~ n_f \bigg( -\dfrac{113366}{81}
                         - \dfrac{56453~ \zeta_2}{405} + \dfrac{21703~ \zeta_2^2}{135}
                         \nonumber\\
                         &
                         + \dfrac{8840~ \zeta_3}{27} - \dfrac{2000~ \zeta_2~ \zeta_3}{3} + \dfrac{6952~ \zeta_5}{9} \bigg)
                         + C_A \bigg \{ n_f^2 \bigg( \dfrac{6914}{81} - \dfrac{7088~ \zeta_2}{405} - \dfrac{2288~ \zeta_2^2}{135}
                         + \dfrac{688~ \zeta_3}{27} \bigg) 
                         \nonumber \\
                         &+ C_F n_f \bigg(-1797 + 96~ \log\dfrac{\mu_R^2}{m_t^2} - \dfrac{4160~ \zeta_2}{9} + 96~ \log\dfrac{\mu_R^2}{m_t^2} ~\zeta_2
                         + \dfrac{176~ \zeta_2^2}{45} + \dfrac{1856~ \zeta_3}{3}
                         \nonumber \\
                         &+ 192~ \zeta_2~ \zeta_3 + \dfrac{3872~ \zeta_5}{9} \bigg) \bigg\} \bigg].
\end{align}

\section{SV coefficients in soft collinear distribution}
\label{appendix:B}

In this section, we present the explicit expressions for $\hat{\phi}_{g}^{SV,\left(i\right)}\left(\varepsilon\right)$ appearing in Eqn. (\ref{eq:phicapg}):
\begin{align}
 \hat{\phi}_{g}^{SV,\left(1\right)}\left(\varepsilon\right) &= 
 -\dfrac{3}{16} C_A \varepsilon^2 \zeta_2^2+\frac{8
    C_A}{\varepsilon^2}+\frac{7 C_A \zeta_3}{3}\varepsilon-3
    C_A \zeta_2,
    \nonumber \\
  \hat{\phi}_{g}^{SV,\left(2\right)}\left(\varepsilon\right) &=   
    \varepsilon \bigg\{\frac{3}{16} \beta_0 C_A \zeta_2^2+\frac{11
    C_A^2 \zeta_2^2}{80}-\frac{203}{6} C_A^2 \zeta_2
    \zeta_3+\frac{707 C_A^2 \zeta_2}{27}+\frac{2077 C_A^2
    \zeta_3}{54}+\frac{43 C_A^2 \zeta_5}{2}
    \nonumber \\
   & -\frac{3644
    C_A^2}{243}-\frac{1}{40} C_A n_f \zeta_2^2-\frac{98
    C_A n_f \zeta_2}{27}-\frac{155 C_A n_f
    \zeta_3}{27}+\frac{488 C_A n_f}{243}\bigg\}
    +\bigg\{ 2
    C_A^2 \zeta_2^2
    \nonumber \\
    &-\frac{469 C_A^2 \zeta_2}{18}-\frac{88
    C_A^2 \zeta_3}{3}+\frac{1214 C_A^2}{81}
    +\frac{35 C_A
    n_f \zeta_2}{9}+\frac{16 C_A n_f
    \zeta_3}{3}-\frac{164 C_A n_f}{81}-\frac{7 \beta_0 C_A
    \zeta_3}{3}\bigg\}
    \nonumber \\
    &  +\dfrac{1}{\varepsilon}\bigg\{6
    \beta_0 C_A \zeta_2
    +\frac{11 C_A^2 \zeta_2}{3}+14
    C_A^2 \zeta_3-\frac{404 C_A^2}{27}-\frac{2 C_A
    n_f \zeta_2}{3}+\frac{56 C_A
    n_f}{27}\bigg\}
    \nonumber \\
    &+\frac{1}{\varepsilon^2}\bigg\{-4 C_A^2 \zeta_2+\frac{134
    C_A^2}{9}-\frac{20 C_A n_f}{9}\bigg\}-\frac{4 \beta_0
    C_A}{\varepsilon^3},
    \nonumber \\
\hat{\phi}_{g}^{SV,\left(3\right)}\left(\varepsilon\right) &=
    \frac{32
    \beta_0^2 C_A}{9 \varepsilon^4}
    +\dfrac{1}{\varepsilon^3}\bigg\{\frac{64}{9} \beta_0 C_A^2
    \zeta_2-\frac{2144 \beta_0 C_A^2}{81}+\frac{320 \beta_0
    C_A n_f}{81}-\frac{8 \beta_1
    C_A}{9}\bigg\}
    \nonumber \\
    &\dfrac{1}{\varepsilon^2}\bigg\{-12 \beta_0^2 C_A \zeta_2-\frac{88}{9} \beta_0
    C_A^2 \zeta_2-\frac{112}{3} \beta_0 C_A^2
    \zeta_3+\frac{3232 \beta_0 C_A^2}{81}+\frac{16}{9} \beta_0
    C_A n_f \zeta_2
    \nonumber \\
    &-\frac{448 \beta_0 C_A
    n_f}{81}+\frac{352 C_A^3 \zeta_2^2}{45}-\frac{2144
    C_A^3 \zeta_2}{81}+\frac{176 C_A^3
    \zeta_3}{27}+\frac{980 C_A^3}{27}+\frac{320}{81} C_A^2
    n_f \zeta_2
    \nonumber \\
    &-\frac{224}{27} C_A^2 n_f
    \zeta_3-\frac{1672 C_A^2 n_f}{243}+\frac{64}{9} C_A
    C_f n_f \zeta_3-\frac{220 C_A C_f
    n_f}{27}-\frac{32 C_A n_f^2}{243}\bigg\}    
    \nonumber \\
    &+\dfrac{1}{\varepsilon}\bigg\{-8 \beta_0 C_A^2
    \zeta_2^2+\frac{938}{9} \beta_0 C_A^2
    \zeta_2+\frac{352}{3} \beta_0 C_A^2 \zeta_3-\frac{4856
    \beta_0 C_A^2}{81}-\frac{140}{9} \beta_0 C_A n_f
    \zeta_2
    \nonumber \\
    &-\frac{64}{3} \beta_0 C_A n_f
    \zeta_3+\frac{656 \beta_0 C_A n_f}{81}+3 \beta_1
    C_A \zeta_2-\frac{352}{15} C_A^3
    \zeta_2^2-\frac{176}{9} C_A^3 \zeta_2
    \zeta_3+\frac{12650 C_A^3 \zeta_2}{243}
    \nonumber \\
    &+\frac{1316
    C_A^3 \zeta_3}{9}-64 C_A^3 \zeta_5-\frac{136781
    C_A^3}{2187}+\frac{32}{5} C_A^2 n_f
    \zeta_2^2-\frac{2828}{243} C_A^2 n_f
    \zeta_2-\frac{728}{81} C_A^2 n_f \zeta_3
    \nonumber \\
    &+\frac{11842
    C_A^2 n_f}{2187}-\frac{32}{15} C_A C_f n_f
    \zeta_2^2-\frac{4}{3} C_A C_f n_f
    \zeta_2-\frac{304}{27} C_A C_f n_f
    \zeta_3+\frac{1711 C_A C_f n_f}{81}
    \nonumber \\
    &+\frac{40}{81}
    C_A n_f^2 \zeta_2-\frac{112}{81} C_A n_f^2
    \zeta_3+\frac{2080 C_A
    n_f^2}{2187}\bigg\}+\bigg\{-\frac{11}{30} \beta_0 C_A^2
    \zeta_2^2+\frac{812}{9} \beta_0 C_A^2 \zeta_2
    \zeta_3
    \nonumber \\
    &-\frac{5656}{81} \beta_0 C_A^2
    \zeta_2
    -\frac{8308}{81} \beta_0 C_A^2
    \zeta_3-\frac{172}{3} \beta_0 C_A^2 \zeta_5+\frac{29152
    \beta_0 C_A^2}{729}+\frac{1}{15} \beta_0 C_A n_f
    \zeta_2^2
    \nonumber \\
    &+\frac{1240}{81} \beta_0 C_A n_f
    \zeta_3-\frac{3904 \beta_0 C_A n_f}{729}+\frac{1}{16}
    \beta_1 C_A \varepsilon \zeta_2^2-\frac{7 \beta_1 C_A
    \zeta_3}{9}+\frac{152 C_A^3 \zeta_2^3}{189}+\frac{1964
    C_A^3 \zeta_2^2}{27}
    \nonumber \\
    &+\frac{11000}{27} C_A^3 \zeta_2
    \zeta_3-\frac{765127 C_A^3 \zeta_2}{1458}+\frac{536
    C_A^3 \zeta_3^2}{9}-\frac{59648 C_A^3
    \zeta_3}{81}-\frac{1430 C_A^3 \zeta_5}{9}+\frac{7135981
    C_A^3}{26244}
    \nonumber \\
    &-\frac{532}{27} C_A^2 n_f
    \zeta_2^2-\frac{1208}{27} C_A^2 n_f \zeta_2
    \zeta_3+\frac{105059}{729} C_A^2 n_f
    \zeta_2+\frac{45956}{243} C_A^2 n_f
    \zeta_3+\frac{148}{9} C_A^2 n_f \zeta_5
    \nonumber \\
    &-\frac{716509
    C_A^2 n_f}{13122}+\frac{152}{45} C_A C_f n_f
    \zeta_2^2-\frac{88}{3} C_A C_f n_f \zeta_2
    \zeta_3+\frac{605}{18} C_A C_f n_f
    \zeta_2+\frac{2536}{81} C_A C_f n_f
    \zeta_3
    \nonumber \\
    &+\frac{112}{9} C_A C_f n_f
    \zeta_5-\frac{42727 C_A C_f n_f}{972}+\frac{32}{27}
    C_A n_f^2 \zeta_2^2-\frac{1996}{243} C_A n_f^2
    \zeta_2-\frac{2720}{243} C_A n_f^2
    \zeta_3
    \nonumber \\
    &+\frac{11584 C_A n_f^2}{6561}-\frac{1}{4} \beta_0^2 C_A
    \zeta_2^2+\frac{784}{81} \beta_0 C_A n_f
    \zeta_2\bigg\}.
\end{align}

\section{Singular NSV coefficients in soft collinear distribution}
\label{appendix:C}

In this section, we present the explicit expressions for the singular coefficients $\varphi_{s,g}^{NSV,\left(i\right)}\left(z,\varepsilon\right)$ appearing in Eqn. (\ref{eq:phiNSV}):
\begin{align}
\varphi_{s,g}^{NSV,\left(1\right)}\left(z,\varepsilon\right) &= -\dfrac{8 C_A}{\varepsilon},
\nonumber \\
\varphi_{s,g}^{NSV,\left(2\right)}\left(z,\varepsilon\right) &= \dfrac{8 \beta_0 C_A}{\varepsilon^2} + \dfrac{1}{\varepsilon}\textbf{\bigg\{}
    C_A^2 \left(8 \zeta_2-\frac{268}{9}\right)+\frac{40
    C_A n_f}{9}+16 C_A^2~\text{log}{\text{$(1-z)$}}\bigg\},
    \nonumber \\
\varphi_{s,g}^{NSV,\left(3\right)}\left(z,\varepsilon\right) &=  
-\dfrac{32 \beta_0^2 C_A }{3 \varepsilon^3}+
\dfrac{1}{\varepsilon^2}\bigg\{\frac{8
    \beta_1 C_A }{3}-\frac{8}{3} \beta_0 \bigg(
    C_A^2 \left(8 \zeta_2-\frac{268}{9}\right)+\frac{40
    C_A n_f}{9}
    \nonumber \\
    & +16 C_A^2~
    \text{log}{\text{$(1-z)$}}\bigg)\bigg\}
    + \dfrac{2}{3 \varepsilon}\bigg\{ C_A^3
    \left(-\frac{176 \zeta_2^2}{5}+\frac{1072 \zeta_2}{9}+\frac{56
    \zeta_3}{3}-166\right)
    \nonumber \\
    & +\frac{80}{3} C_A^2 n_f T_f
     +C_A^2 n_f \left(-\frac{160 \zeta_2}{9}+\frac{112
    \zeta_3}{3}+\frac{548}{27}\right)+16 C_A C_f n_f
    T_f
    \nonumber \\
    & +C_A C_f n_f \left(\frac{86}{3}-32
    \zeta_3\right) +\frac{16 C_A n_f^2}{27}+\text{log}{\text{$(1-z)$}}
    \bigg(C_A^3 \left(\frac{2144}{9}-64 \zeta_2\right)
    \nonumber \\
    & -\frac{320
    C_A^2 n_f}{9}\bigg)\bigg\}.
\end{align}

\end{appendices}

%%%%%%%%%%%%%%%%%%%%%%%%%%%%%%%%%%%%%%%%
%\clearpage
%\newpage
%\mbox{~}

\printbibliography

\end{doublespace}

\end{document}